\documentclass{aa} %
  
\usepackage{graphicx}
\usepackage{txfonts}
\usepackage{url}
\usepackage{natbib}
\bibpunct{(}{)}{;}{a}{}{,} % to follow the A&A stylefile:///home/marco/Dropbox/paper_cromo/aa.dem
\usepackage[utf8]{inputenc}
\usepackage{subfigure}
\usepackage{verbatim} 

\begin{document}

\title{Plasma flows and magnetic field interplay \\
during the formation of a pore }
%\subtitle{I. Observations and analysis of photospheric data}

\author{I. Ermolli\inst{1}
                \and
                A. Cristaldi\inst{1}
                \and
                F. Giorgi\inst{1}
                \and
                F. Giannattasio\inst{1}\thanks{Current address: INAF Istituto Nazionale di Astrofisica - Istituto di Astrofisica e Planetologia Spaziali, Via del Fosso del Cavaliere, 100, 00133 Roma, Italy}
                \and
                M. Stangalini\inst{1}
                \and
                P. Romano\inst{2}
                \and
                A. Tritschler\inst{3}
                \and
                F. Zuccarello\inst{4}
   % \and
   % S. Criscuoli\inst{2}
   }
 
 \offprints{I. Ermolli}

\institute{INAF  Istituto Nazionale di Astrofisica -- Osservatorio Astronomico di Roma, Via Frascati 33,
                I-00040 Monte Porzio Catone, Italy
                                \and
                                INAF  Istituto Nazionale di Astrofisica -- Osservatorio Astrofisico di Catania, Via S. Sofia 78,
                I-95123 Catania, Italy
                                \and
                NSO  National Solar Observatory, Sacramento Peak Box 62, Sunspot, NM 88349, USA
                \and
                Dipartimento di Fisica e Astronomia -- Sezione Astrofisica, Universit\`a di Catania, Via S. Sofia 78, I-95123 Catania, Italy
          }

\date{Received ...; accepted ...}

\abstract
% context heading (optional)
{} %leave it empty if necessary
{
Recent %3D radiative MHD 
simulations of solar magneto-convection has offered new levels of understanding of  the interplay between plasma motions and magnetic fields in evolving active regions. We aim at verifying some aspects of the formation of magnetic regions derived from  recent numerical studies in  observational data. 
}
%}
%% methods heading (mandatory)
{
We studied  the formation of a pore in the active region (AR) NOAA 11462. We analysed data  obtained with the Interferometric Bidimensional Spectrometer (IBIS) at the Dunn Solar Telescope  on April 17, 2012,   consisting of  full Stokes measurements of the Fe I 617.3 nm lines.  Furthermore, we analysed   SDO/HMI observations in the continuum and vector magnetograms  derived from the  Fe I 617.3 nm line data taken from April 15 to 19, 2012.  We estimated  the  magnetic field strength and vector components  and the line-of-sight (LOS) and  horizontal motions in the photospheric region hosting the pore formation. 
%Unlike previous studies, the above quantities were measured  during the pore formation process. 
We discuss our results  in light of  other observational studies and recent advances  of  numerical simulations. 
%, attempting to provide observational constraints on current numerical simulations and models of sunspot evolution. . 
}
%% results heading (mandatory)
{
%The studied pore formation occurs through three main stages that are characterized by a progressive coalescence of the emerged fields  {\bf by plasma motions}. 
The pore formation occurs in less than 1 hour  in the leading region of the AR. We observe that  the evolution of the flux patch in the leading part of the AR  is faster (< 12 hour) than the evolution (20-30 hour) of the more diffuse and smaller scale flux patches in the trailing region. During the pore formation, the   ratio between magnetic and dark area decreases from 5 to 2. We observe strong  downflows  at the forming pore boundary and    diverging proper  motions  of plasma in the vicinity of the evolving feature that are  directed towards  the forming pore. 
The average values  and trends of the various quantities estimated in the AR are in agreement with  results of former observational studies of steady pores and with their modelled counterparts, as seen in recent numerical simulations of a rising-tube process. The  agreement with the outcomes of the numerical studies    holds 
for  both the signatures of the flux emergence process (e.g. appearance of small-scale mixed polarity patterns and elongated granules) and the evolution of the region.  The processes driving the formation of the pore are identified with  the emergence of a magnetic flux concentration and the subsequent reorganization of the emerged flux, by the combined  effect of  velocity and magnetic field,  in and around the evolving structure. 
}
%% conclusions heading (optional), leave it empty if necessary
{}
%The results presented in this study thus show that most recent  MHD simulations are  very successful not only in explaining the average properties of the formed regions, 
%but also in depicting most of the mechanisms  that drive  the evolution of the magnetic regions emerged in the solar photosphere. 
%}


\authorrunning{I. Ermolli et al.}
\titlerunning{Velocity and magnetic fields  in a forming pore}

\maketitle

\section{Introduction}

Sunspots represent the  best-known manifestation  of solar magnetism \citep[for a review see e.g.][]{Solanki_2003,Rempel_schliche2011}. The structure and dynamics  of sunspots have been investigated for a long time, but their evolution after emergence in the solar atmosphere cannot be predicted by current knowledge as yet. %of sunspots  %of the processes involved
%does not yet allow predicting the evolution of sunspots after their emergence  in the solar atmosphere.  
In addition to  advance science,  the ability to predict the evolution of sunspots presently entails economic and ethical consequences, by allowing efficient protection   of technological systems and human life from space weather events \citep[][]{Hapgood_2012}. 
 
Pores constitute the first stage of the evolution of sunspots from which they mainly differ in size,  and in  strength and orientation of the magnetic field. 
High-resolution observations  of isolated pores in the photosphere \citep[e.g.][and references therein]{Sainz_etal2012,Sobotka_etal2012,Guglielmino_zuccarello2011,Giordano_etal2008} and chromosphere  \citep[e.g.][]{Sobotka_etal2013}, and  the study of   samples of pores   \citep[e.g.][]{Cho_etal2010,Cho_etal2013,Vargas_etal2010,Toriumi_etal2014}, show that 
these features  are small sunspots  without penumbra and with a prevailing vertical magnetic field that can  reach  up to 1-2 kG strength in the photosphere \citep[e.g.][]{Sobotka_etal2012}.  %Observations  reveal   
%have unveiled 
%regular  plasma motions inside and around pores. In particular, 
The periphery of pores displays strong downflows with plasma velocities that decrease with the  atmospheric height;   
supersonic flows  were also reported
in the chromosphere \citep[e.g.][and references therein]{Lagg_etal2007,Sobotka_etal2012,Cho_etal2013,Sobotka_etal2013}. Converging horizontal motions appear around pores with  velocities  twice as high as those  found inside them and the highest values  near the border of the pores \citep[e.g.][and references therein]{Vargas_etal2012,Verma_etal2014}. 
%,  due to their contamination by convective flows.
%High-resolution o
Like sunspots, pores  host 
 fine bright features  and  several types of waves \citep[e.g.][]{Balthasar_etal2000,Bogdan_2006,Stangalini_etal2011,Stangalini_etal2012}.

High-resolution observations show  that 
pores  are formed by the merging
of small magnetic elements  dragged together  %motions of which being driven
by %sub- and surface 
plasma  motions (e.g. \citet{Sobotka_2003,Yang_etal2003}, and references therein). 
However, until recent times, limited diagnostic capabilities  have impeded the understanding of 
%the processes involved in the formation of pores and larger-scale magnetic structures. For example, it is still unclear 
whether  the above process results from 
emergence through the photosphere of a magnetic field generated by global dynamo mechanisms in the solar interior, or from amplification and structuring of a magnetic field generated by local convective dynamo due to plasma motions.  
Current spectro-polarimetric instruments and methods, and recent numerical simulations promise to offer new 
 levels of understanding  of the processes involved in the formation of pores and larger scale magnetic structures; see for example
 the observational studies by \citet{Schlichenmaier_etal2010}, \citet{Rezaei_etal2012}, \citet{Bello_etal2012}, \citet{Romano_etal2013,Romano_etal2014}, and \citet{Watanabe_etal2014} concerning the conditions that lead to the formation of penumbral regions, and the three-dimensional (3D) radiative magneto-hydro-dynamic (MHD) simulations of flux emergence by \citet{Rempel_cheung2014}.

  \begin{figure}%[ht!]
 % \centering{\hspace {0.7cm} 1                       \hspace{3.cm}                                     2            \hspace{3.cm}                                                 3     \hspace{3cm}                        4        }\\
\centering
{
\includegraphics[height=2.8cm,trim=0.5cm 2.5cm 4.7cm 1.3cm,clip=true]{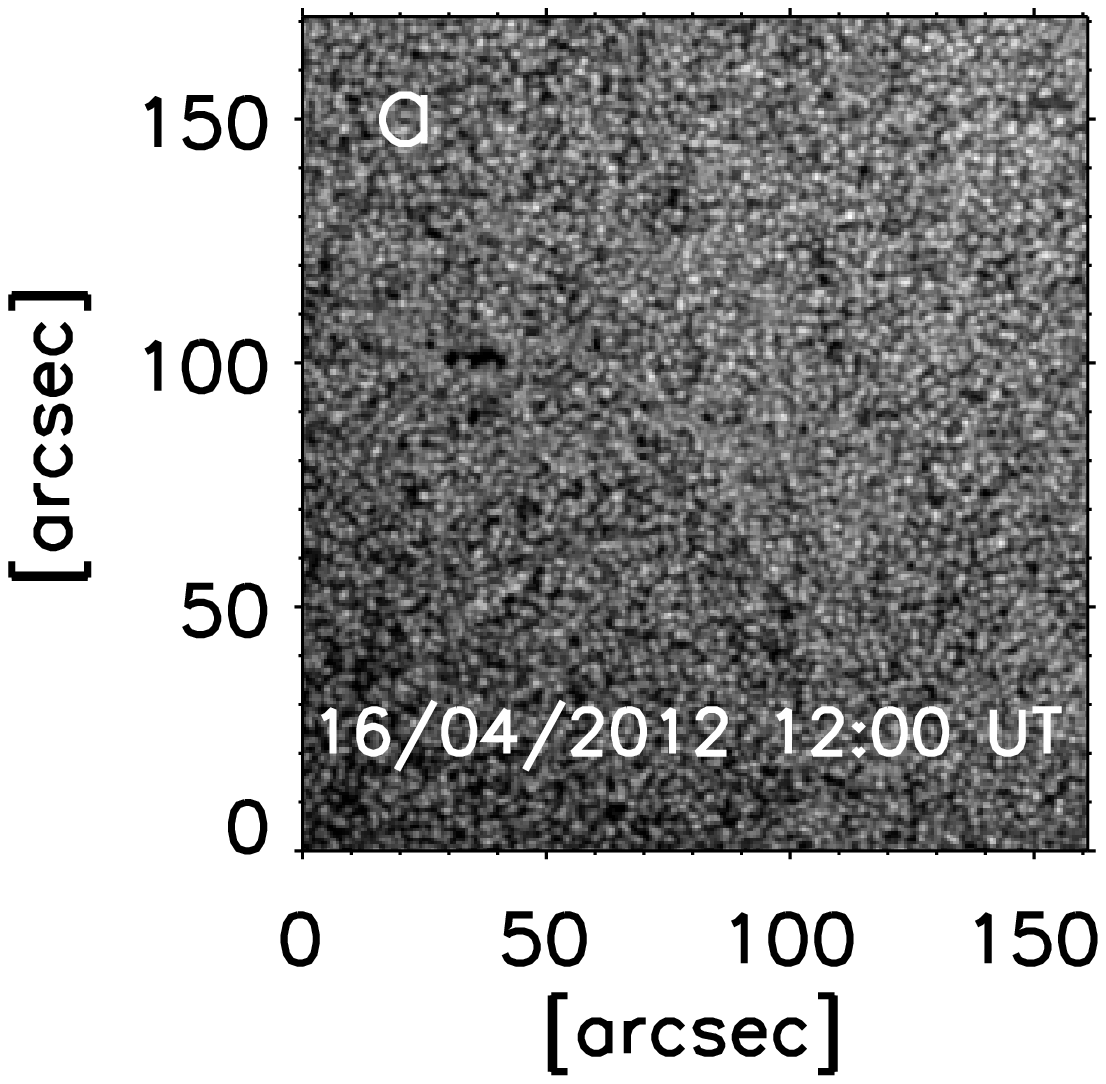}\includegraphics[height=2.8cm,trim=4.2cm 2.5cm 4.7cm 1.3cm,clip=true]{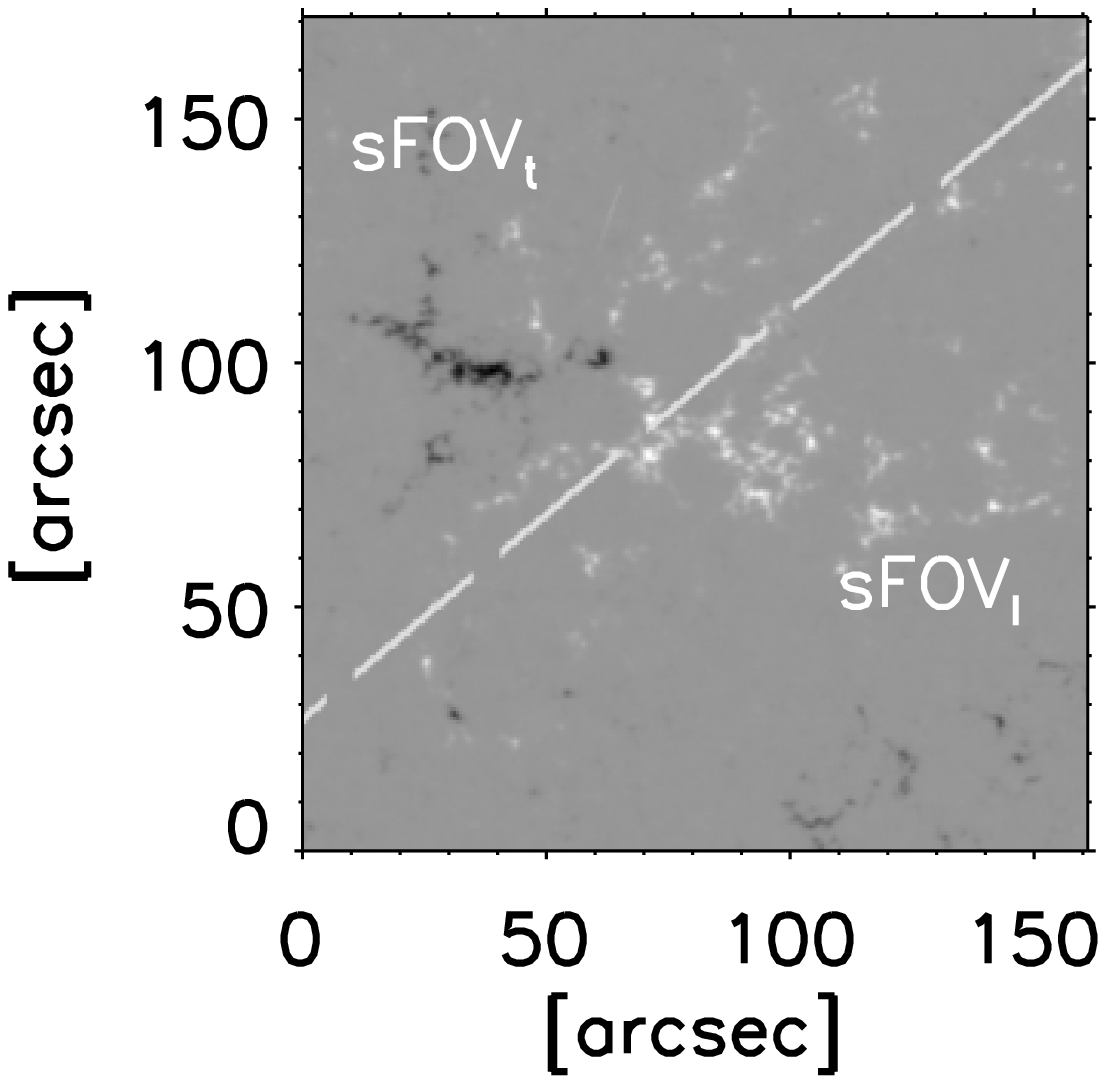}
\includegraphics[height=2.8cm,trim=0.5cm 2.5cm 4.7cm 1.3cm,clip=true]{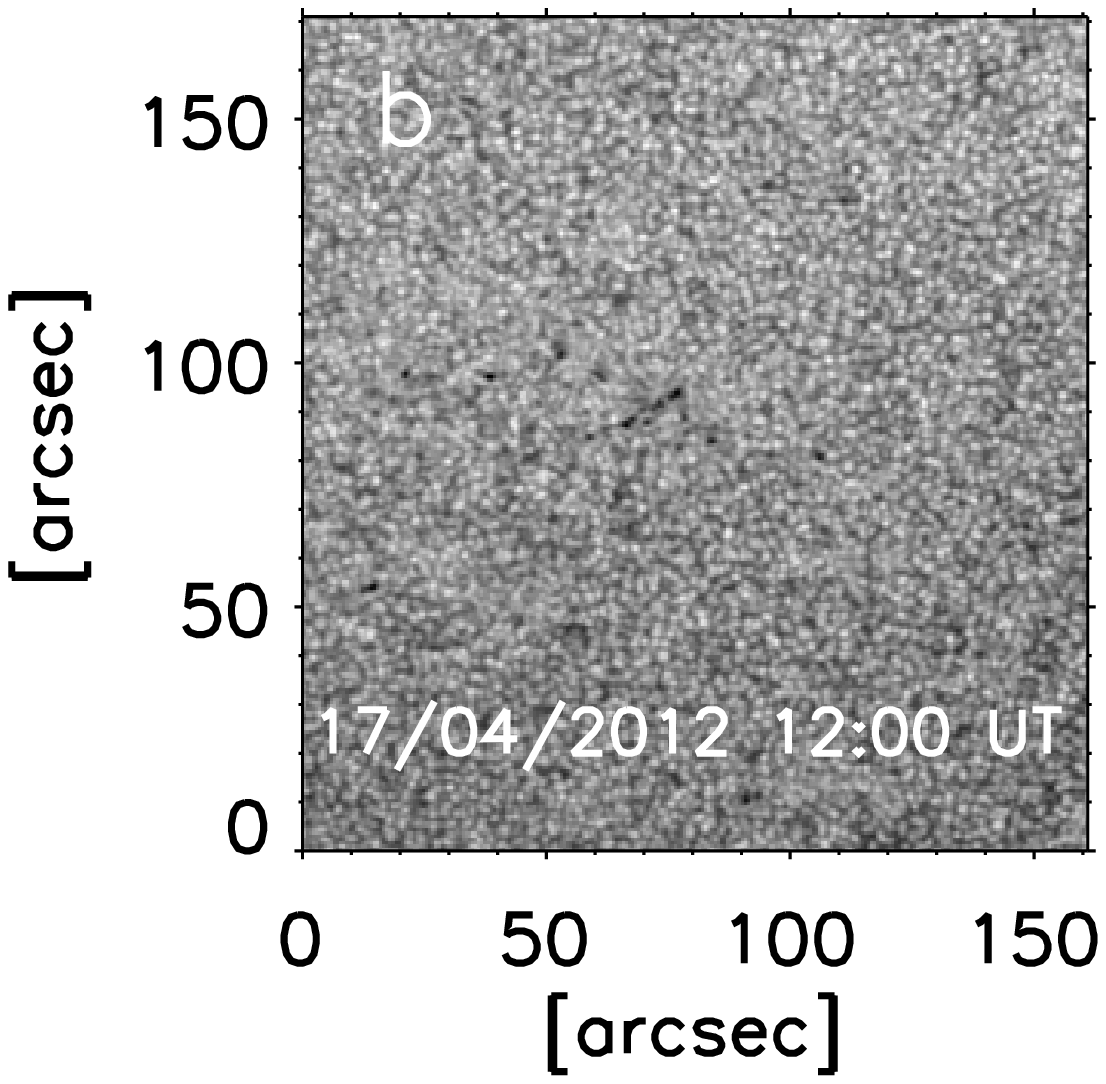}\includegraphics[height=2.8cm,trim=4.2cm 2.5cm 4.7cm 1.3cm,clip=true]{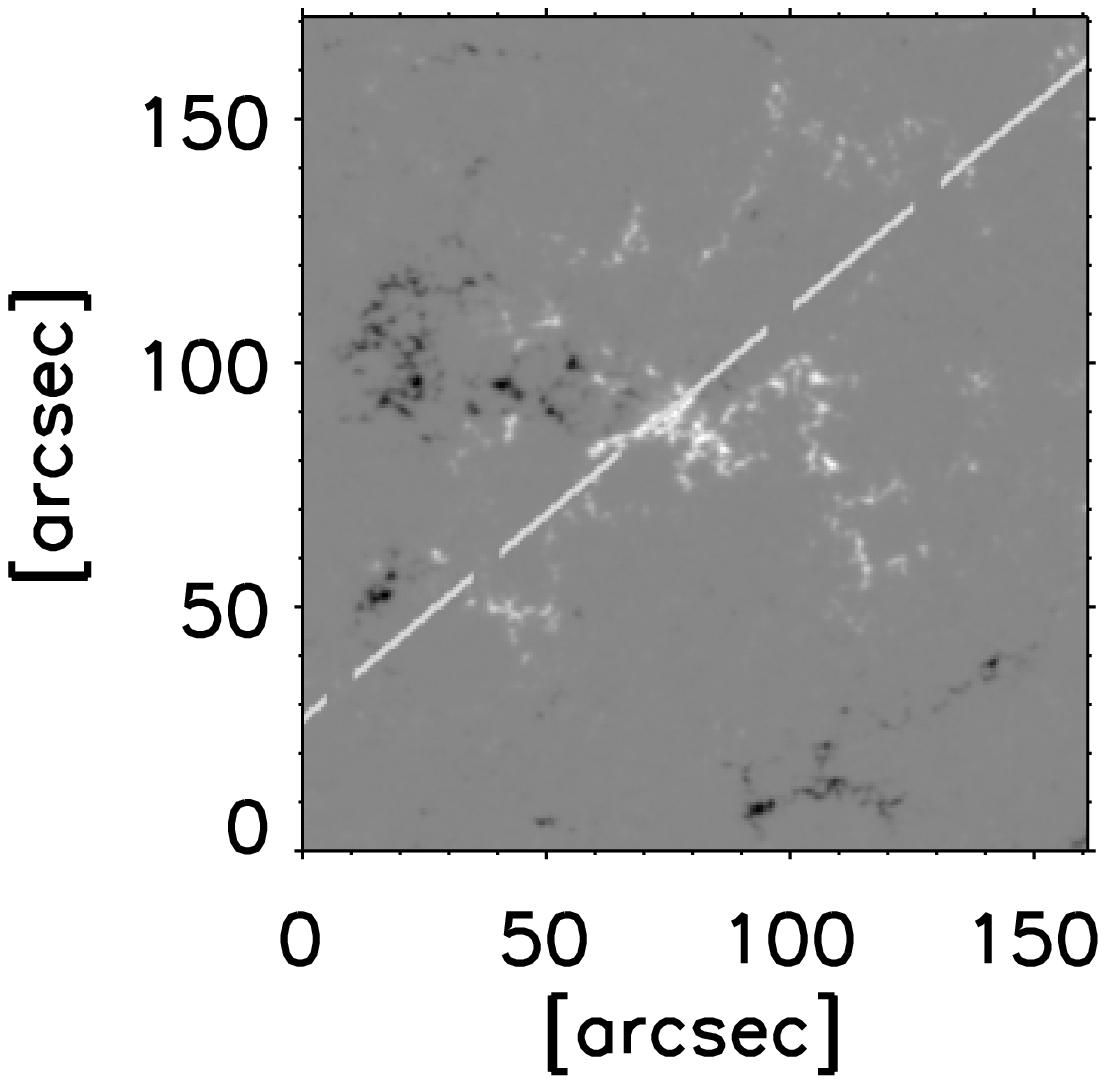}
\includegraphics[height=2.8cm,trim=0.5cm 2.5cm 4.7cm 1.3cm,clip=true]{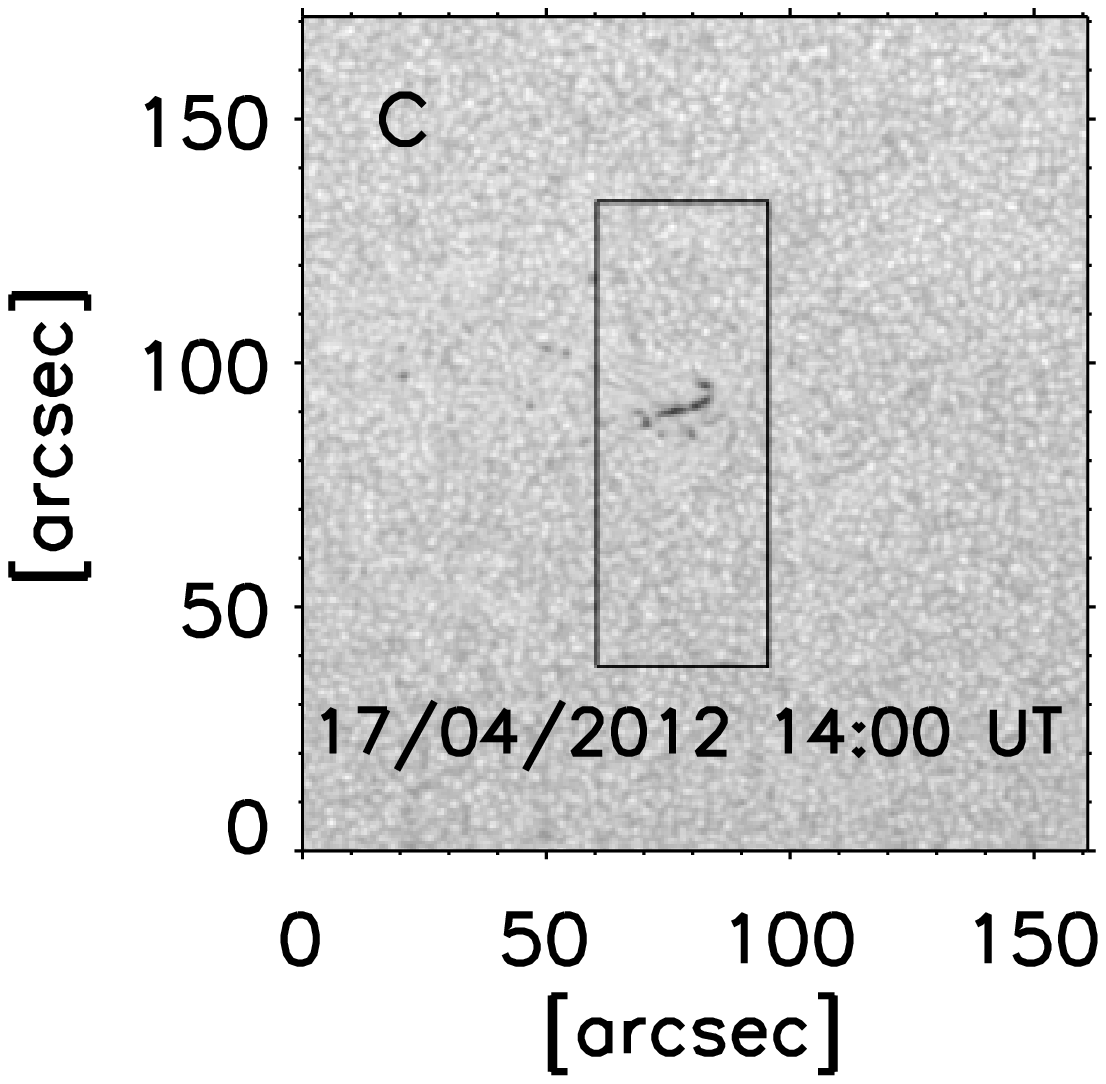}\includegraphics[height=2.8cm,trim=4.2cm 2.5cm 4.7cm 1.3cm,clip=true]{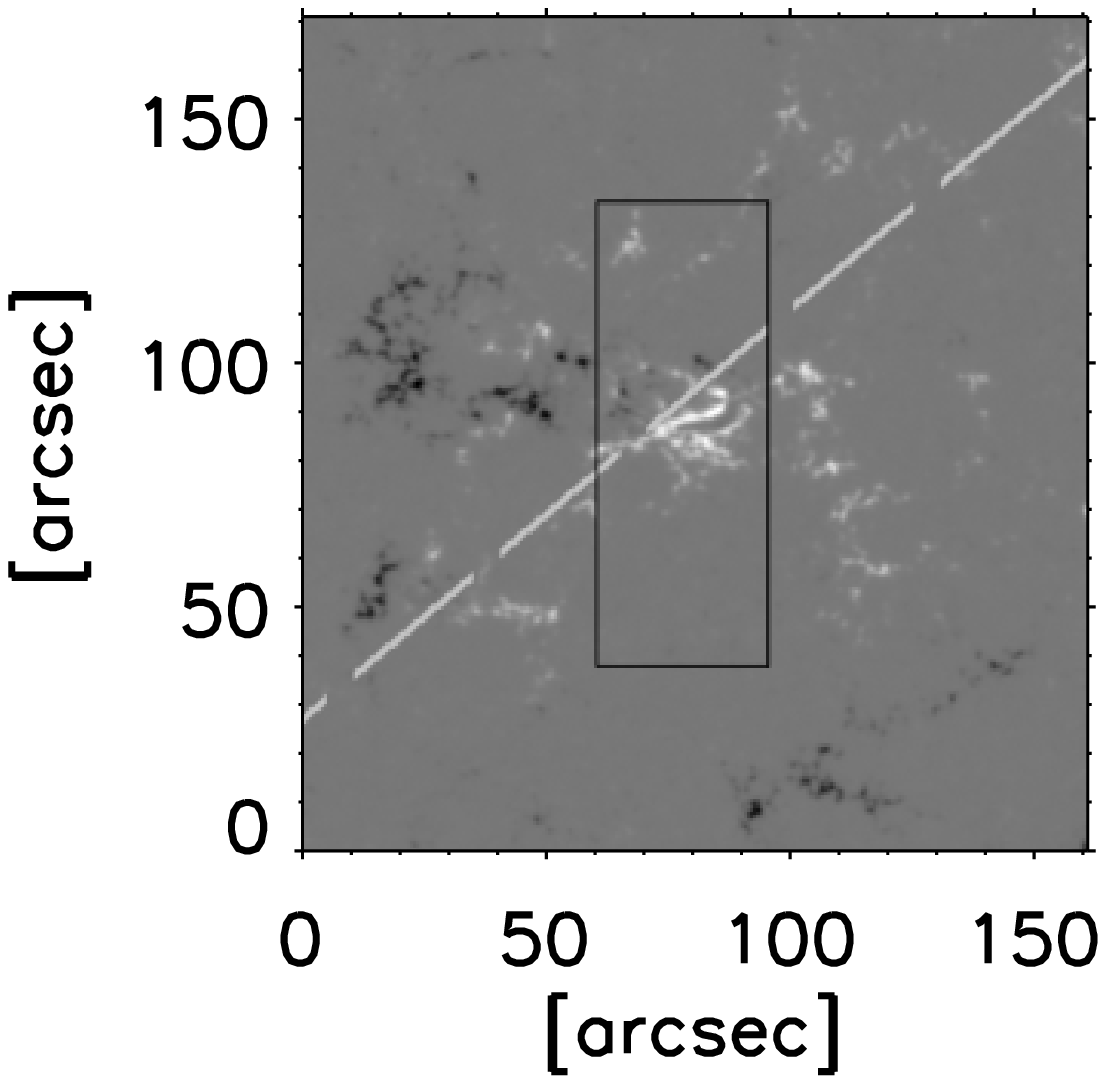}
\includegraphics[height=2.8cm,trim=0.5cm 2.5cm 4.7cm 1.3cm,clip=true]{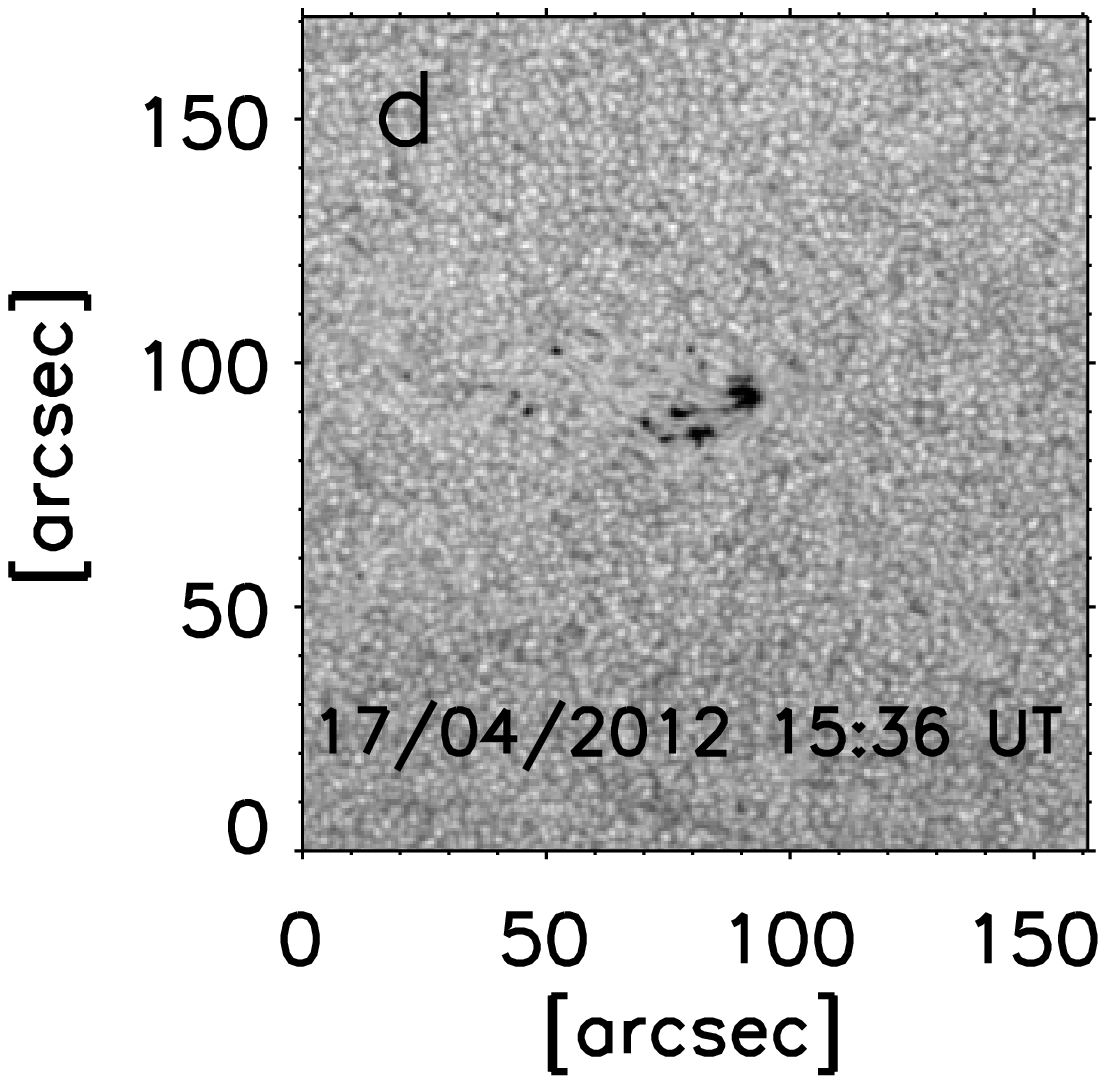}\includegraphics[height=2.8cm,trim=4.2cm 2.5cm 4.7cm 1.3cm,clip=true]{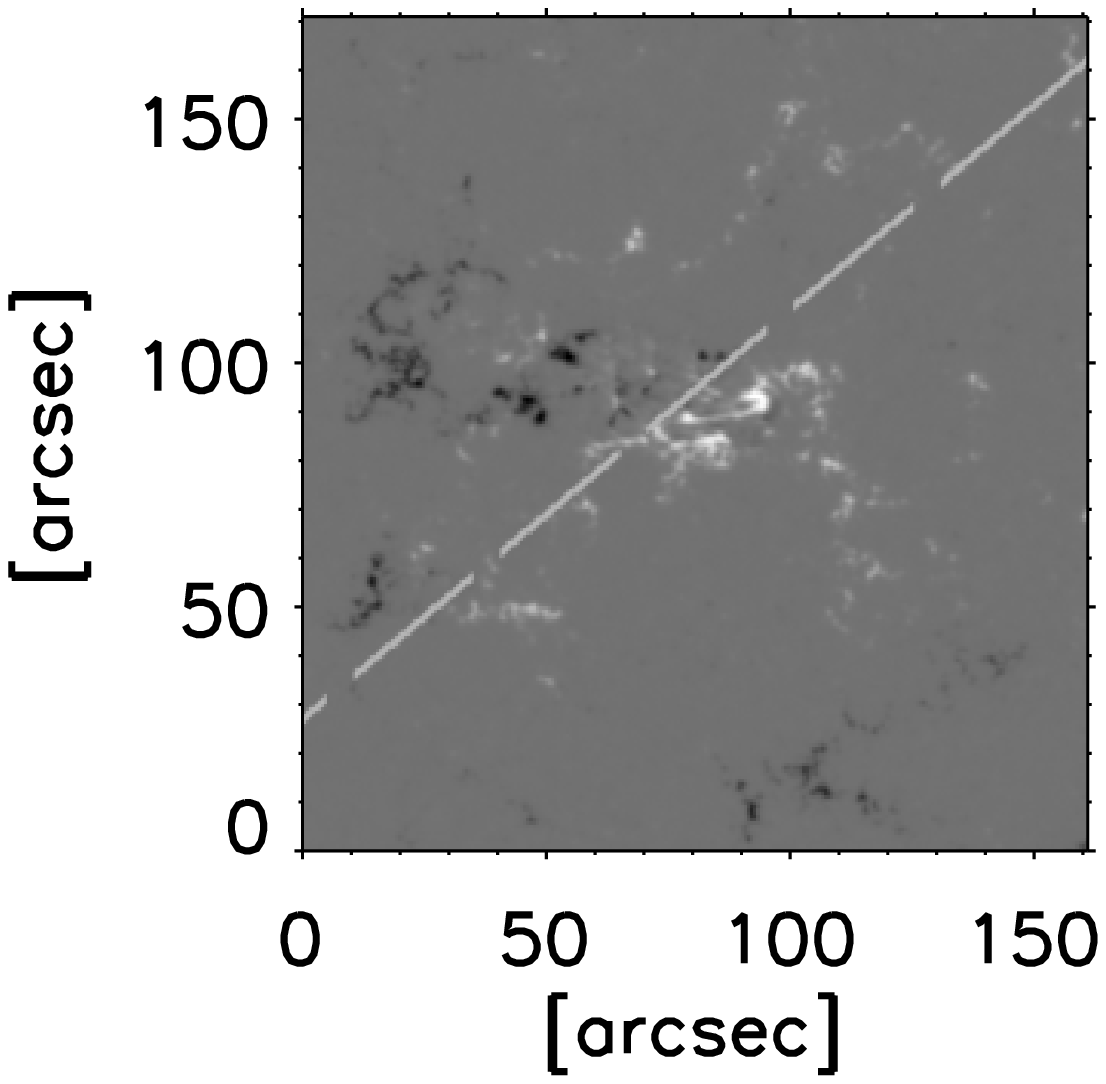}
\includegraphics[height=2.8cm,trim=0.5cm 2.5cm 4.7cm 1.3cm,clip=true]{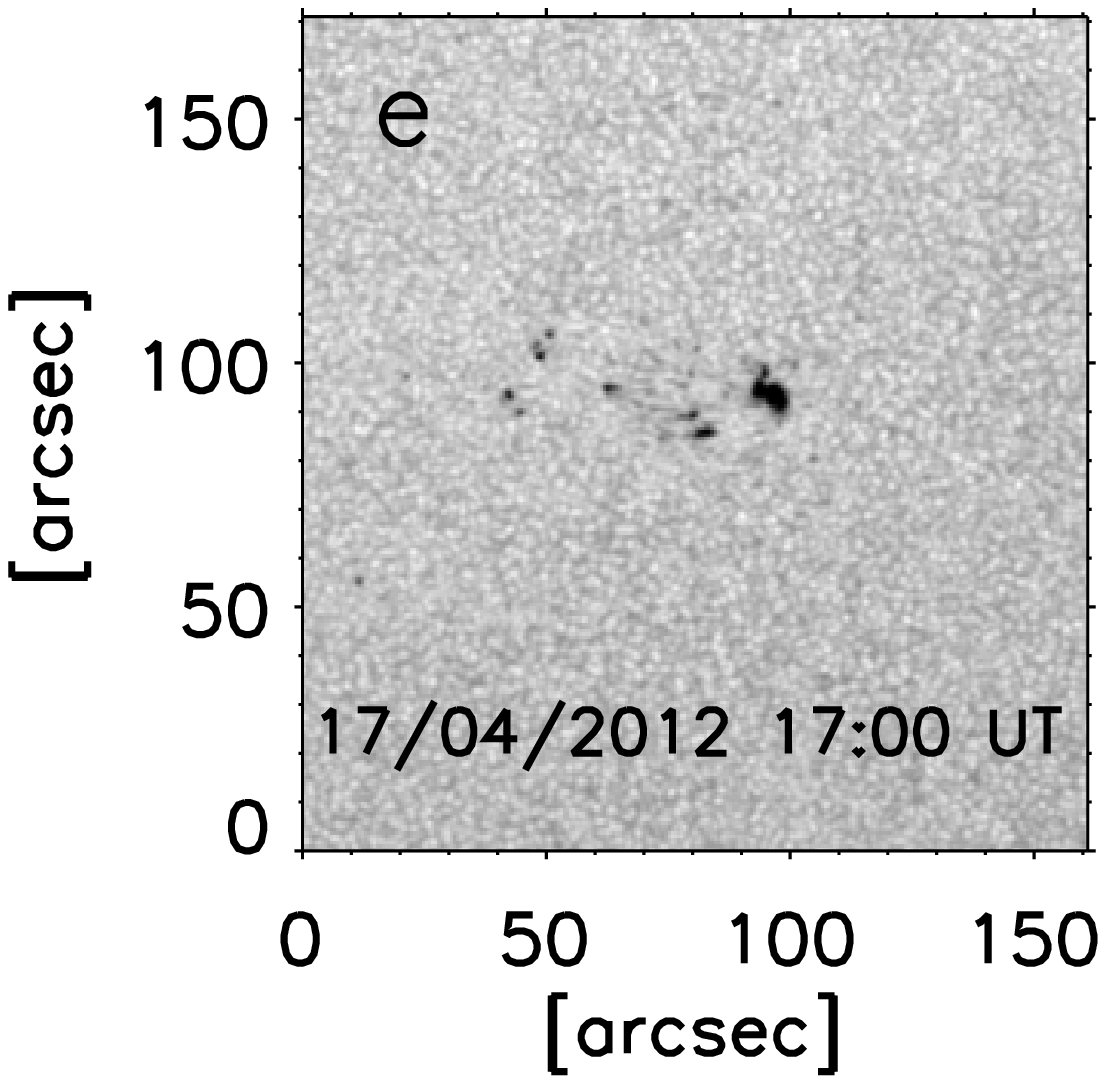}\includegraphics[height=2.8cm,trim=4.2cm 2.5cm 4.7cm 1.3cm,clip=true]{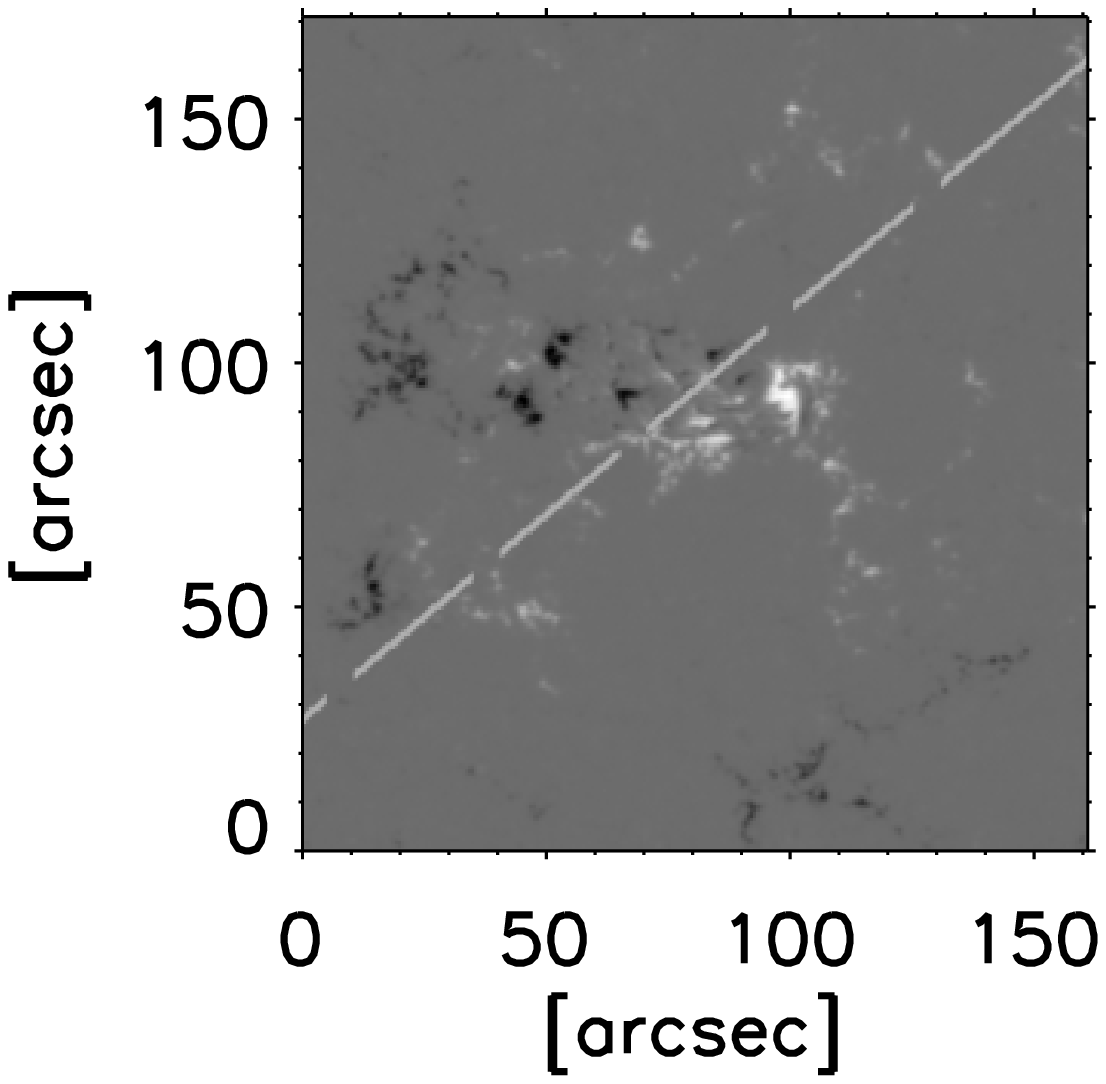}\
\includegraphics[height=2.8cm,trim=0.5cm 2.5cm 4.7cm 1.3cm,clip=true]{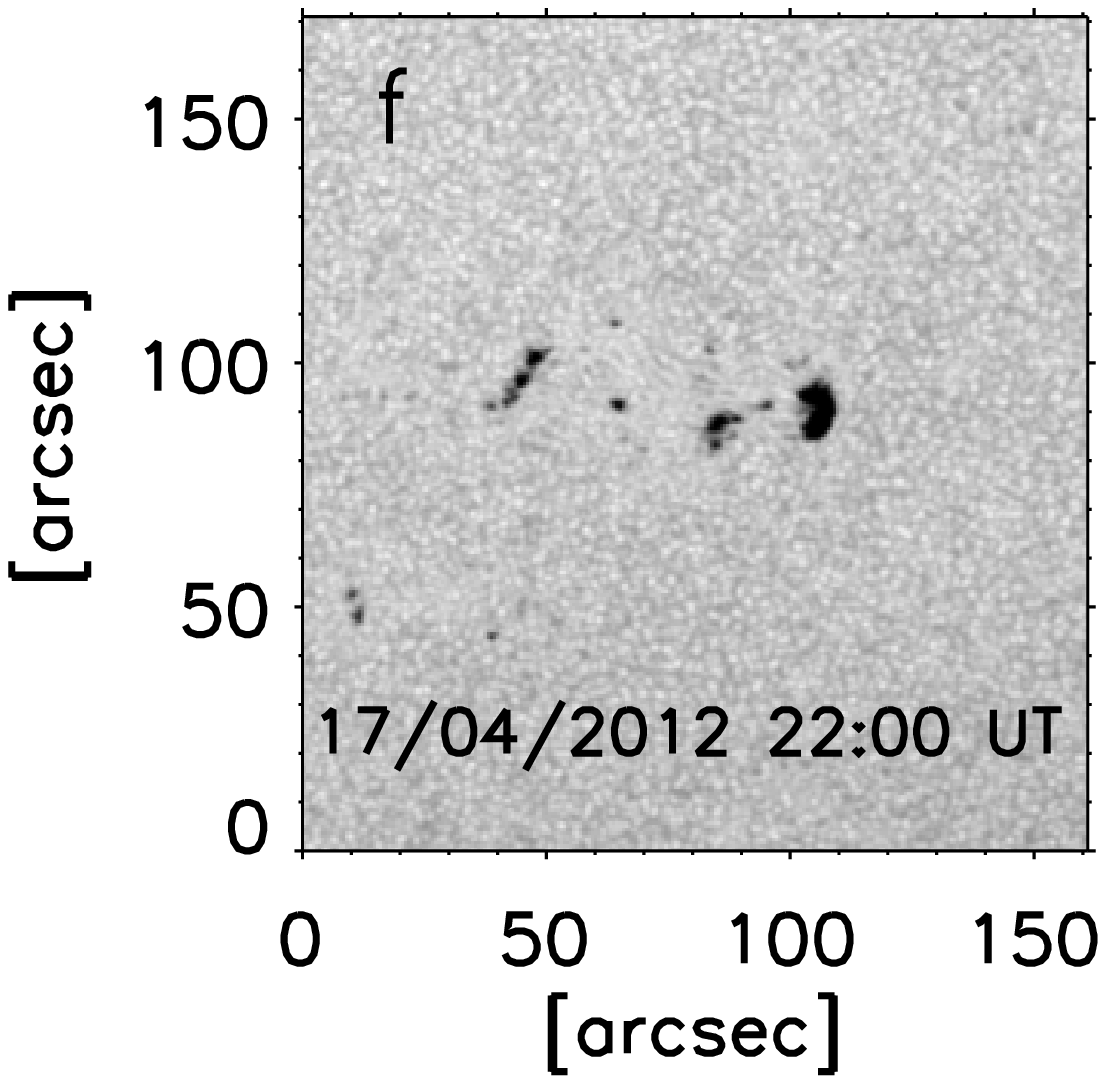}\includegraphics[height=2.8cm,trim=4.2cm 2.5cm 4.7cm 1.3cm,clip=true]{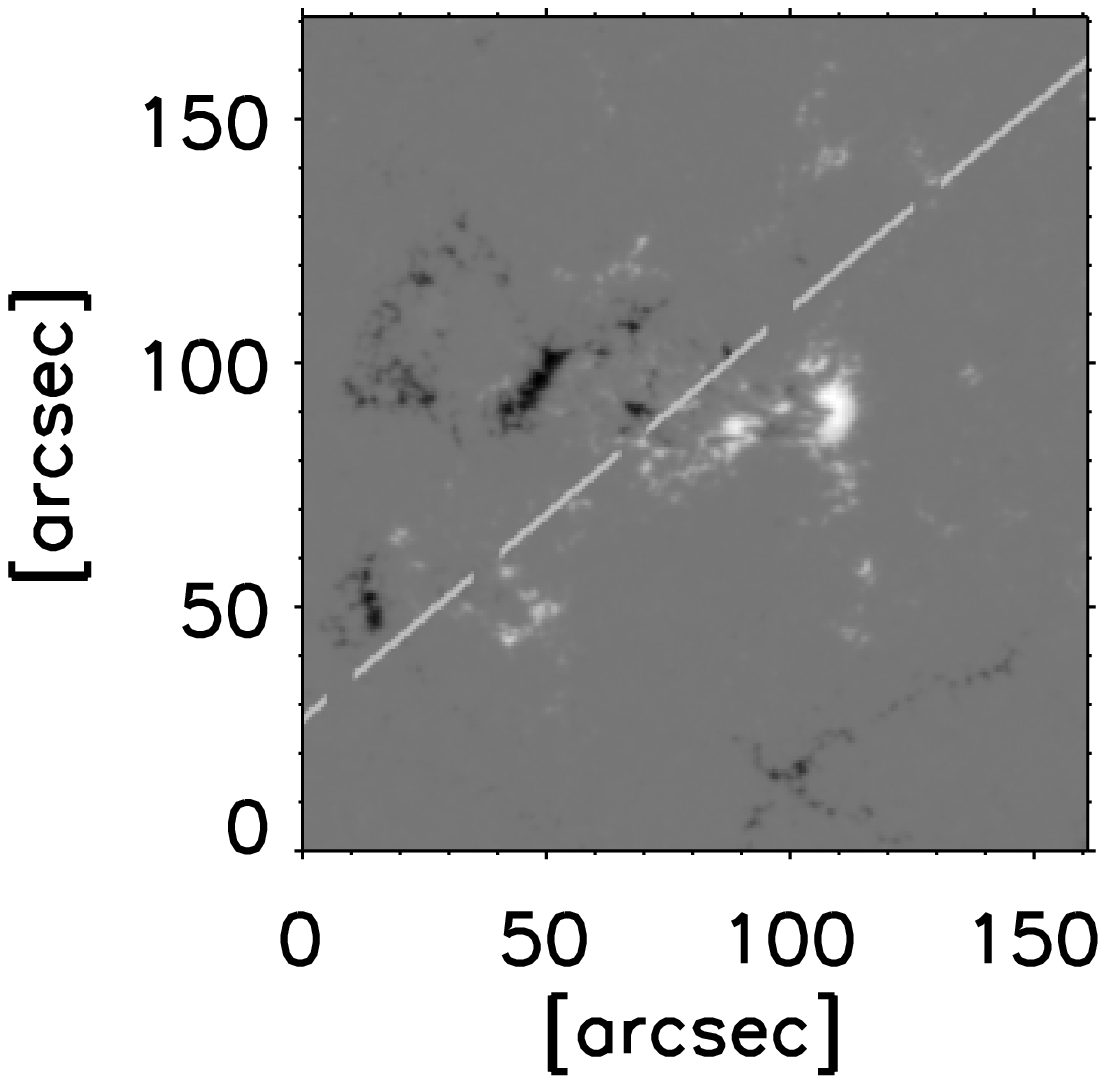}
\includegraphics[height=2.8cm,trim=0.5cm 2.5cm 4.7cm 1.3cm,clip=true]{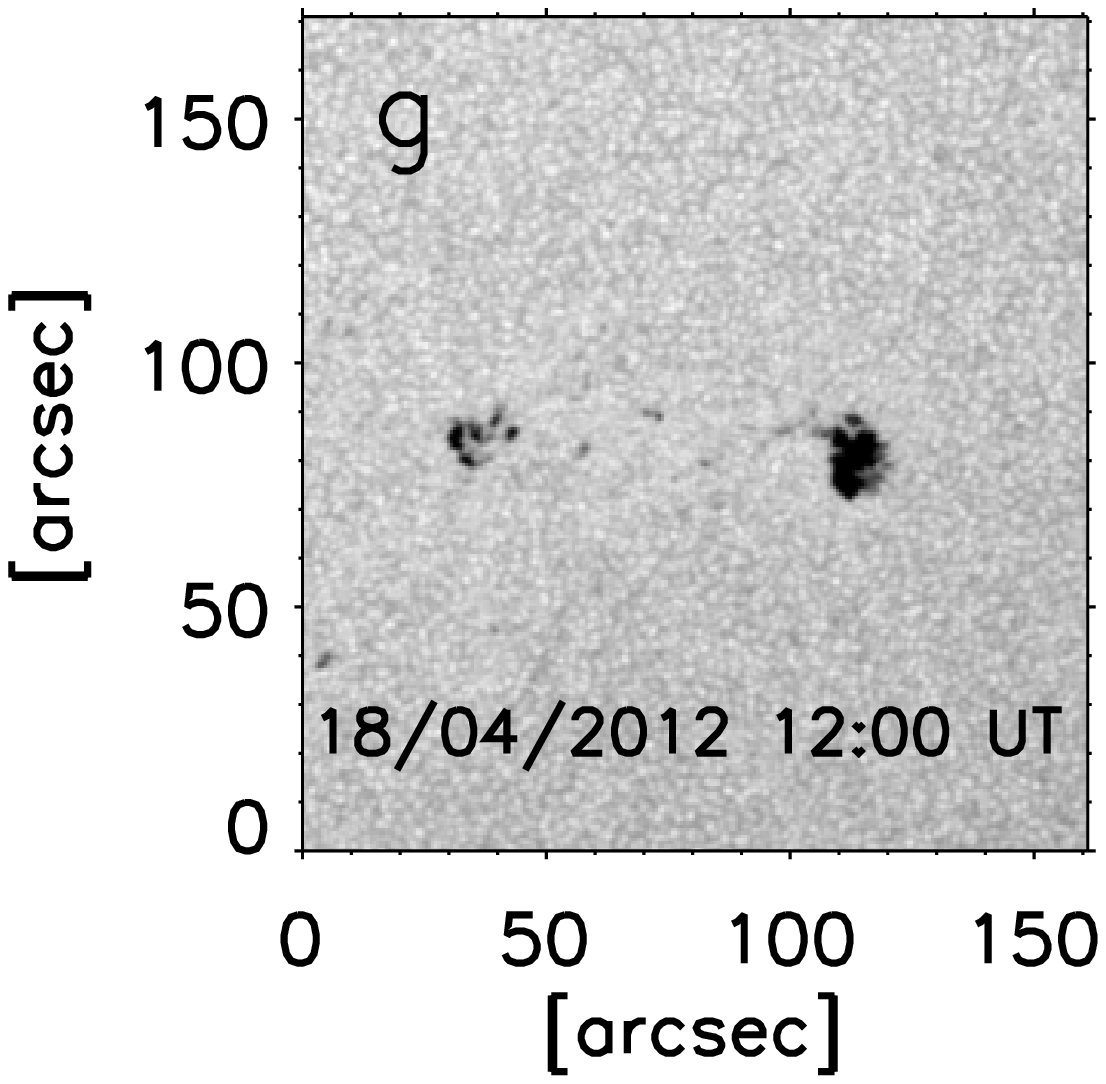}\includegraphics[height=2.8cm,trim=4.2cm 2.5cm 4.7cm 1.3cm,clip=true]{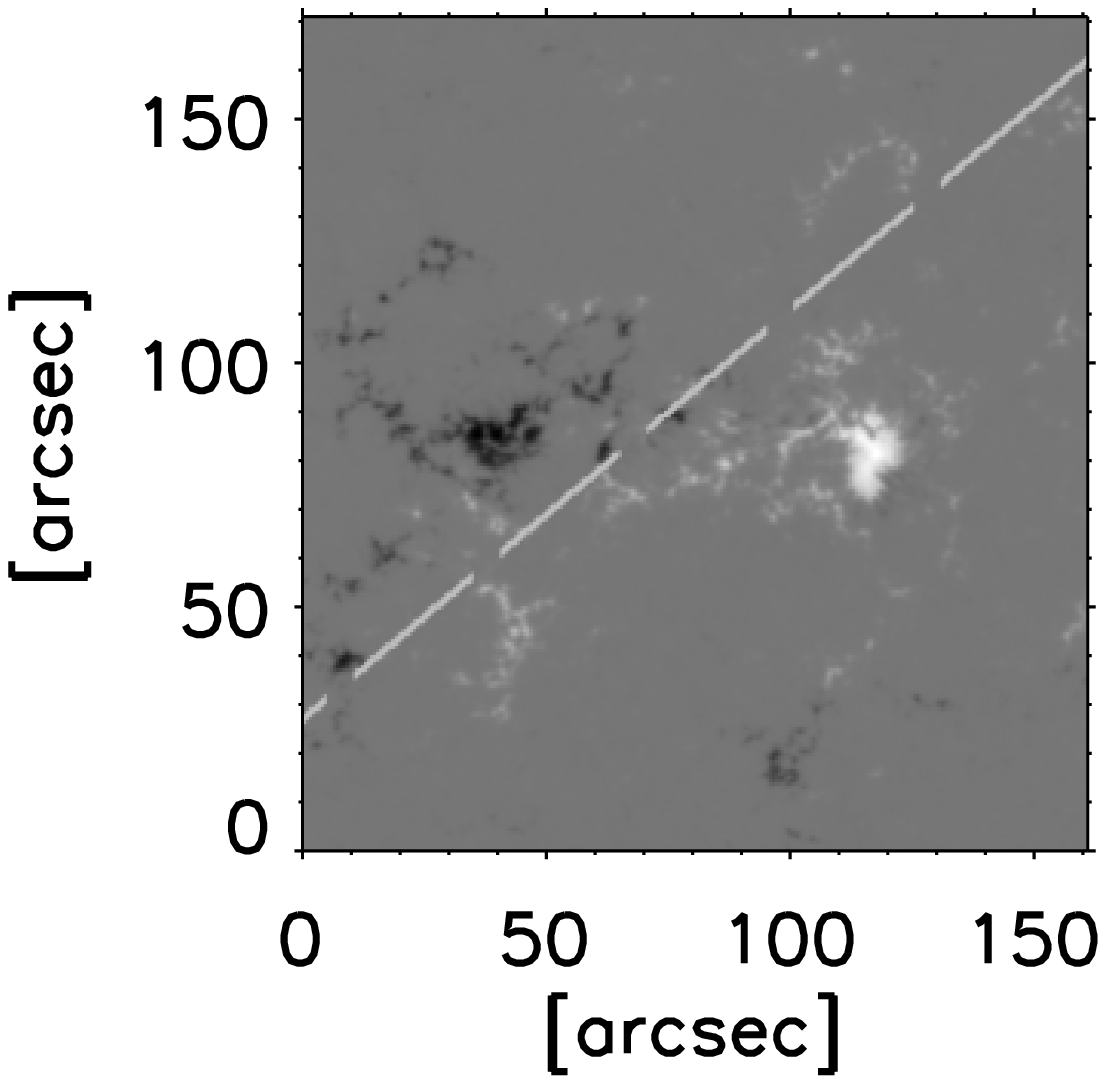}
\includegraphics[height=3.45cm,trim=0.5cm 0.5cm 4.7cm 1.3cm,clip=true]{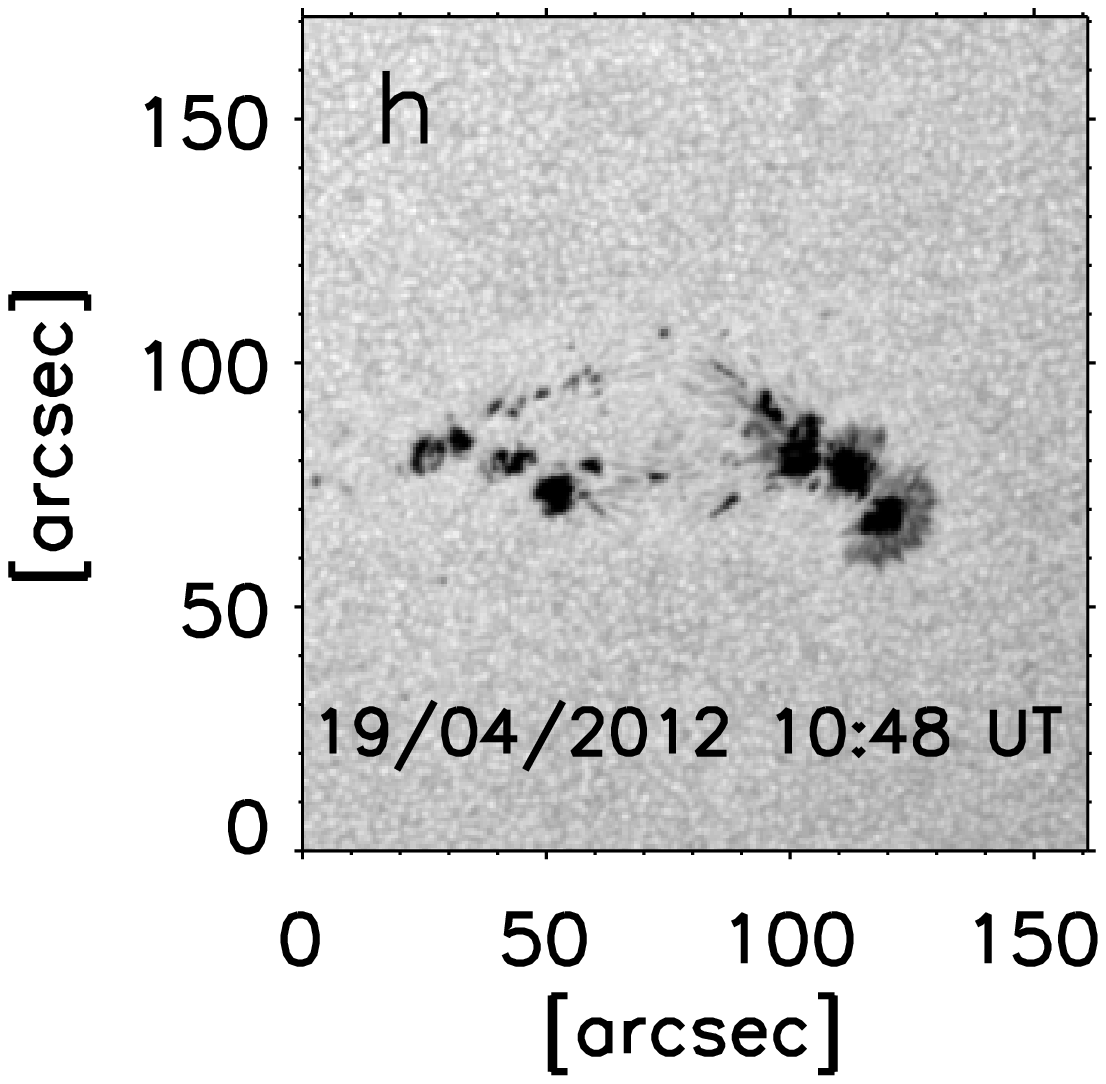}\includegraphics[height=3.45cm,trim=4.2cm 0.5cm 4.7cm 1.3cm,clip=true]{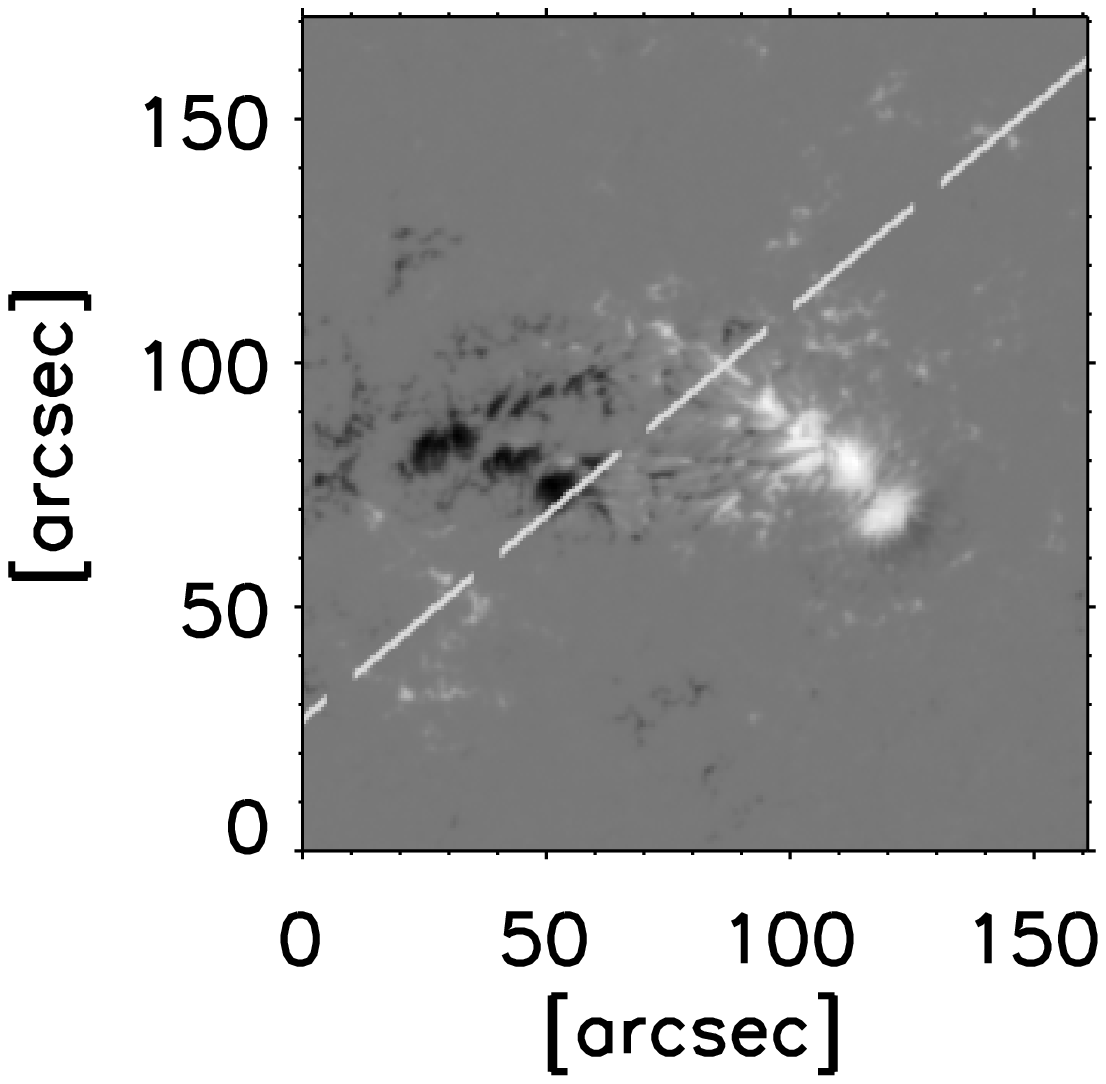}
}
\caption{\footnotesize{AR  11462 as seen in the SDO/HMI continuum filtergrams (left) and LOS magnetograms (right) %during its evolution  %t the times shown on the continuum image.  % (from top to bottom) at different times, 
from April 16, 2012, 12:00 UT, to April 19, 2012, 10:48 UT. %
%From top to bottom, with letters between brackets indicating line label: the images correspond to data taken on  April 16 12:00 UT (a), April 17 12:00 UT (b), 14:00 UT (c), 15:00 UT (d), 15:36 UT (e), 16:00 UT (f), and 16:36 UT (g), 2012. 
The box  in the April 17, 2012, 14:00 UT data show  the FOV of IBIS data.  More details in Sect. 3.1.
        The magnetic field in the background magnetogram is shown in the range of values [-1.5,1.5] kG. The full temporal evolution of the analysed data is shown in a movie available online. 
}}
\label{f1b}
\end{figure}

 In this framework,   we hereby study the evolution of a magnetic region that  led a  diffuse dipolar flux patch to form a pore, and, at a later stage, dipolar sunspot groups in AR 11462. We analyse a  unique  ground-based spectro-polarimetric data set  acquired during the pore formation, and complement these observations 
%Besides, we analyze 
with space-borne data  from the SDO mission. 
%with higher spatial, temporal, and spectral resolution than reported in the literature so far.  % operated at the NSO/Dunn Solar Telescope. 
We focus on the photospheric processes  %. 
%In particular, the present paper focuses on the photospheric processes 
 related to the pore formation by investigating the physical conditions  in the whole evolving region  as observed on timescales longer than about 10 minutes. The data and methods employed are presented in Sect. 2, while the results obtained are described in Sect. 3. We discuss our findings in light of  other observational studies  and  the minute outcomes of recent simulations, and then draw our conclusions in Sect. 4.

\section{Data and methods}

  \subsection{Observations}

The region analysed  in our  study was observed with the Interferometric Bidimensional Spectrometer  \citep[IBIS;][]{Cavallini_2006}  at  the Dunn Solar Telescope of the National Solar Observatory (NSO/DST) on April 17, 2012,   from 13:58 UT to 20:43 UT, within the AR  NOAA  11462 (hereafter referred to as AR), at initial disk position [S24, E0.3].   
The observations were assisted by the  adaptive optics  system
of the NSO/DST \citep{Rimmele_etal2004}, under fair conditions of atmospheric seeing. 
% allowed for acquisition of data with rather stable spatial resolution  and contrast over a $\approx$ 4.5 hr  time interval.   
No data were acquired between %There was no data taken of the  AR  in  the time interval between 
16:30 UT and 18:30 UT because of a worsening of the seeing started at about 16:15 UT.

{\bf  The IBIS data consist of 223 sequences, each containing narrowband filtergrams derived from a 24-, 30-, and 25- point  scan of the Fe I 617.3 nm, Fe I 630.2 nm, and Ca II 854.2 nm lines, respectively,  over a field of view (FOV) of  $\approx$ 40$\times$90 arcsec$^2$  with a  cadence of 67 s. 
 The Fe I  data include  sequential  measurements of the six polarization  states  [I+Q, I+V, I-Q, I-V, I-U, I+U] at  each wavelength position of the line sampling (FWHM 2 pm, step 2 pm), while the Ca II data (FWHM 4.4 pm, step 4.4 pm) include only Stokes I measurements.  } %It is worth mentioning that, although measurements from three modulation cycles would suffice the  estimation of Stokes parameters, 
 %The above six measurements are usually acquired with the IBIS to compensate observations for  instrumental effects more accurately than achieved with less  states.
  %, e.g. flat field induced cross talk.} 
 Each measurement consists of a  single filtergram taken with an integration time of 60 ms and pixel scale of  $\approx$ 0.09 arcsec.  The above data are complemented with  simultaneous broadband filtergrams obtained  at  633.32 $\pm$ 5 nm with the same exposure time and FOV of the narrowband data for the post-facto image restoration.

%  QUI FIGURA
In this study we focus on the available Fe I 617.3 nm line data; according to \citet[][]{Norton_etal2006}, the excitation potential, %($\chi$), 
effective Land\'e factor,  %(g$_{eff}$ ), 
and  average line-formation height  %(h) 
of the 
  Fe I 617.3 nm  line %, Fe I 630.25 nm
are  2.22 eV, 2.5, and 250-350 km, respectively. %We defer  the analysis of other available IBIS data  to future work. 
%; following  \citet[][]{Faurobert_etal2013} the same quantities for the  Fe I 630.25 nm line data are 
%3.68, 2.5, and 20-210 km, respectively.

The AR evolution was also studied  by analysing the   data 
obtained  with the Helioseismic and Magnetic Imager \citep[HMI;][]{Scherrer_etal2012,Schou_etal2012}   aboard the Solar Dynamics Observatory \citep[SDO;][]{Pesnell_etal2012}. 
In particular, we analysed the  Level 1.5 SDO/HMI photospheric  full-disk
filtergrams and  magnetograms 
 at the Fe I 617.3 nm line taken from April 15 to April 19, 2012, when the AR longitudinal distance was within $\pm$ 30$^{\circ}$ of the central meridian.  
 The data consist of $\approx$ 500  images, each 4096 $\times$ 4096 pixels, with pixel size of 0.505 arcsec and cadence of 720 s.  Additional information about the  AR  was derived from  the SDO/HMI Space-weather Active Region Patches \citep[SHARP;][]{Hoeksema_etal2014,Bobra_etal2014} maps obtained during the  same time interval of the other SDO data. %{\bf We analysed 

\subsection{Methods} 
The IBIS observations were processed with the standard reduction pipeline\footnote{http://nsosp.nso.edu/dst$-$pipelines} to compensate data   for the dark and flat-field response of the CCD devices,  
instrumental blueshift, and instrument- and telescope-induced polarizations.  Besides, they  were also restored for seeing-induced  degradations, using the 
 Multi-Frame Blind Deconvolution technique \citep[MFBD;][and references therein]{vanNoort_etal2005}.  
%The  IBIS data were further  processed as described in the following.

To get quantitative estimates of the physical parameters in the observed region, we performed spectro-polarimetric inversions of a subset  of the  IBIS   measurements  with the SIR code \citep[][]{Ruizcobo_deltoro1992,Bellotrubio_2003}. %In particular, 
%Since we aimed at investigating physical conditions and processes occurring during the whole pore formation on time scales  longer than 10 minute, 
 We selected   20 time intervals during the IBIS observations that  track the evolution of the region under good and stable  seeing conditions.  We processed, with the SIR code, a 
sub-array of 360$\times$350  pixels  extracted from the selected data  and  centred on the evolving feature.

Following common approaches  (see e.g. \citet[][]{asensio_etal2012,Requerey_etal2015,Buehler_etal2015}, and references therein),  we performed the  data inversion by considering  one component plus a stray-light component of unspecified amplitude.  This latter component thus acts as a free parameter.  Depending on the amount of polarization in each  pixel, we  considered the first component to be magnetic or  quiet. We defined the  magnetized pixels  as those in which the total circular polarization signal is $\ge$ 2 times the  standard deviation of the signal estimated over the sub-array.  Examples of the identified regions are given in  Fig. \ref{a1}, which is  available online.  
{\bf We  considered  
 the  Harvard$-$Smithsonian Reference Atmosphere \citep[HSRA;][]{Gingerich_etal1971}   as an initial guess model  for  the quiet  regions and the same model, but modified with an initial value for the magnetic field strength of 0.2 kG, %consisting of the former and the cool model by \citet[][]{Collados_etal1994}  
for the magnetized regions.  
We  performed  the data inversion by applying  two computational cycles, each one %a total of two times  with each inversion 
including  
up to 30  iterations. At the first  cycle, we  considered  all modelled quantities to be constant along the LOS and assigned them one node.  At the second cycle, we added one node in  the temperature. The magnetic field strength, inclination,  azimuth,  LOS velocity, and micro-turbulent velocity, however, were   considered  to be constant with height, i.e. the temperature was assigned two nodes, and the other quantities were given one node. The two  nodes were set 
at log($\tau$) = 1.4 and 
log($\tau$) = -4. 
 Since we performed one-component inversions,  the magnetic filling factor  is equal to unity; we set  the macroturbulent velocity   to 0.75 km/s.  
For quiet Sun region pixels, we gave the I measurements four times the statistical weight of the Q, U, V profiles; for magnetized region pixels, we did the opposite. 
% ???? due to the high spatial resolution of the data.
At each iteration,  the synthetic profiles derived from the solution of the radiative transfer equation were also convolved with the  spectral response function  of IBIS %,  
\citep[][]{Reardon_cavallini2008} and %. 
%The obtained spectra were also 
weighted by considering  the stray-light contamination on the data. We estimated the latter quantity    by averaging Stokes I spectra in a region characterized by low  polarization degree over the inverted  sub-array, as in for example \citet[][]{Bellotrubio_etal2000}.

}

We tested the accuracy of the results obtained  
for different initializations of the inversions and chose the initialization that produced 
the best fit between the synthesized profile and measurement over each pixel and the  
largest physical consistency of the estimated quantities over the whole FOV inverted.  
We found that increasing the number of inversion cycles did not further improve the result of our calculation;  on average less than 12-15 iterations allowed the  computational convergence.    Examples of the results obtained and comparisons between the observed and inverted profiles  at 20 positions on the analysed sub-FOVs are given in Figs. \ref{a1}-\ref{a8}, which are  available online.  Values of the stray-light fraction, LOS magnetic field strength, field inclination and azimuth, and LOS velocity  derived  from our data analysis  at the same positions of the compared profiles are listed in Tables \ref{tablea2}-\ref{tablea8}, which are available online.

 We then  transformed
the magnetic field inclination and azimuth derived from the data inversions  into the local solar frame (LSF). 
We resolved the 180 degree ambiguity of the azimuth angle via the NPFC code \citep{Georgoulis_2005}.

In order to describe   the  horizontal proper motions   in the IBIS FOV  and estimate their velocity, $v_{H}$,  we applied  the Fourier local correlation tracking method \citep[FLCT;][and references therein]{Fisher_welsch2008} to the available line-continuum data.  
We set the  FWHM of the Gaussian tracking
window  to 0.5 arcsec %approximately a half of the granule
%size, to %. This was done to 
to properly track magnetic structures with spatial scales smaller than the typical granular size; we made the  %, which are very common in the analysed data. 
 temporal integration  over a 13 minute  time interval. 
Finally, we computed the plasma LOS 
 velocity, v$_{LOS}$,  by the Doppler shifts of line cores with respect to the average quiet Sun line
centre position in the IBIS FOV. We computed the  reference value for each filtergram of the analysed series. The series were previously processed with a subsonic filtering with a phase$-$velocity cut-off set to 5 km/s.

We processed 
the time series of the SDO/HMI  data  %analysed in our study 
according to \citet[][]{ermolli_etal2014}, by extracting from each image the 512$\times$512 pixel$^2$ sub-array centred on the AR baricentre.  
Besides, we produced photospheric velocity maps of the horizontal plasma motions  via the   differential affine velocity estimator method for vector magnetograms \citep[DAVE4VM;][]{Schuck_2006} on the SDO/HMI SHARP data. In particular, 
following  \citet[][]{Schuck_2008}, we derived persistent plasma motions by comparing magnetograms 
taken 24 minute apart with a 5.5 arcsec FWHM  apodization window.

\begin{figure}%[ht!]
\centering{
\includegraphics[width=8.7cm,trim=1.cm 0.7cm 0.7cm 0.6cm,clip=true]{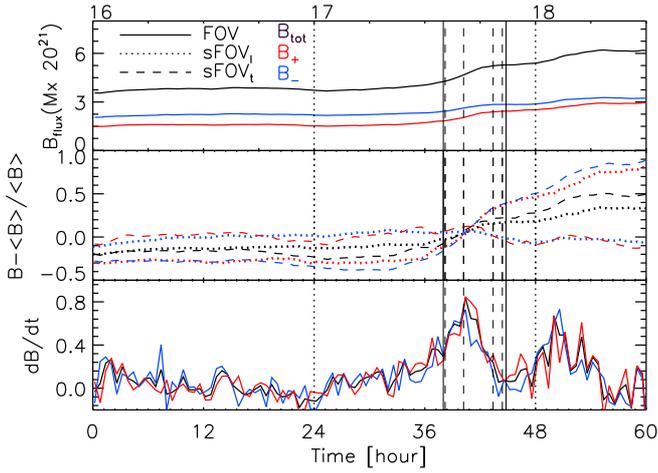} %\\
}
\caption{\footnotesize{Evolution of the magnetic flux  (top, middle) and flux derivative (bottom) in AR  11462 from the SDO/HMI LOS magnetograms acquired from April 16, 2012, 00:00 UT, to April 18, 2012, 12:00 UT, by analysing the subfield  shown in Figs. \ref{f1b}  (solid line),  and the leading and trailing parts of the same subfield separately (dotted and dashed  lines, respectively). The black, red, and blue lines  indicate the total, positive, and negative magnetic flux, respectively. The vertical lines  show the time interval of the IBIS observations (solid), the times  the data shown in Fig. \ref{f2}  were taken (dashed), and the times of 00:00 UT from April 16 to April 18, 2012 (dotted). 
The additional axis  indicates calendar days at 00:00 UT. For clarity, flux values are only shown from data taken with a cadence of 36 minute.
}}
\label{f1}
\end{figure}

\begin{figure}
\centering{
\includegraphics[height=5.cm,trim=0.7cm 1.9cm 10.6cm 0cm,clip=true,keepaspectratio=true]{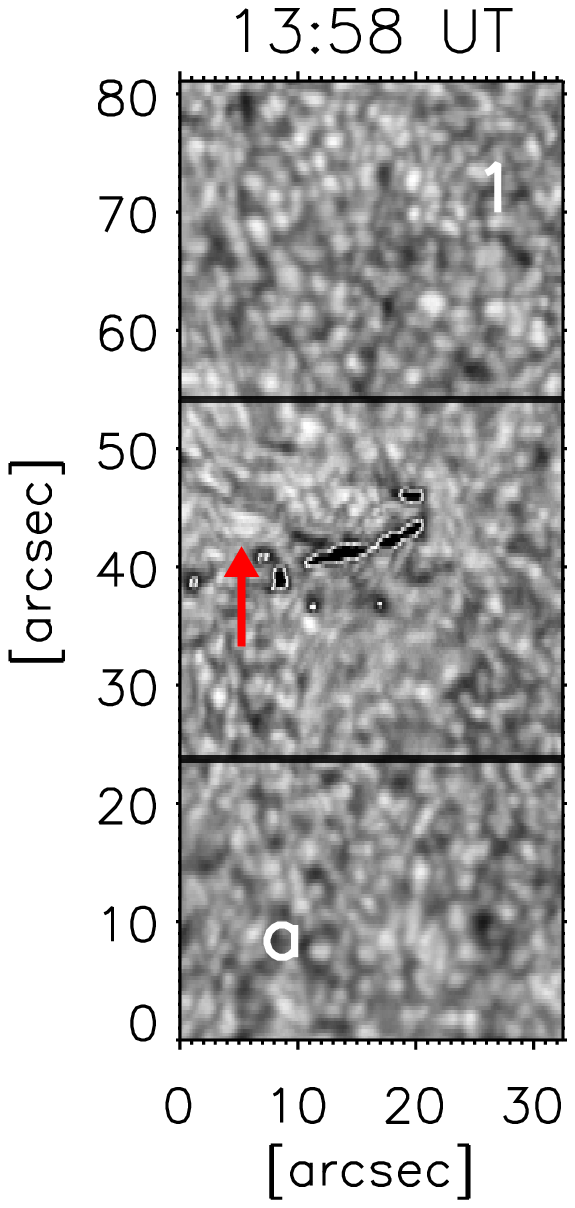}\includegraphics[height=5.cm,trim=2.9cm 1.9cm 10.6cm 0cm,clip=true,keepaspectratio=true]{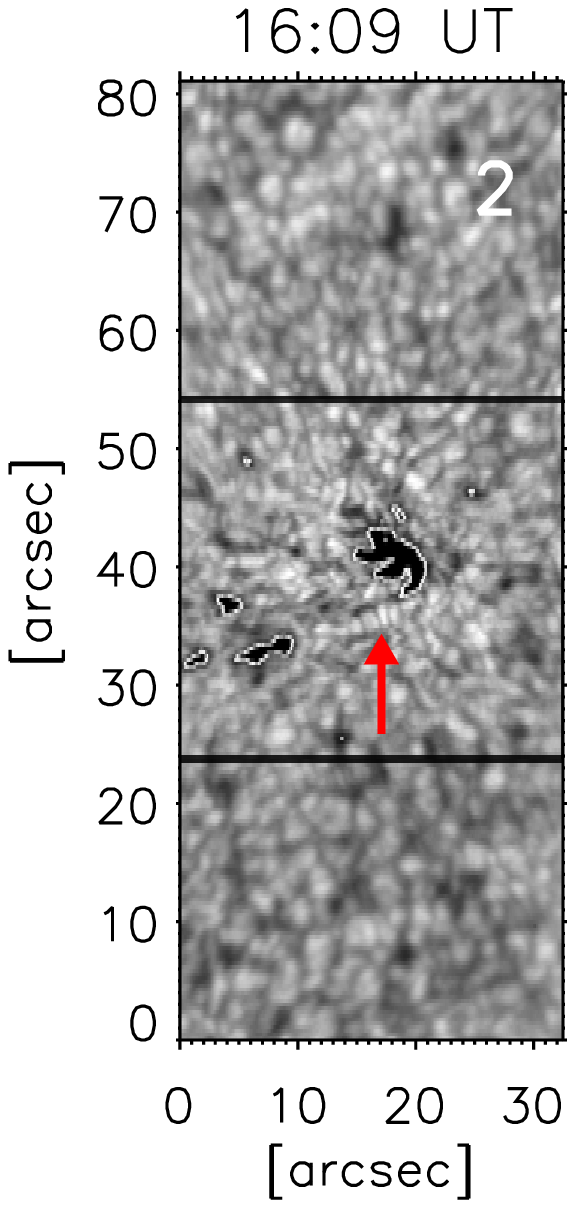}\includegraphics[height=5.cm,trim=2.9cm 1.9cm 10.6cm 0cm,clip=true,keepaspectratio=true]{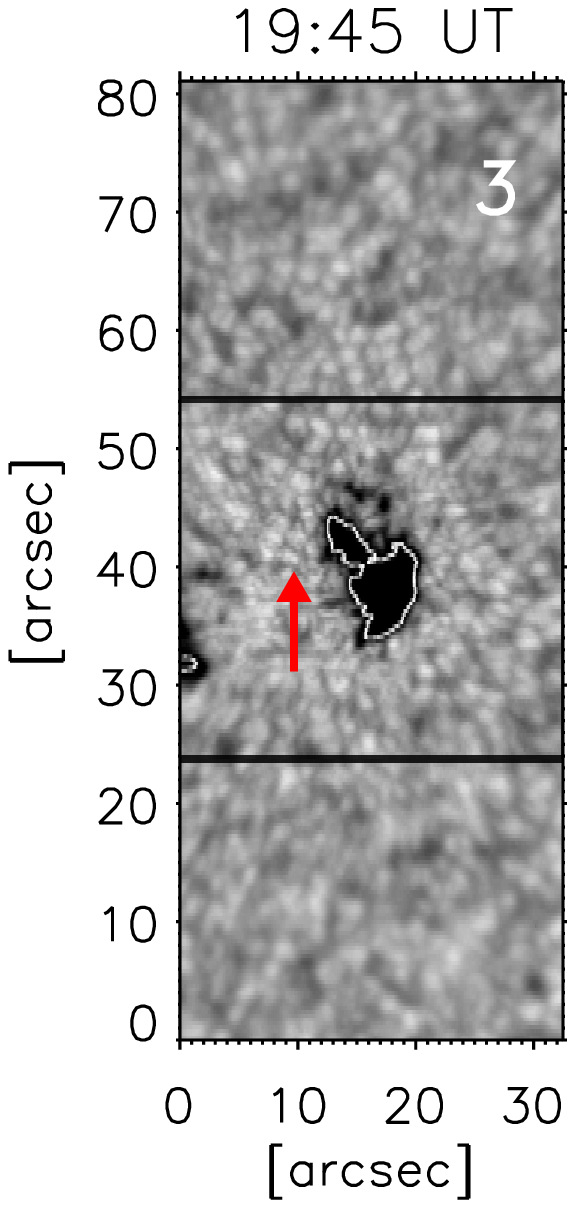}\includegraphics[height=5.cm,trim=2.9cm 1.9cm 10.6cm 0cm,clip=true,keepaspectratio=true]{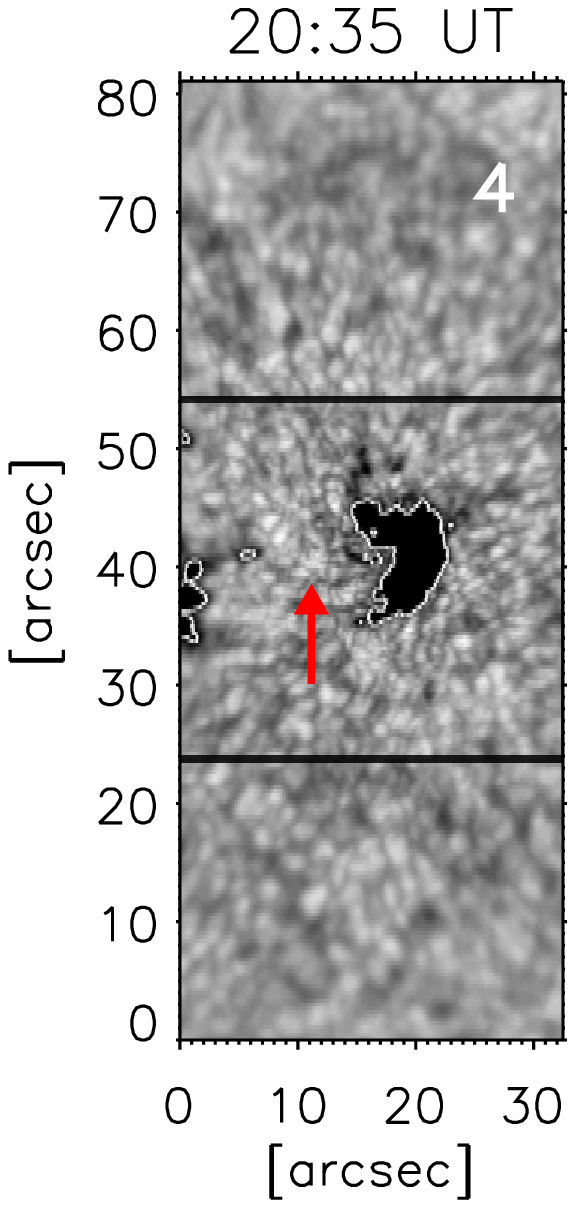}\\
\includegraphics[height=4.6cm,trim=0.7cm 1.9cm 10.6cm 0.9cm,clip=true,keepaspectratio=true]{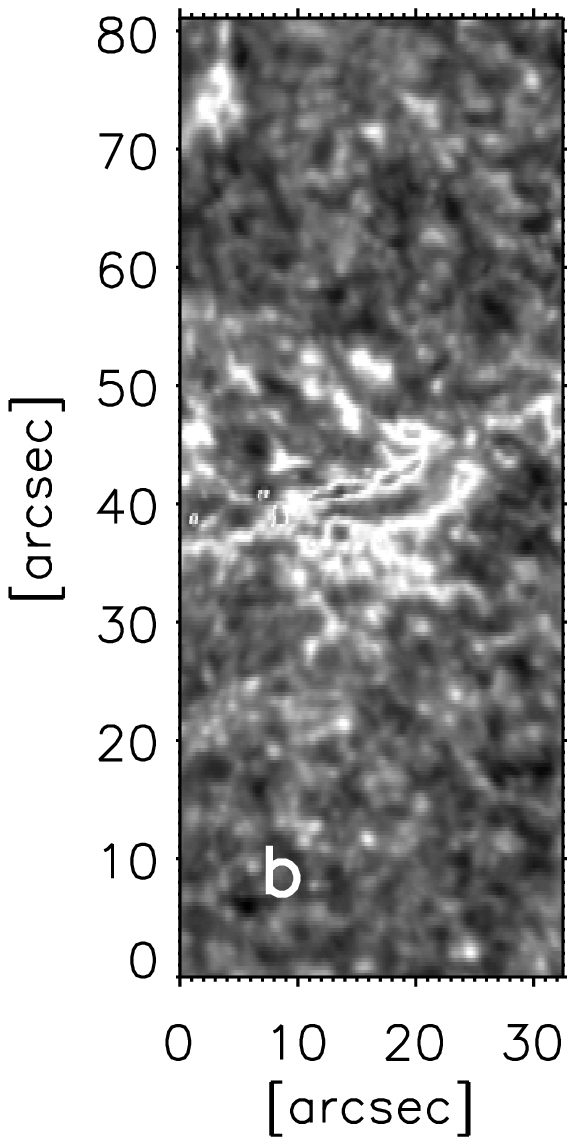}\includegraphics[height=4.6cm,trim=2.9cm 1.9cm 10.6cm 0.9cm,clip=true,keepaspectratio=true]{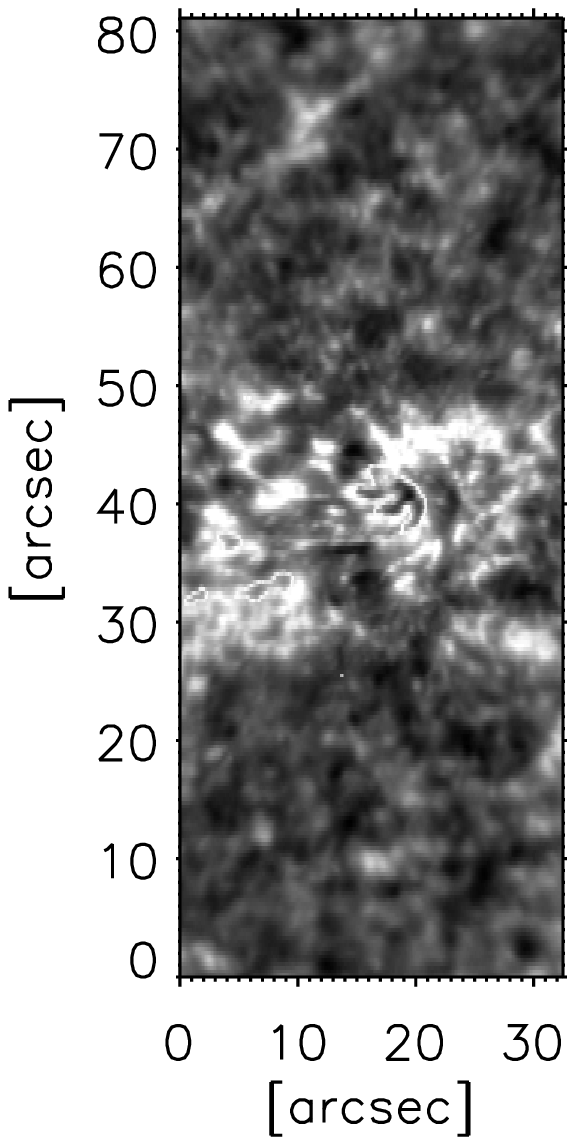}\includegraphics[height=4.6cm,trim=2.9cm 1.9cm 10.6cm 0.9cm,clip=true,keepaspectratio=true]{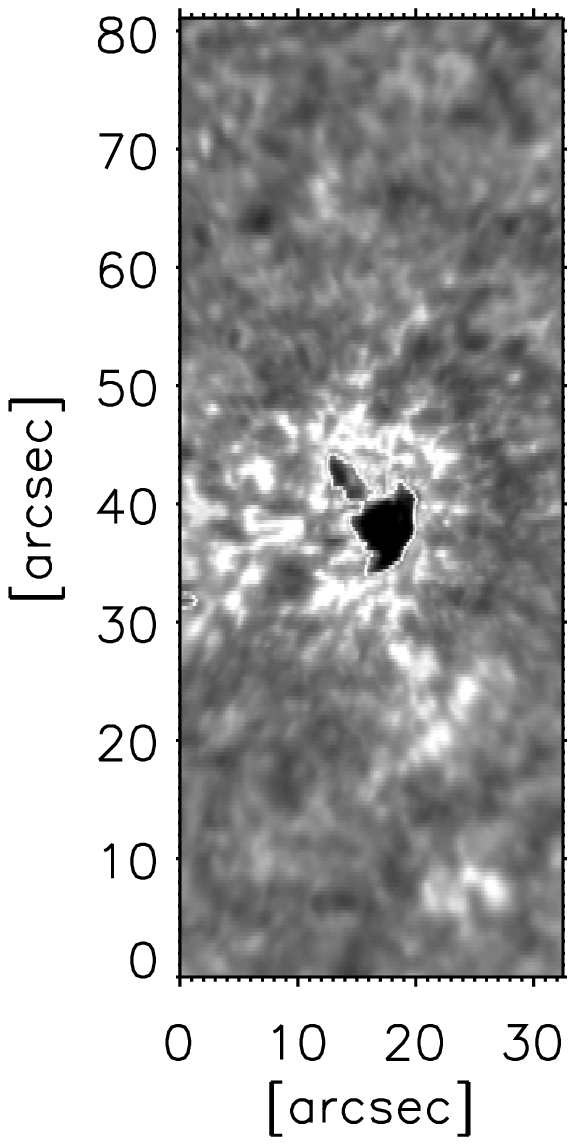}\includegraphics[height=4.6cm,trim=2.9cm 1.9cm 10.6cm 0.9cm,clip=true,keepaspectratio=true]{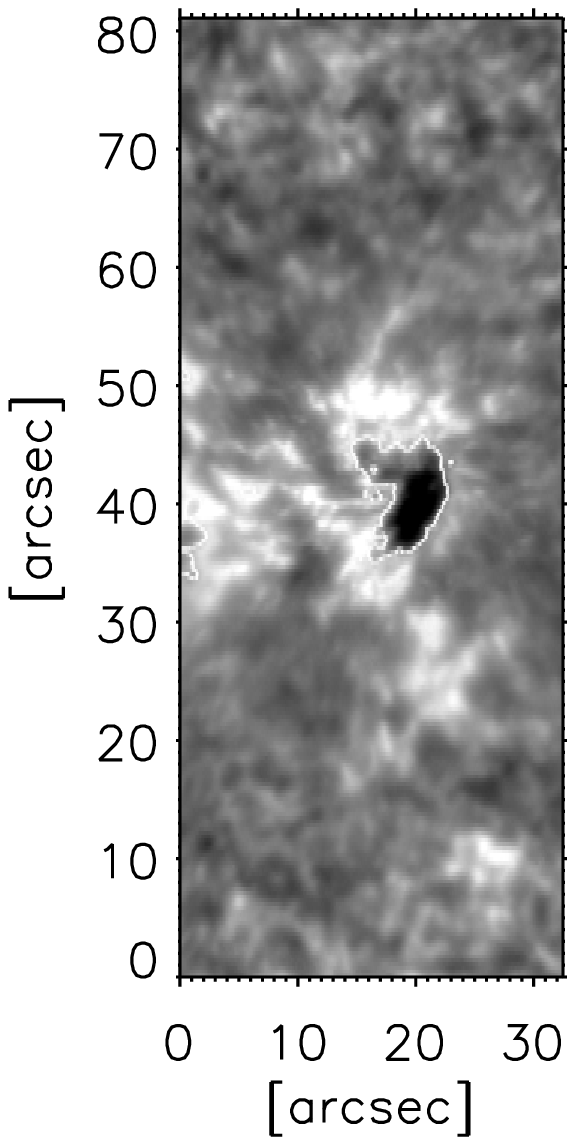}\\
\includegraphics[height=5.4cm,trim=0.7cm 0.3cm 10.6cm 0.9cm,clip=true,keepaspectratio=true]{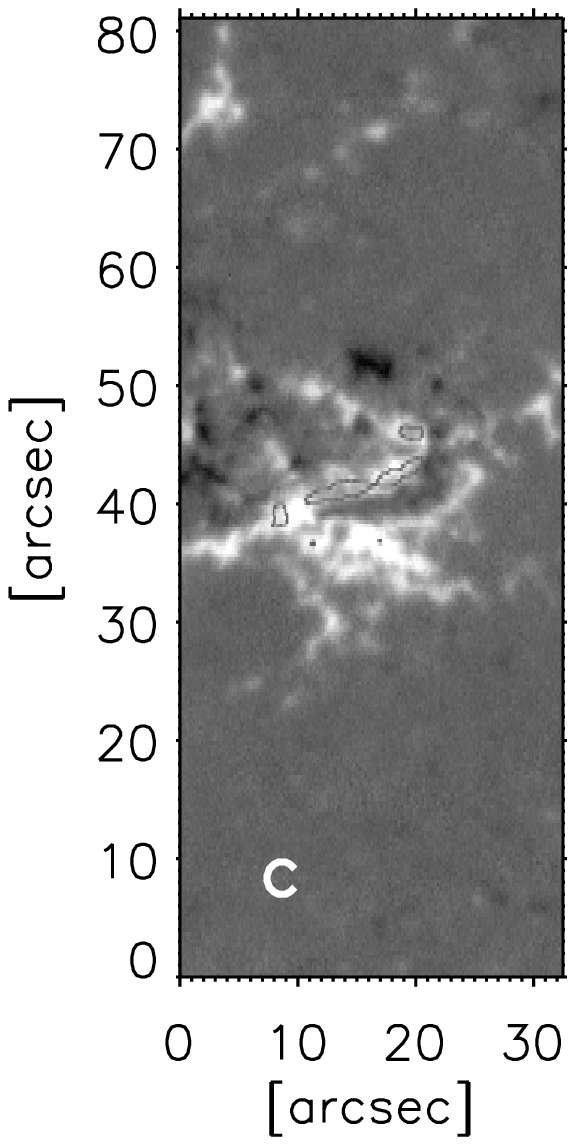}\includegraphics[height=5.4cm,trim=2.9cm 0.3cm 10.6cm 0.9cm,clip=true,keepaspectratio=true]{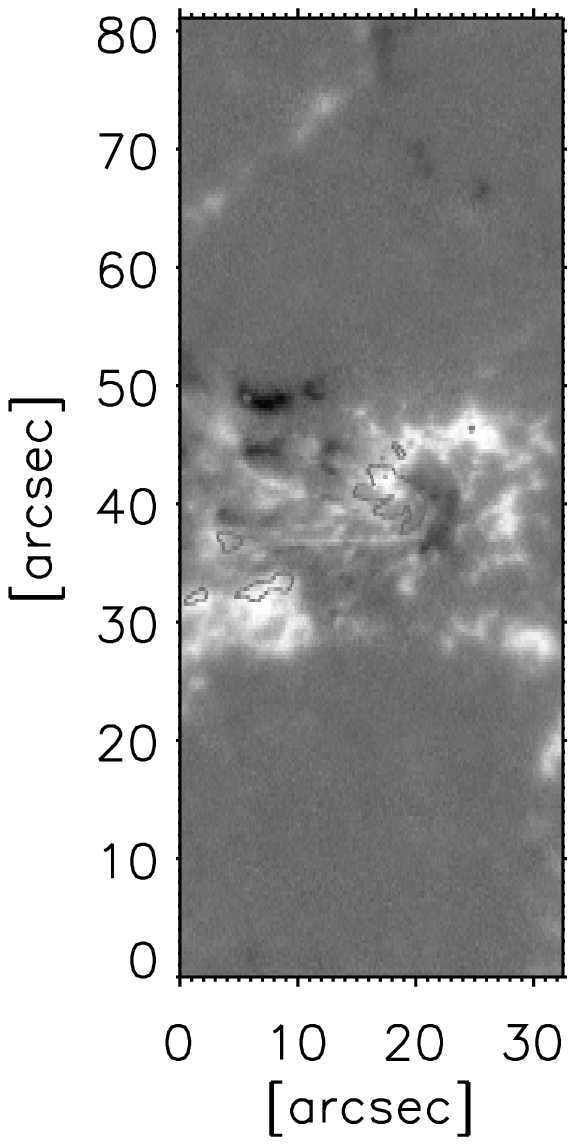}\includegraphics[height=5.4cm,trim=2.9cm 0.3cm 10.6cm 0.9cm,clip=true,keepaspectratio=true]{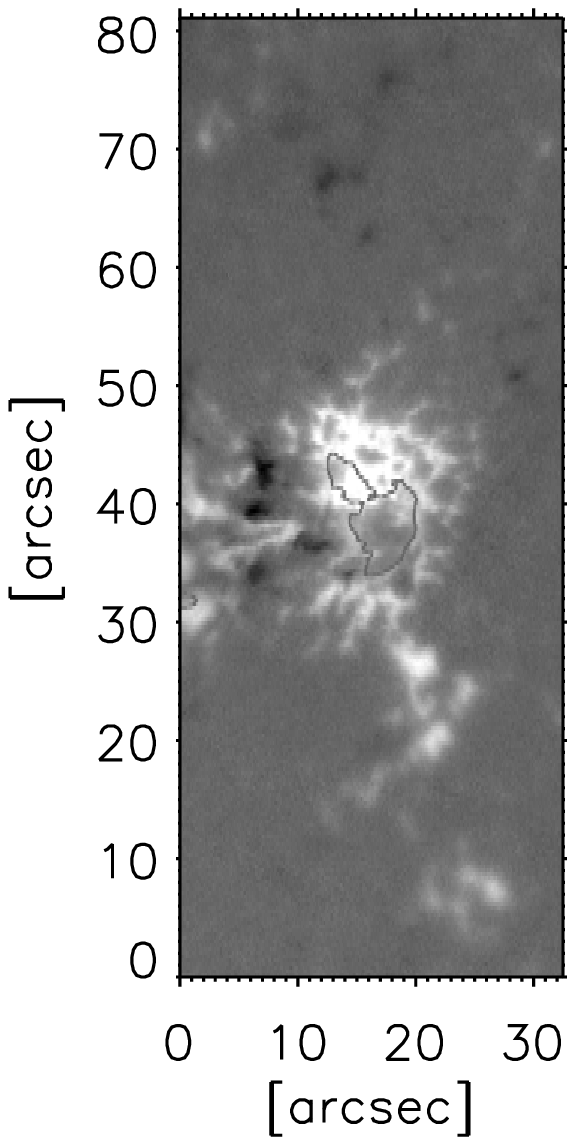}\includegraphics[height=5.4cm,trim=2.9cm 0.3cm 10.6cm 0.9cm,clip=true,keepaspectratio=true]{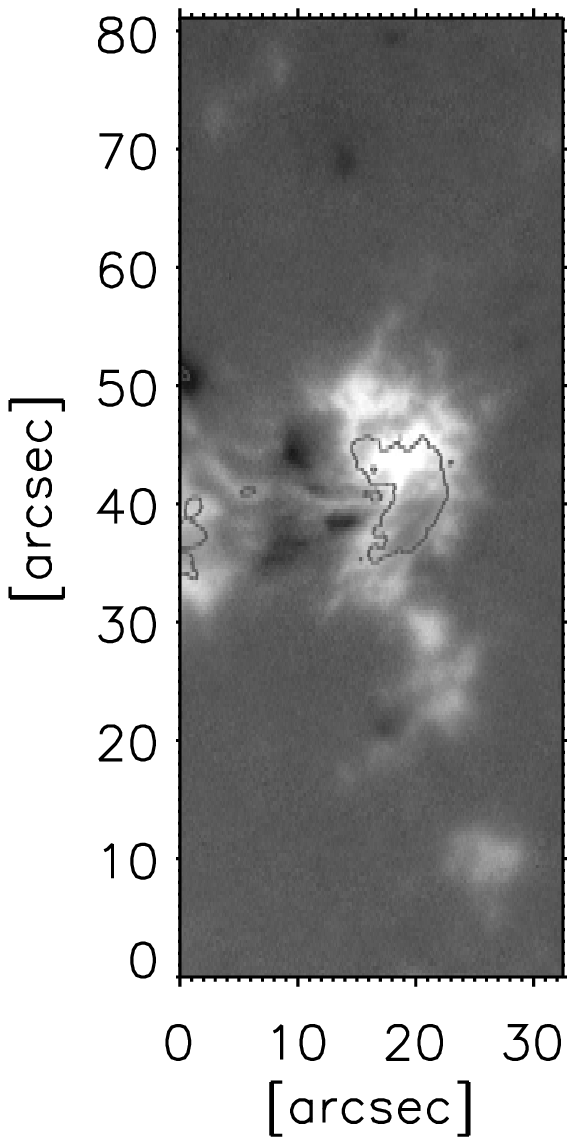}\\
}
\caption{\footnotesize{Examples of the IBIS Fe I 617.3 nm line data  analysed in our study; letters (numbers) between brackets indicate the line (column)  label. (a)  Stokes I in  the continuum near the line, (b) Stokes I in the line core, (c) Stokes V in the blue wing, near the line core, at  four  stages of pore evolution, from data taken  at (1) 13:58 UT, (2) 16:09 UT, (3)  19:45 UT, and (4)  20:35 UT. North (west) is at the top (right). The black box in the top panels  shows the subfield inverted with the SIR code and shown in Figs. \ref{f3} and \ref{f5}. The contours overplotted in each panel indicate the location of the evolving structure, as singled out by applying an intensity threshold criterion, I$_c$ <0.9 I$_{qs}$, where $I_{qs}$ is the average quiet Sun intensity. More details are provided in Sect. 3.1.}}
\label{f2}
\end{figure}

\begin{figure}%[h!]
\centering{
\includegraphics[width=8.5cm,trim=.2cm 4.5cm 0.cm 16.4cm,clip=true]{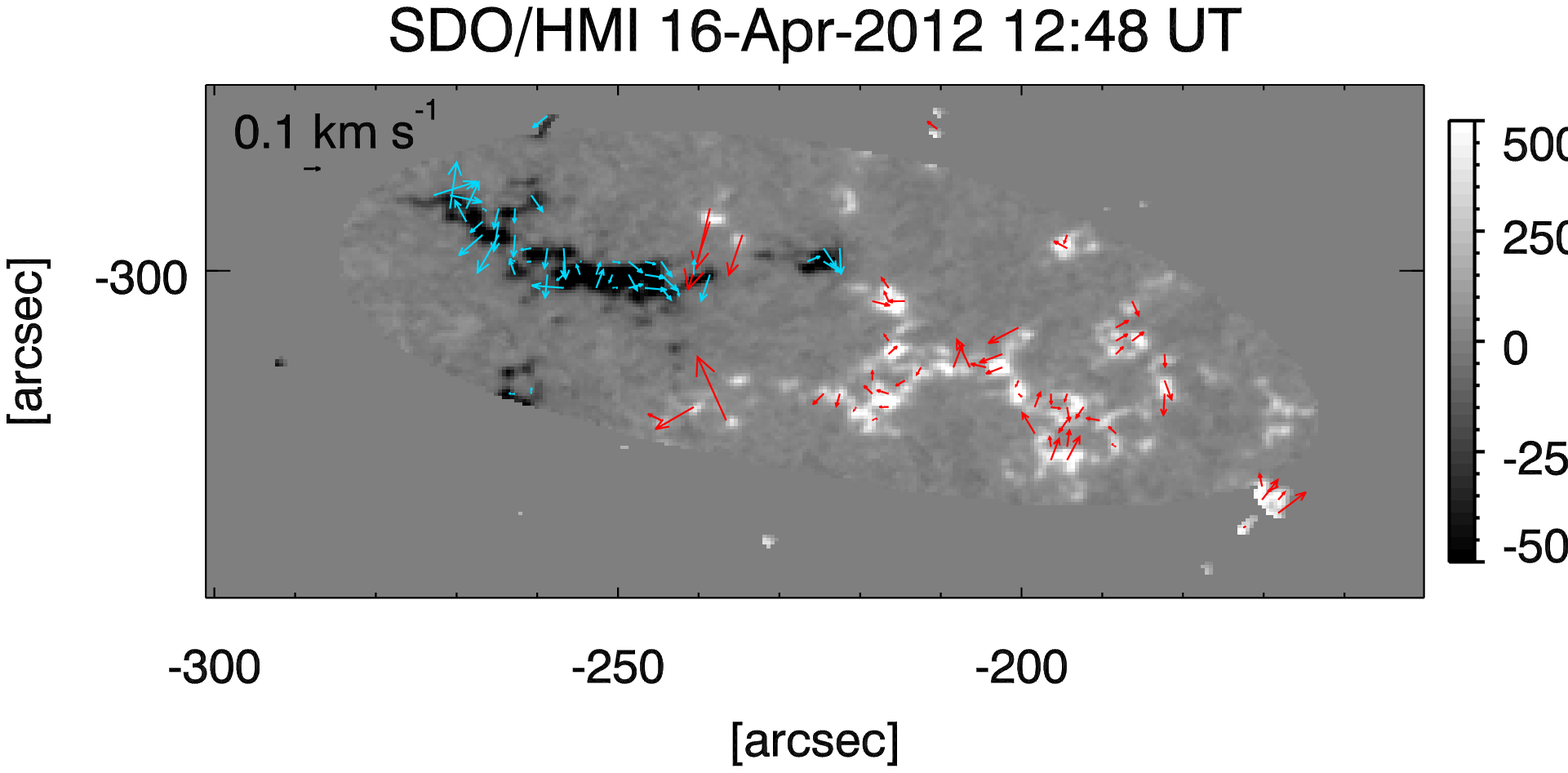}\\
\includegraphics[width=8.5cm,trim=.2cm 4.5cm 0.cm 16.4cm,clip=true]{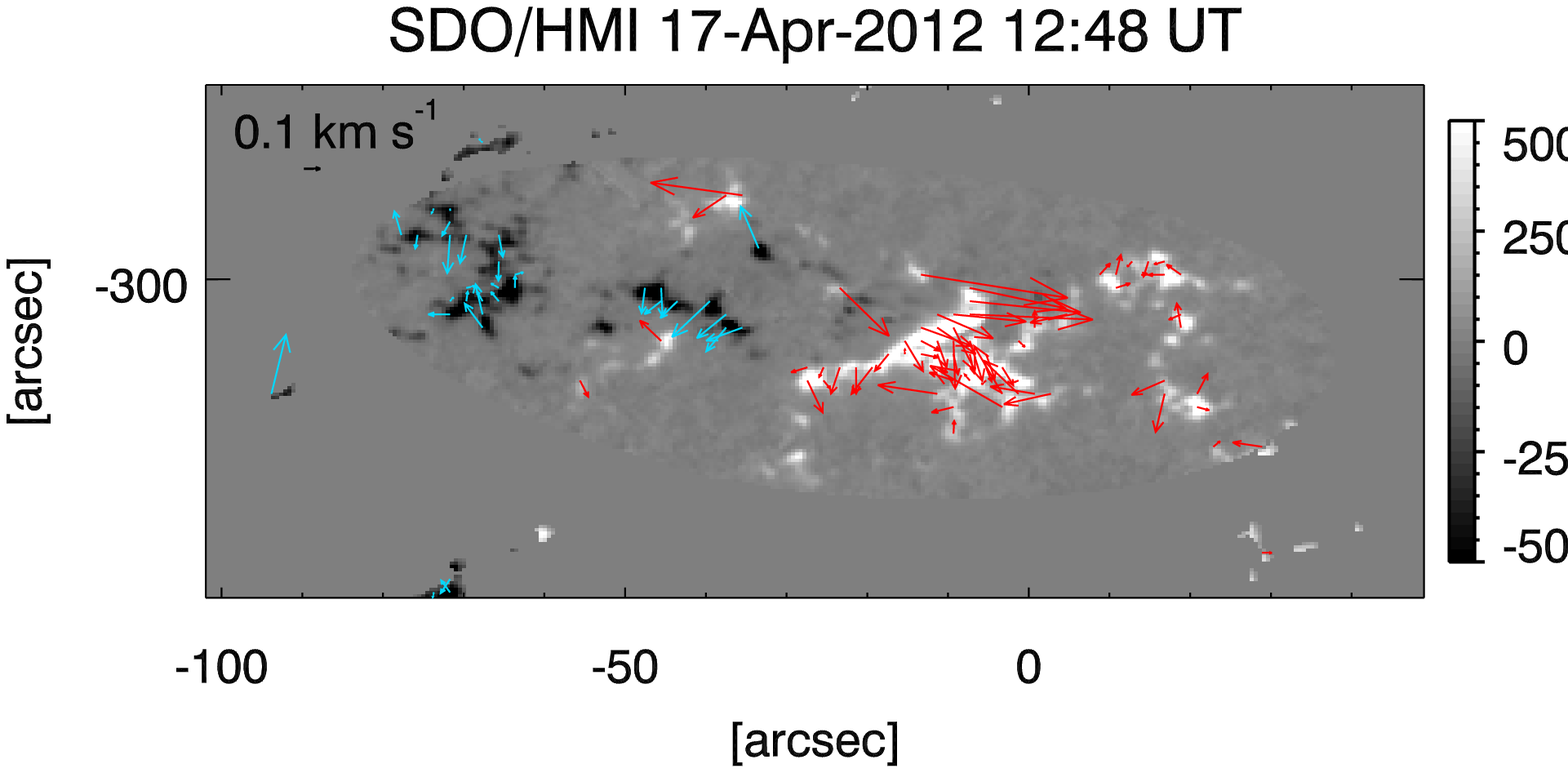}\\
\includegraphics[width=8.5cm,trim=.2cm 4.5cm 0.cm 16.4cm,clip=true]{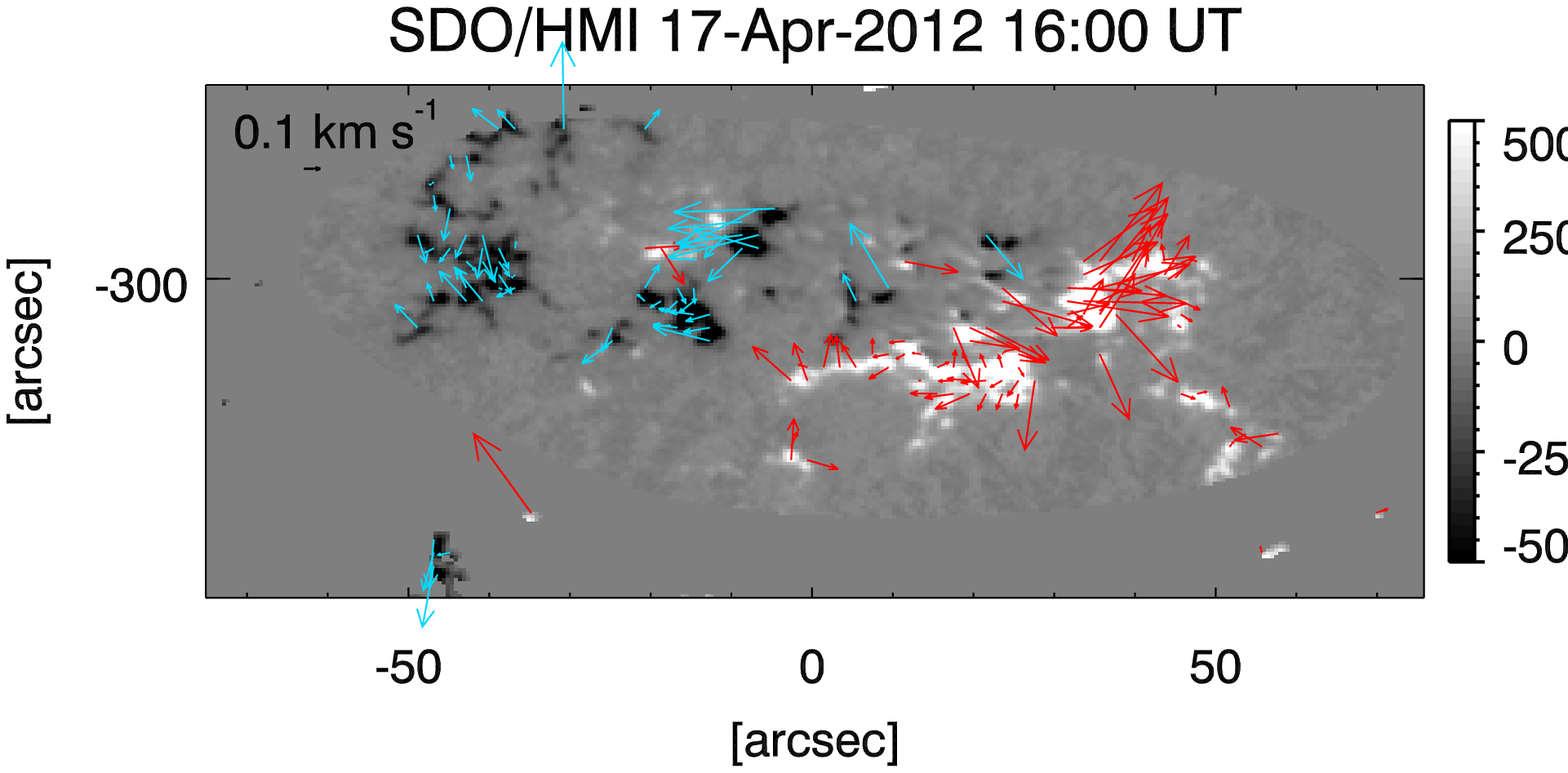}\\
\includegraphics[width=8.5cm,trim=.2cm 4.5cm 0.cm 16.4cm,clip=true]{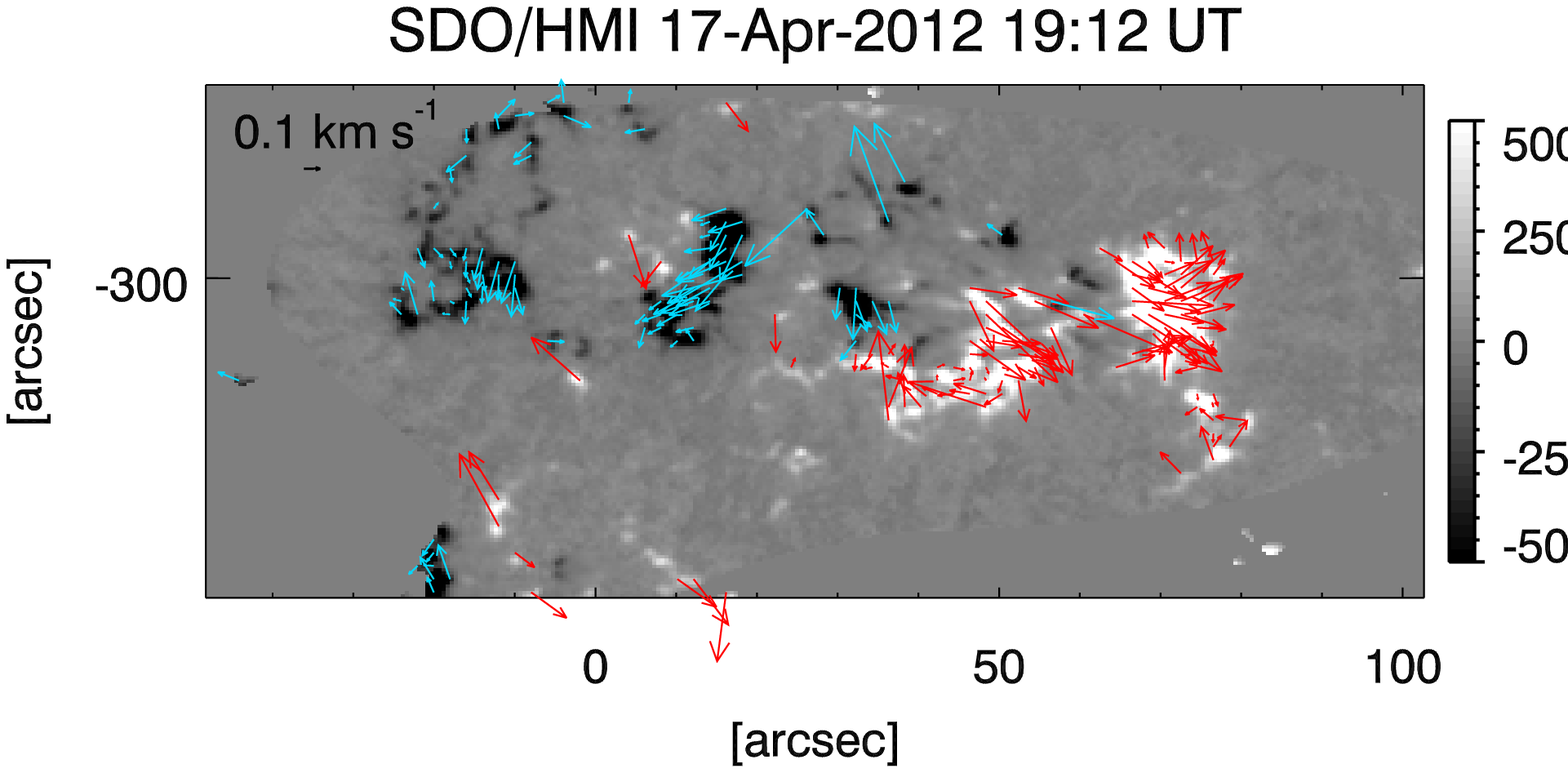}\\
\includegraphics[width=8.5cm,trim=.2cm 4.5cm 0.cm 16.4cm,clip=true]{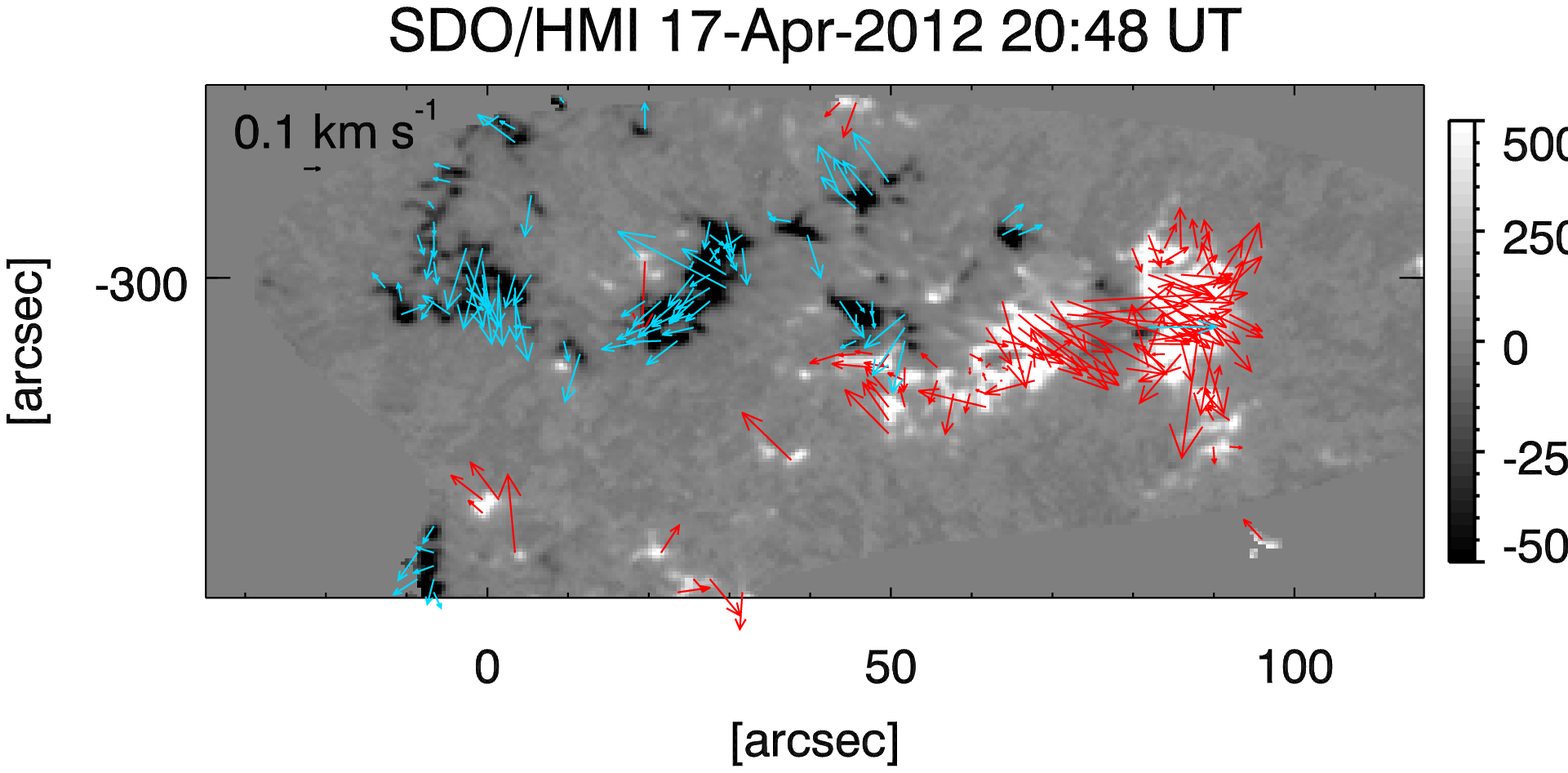}\\
\includegraphics[width=8.5cm,trim=.2cm 3.5cm 0.cm 16.4cm,clip=true]{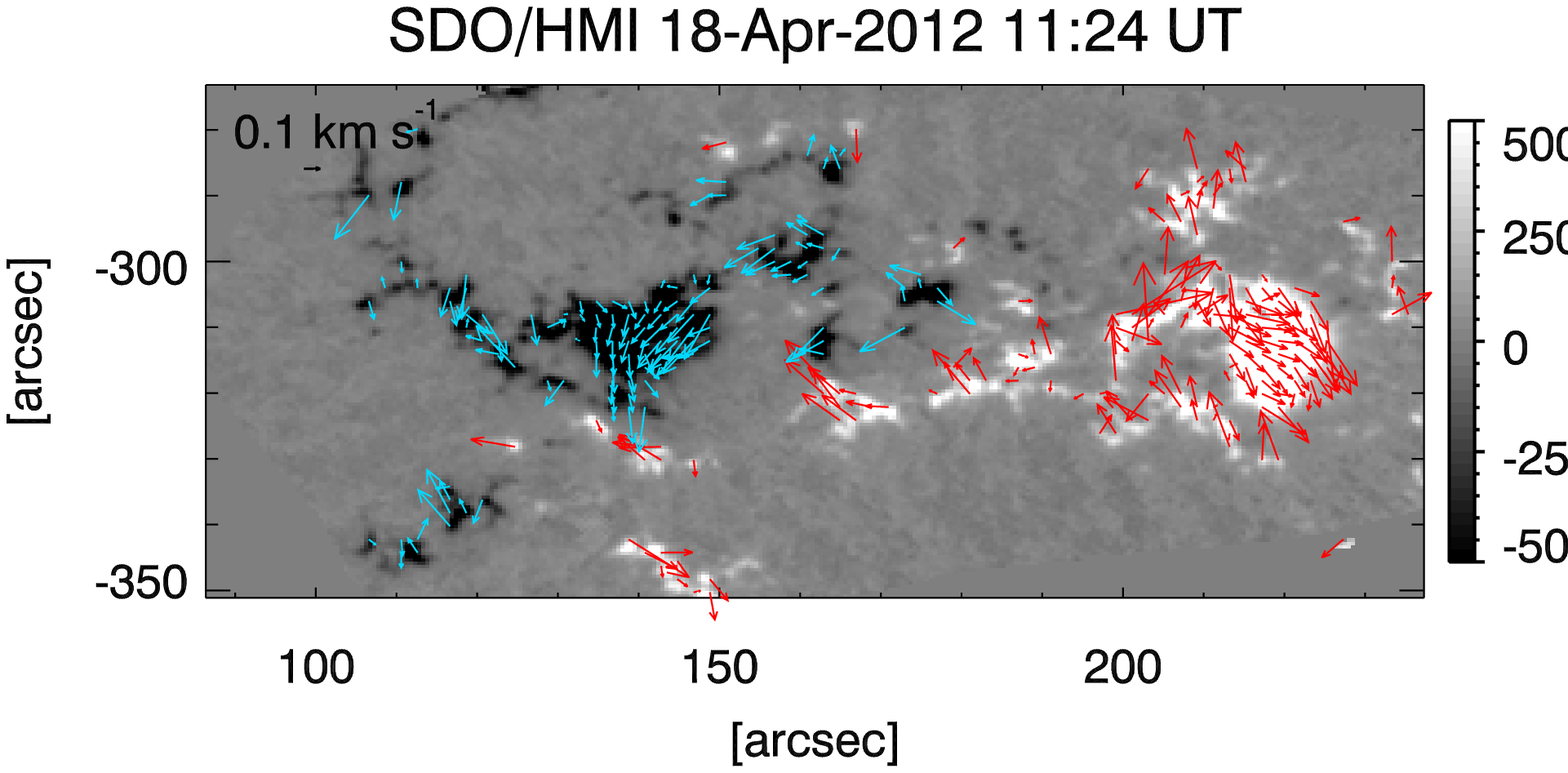}\\
} 
\caption{\footnotesize{Examples of the photospheric horizontal velocity maps derived from the SDO/HMI SHARP data  %by applying  the DAVE4VM method to observations %. From top to bottom: results derived from the 
%data 
taken from %24 hour before the start of the  pore formation (
Apr. 16, 2012, 12:48 UT  to %at the beginning (Apr. 17, 2012, 12:48 UT) and 
%during the pore formation in the time interval of the IBIS observations (Apr. 17, 2012, 16:00 UT, 19:12 UT, 
%20:48 UT), and several hour after the formation of the funnel-shaped  structure (
Apr. 18, 2012, 11:24 UT, which is from 24 hour before the start of the pore formation to several hour after. The arrows indicate the horizontal velocity; red (light blue) shows  the leading (trailing), positive (negative) polarity. In all panels but the first, the studied pore coincides with the largest size structure in the leading region.  The magnetic field in the background magnetogram is shown in the range of values specified in the colour bars. The full temporal evolution of all computed maps is shown in a movie available online.}}
\label{f5b}
\end{figure}

\section{Results}

\subsection{AR evolution and pore formation}

The AR was observed on the solar disk from April 16 to
April 22, 2012, when it reached  the western limb.
Figure \ref{f1b} shows the AR as seen in the  SDO/HMI  observations  taken at given times from April 16, 2012, 12:00 UT, to April 19, 2012, 10:48 UT.  In each panel, we show the subfield of $\approx$ 160$\times$170 arcsec$^2$, centred on the AR,  %. This sub-field is  
that was analysed   to describe the evolution of the magnetic field and radiative flux in the AR. We also show (Fig. \ref{f1b}  column 2) the two parts (sFOV) of the above subfield that were   considered   to describe the  evolution of the trailing (negative) and leading (positive) polarity regions in the AR, sFOV$_t$ and   sFOV$_l$, respectively.  The full temporal evolution of the  analysed subfield 
is shown in the movie attached to Fig.  \ref{f1b}.

 According to the 
NOAA/USAF active region summary, in the early stages of its evolution, the AR consists of seven tiny  field features organized to form a diffuse dipolar flux region,  which is seen  for example in Fig.  \ref{f1b} (lines a-b). The outermost available observations of the  AR consists  of several sunspots and pores, which are already seen in the observations taken on April 19, 2012 and shown, for example in Fig. \ref{f1b} (line h).

Figure \ref{f1}   shows the evolution of the magnetic flux (top and middle panels) and flux derivative (bottom panel) in the AR, from the SDO/HMI LOS magnetograms taken from April 16 00:00 UT to April 18 12:00 UT, 2012, over the subfield and  the two sFOVs shown in Fig. \ref{f1b}  and in the attached movie.   The  total unsigned magnetic  flux (B$_{tot}$) emerged in the AR   (Fig. \ref{f1}, top panel)  remains almost constant with values below 4$\times$10$^{21}$ Mx till  April 17, 2012, $\approx$ 12:00 UT, then it undergoes an increase.

 The flux evolution   %of the flux derivative ($dB/dt$) 
 points out an increased flux of comparable magnitude in both the sFOV$_l$ and sFOV$_t$ on April 17, 2012,   between 9:30 UT and 12:30 UT,
  before the start of the IBIS measurements  (Fig. \ref{f1}, middle and bottom panels). This flux increase, which is consistent with the flux increase by a rising-tube process, %.  and This flux increase 
  precedes the formation of a filamentary, weak S-shaped structure and occurred in sFOV$_l$ between $\approx$ 12:30 UT and 13:30 UT.   
The IBIS measurements targeted the evolution that leads the  above filamentary  structure, of positive polarity flux, to form  a pore, on April 17, 2012 from 13:58 UT to 20:43 UT. 
Figure \ref{f1b} (line c) shows the FOV of the IBIS data, whose examples are given  in Fig. \ref{f2}.

{\bf 
The time interval of the IBIS measurements is characterized by a steep increase of flux in the AR lasting until April 17, 2012, $\approx$ 17:00 UT (Fig. \ref{f1}, middle and bottom panels)   and small flux changes afterwards. %{\bf In addition, we note that the amplitude of flux variations increases with time.} 
Available observations show that on April 17, 2012 between 15:15 UT and 15:30 UT the right end of the initial S-shaped structure breaks away at roughly one-third of the length of the structure. After the break, the right end of the structure evolves to form the pore, which  at some first stages resembles a tiny 
U-shaped  (e.g. SDO/HMI observations at 15:36 UT), then a trident-shaped structure (e.g. 16:09 UT,  IBIS data in Fig. \ref{f2}, column 2). In the early stages of its evolution, the S-shaped   structure  is aligned to about 10 degree to the east-west direction (Fig. \ref{f2}, column 1).   %
The IBIS  observations  show  elongated granules near  the evolving structure, mostly located east of this structure in agreement with, for instance \citet[][]{Centeno_2012} and \citet[][]{Verma_etal2016}; these granules   are oriented almost perpendicularly to the axis of the evolving feature at the early stages of its evolution and parallel to it later on (see e.g. the IBIS observations in Fig. \ref{f2} line a, granules indicated by the red bar in each panel).
%, likely reflecting the field orientation below the solar surface. 
%The evolution from the initial  filamentary structure to the pore occurs rapidly between $\approx$ 15:15 UT and $\approx$ 16:30 UT.

The  formation of the positive polarity pore is accompanied by the emergence of small-scale magnetic features of opposite polarity. Indeed, 
since  the 
 early stages of the pore formation,  i.e. between 10:00 UT and 12:00 UT and more extensively after 14:00 UT, the SDO/HMI observations  show %small-scale  ($\leq$ 10 arcsec)  flux features  Eastward of sFOV$_{t}$, on the trailing (negative polarity) region of the AR,  at the time the  pore formation starts on the leading (positive polarity) flux area.  {\bf 
%At the early stages of the pore formation,  this 
a tiny magnetic patch of negative polarity flux  that emerges   north-east of the evolving, positive flux  structure, at  about 100 degree to the east-west direction  (see e.g. the movie attached to Fig.  \ref{f1b}).  This diffuse negative polarity patch mostly  lies outside the FOV of the IBIS observations. It nearly  preserves  the same size during the pore formation, but similar sized negative flux features  appear south of it  (e.g.  the SDO/HMI observations at 16:12 UT in the movie attached to Fig.  \ref{f1b}). The negative polarity features show clear links to the evolving, positive polarity region. This suggests that they  belong to the same emerging flux loop from which  the AR evolution  could be ensued. The negative patches are 
  clearly seen on the available 
data at the time the pore has already increased significantly in size. %, but they can be found  in the other available data. 
Near the evolving positive flux structure, the SDO/HMI magnetograms also show small-scale,  mixed polarity  features that counterstream;  as seen in the movie attached to Fig.  \ref{f1b},  the negative flux moves  towards the sFOV$_{t}$, while the positive polarity flux moves towards the 
sFOV$_{l}$.  % (Fig. \ref{f1b} bottom panel and Fig. \ref{f1b2} first and second  panels). 
This arrangement of opposite polarity flux resembles  the footpoints of an emerging loop. 
The SDO/HMI magnetograms show the growth of flux regions by coalescence of smaller scale, same polarity features. This process  also shows up as short-lived, small-scale light features observed in the photospheric available observations, in both the SDO/HMI and IBIS data, near the eastward side (leftmost)  of the evolving structure.
}

Both the SDO/HMI and IBIS  observations reveal a counterclockwise rotation of the growing pore, which is clearly seen as a swirling motion of the plasma immediately outside the western (rightmost) border of the evolving structure.  
%, by advection of the surrounding plasma motions.  
  Figure \ref{f5b}   shows  some maps of the horizontal motions  derived from the SDO/HMI data;   the full evolution of all the computed maps is shown in the movie attached to Fig.  \ref{f5b}. In these maps,  the arrows were plotted to show persistent motions (> 12 min) at moderate resolution (> 2 arcsec).  Figure \ref{f5b}  and the attached movie show the above-mentioned rotation %plasma rotation inferred from  the visual inspection of  the observationsdata. In  Fig. \ref{f5b} we present the maps obtained  from the
 %SDO/HMI data, which allow us to depict the plasma motions 
 on larger size regions of both polarities in the  AR with  velocities up to  about 1 km/s (see e.g. the maps at 16:00 UT and 19:12 UT).  
In addition, they also  %to the plasma rotation clearly seen in the leading region of the AR that hosts the pore formation, these maps 
display an outwards motion of the formed pore,  with respect to the primary flux patch, and a drift of the negative polarity flux  in the opposite direction  (see e.g. the maps at 19:12 UT and 20:48 UT).  As seen on the movies available online, all together these plasma motions seem to foster the accumulation of flux.  
%During the pore formation, the  negative polarity patch of the AR  shows a counterclock rotation.}
Figure  \ref{f5b}  also shows that,
after the pore formation, the %leading area of the AR shows ordered proper  
average  velocity of the horizontal plasma motions at the evolving region reduces to about  0.3-0.4 km/s 
(Fig. \ref{f5b}, bottom panel).

Throughout the duration of the IBIS observations, 
the coherent structure resulting from the evolution of the  S-shaped feature  displays  a fragmented, changing core in the IBIS 617.3 nm data and lack of  penumbra around its  entire perimeter. 
 The formed pore shows a rather symmetric funnel-shaped structure (Fig. \ref{f2}, column 4) that lasts about  9  hours.  During the same time interval,  the field aggregation in the trailing (negative) polarity region   forms in a smaller scale   (< 10 arcsec)   and more diffuse features.  
 While the 
 pore is formed  in the leading  region in less than  1 hour,  the  more diffuse and smaller scale magnetic structures on the trailing  region show a more gradual increase in flux over a time interval of 
 20-30 hours.

%%%%%QUI2

 % and that the flux emergence occurs in the central part of the pre-exiting flux system,  in the form of small-scale dipoles.  
 On April 18-22, 2012,   the  total  magnetic  flux estimated in the evolving region slightly increases by remaining almost constant for about 24 hour, and then it increases up to  values close to 9$\times$10$^{21}$ Mx  (not shown in Fig. \ref{f1}). This second flux increase   follows the formation of  a partial penumbra around the formed pore (Fig. \ref{f1b}, line g). This flux 
leads, in about 10 hours, to the  final configuration of the region, which  consists  of a spot in the leading polarity region of the AR, two smaller   spots in the same area, and several pores in the trailing  flux region  of the AR  (Fig. \ref{f1b}, line h).  The leading spot shows a partial penumbra  on the side opposite from the following polarity of the AR,  as reported, for example by \citet[][]{Schlichenmaier_etal2010}. The steepest change of the total flux during the analysed time interval occurs during the  formation of coherent magnetic structures eastward of the mature leading spot in the leading polarity region.

% The detailed evolution of the AR in the SDO/HMI continuum intensity images and LOS magnetograms, and the evolution of the horizontal plasma motions derived from the SDO/HMI SHARP data {\bf are shown in the movie available online}.
 %,  can be seen in  the supplementary 
%movies 1-3. 

%%%%%%%%%QUIQUIQUIQ

\begin{figure*}%[ht!]
%  \centering{\hspace {1cm} 1                       \hspace{2.5cm}                                     2            \hspace{2.5cm}                                                 3     \hspace{2.5cm}                        4  \hspace{2.5cm}                        5 \hspace{2.5cm}                        6     \hspace {1cm}    }\\
%
\centering{
\hspace{.7cm}
\includegraphics[height=0.68cm,trim=3.5cm 9.75cm 5.8cm 1.cm,clip=true,keepaspectratio=true]{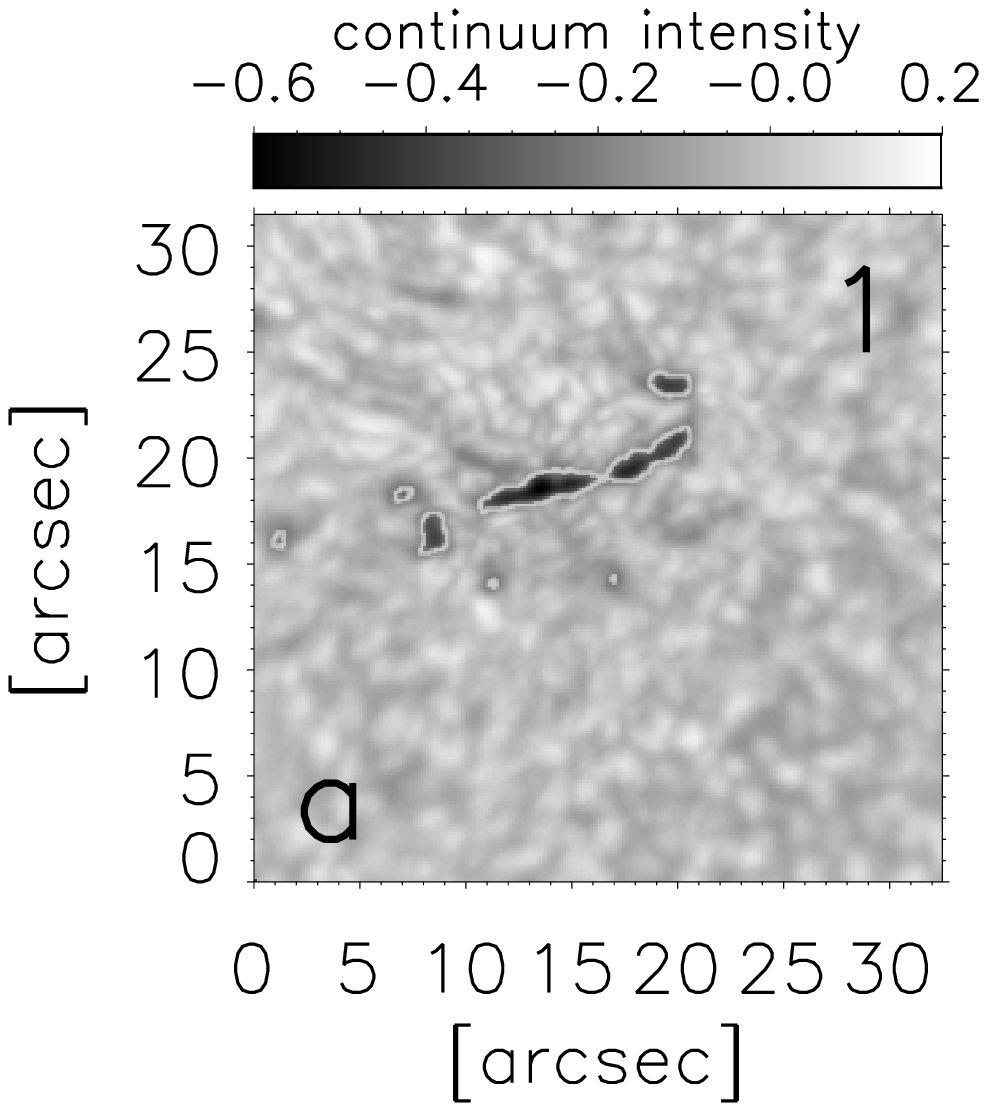}  %5.8
\includegraphics[height=0.68cm,trim=3.5cm 9.75cm 5.8cm 1.cm,clip=true,keepaspectratio=true]{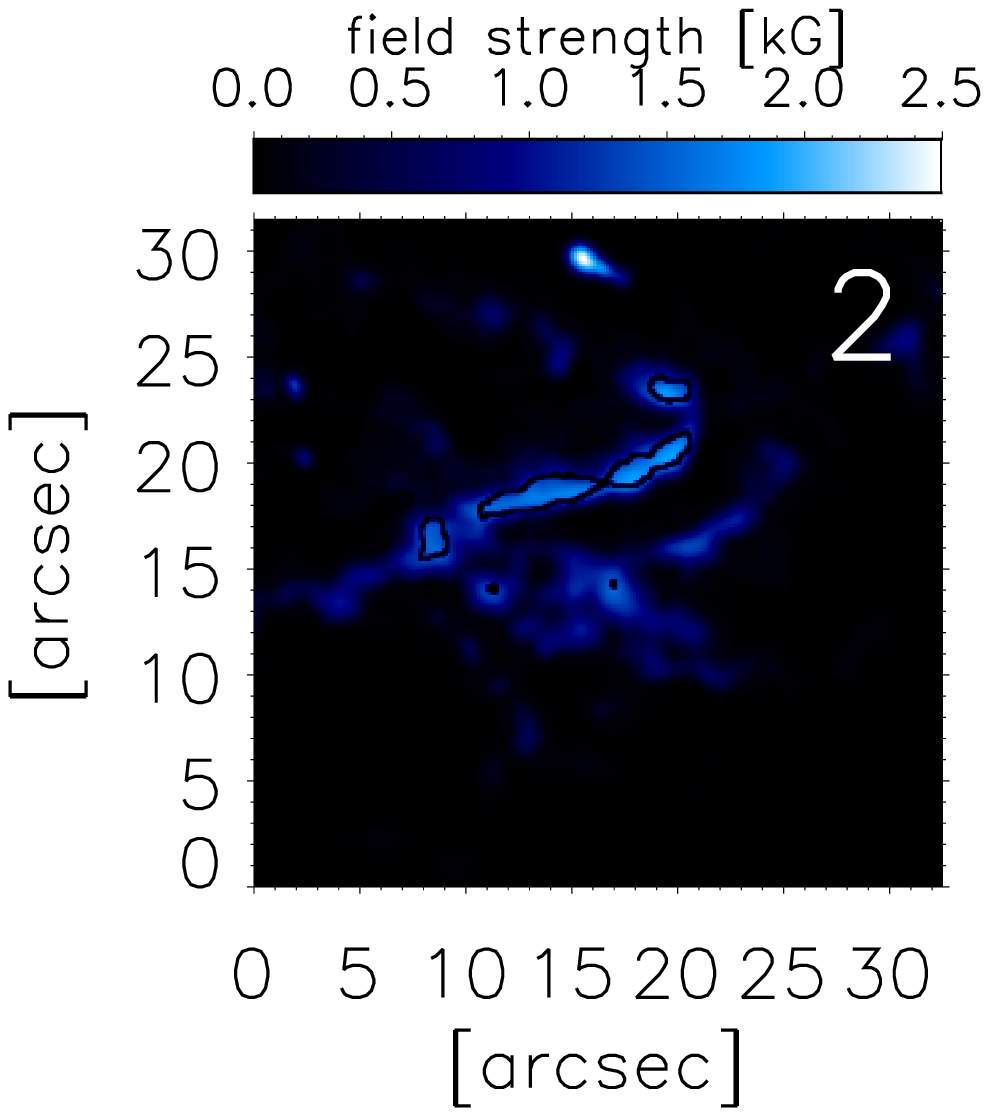}
\includegraphics[height=0.68cm,trim=3.5cm 9.75cm 5.8cm 1.cm,clip=true,keepaspectratio=true]{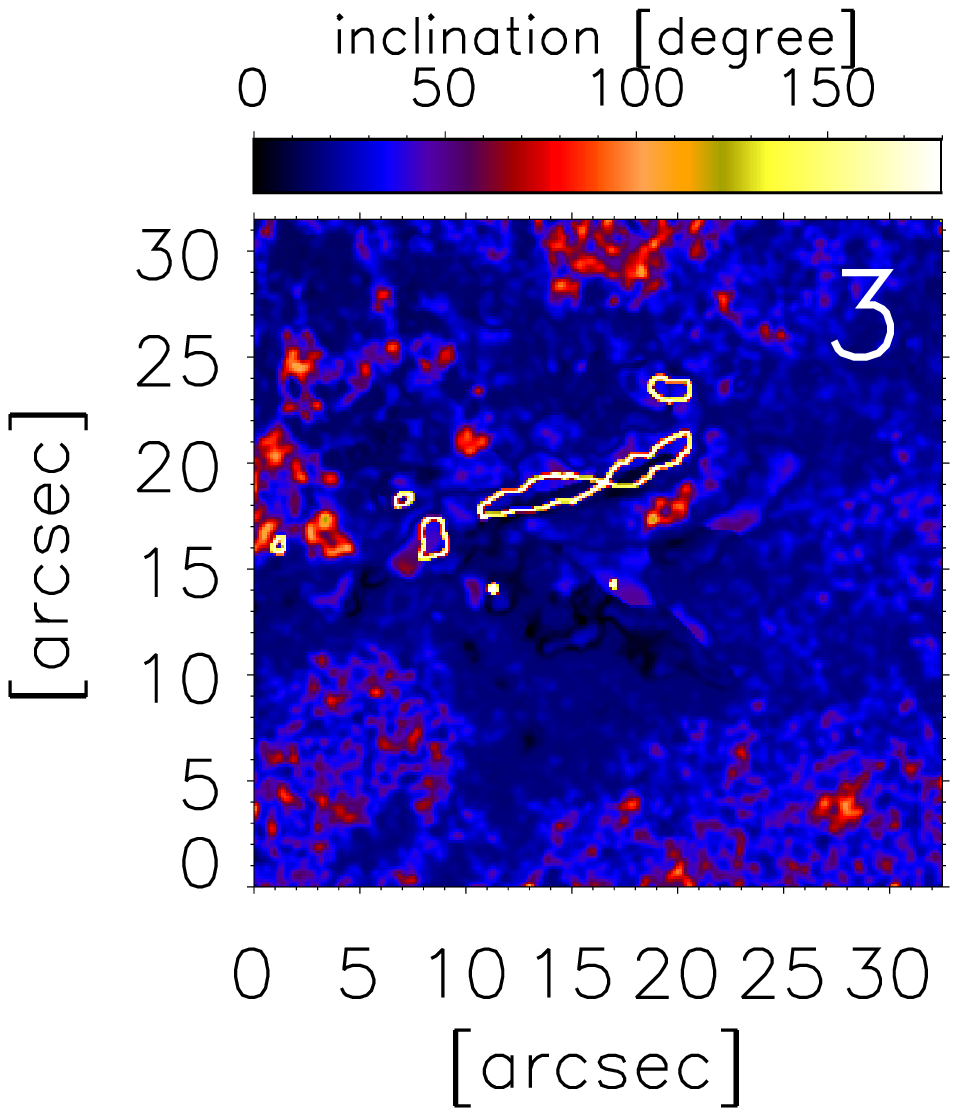}
\includegraphics[height=0.68cm,trim=3.5cm 9.75cm 5.8cm 1.cm,clip=true,keepaspectratio=true]{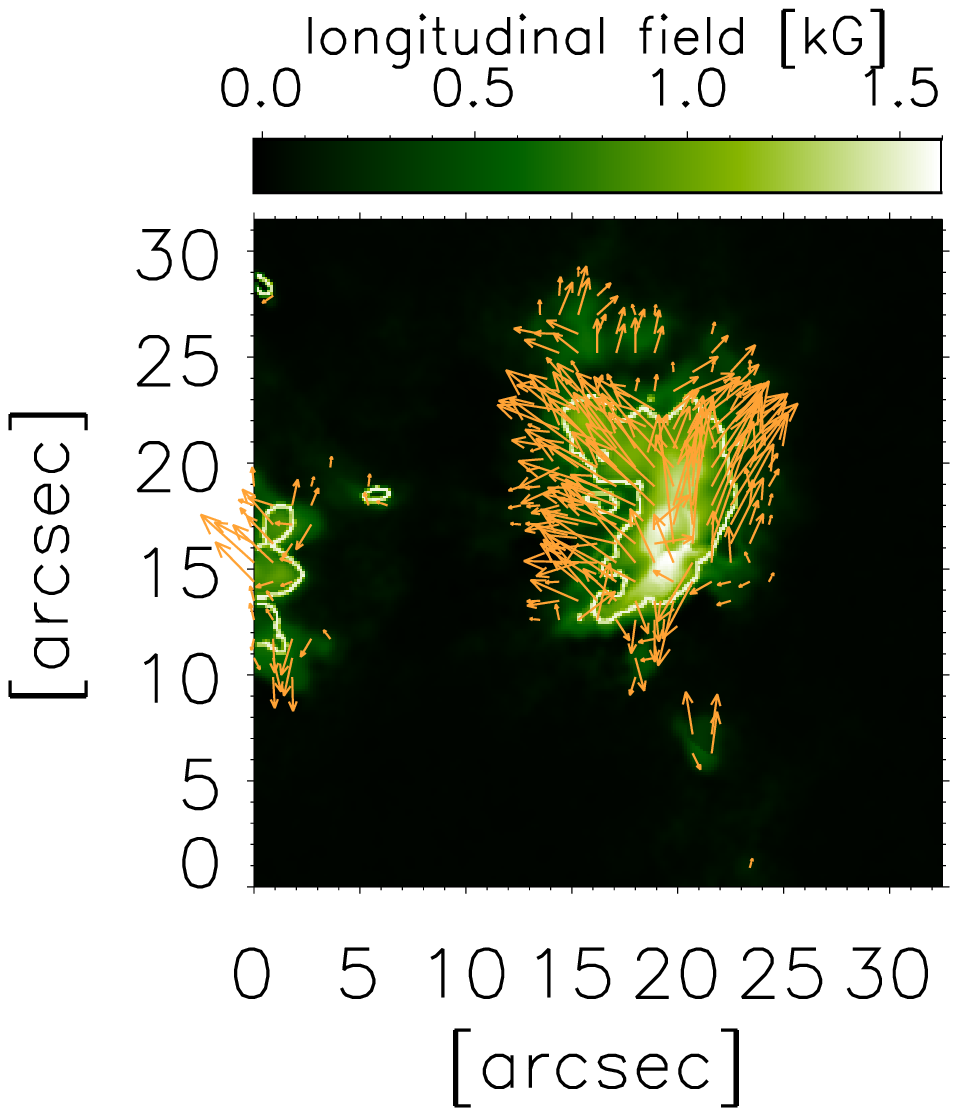}
\hspace{15.9cm}
}\\
\centering{
\includegraphics[height=3cm,trim=1.5cm 2.3cm 4.35cm 1.cm,clip=true,keepaspectratio=true]{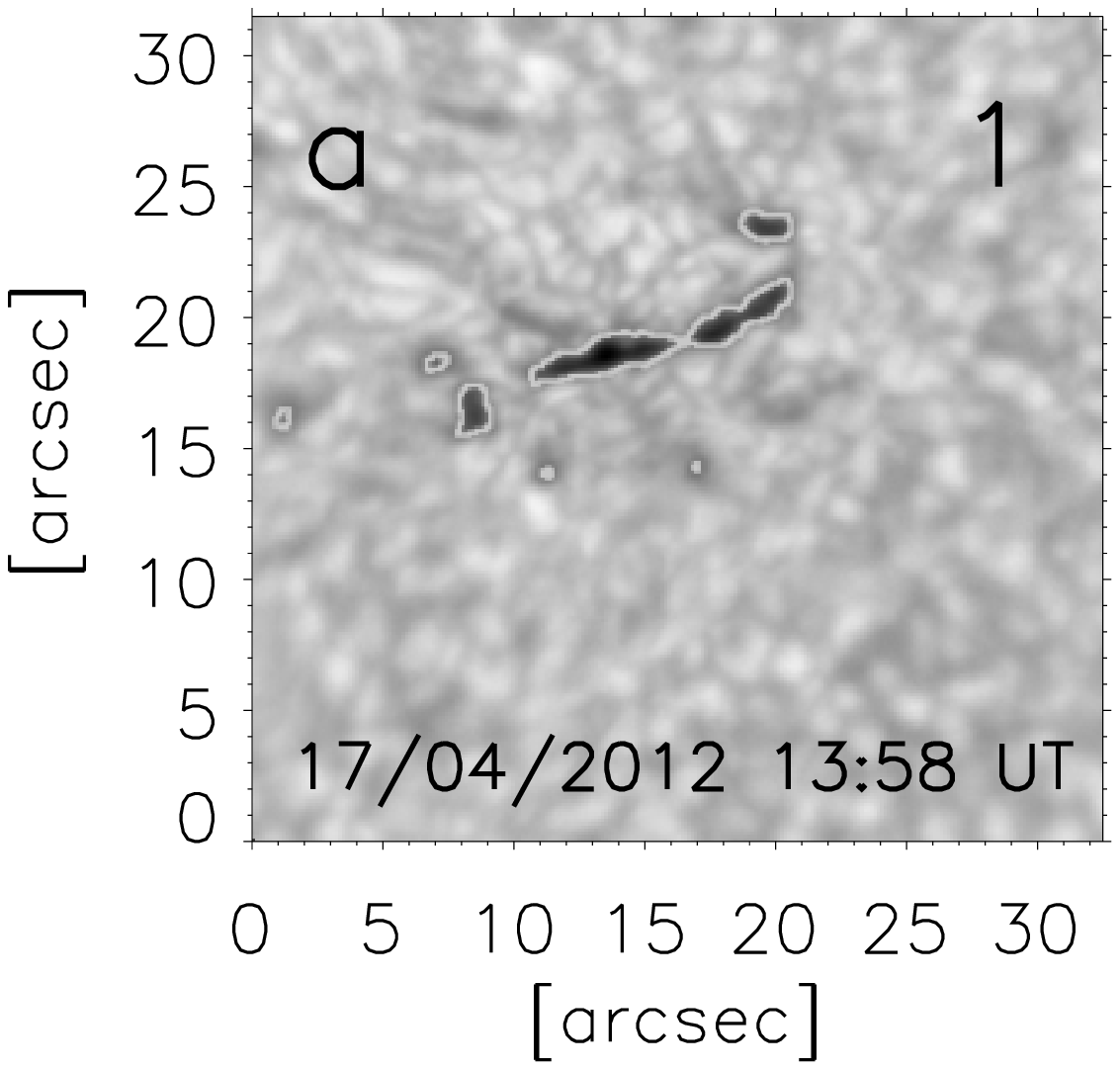}\includegraphics[height=3cm,trim=4.2cm 2.3cm 4.35cm 1.cm,clip=true,keepaspectratio=true]{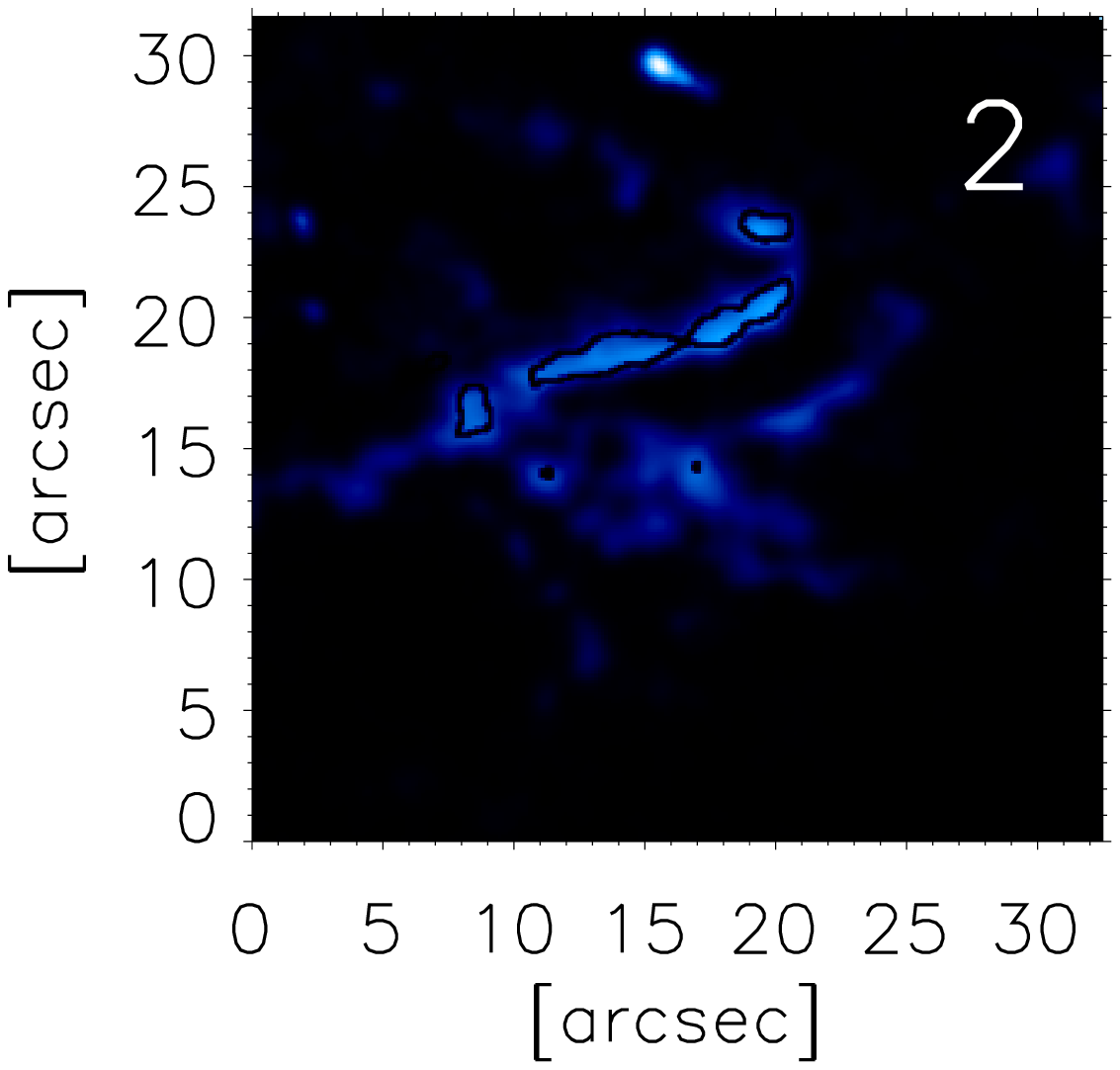}\includegraphics[height=3cm,trim=4.2cm 2.3cm 4.35cm 1.cm,clip=true,keepaspectratio=true]{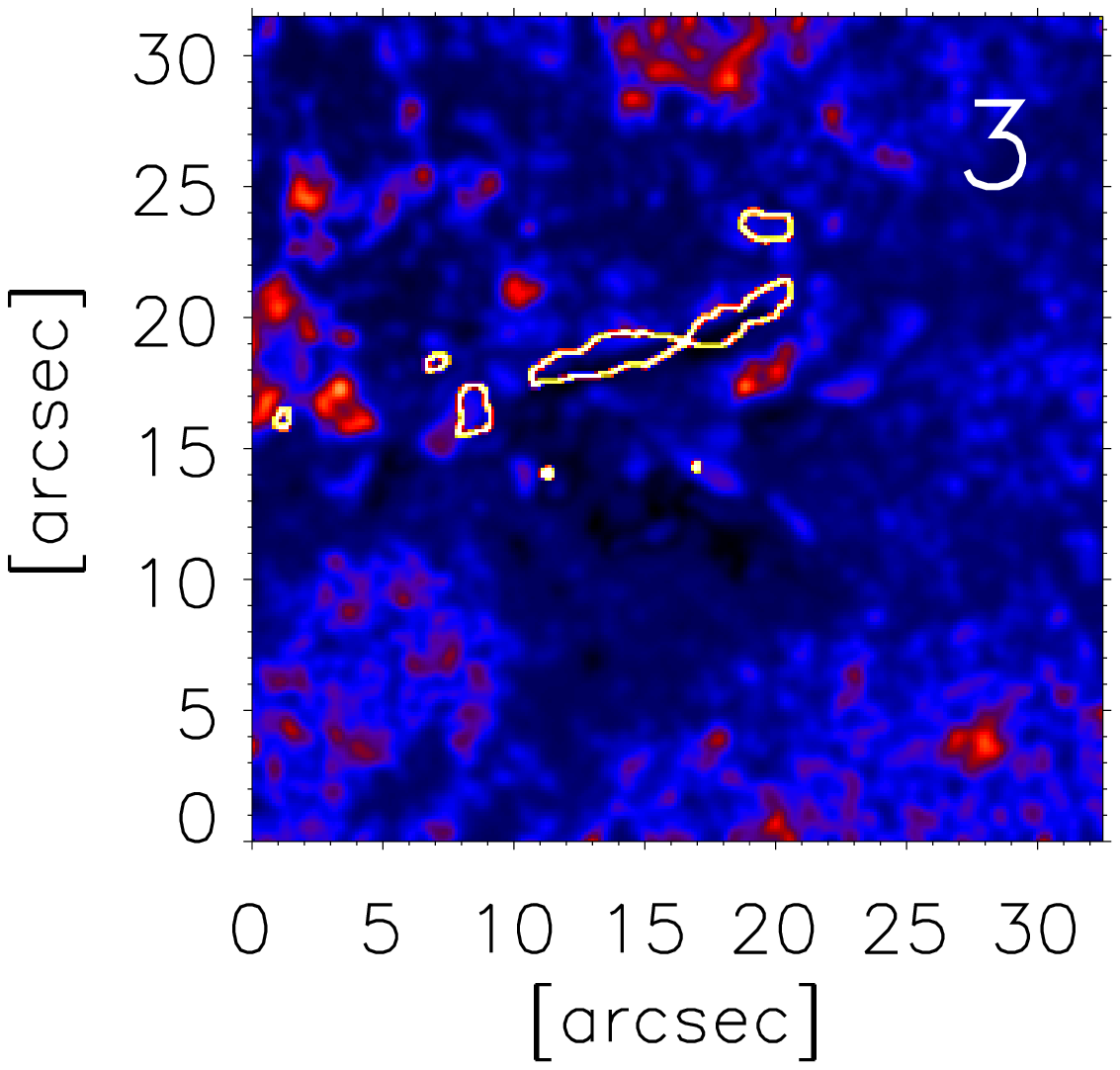}\includegraphics[height=3cm,trim=4.2cm 2.3cm 4.35cm 1.cm,clip=true,keepaspectratio=true]{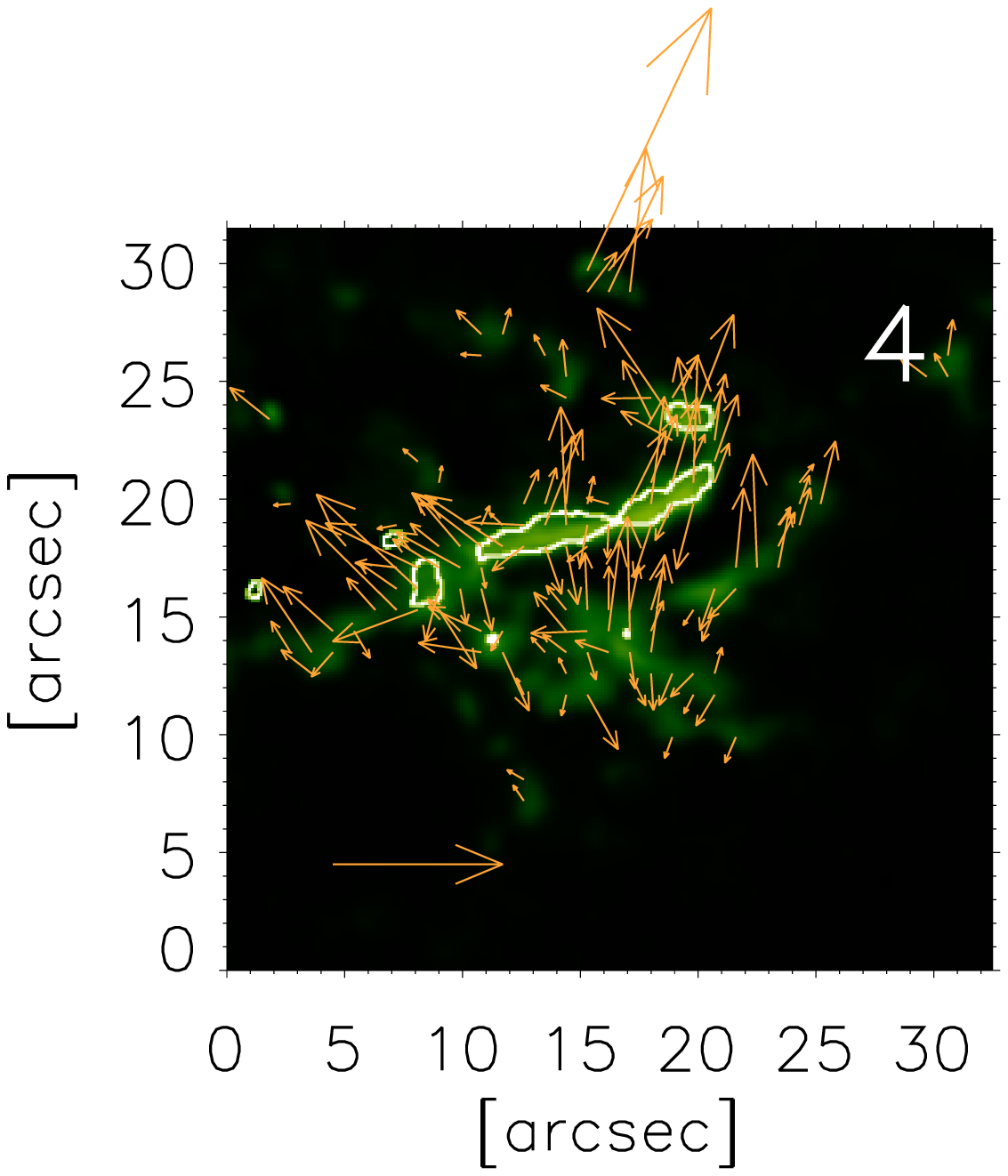}\\
\includegraphics[height=3cm,trim=1.5cm 2.3cm 4.35cm 1.cm,clip=true,keepaspectratio=true]{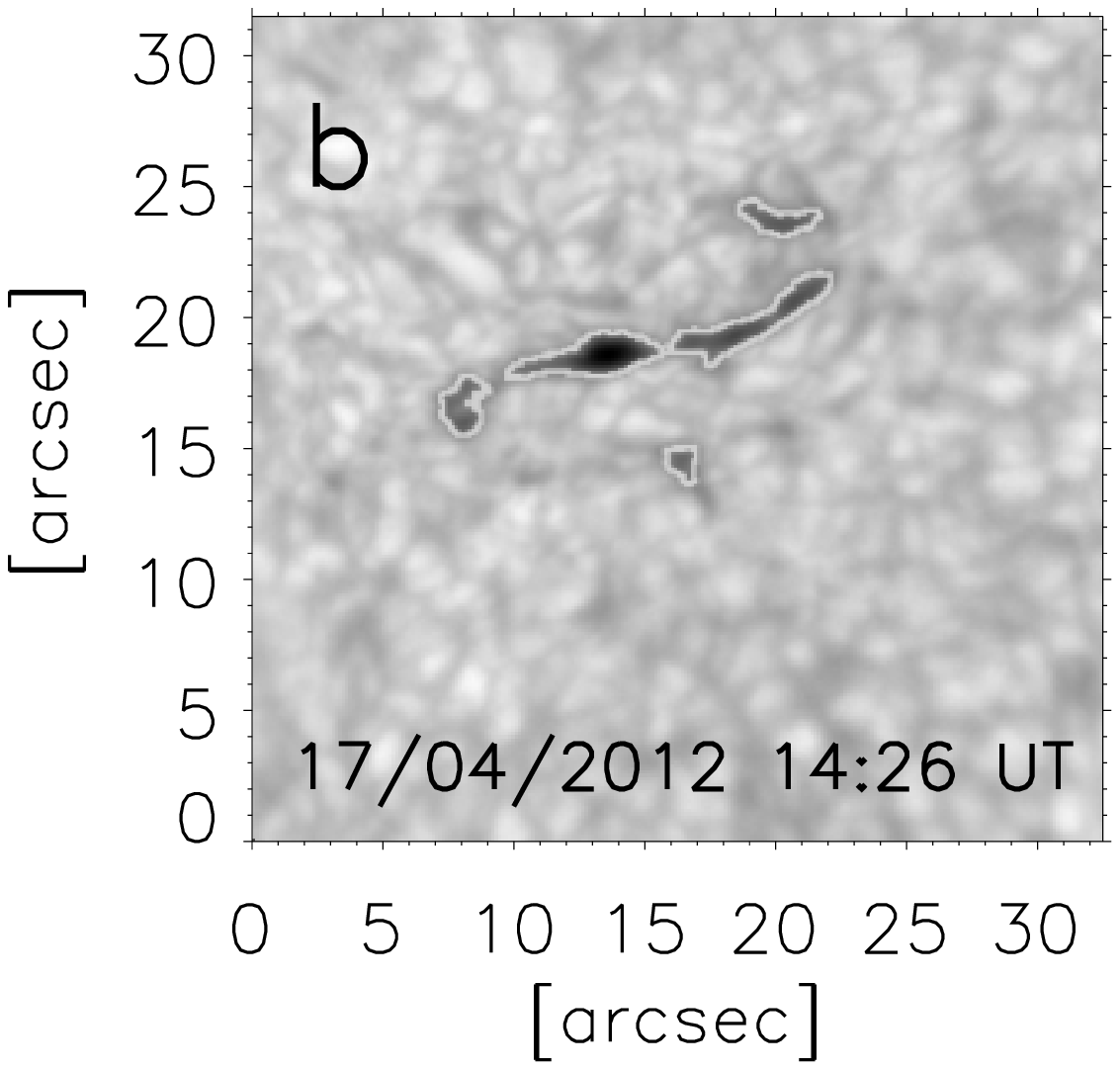}\includegraphics[height=3cm,trim=4.2cm 2.3cm 4.35cm 1.cm,clip=true,keepaspectratio=true]{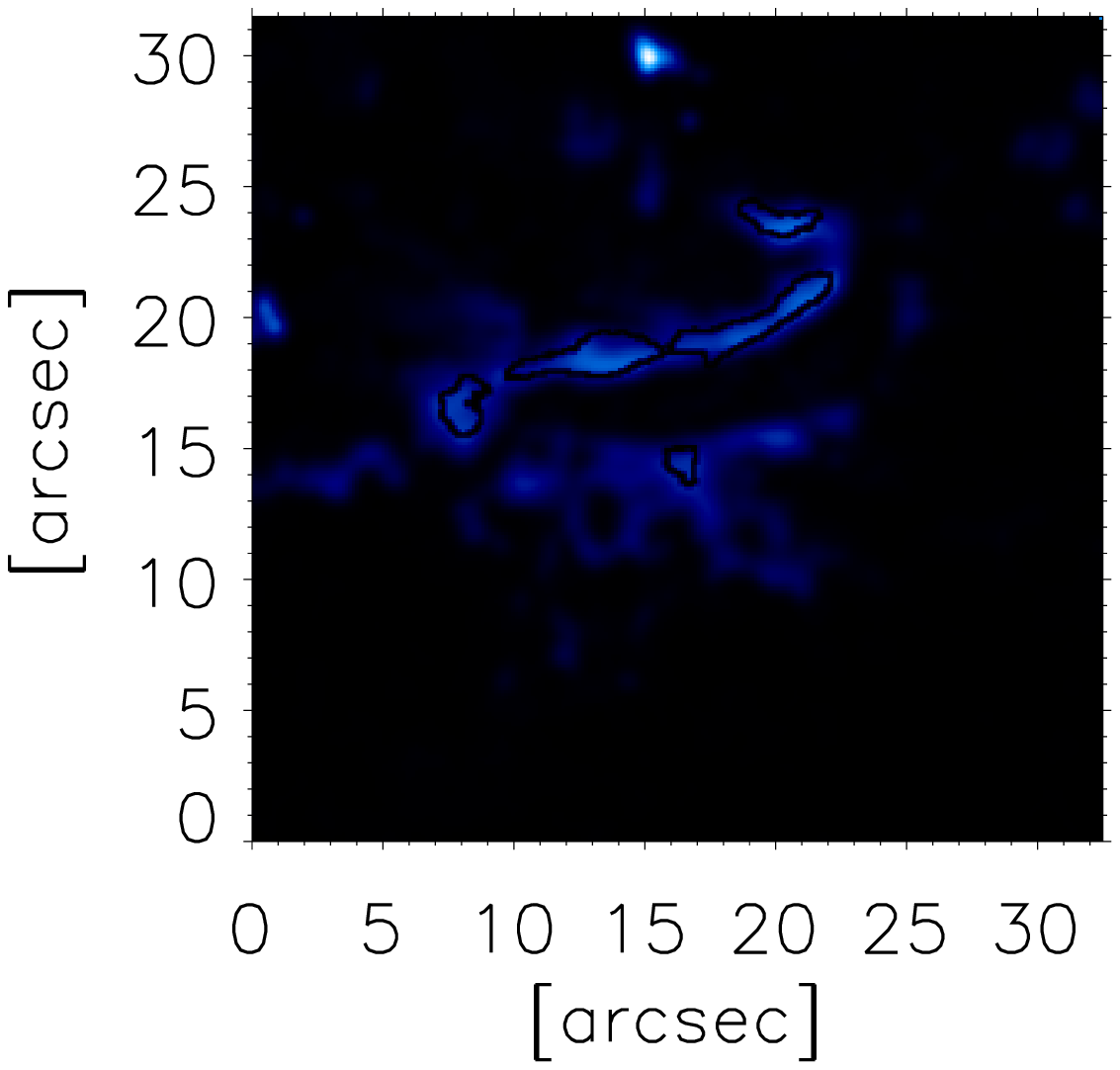}\includegraphics[height=3cm,trim=4.2cm 2.3cm 4.35cm 1.cm,clip=true,keepaspectratio=true]{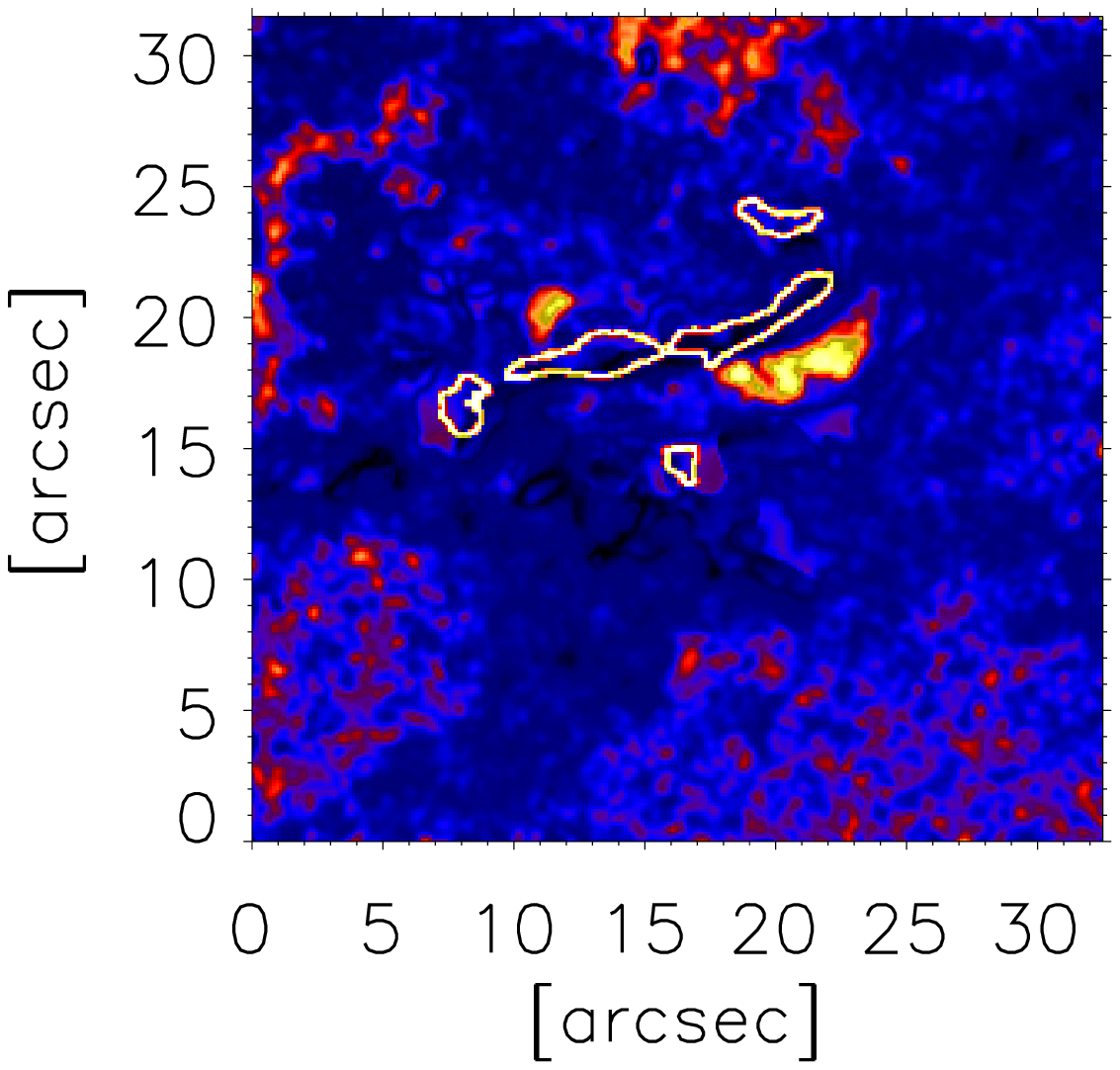}\includegraphics[height=3cm,trim=4.2cm 2.3cm 4.35cm 1.cm,clip=true,keepaspectratio=true]{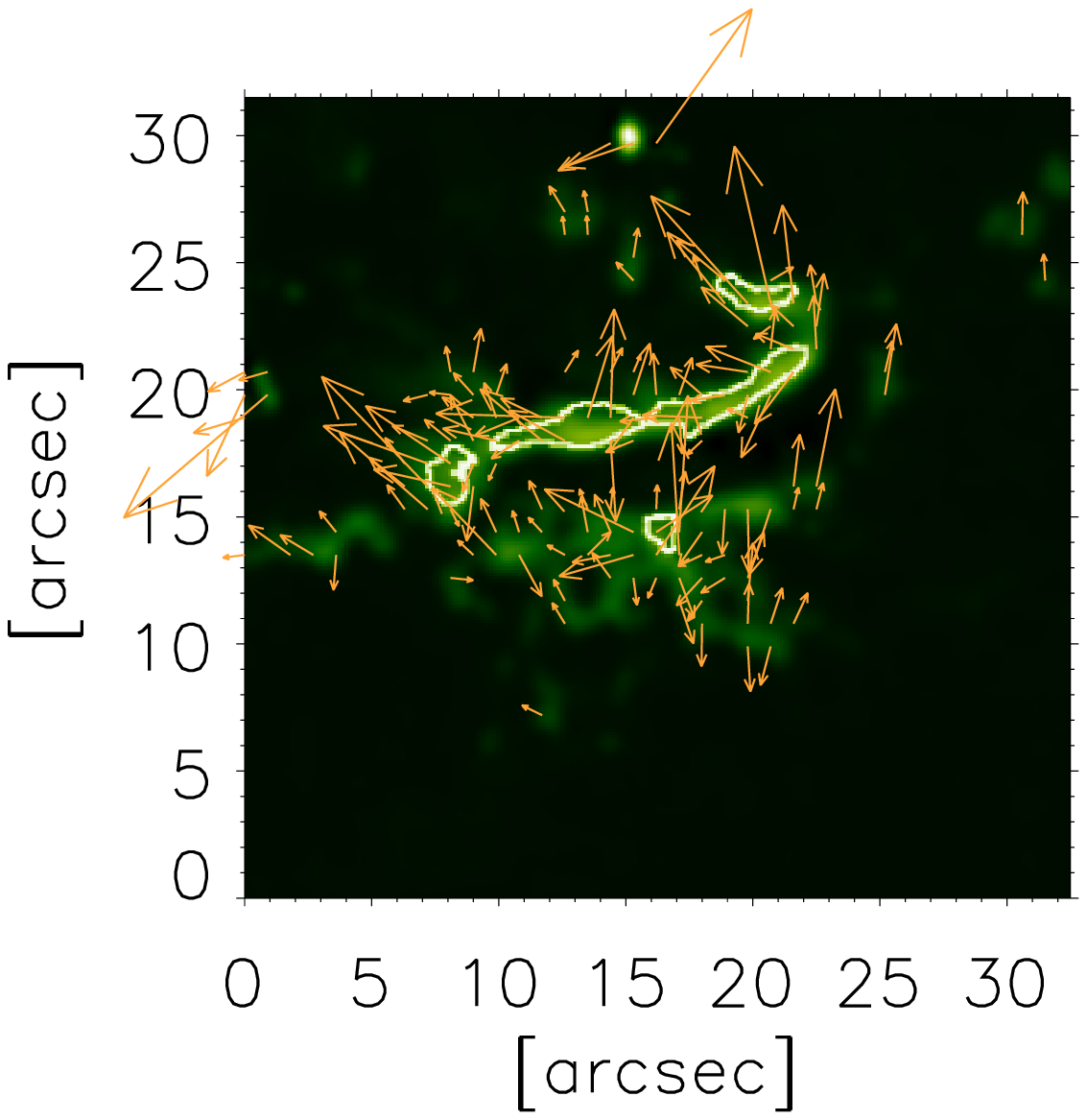}\\
\includegraphics[height=3cm,trim=1.5cm 2.3cm 4.35cm 1.cm,clip=true,keepaspectratio=true]{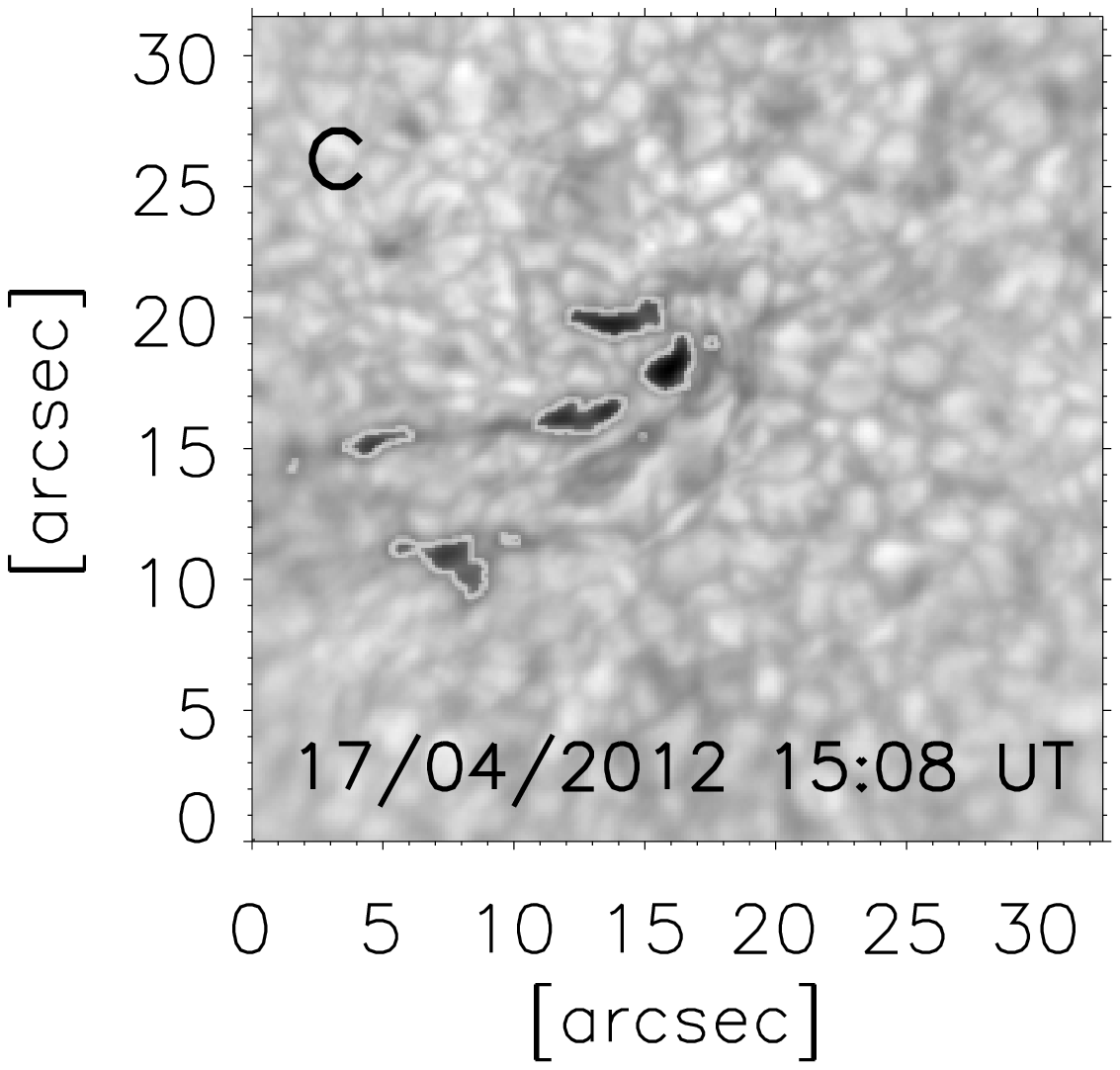}\includegraphics[height=3cm,trim=4.2cm 2.3cm 4.35cm 1.cm,clip=true,keepaspectratio=true]{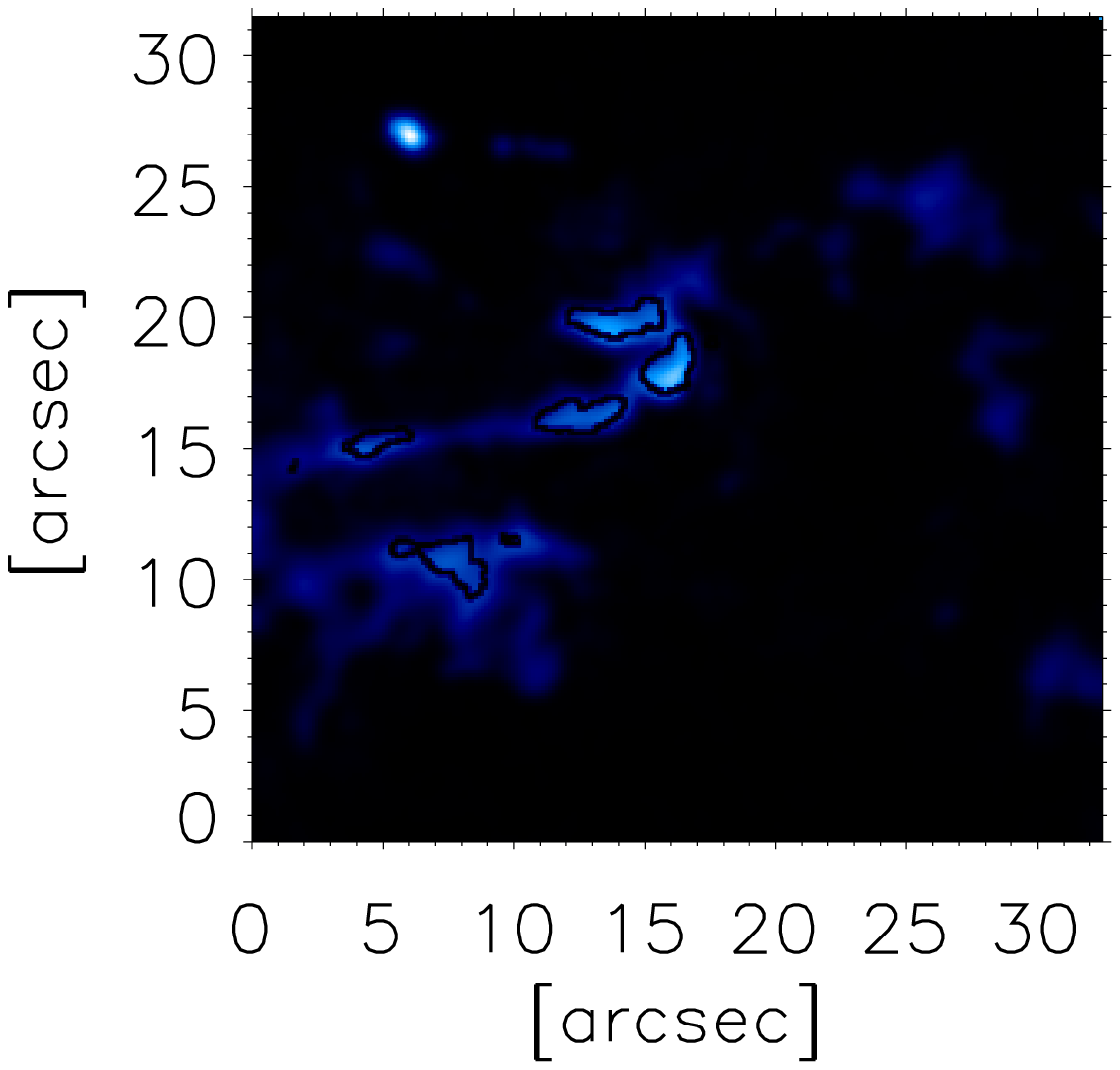}\includegraphics[height=3cm,trim=4.2cm 2.3cm 4.35cm 1.cm,clip=true,keepaspectratio=true]{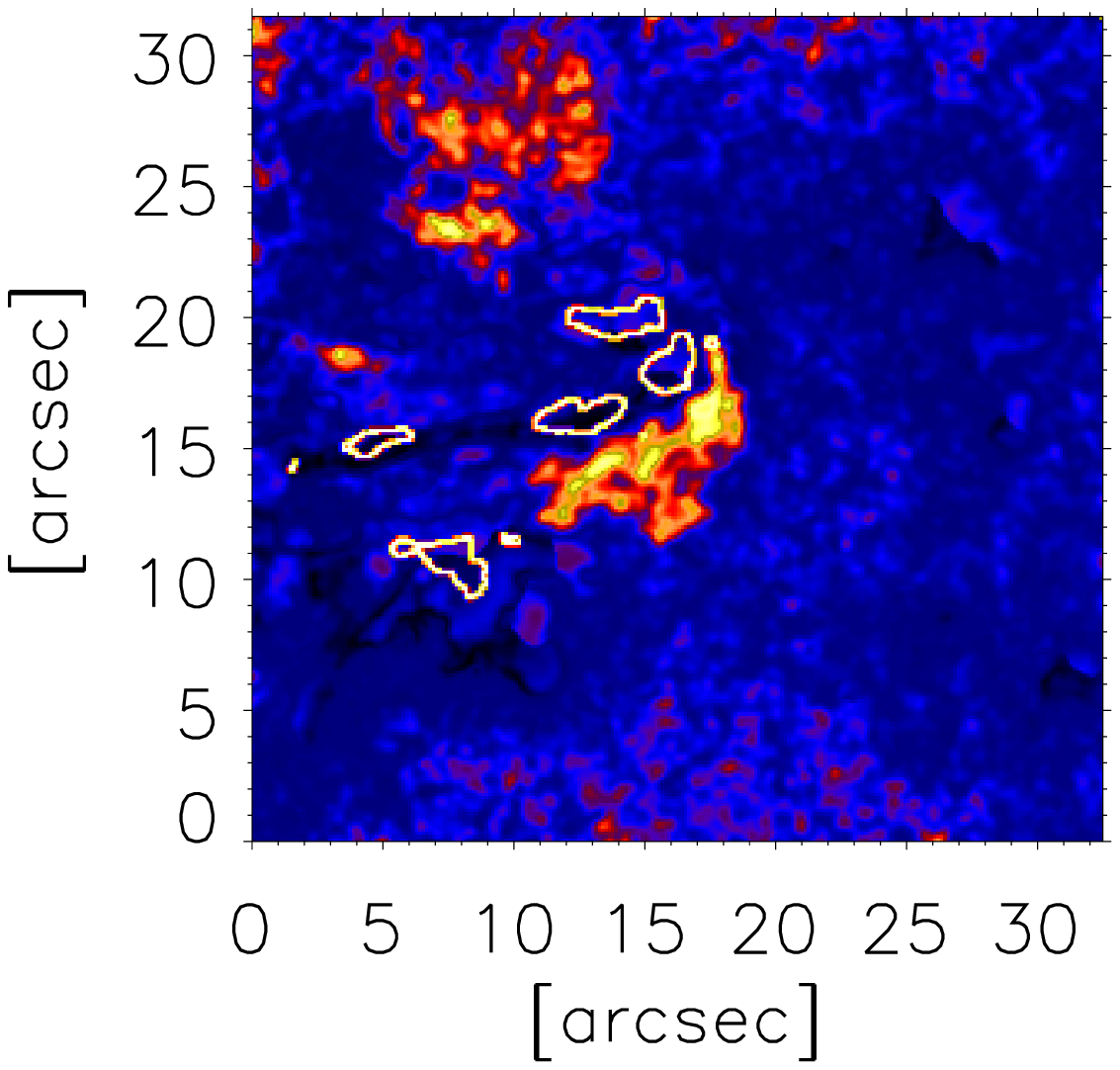}\includegraphics[height=3cm,trim=4.2cm 2.3cm 4.35cm 1.cm,clip=true,keepaspectratio=true]{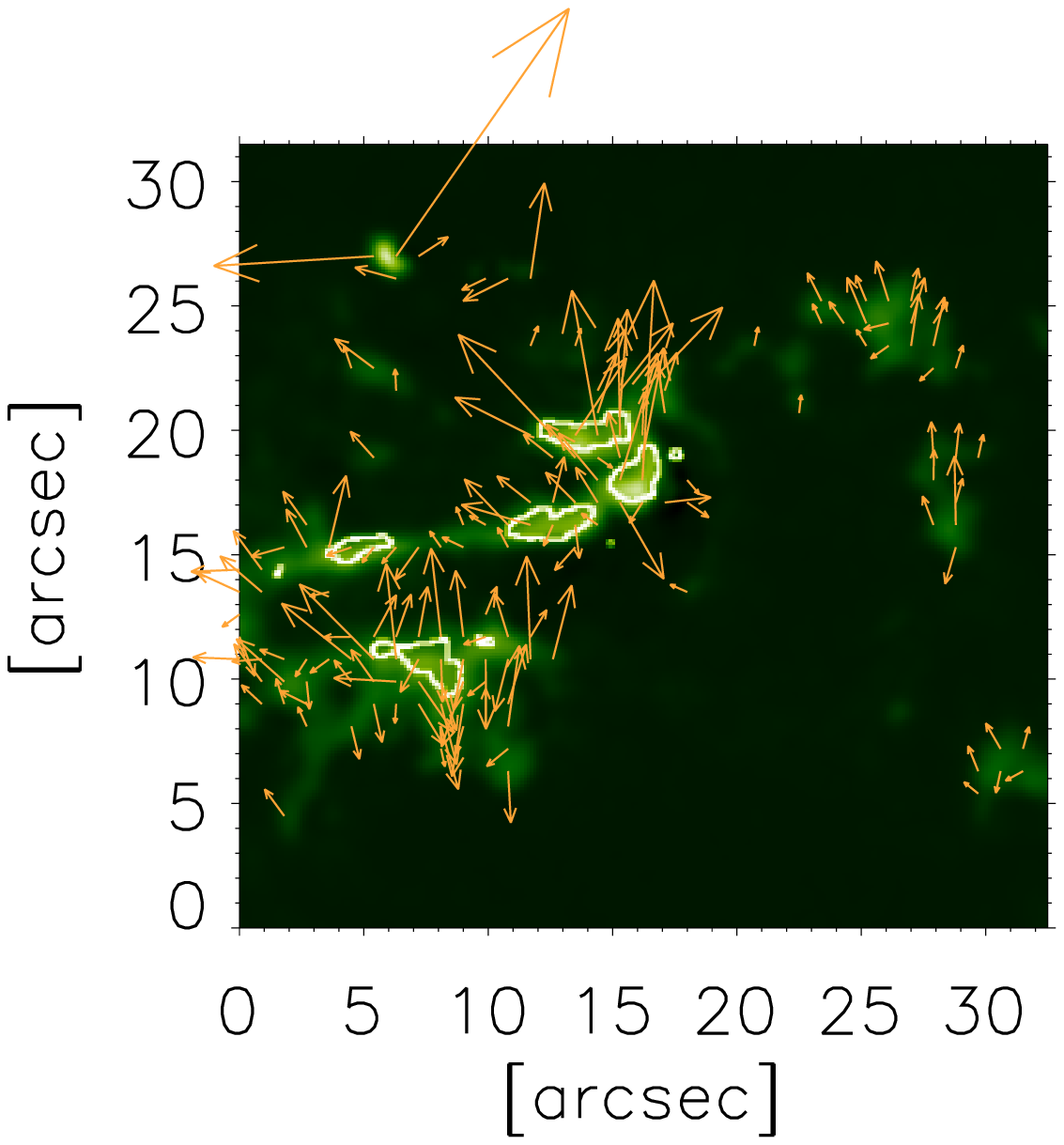}\\
\includegraphics[height=3cm,trim=1.5cm 2.3cm 4.35cm 1.cm,clip=true,keepaspectratio=true]{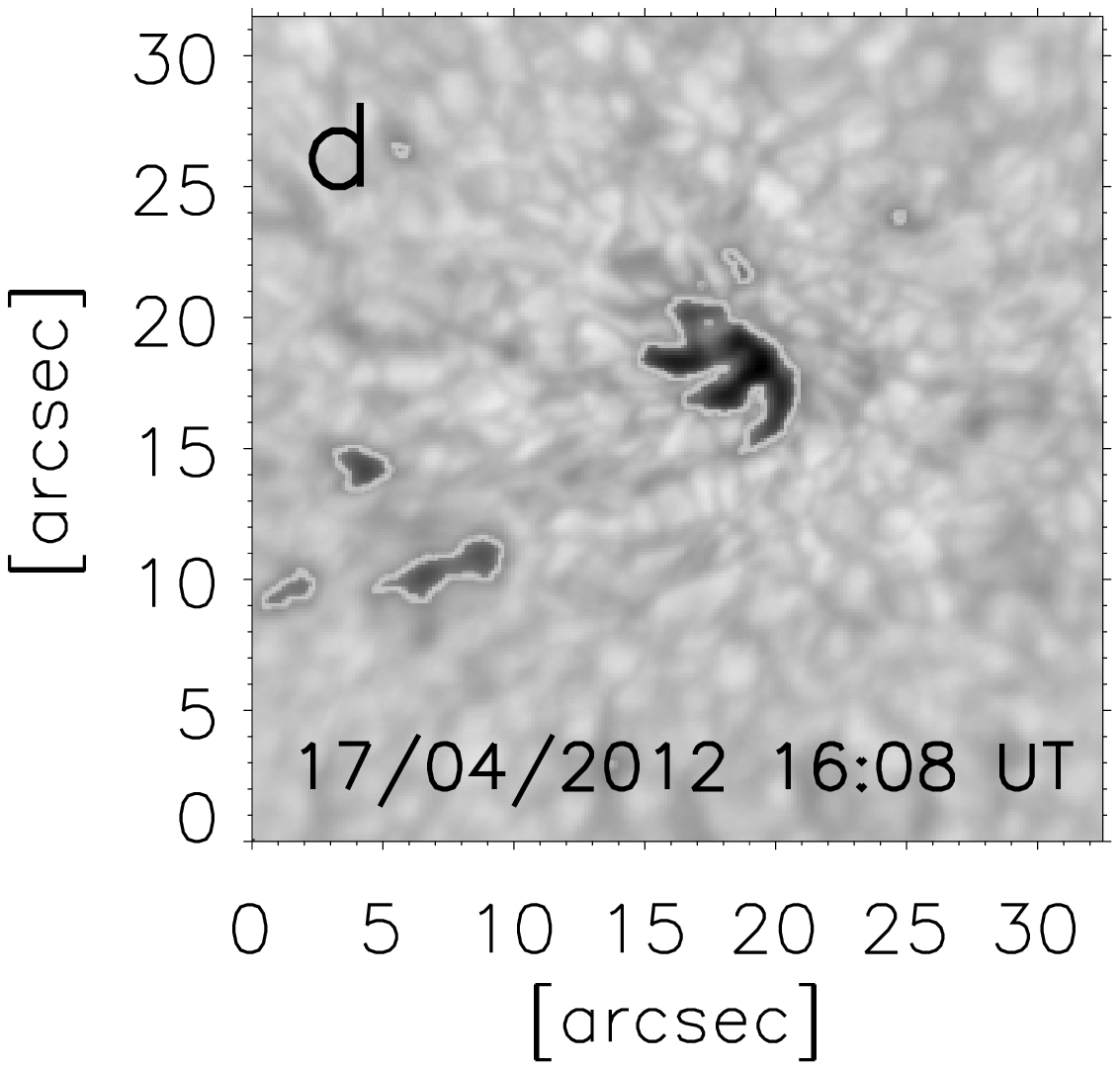}\includegraphics[height=3cm,trim=4.2cm 2.3cm 4.35cm 1.cm,clip=true,keepaspectratio=true]{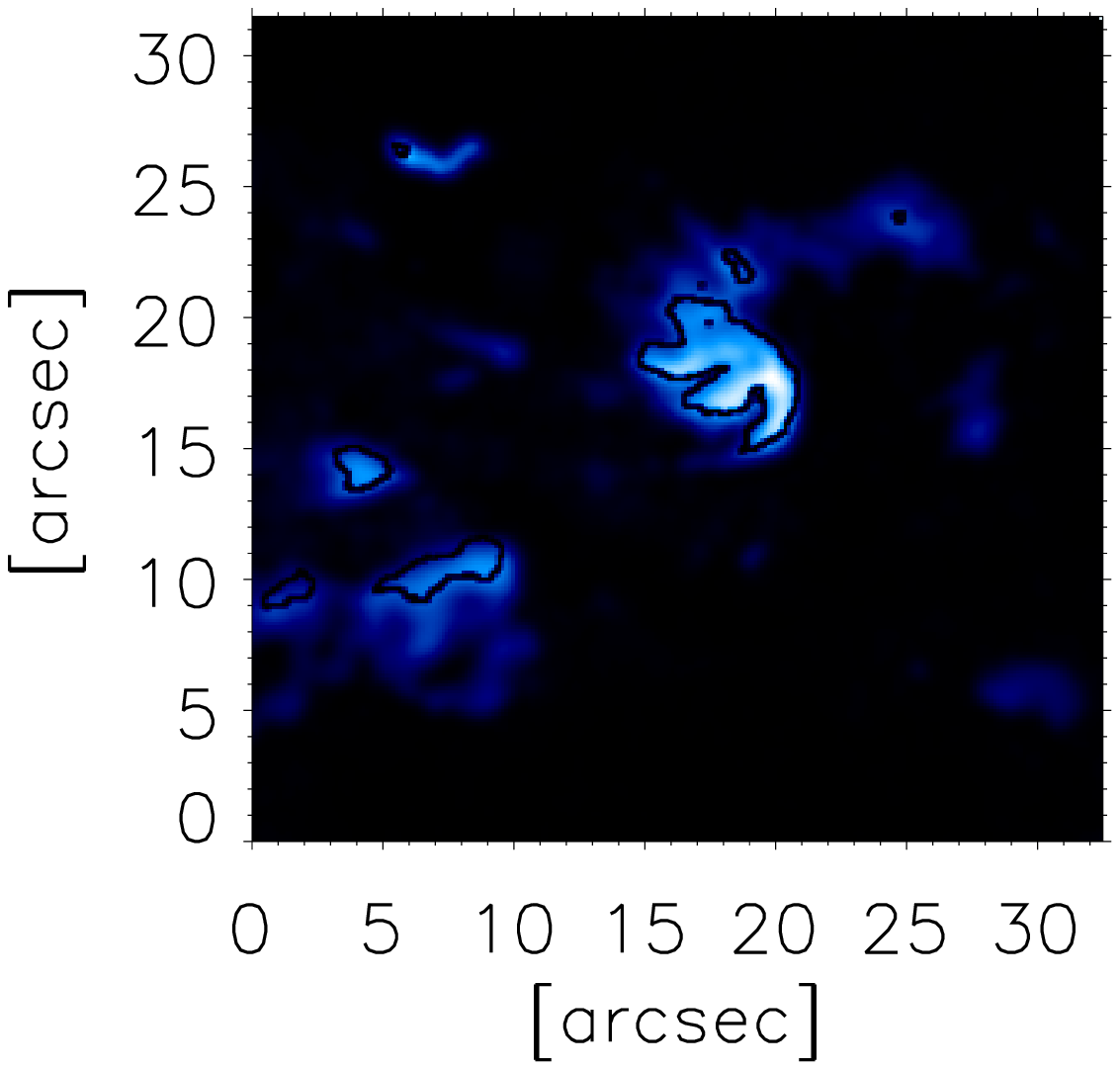}\includegraphics[height=3cm,trim=4.2cm 2.3cm 4.35cm 1.cm,clip=true,keepaspectratio=true]{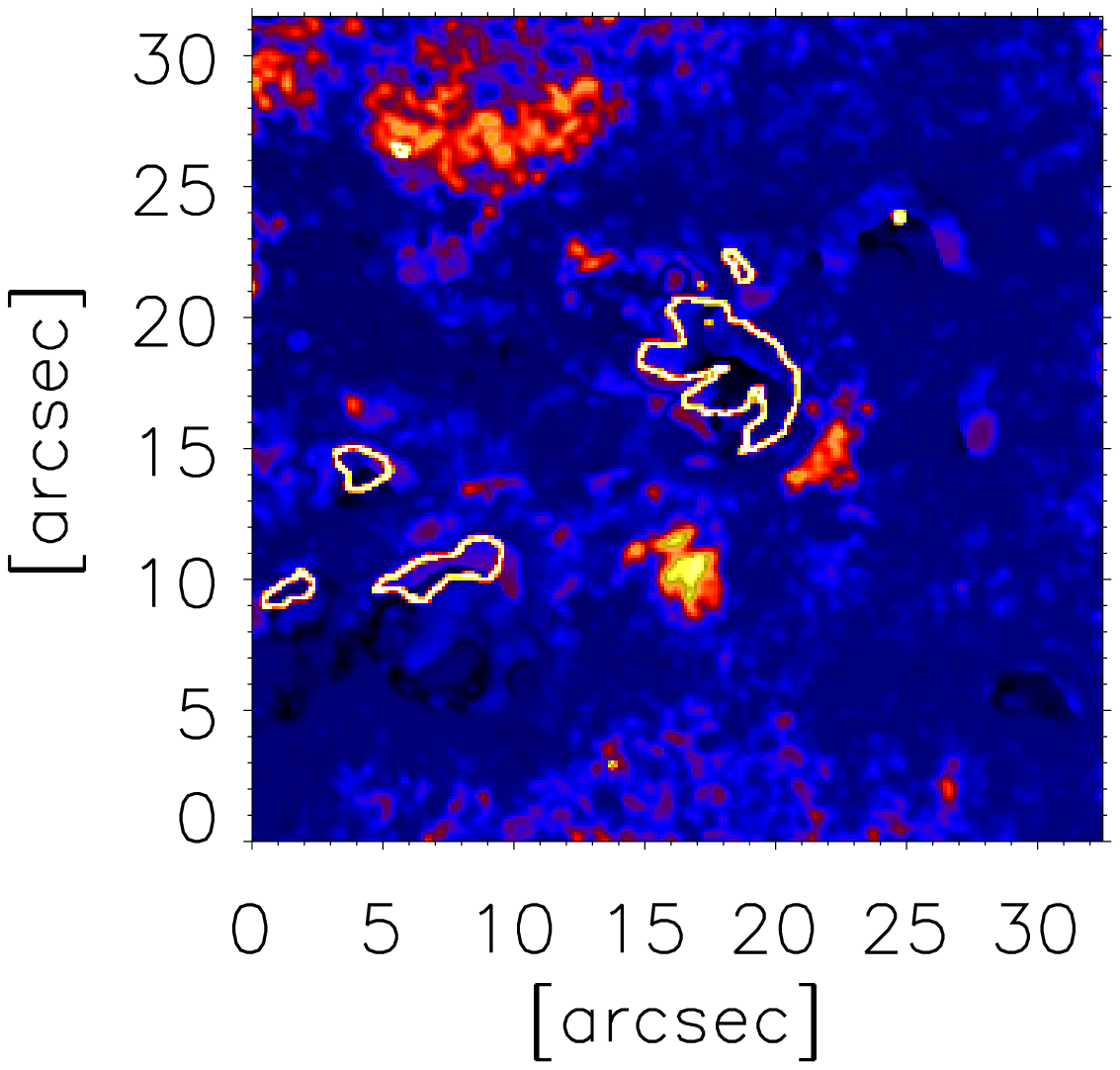}\includegraphics[height=3cm,trim=4.2cm 2.3cm 4.35cm 1.cm,clip=true,keepaspectratio=true]{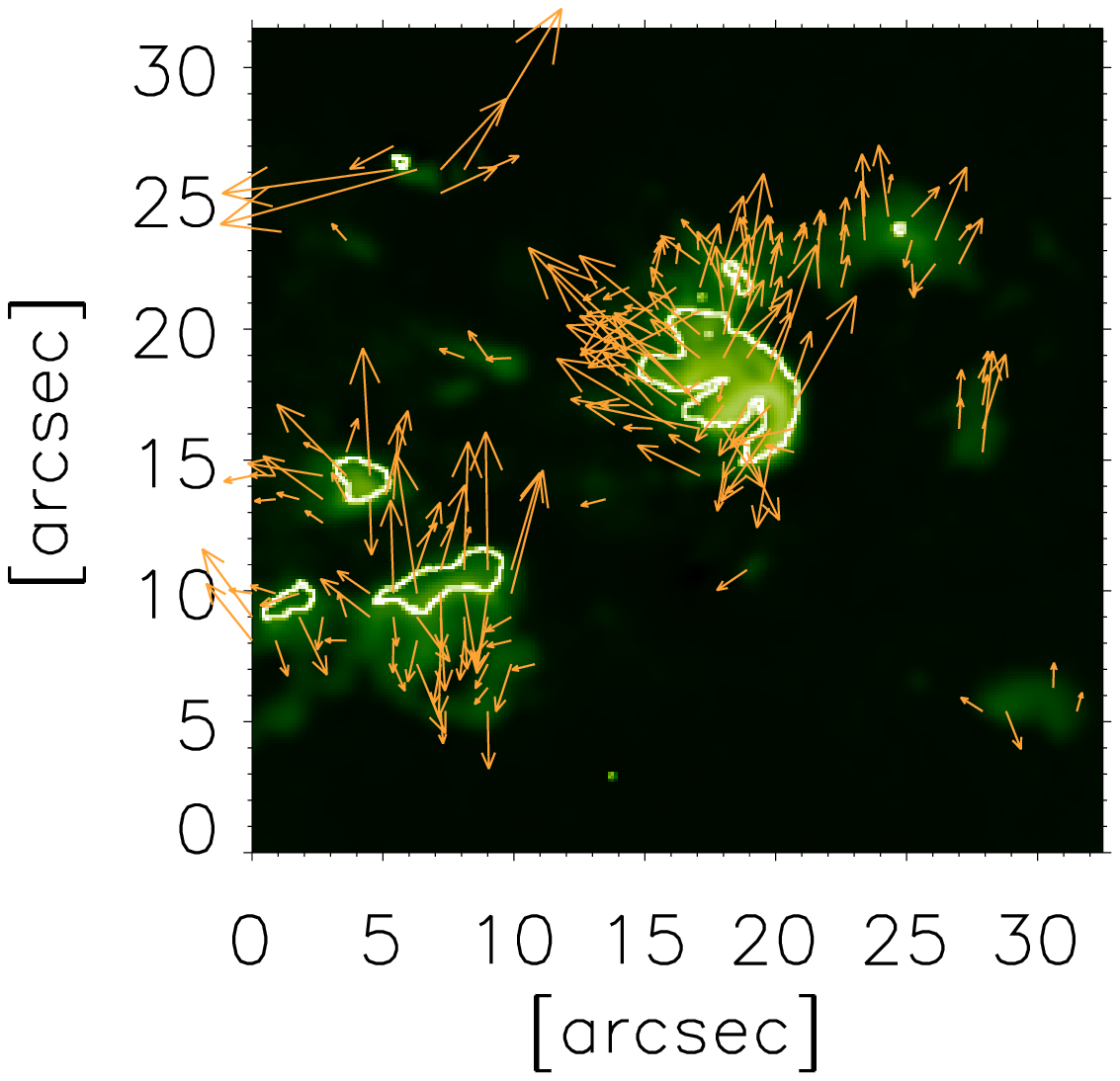}\\
\includegraphics[height=3cm,trim=1.5cm 2.3cm 4.35cm 1.cm,clip=true,keepaspectratio=true]{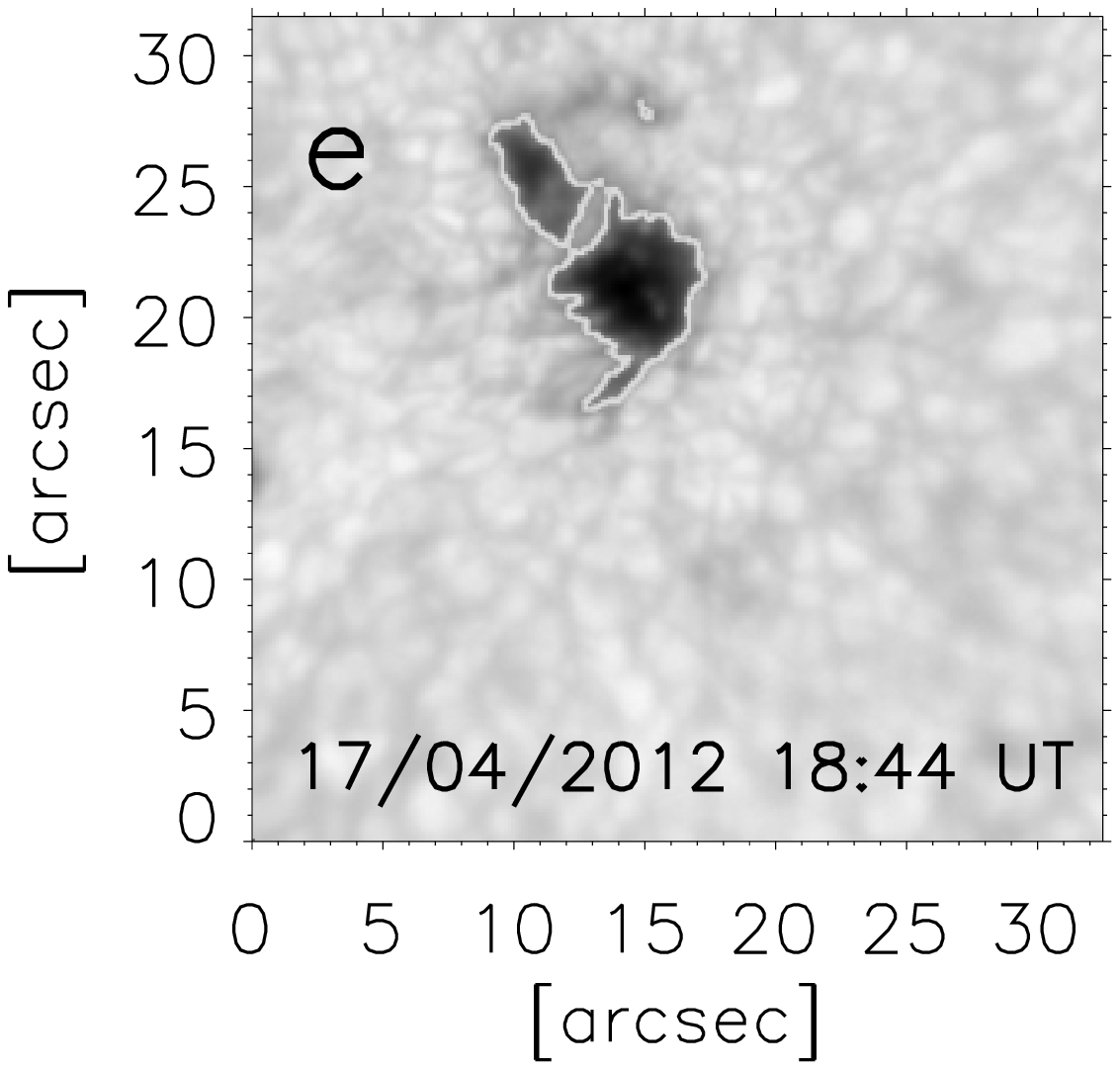}\includegraphics[height=3cm,trim=4.2cm 2.3cm 4.35cm 1.cm,clip=true,keepaspectratio=true]{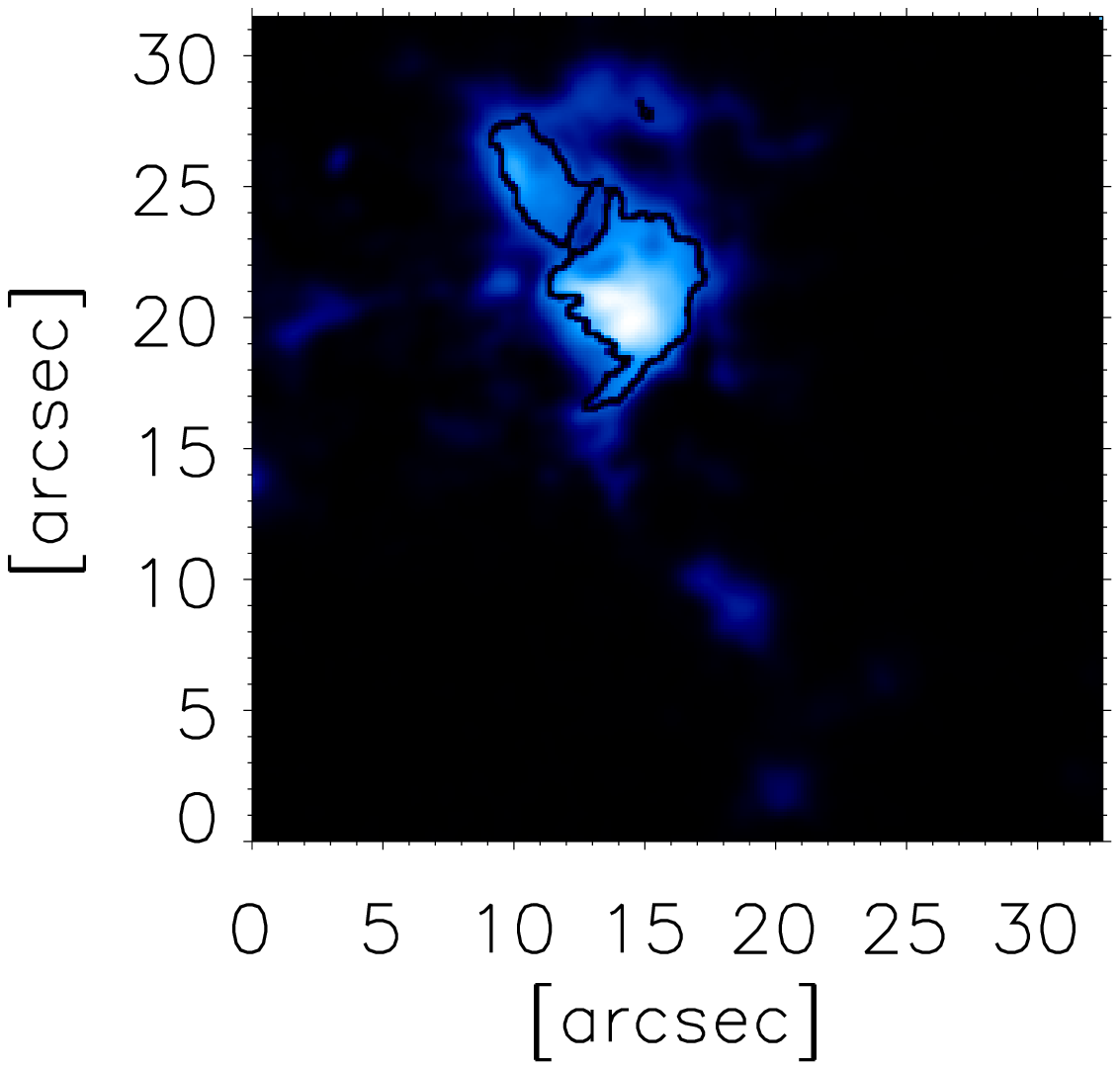}\includegraphics[height=3cm,trim=4.2cm 2.3cm 4.35cm 1.cm,clip=true,keepaspectratio=true]{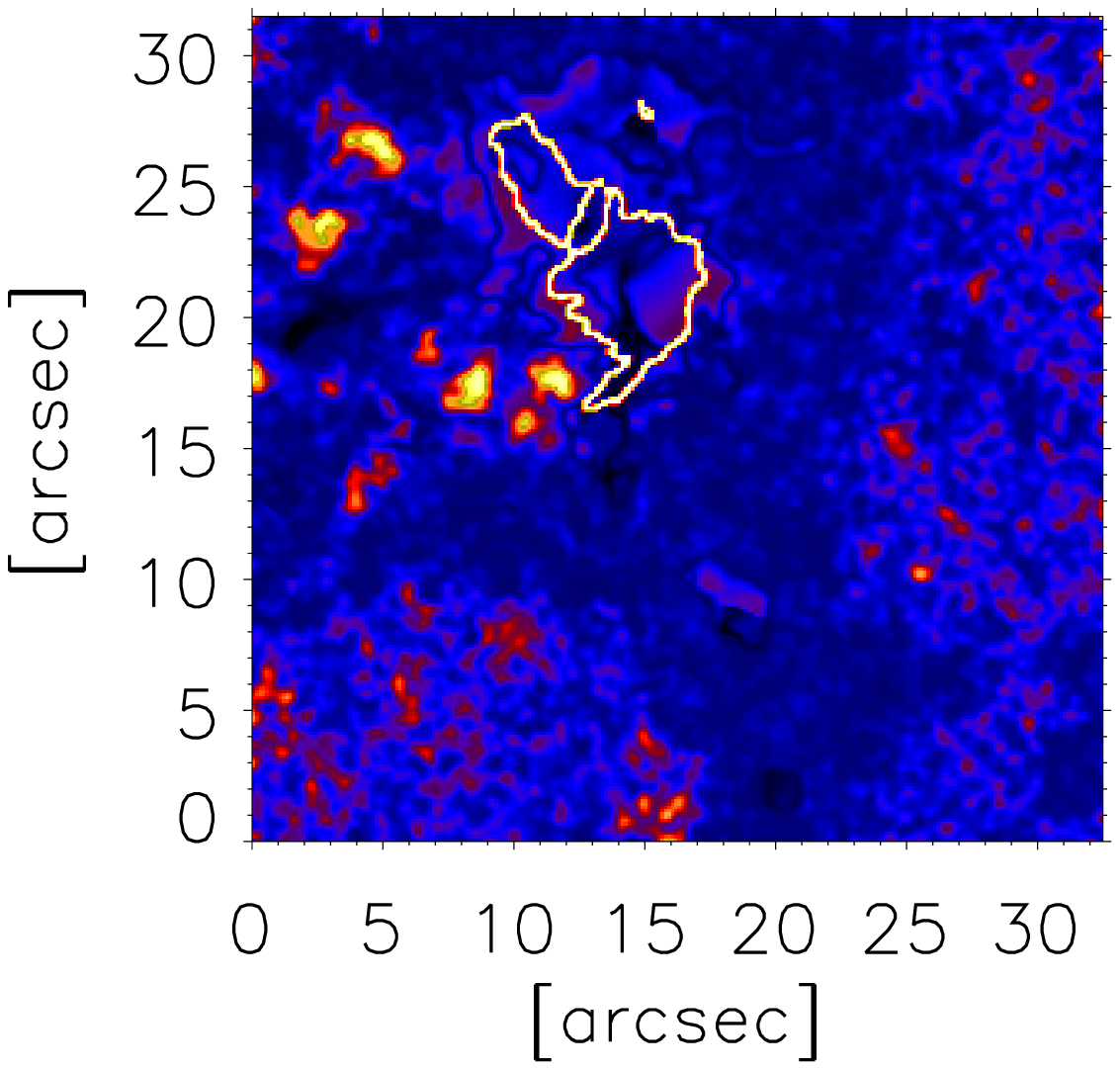}\includegraphics[height=3cm,trim=4.2cm 2.3cm 4.35cm 1.cm,clip=true,keepaspectratio=true]{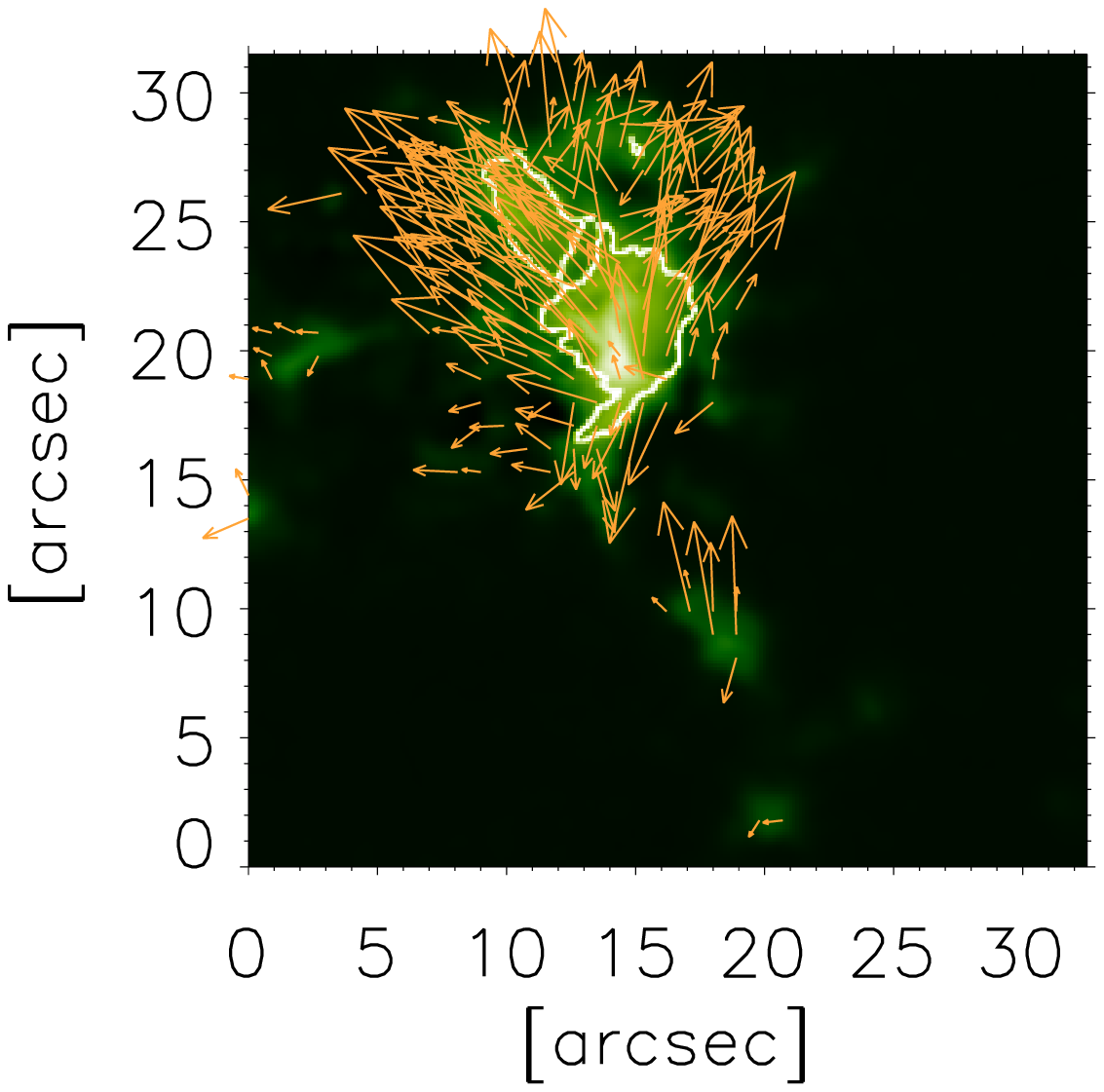}\\
\includegraphics[height=3cm,trim=1.5cm 2.3cm 4.35cm 1.cm,clip=true,keepaspectratio=true]{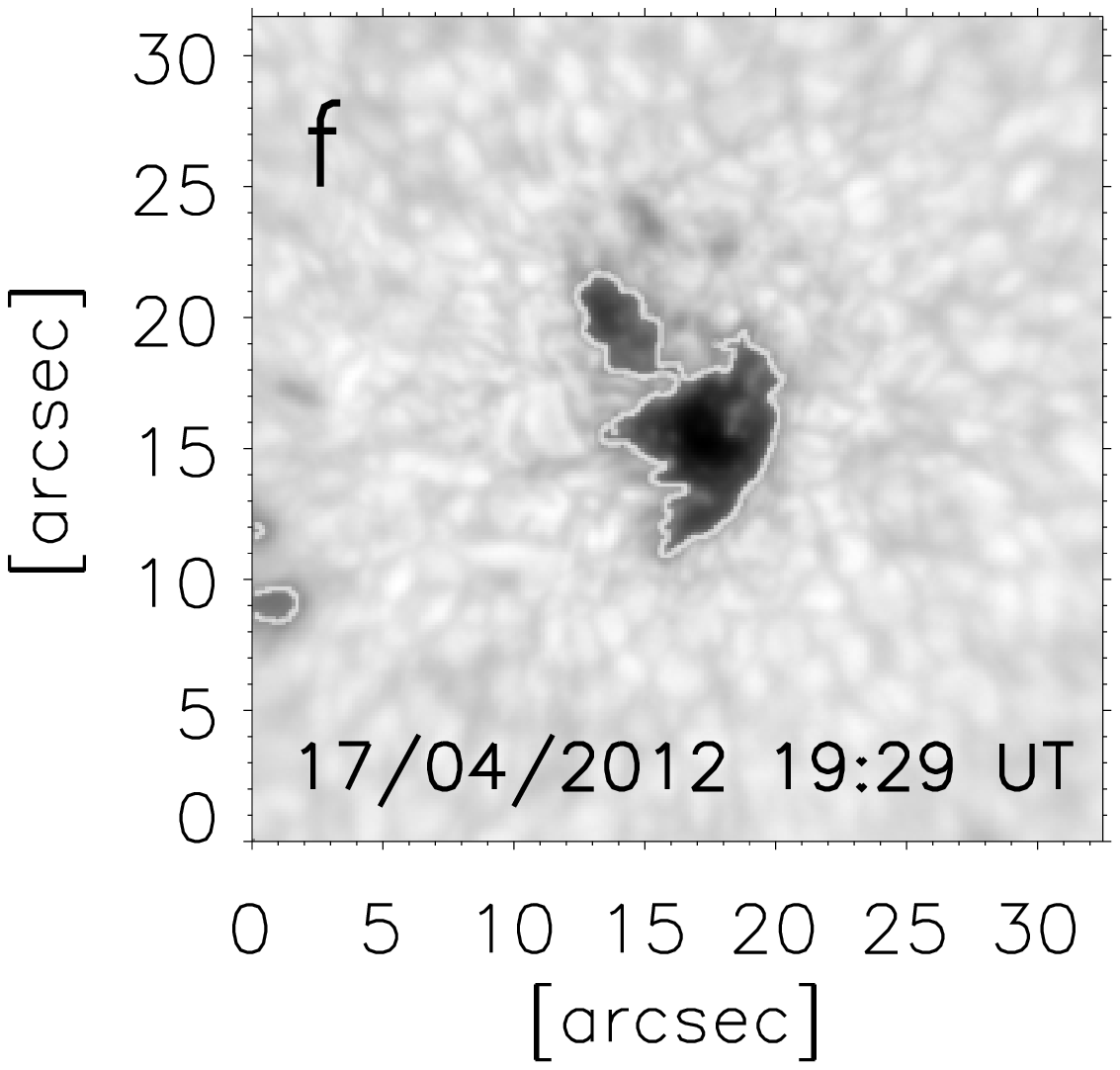}\includegraphics[height=3cm,trim=4.2cm 2.3cm 4.35cm 1.cm,clip=true,keepaspectratio=true]{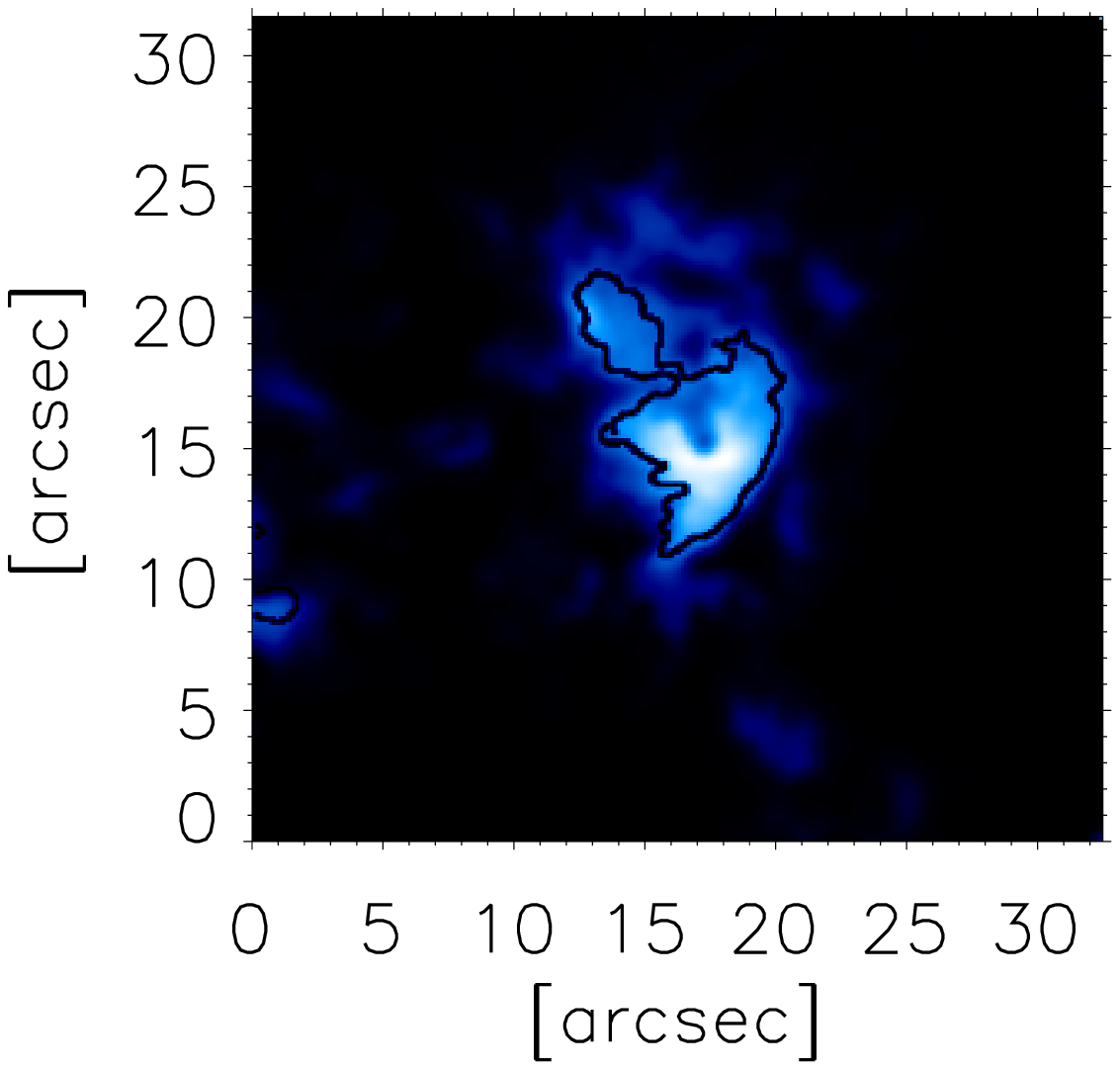}\includegraphics[height=3cm,trim=4.2cm 2.3cm 4.35cm 1.cm,clip=true,keepaspectratio=true]{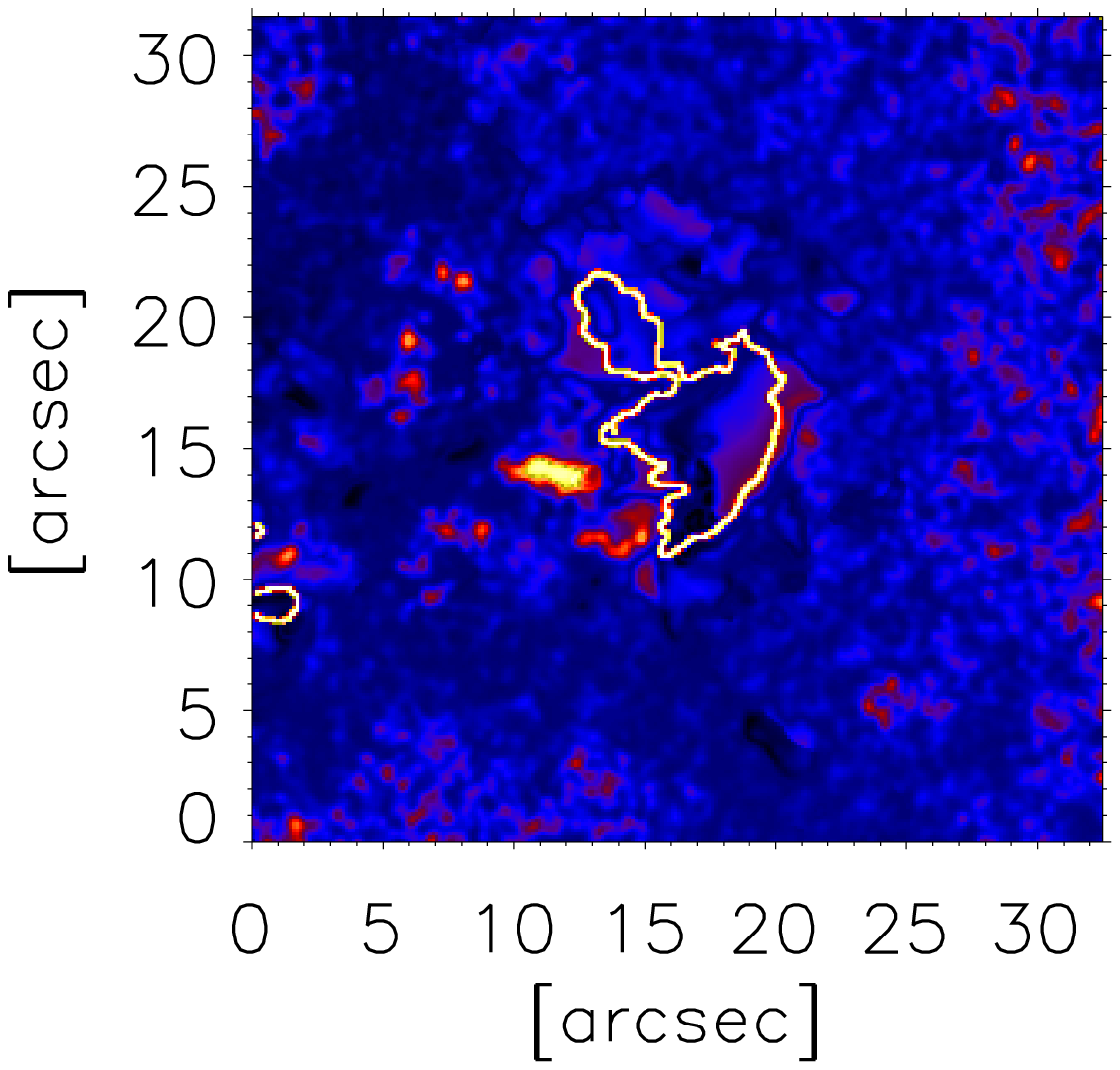}\includegraphics[height=3cm,trim=4.2cm 2.3cm 4.35cm 1.cm,clip=true,keepaspectratio=true]{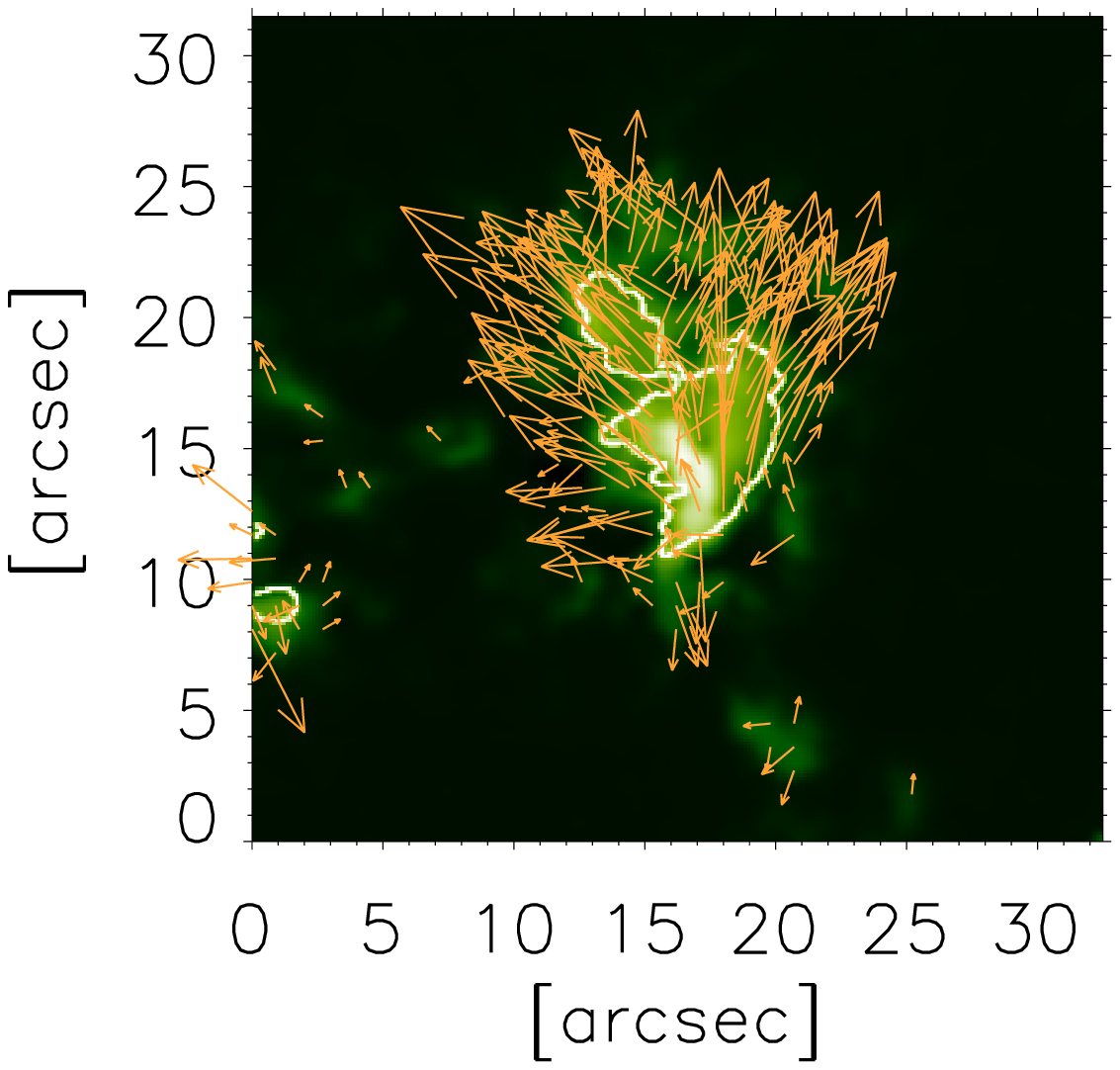}\\
\includegraphics[height=3.6cm,trim=1.5cm .5cm 4.35cm 1.cm,clip=true,keepaspectratio=true]{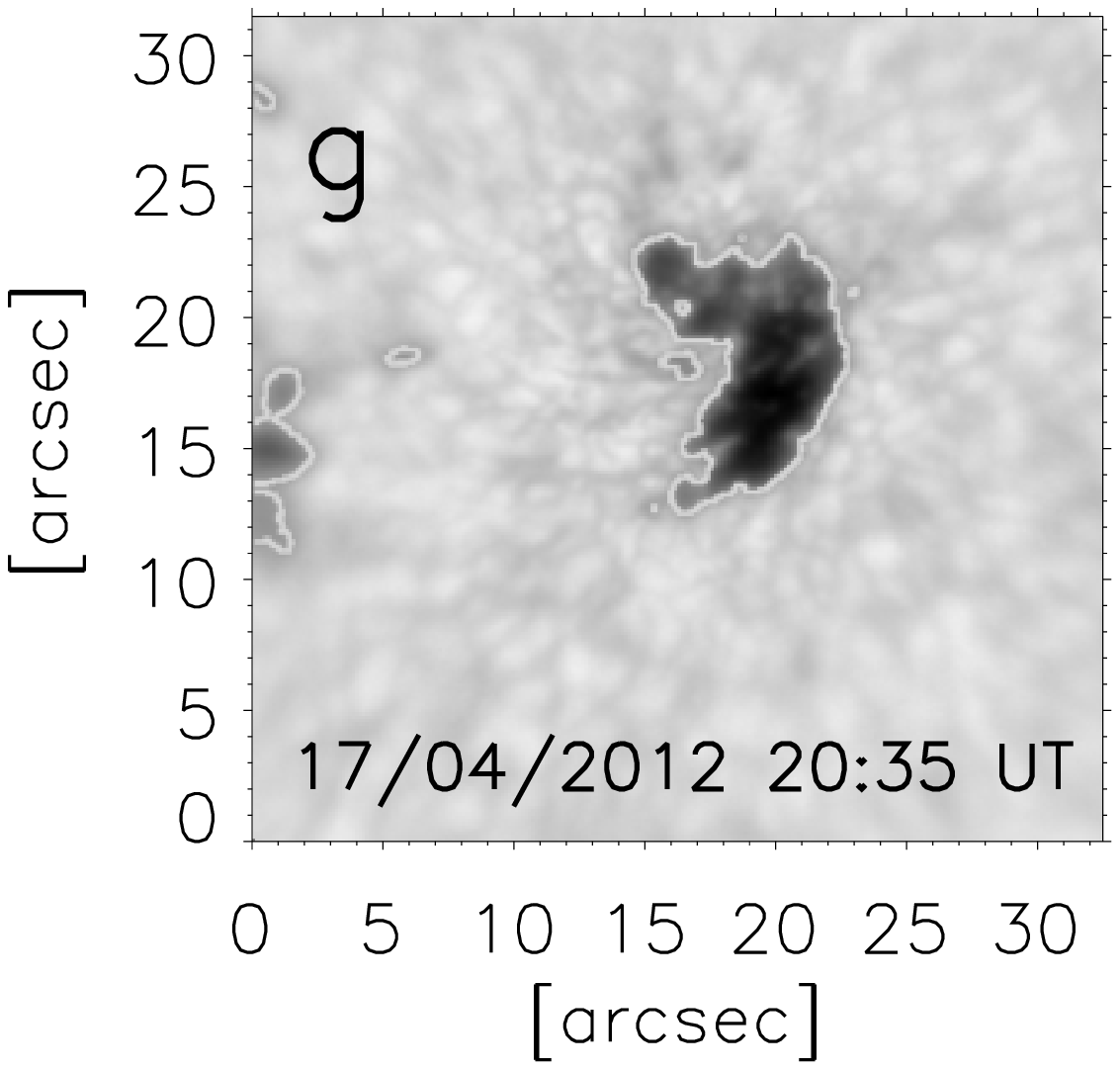}\includegraphics[height=3.6cm,trim=4.2cm .5cm 4.35cm 1.cm,clip=true,keepaspectratio=true]{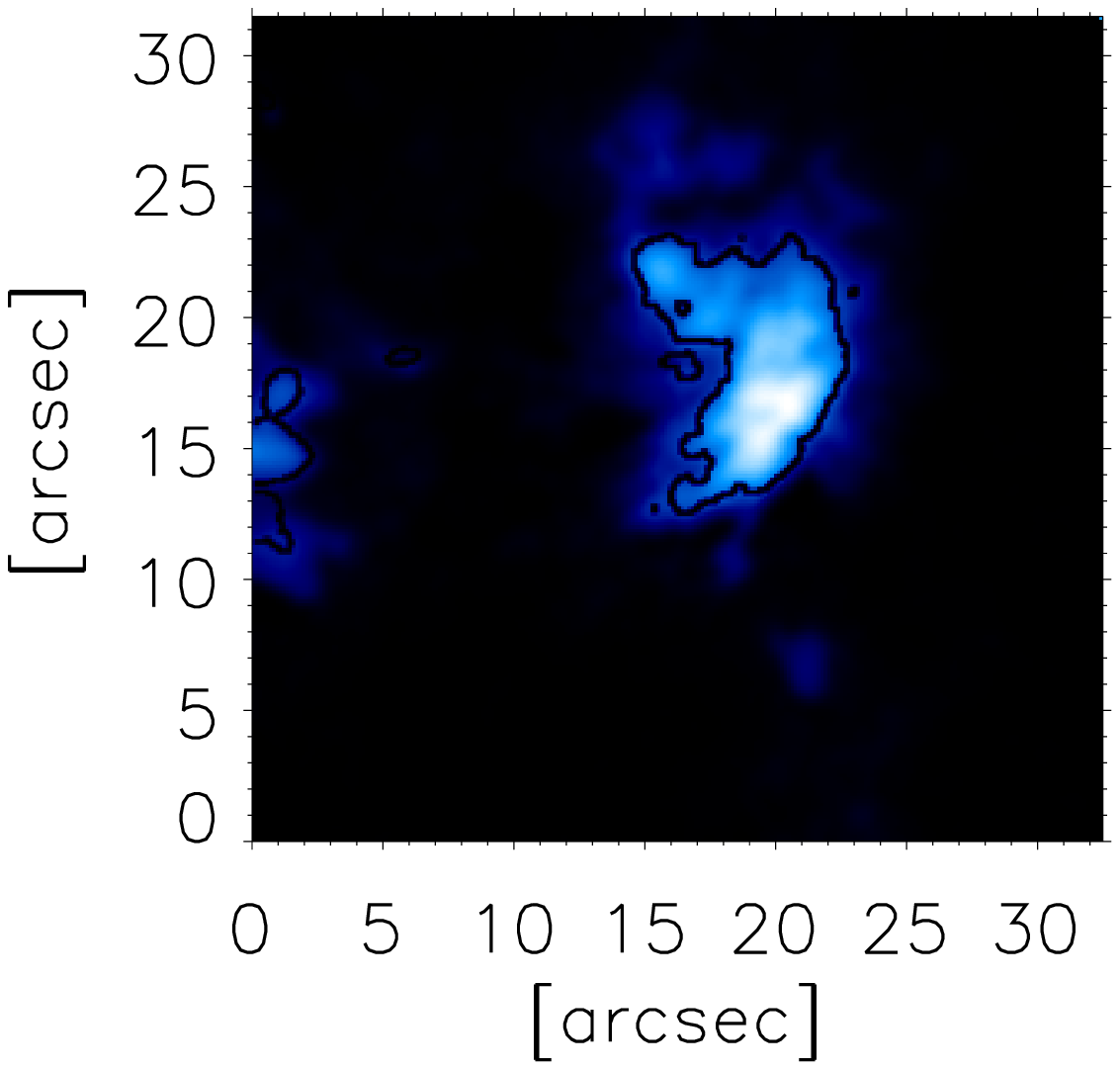}\includegraphics[height=3.6cm,trim=4.2cm .5cm 4.35cm 1.cm,clip=true,keepaspectratio=true]{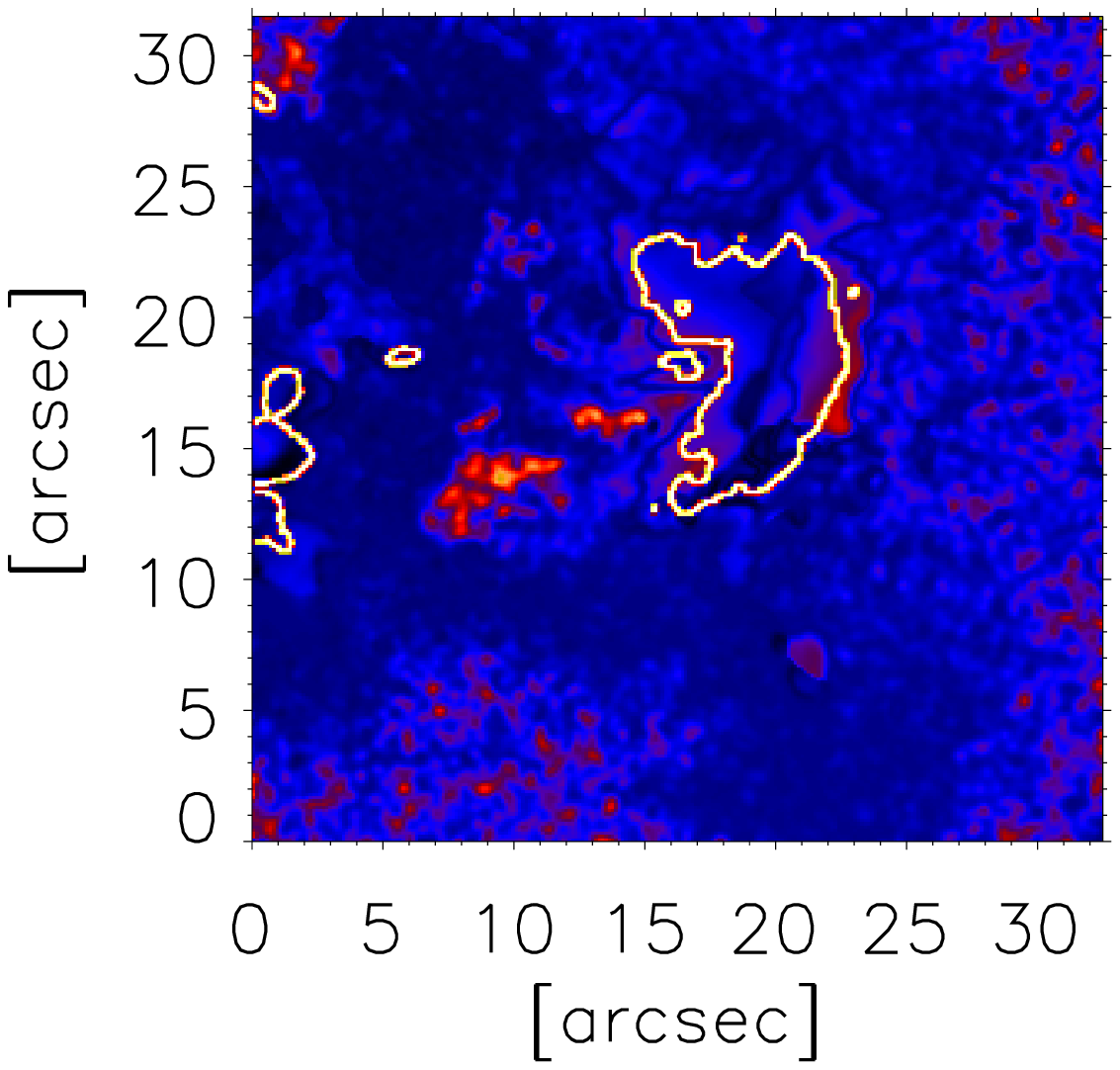}\includegraphics[height=3.6cm,trim=4.2cm .5cm 4.35cm 1.cm,clip=true,keepaspectratio=true]{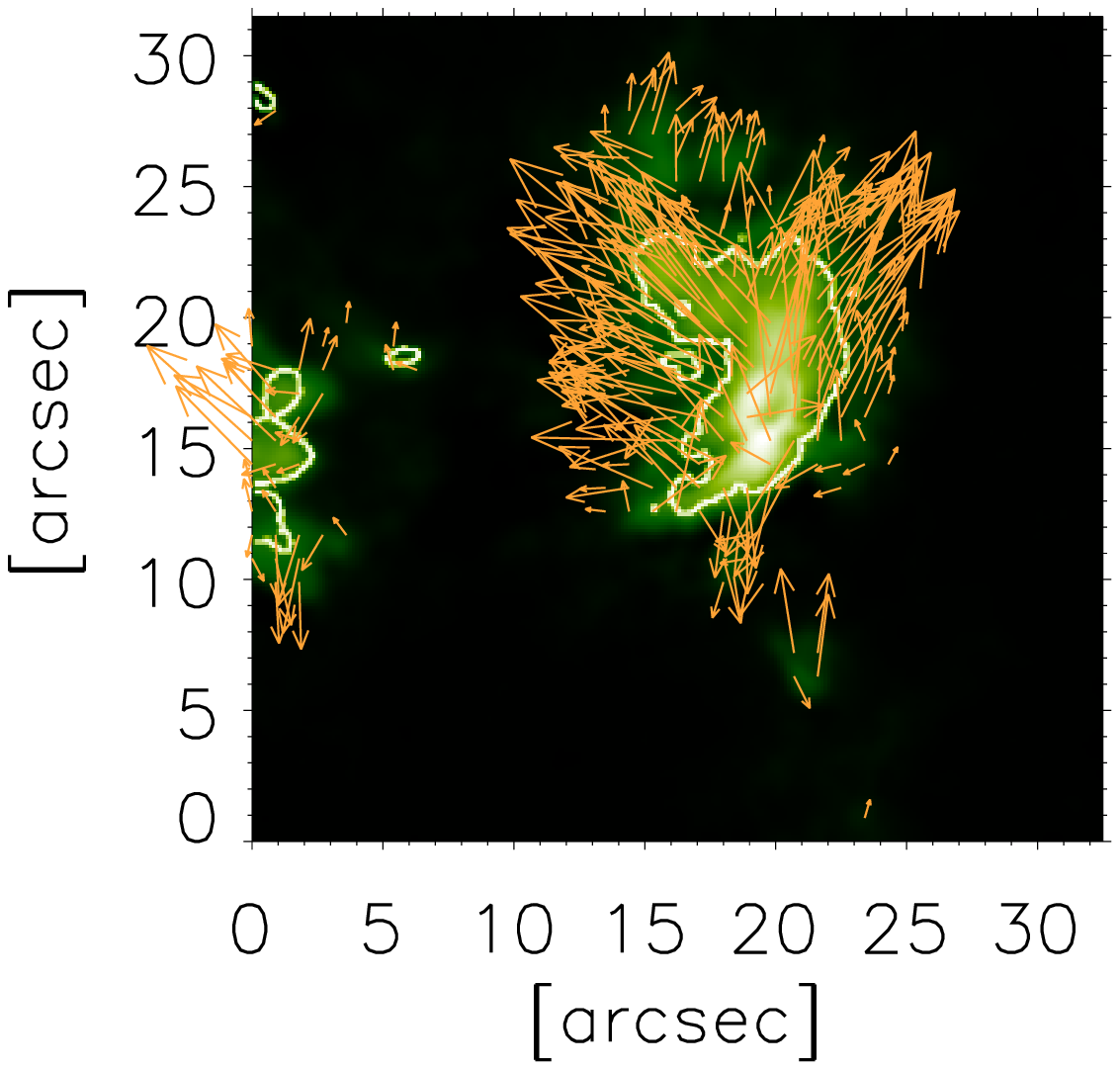}
}
\caption{\footnotesize{From left to right, top to bottom, numbers and letters within brackets indicate the column and line labels, respectively.  Examples of  (1)  continuum intensity, (2)  magnetic field strength,  (3) inclination, (4) B$_l$ longitudinal (background image), and B$_t$ transverse (overplotted vector field) components of the magnetic field in the LSF   derived  from the SIR inversion of the IBIS Fe I 617.3 nm line data,  at optical depth log$\tau _{500}=1$, at seven stages of the pore formation, at (a) 13:58 UT, (b) 14:26 UT, (c) 15:08 UT, (d) 16:08 UT, (e) 18:44 UT, (f) 19:29 UT, and (g) 20:35 UT. North is at the top, and west is to the right. The arrow at the bottom left on panel 4a represents a horizontal field of 1 kG; transverse field components lower than 0.2 kG are not shown. The contour line in each panel shows the location of the evolving structure singled out in the continuum data, as specified in Fig. \ref{f2}. 
}}
\label{f3}
\end{figure*}

%\subsubsection{Magnetic fields} 

\subsection{Magnetic and velocity fields  during the pore formation}

\subsubsection{Magnetic field}

Figure \ref{f3} shows examples of the results  derived from the SIR inversion  of the IBIS  data. In particular, we show the maps of  the magnetic field strength ($B$), and of the field inclination ($\theta$), transverse ($B_t$) and longitudinal ($B_l$) field components in the LSF,  %, and LOS velocities ($v_{LOS}$) 
% the maps of the $B_{LOS}$, total linear polarization $LP$, total circular polarization $CP$, and field inclination $\theta$
  at seven  representative stages of the pore formation.  %corresponding to the main stages displayed in Fig. \ref{f2}. 

%From an inspection of the computed maps, we deduce that the intensity of the magnetic field in the evolving feature  is on average xxx kG, i.e., not as low as it appears in the region corresponding to the segmented LB (approx 900G). 

The $B$ maps  (Fig. \ref{f3}, column 2) indicate that the field extends beyond the visible outline of the evolving feature %as previously reported in the literature \citep[e.g.][and references therein]{Sobotka_etal2012}. Figure \ref{f3}  (second to fourth columns) shows that this holds 
during the whole pore formation;  %The magnetic field strength measured in the evolving structure
the field reaches  values up to   1-2 kG   during the entire  interval analysed and slightly increases  over time.  %This field pattern and strength are consistent with results derived from observations of pores \citep[e.g.][and references therein]{Sobotka_etal2012}. 
%from 1.6 $\pm$ 0.1 to 2.3 $\pm$ 0.4  kGauss.  %during the evolution of the region. 
At the initial stages, 
%{\bf At} the initial stages, 
 the most intense field concentrations are found at the edges of the evolving region; these strong fields are located near  magnetic fields  with lower strength and  opposite polarity than  those found  in the evolving feature (Fig. \ref{f3} lines b, c), from which they are seen to move away  (Fig. \ref{f3} lines e-g). At the initial stages, the magnetic area is 4-5 times larger than the photometric area, as  reported from analysis of an evolving small AR, for example by  \citet[][]{Verma_etal2016}, while   
after the  formation of the coherent feature, the magnetic region is  about 2 times larger than the dark structure.
At the latter stages, the most intense fields are detected in the central-southern  region of the pore.  %The $B_l$ and $B_t$ maps   (e.g. Fig. \ref{f3}, column 4) show that t
The  field is almost vertical with respect to the photosphere in the central section of the evolving feature, and rather inclined outside it (e.g. $B_l$ and $B_t$ maps   in Fig. \ref{f3}, column 4).  
Indeed, the $\theta$ maps (Fig. \ref{f3}, column 3) show  a large patch  characterized by 
values of $\approx$ 45$^{\circ}$ over an area extending well beyond that of the evolving feature,  which  hosts fields with $\theta$ < $\approx$ 25$^{\circ}$.  
The $\theta$   maps also show small patches of magnetic field characterized by an inclination of about 90-180$^{\circ}$, which correspond to the opposite polarity features observed outside the evolving pore, north-east of it. 

After the formation of the coherent, funnel-shaped feature, the  $B_t$  maps (Fig. \ref{f3}, column 4, lines e-g)  show magnetic fields that are  aligned along the direction of the opposite polarity patches in the evolving region facing each other. %The azimuth pattern in opposite polarities   also results along the axis of the elongated granules and plage features seen in the photospheric and chromospheric data, respectively. 
Comparison between the maps derived from the data taken at the initial  (Fig. \ref{f3}, lines a-d) and final (Fig. \ref{f3}, lines e-g) evolutionary stages  shows the same azimuth pattern on both the  fragmented positive polarity patches of the forming  pore and the later formed pore. Therefore, the fields in the   smaller scale, the same polarity  features, and those in the formed pore are coaligned. Since the former features are seen to coalesce and form the pore, they 
show  up as  the leading footpoints of an emerging dipole.

\subsubsection{Line-of-sight and horizontal motions}

\begin{figure}%[ht!]
%  \centering{\hspace {1cm} 1                       \hspace{2.5cm}                                     2            \hspace{2.5cm}                                                 3     \hspace{2.5cm}                        4  \hspace{2.5cm}                        5 \hspace{2.5cm}                        6     \hspace {1cm}    }\\
\centering{
\hspace{1.3cm}
\includegraphics[height=0.8cm,trim=3.5cm 9.75cm 5.8cm 0.7cm,clip=true,keepaspectratio=true]{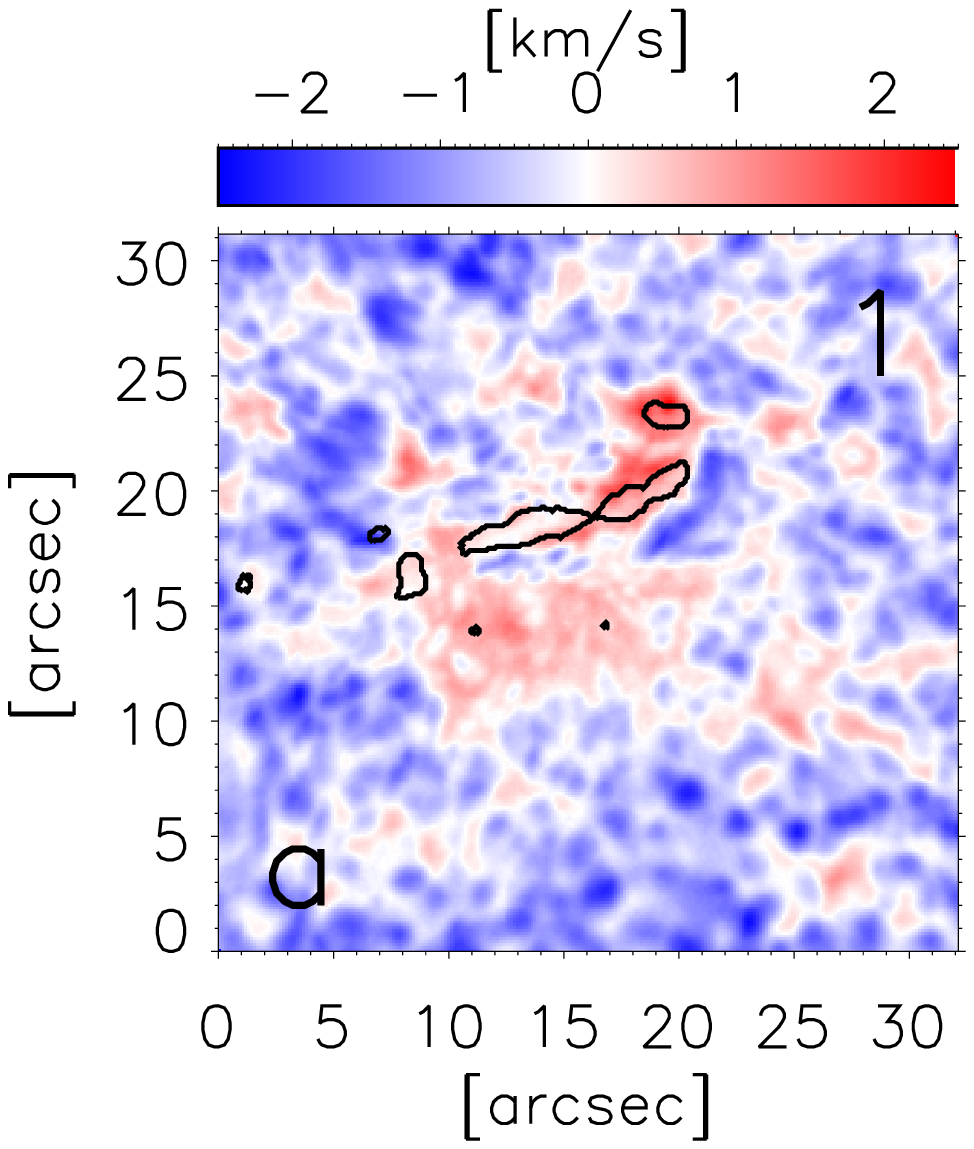}
\includegraphics[width=3cm,trim=3.5cm 12.9cm 5.8cm 0.7cm,clip=true,keepaspectratio=true]{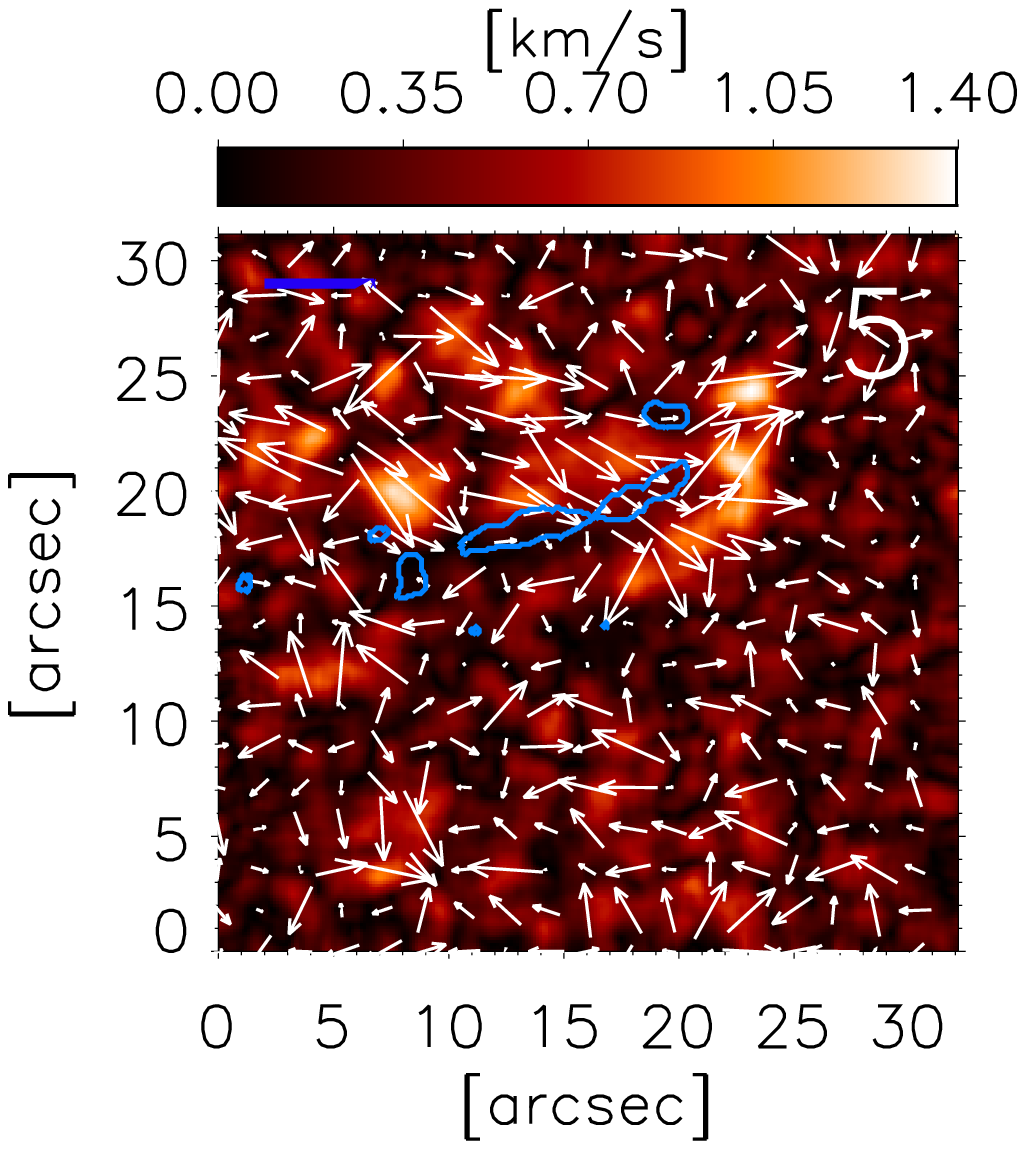}  %3.5 12.9
\hspace{5.9cm}
}\\  
\centering{
\includegraphics[height=3cm,trim=1.5cm 2.3cm 4.35cm 1.1cm,clip=true,keepaspectratio=true]{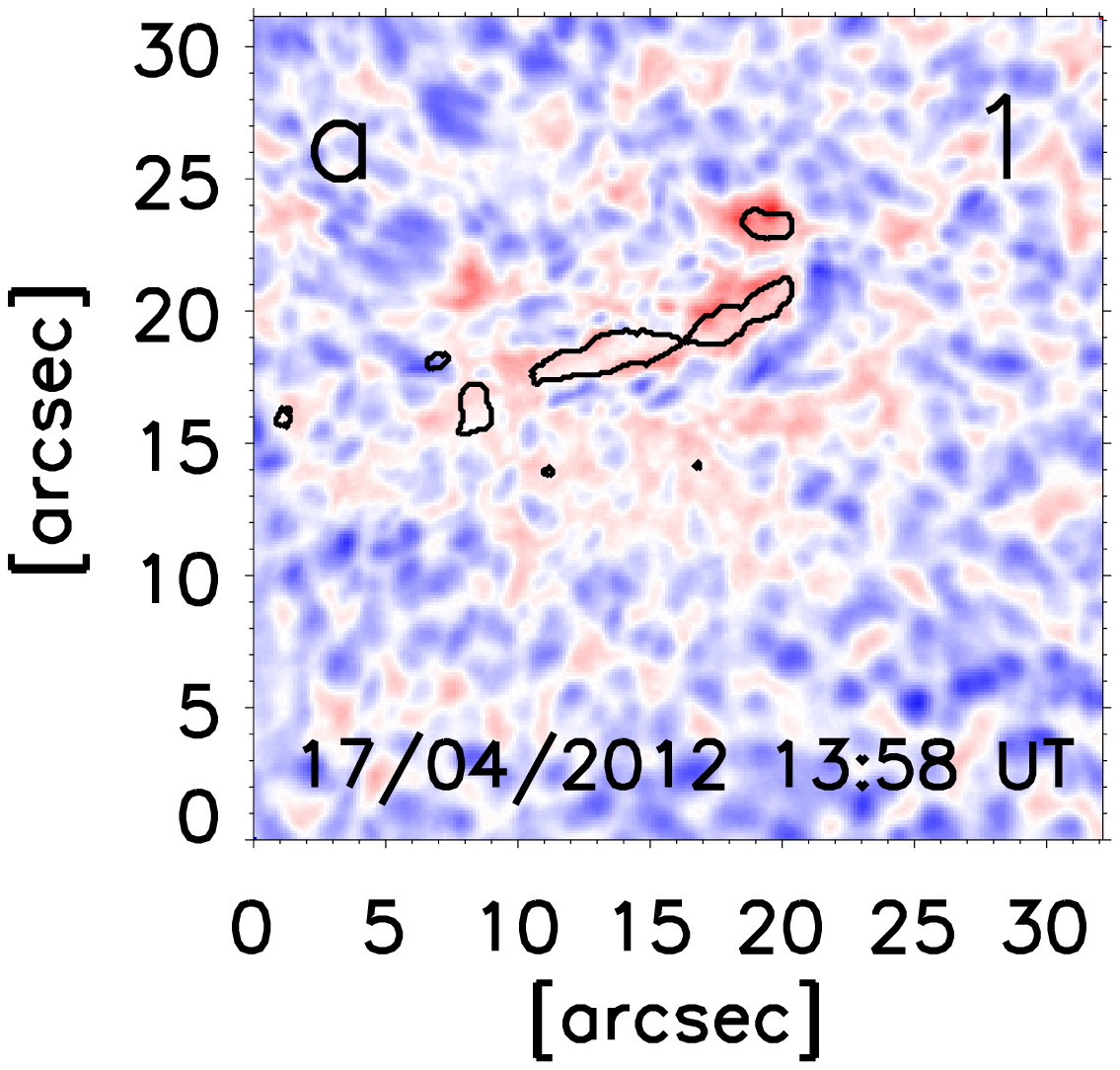}\includegraphics[height=3cm,trim=4.2cm 2.3cm 4.35cm 1.1cm,clip=true,keepaspectratio=true]{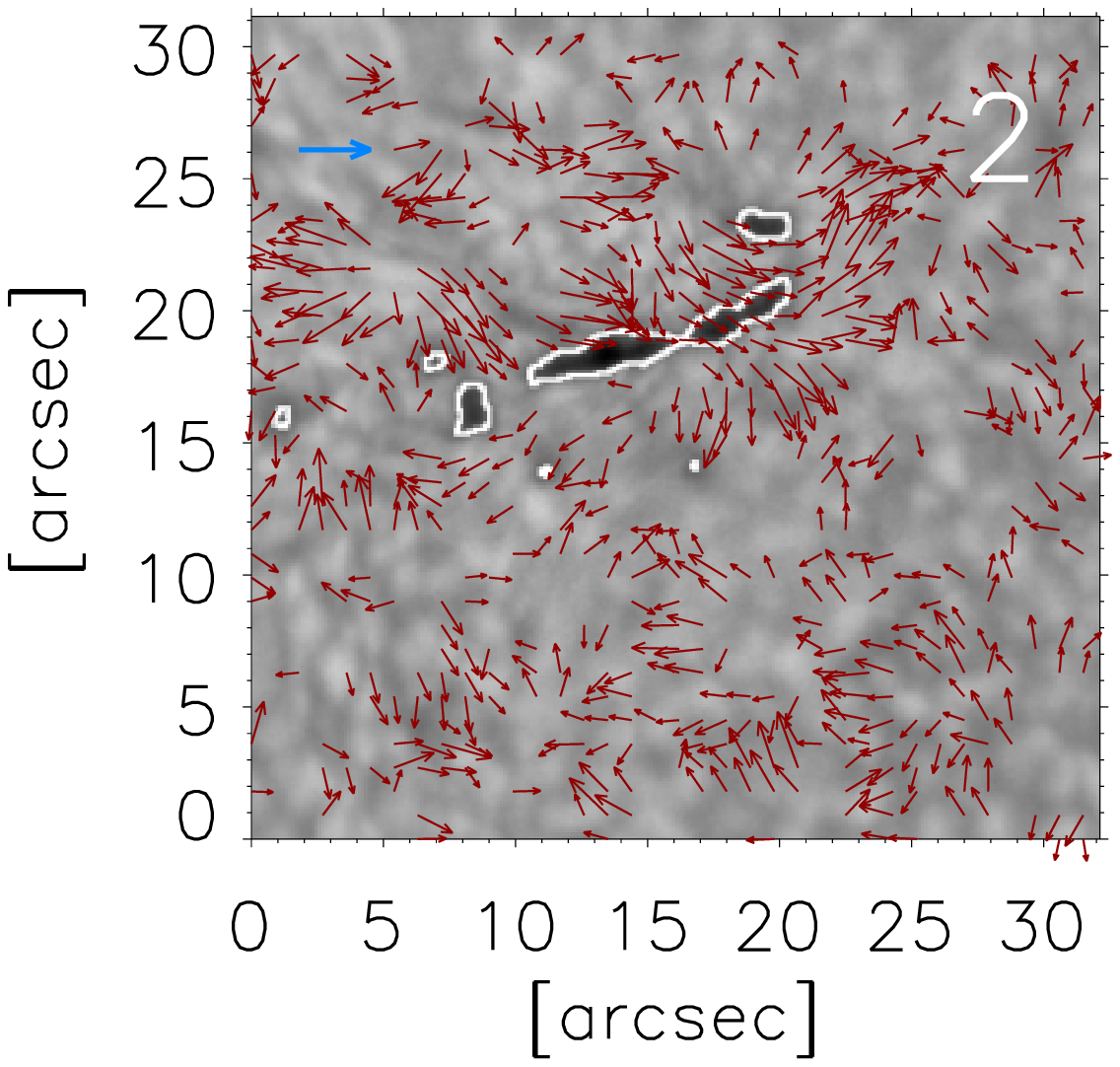}\\
\includegraphics[height=3cm,trim=1.5cm 2.3cm 4.35cm 1.1cm,clip=true,keepaspectratio=true]{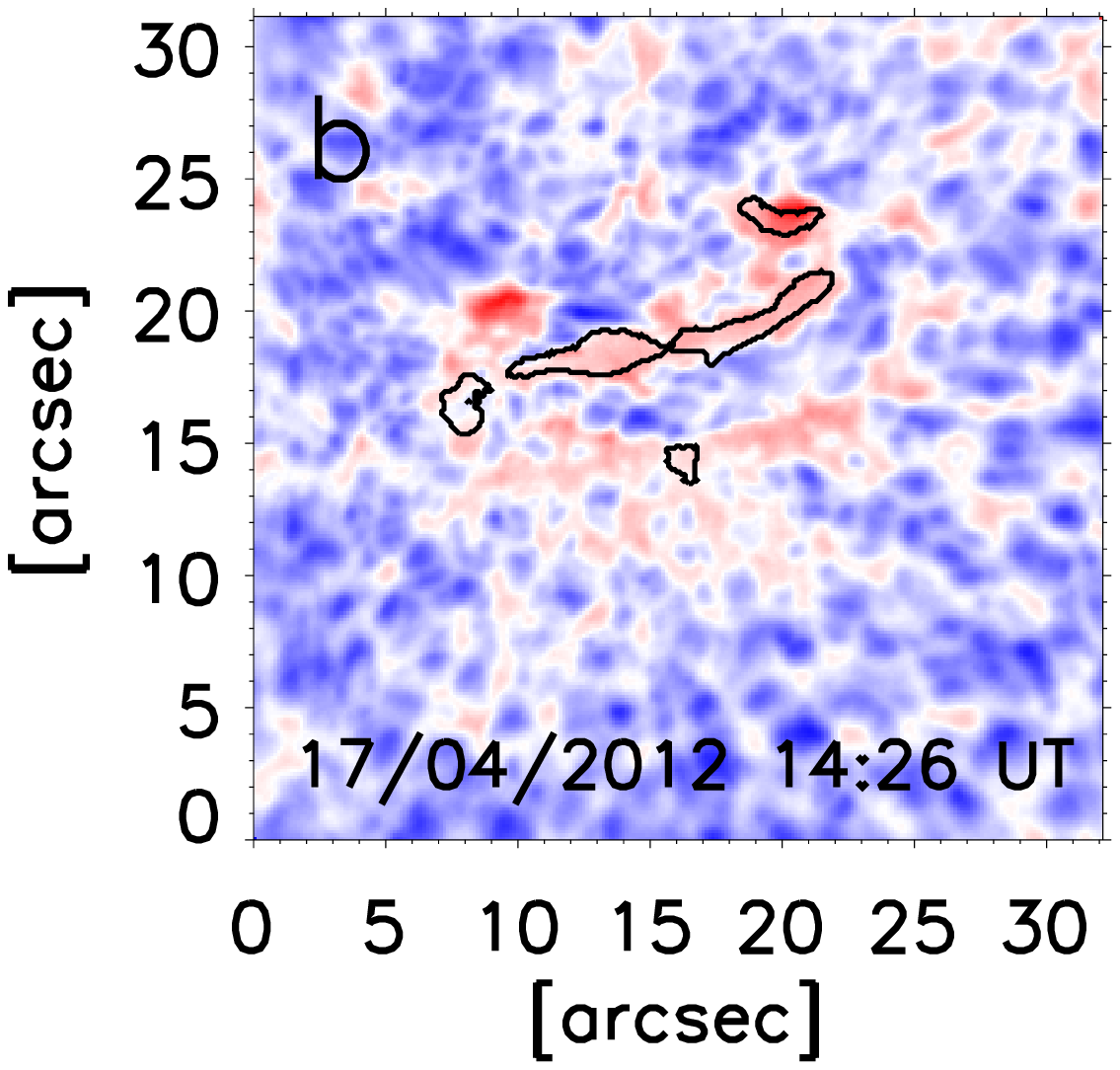}\includegraphics[height=3cm,trim=4.2cm 2.3cm 4.35cm 1.1cm,clip=true,keepaspectratio=true]{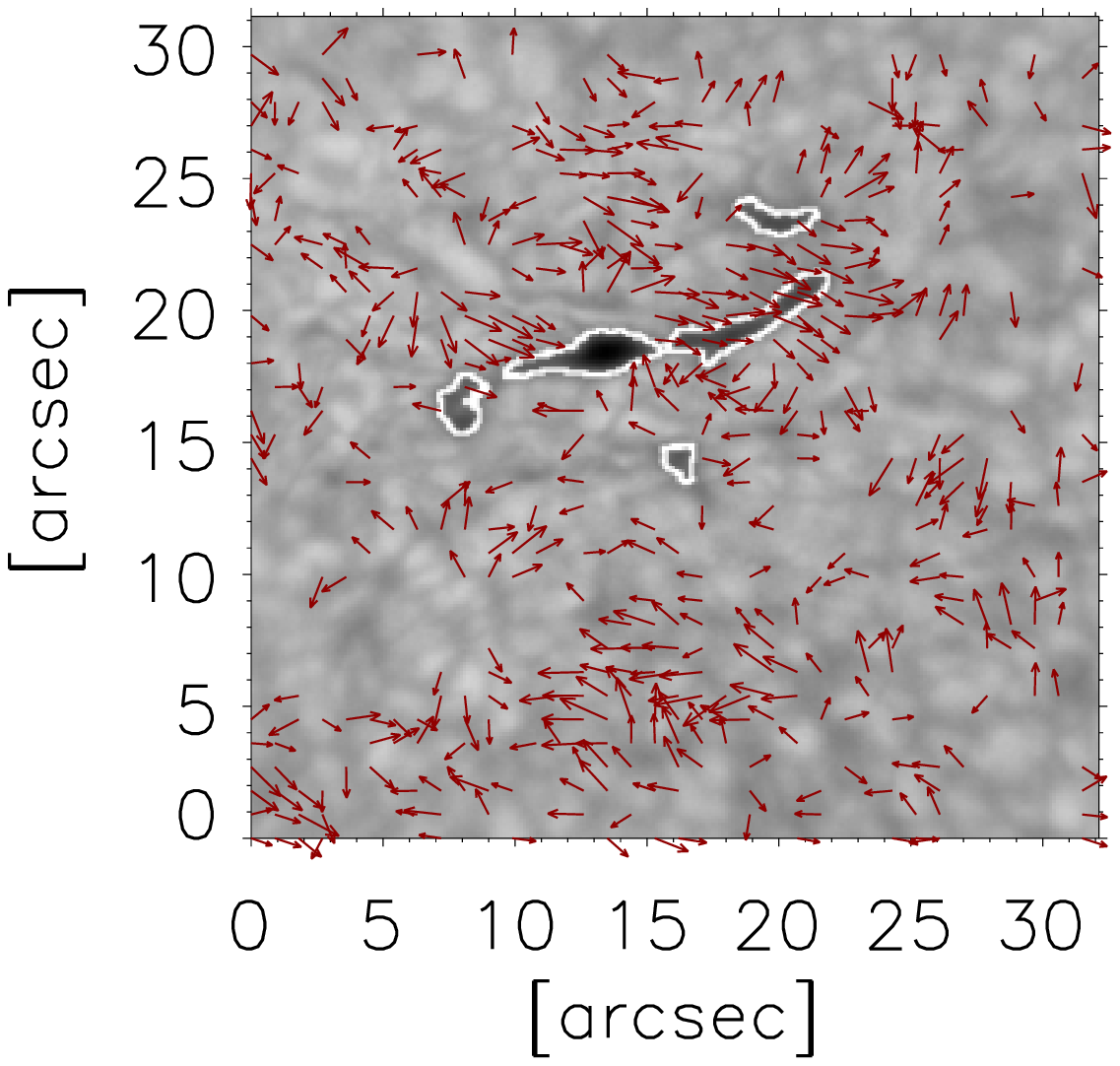}\\
\includegraphics[height=3cm,trim=1.5cm 2.3cm 4.35cm 1.1cm,clip=true,keepaspectratio=true]{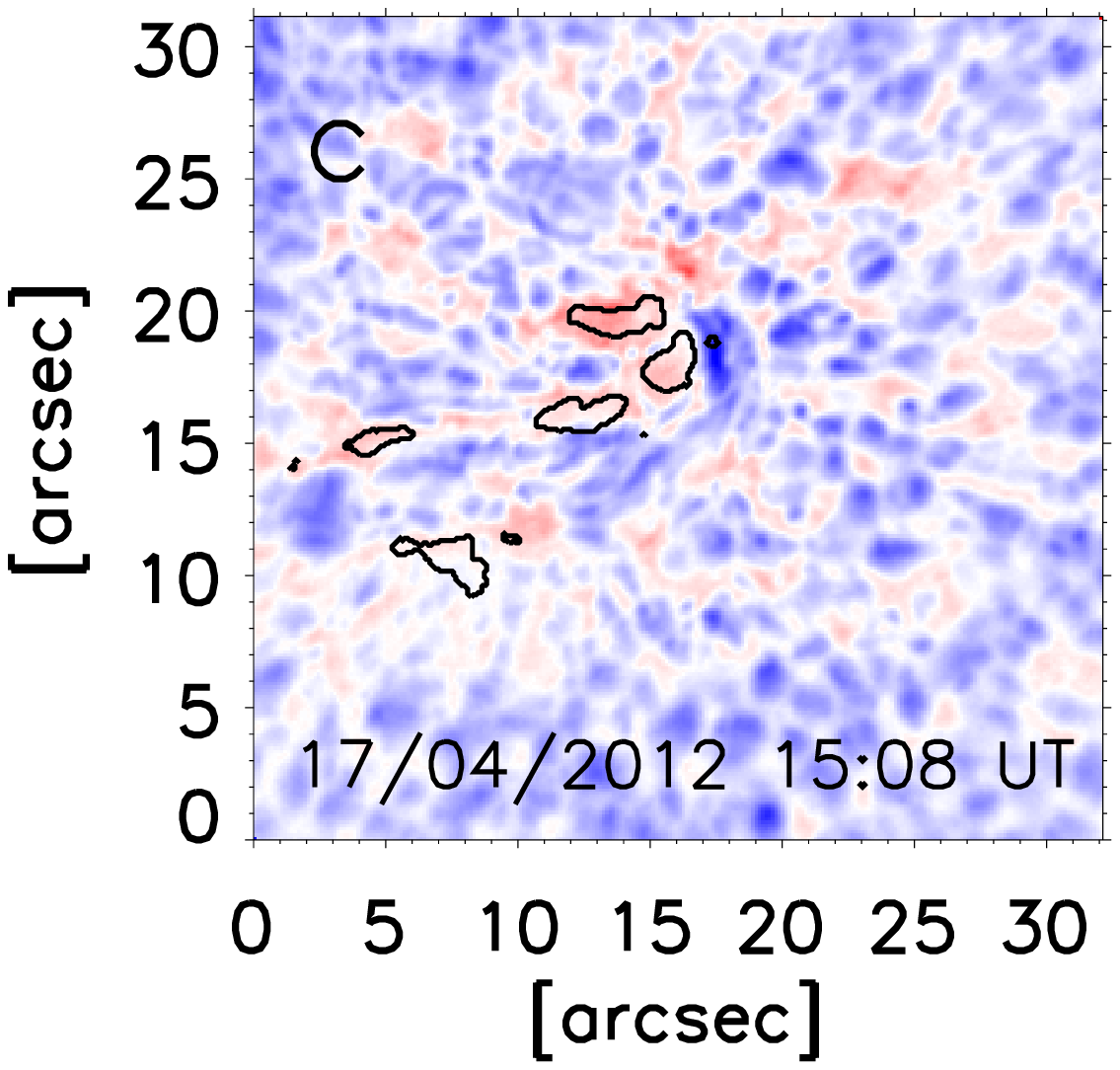}\includegraphics[height=3cm,trim=4.2cm 2.3cm 4.35cm 1.1cm,clip=true,keepaspectratio=true]{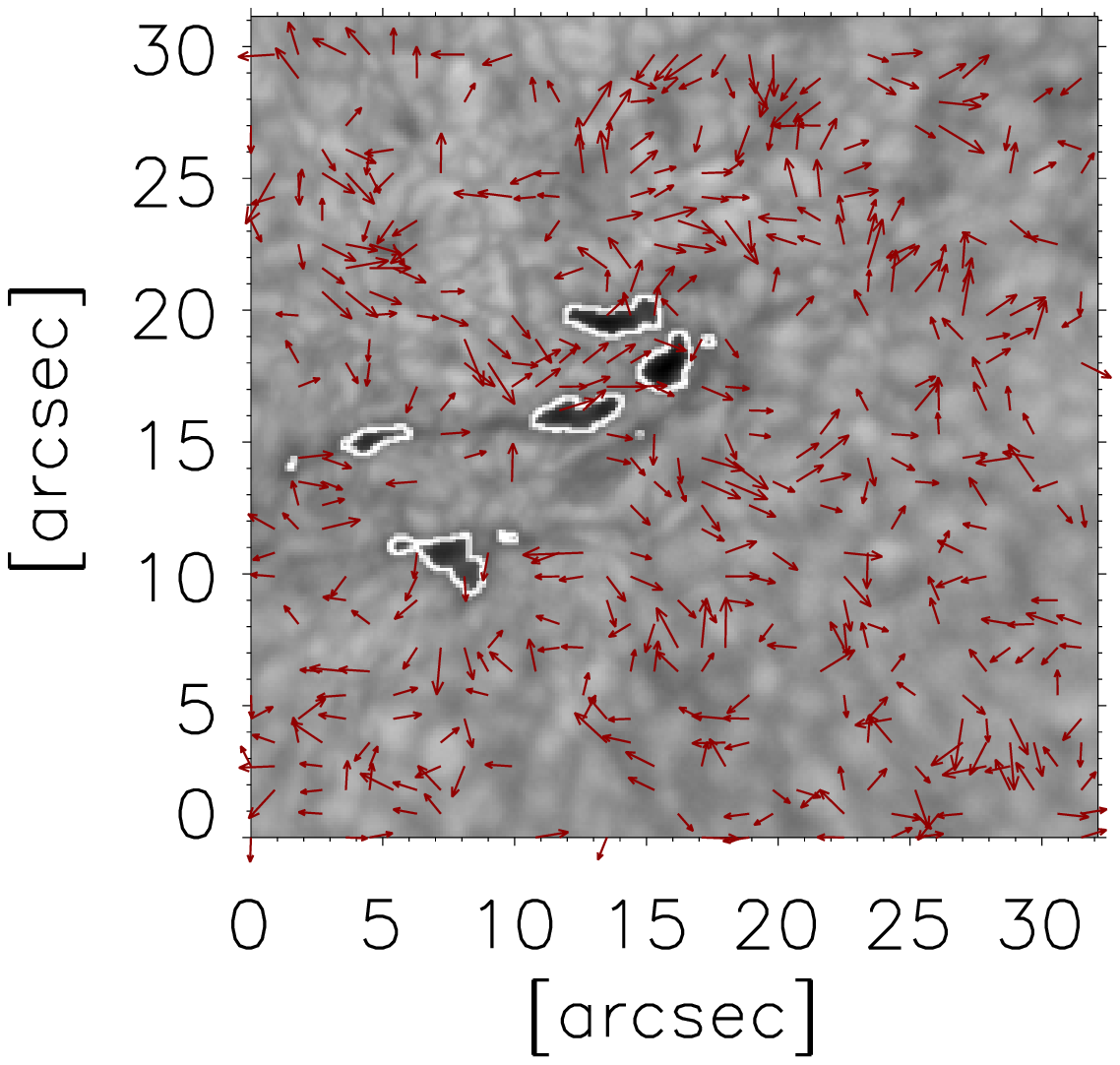}\\
\includegraphics[height=3cm,trim=1.5cm 2.3cm 4.35cm 1.1cm,clip=true,keepaspectratio=true]{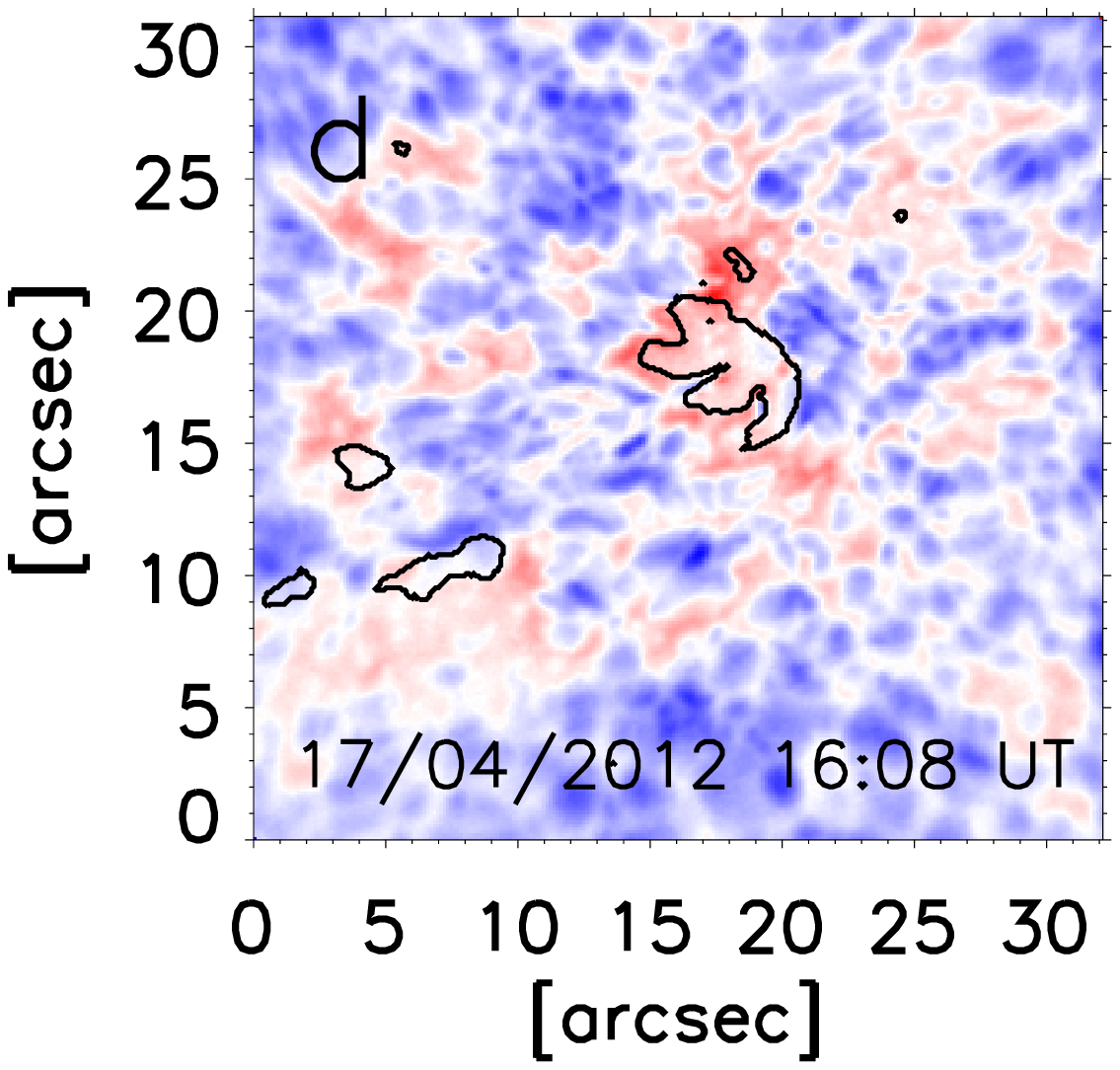}\includegraphics[height=3cm,trim=4.2cm 2.3cm 4.35cm 1.1cm,clip=true,keepaspectratio=true]{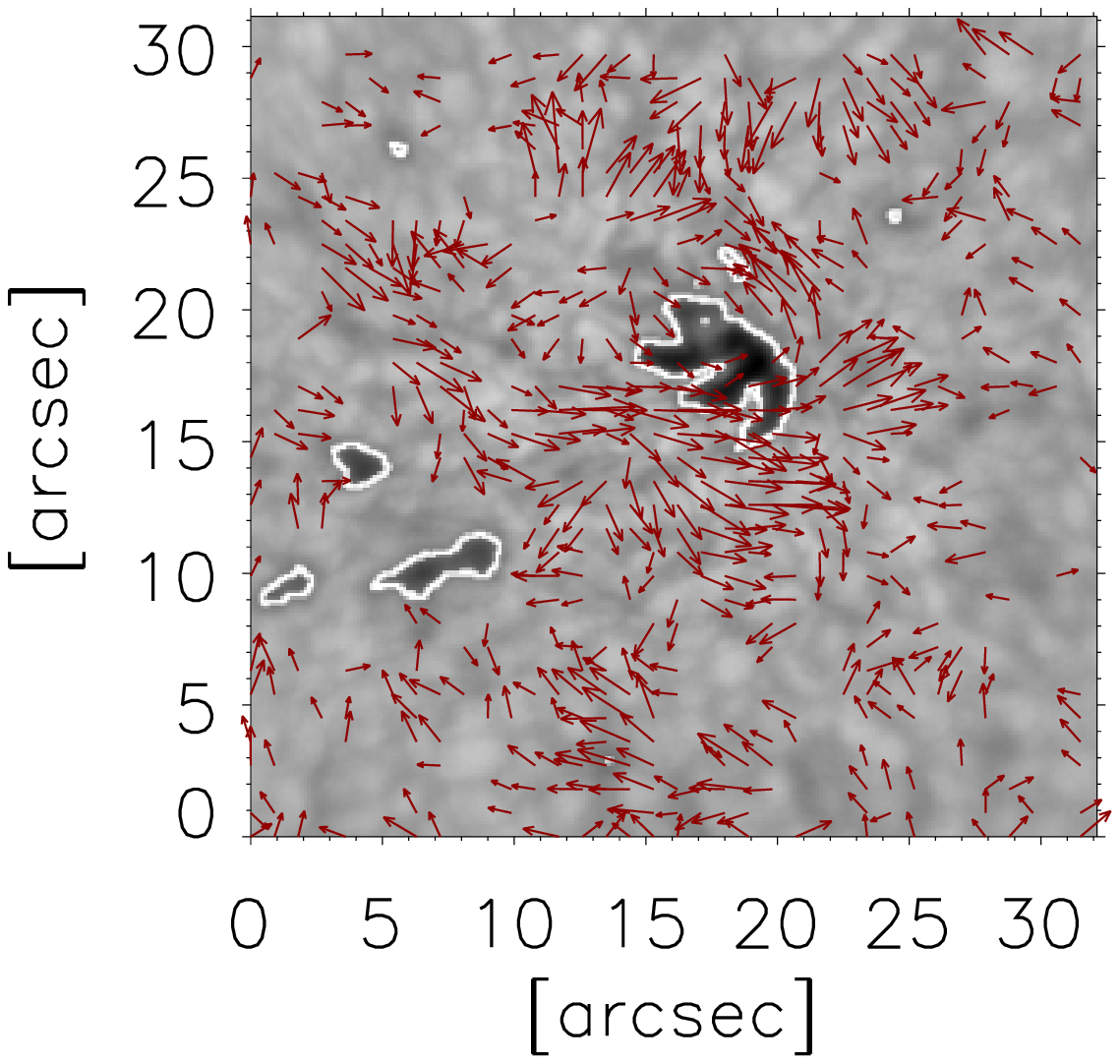}\\
\includegraphics[height=3cm,trim=1.5cm 2.3cm 4.35cm 1.1cm,clip=true,keepaspectratio=true]{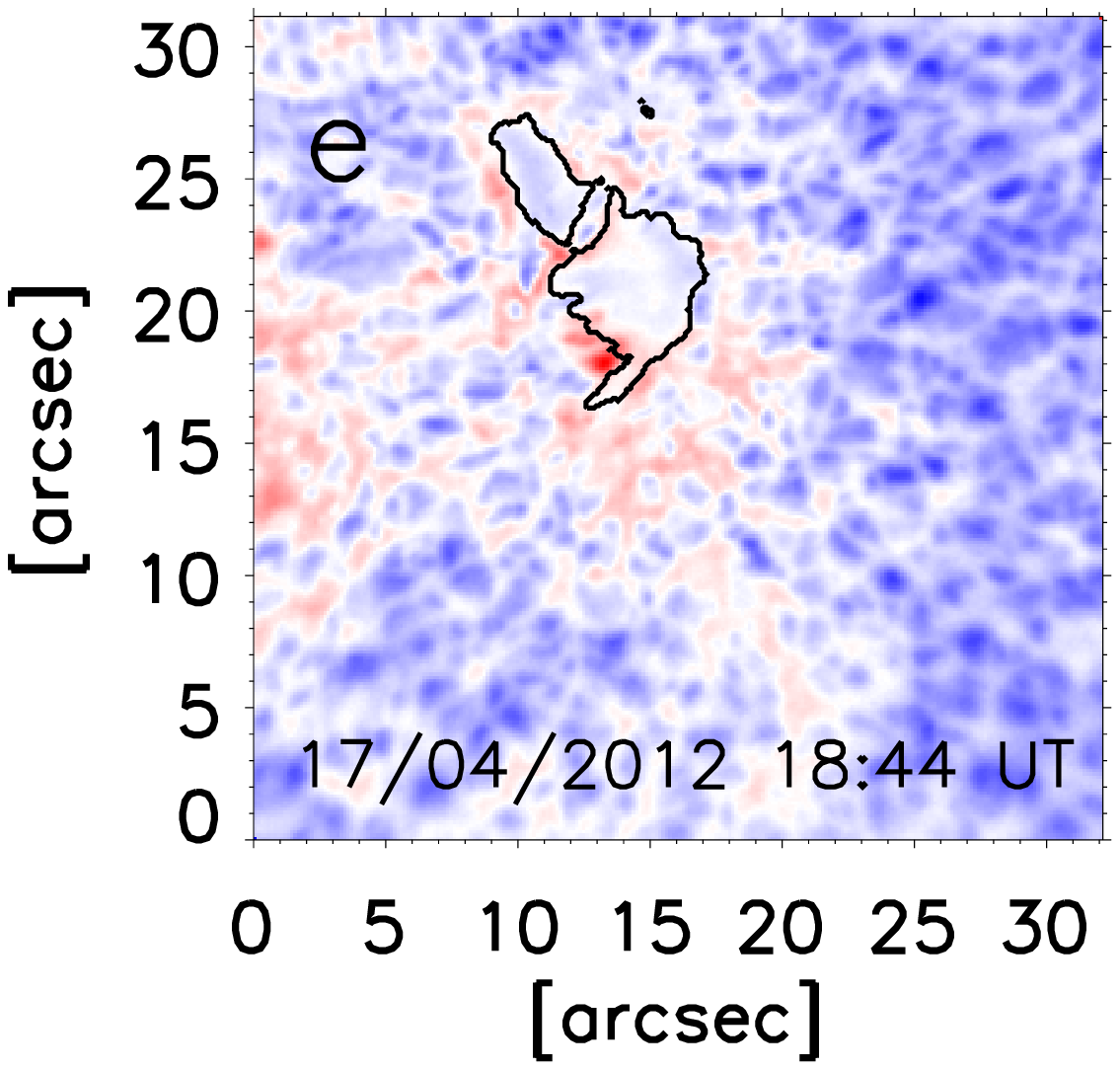}\includegraphics[height=3cm,trim=4.2cm 2.3cm 4.35cm 1.1cm,clip=true,keepaspectratio=true]{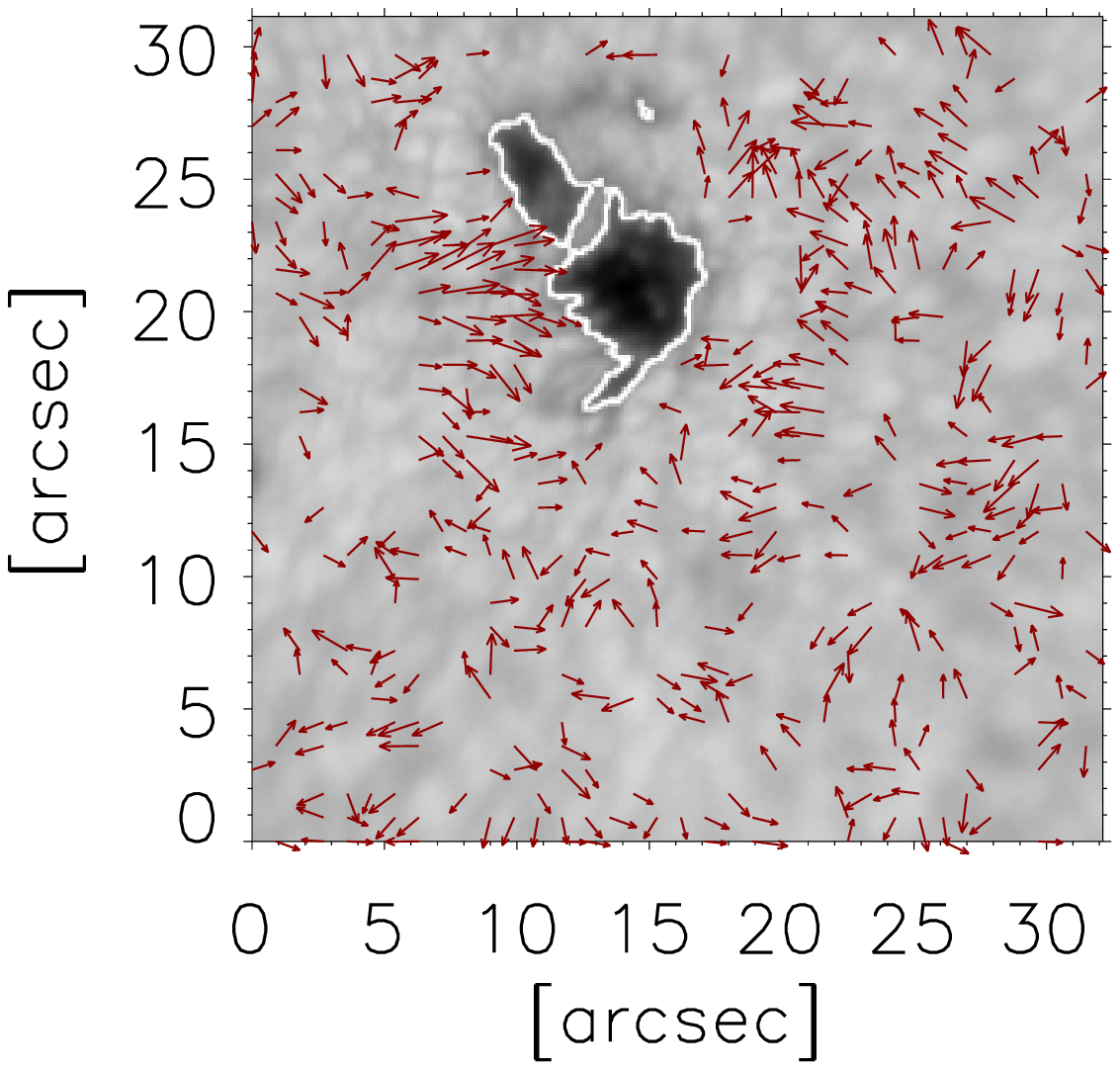}\\
\includegraphics[height=3cm,trim=1.5cm 2.3cm 4.35cm 1.1cm,clip=true,keepaspectratio=true]{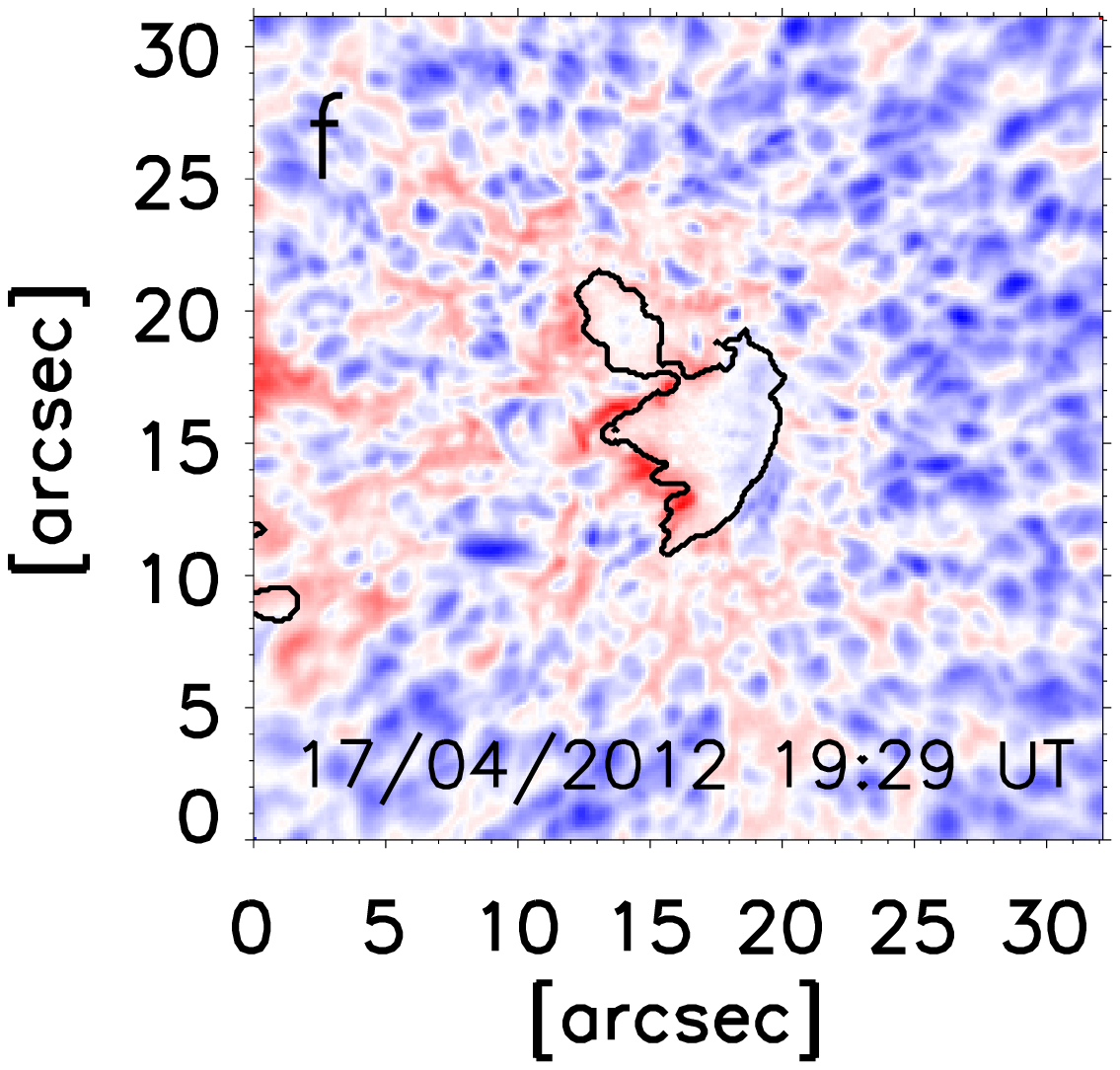}\includegraphics[height=3cm,trim=4.2cm 2.3cm 4.35cm 1.1cm,clip=true,keepaspectratio=true]{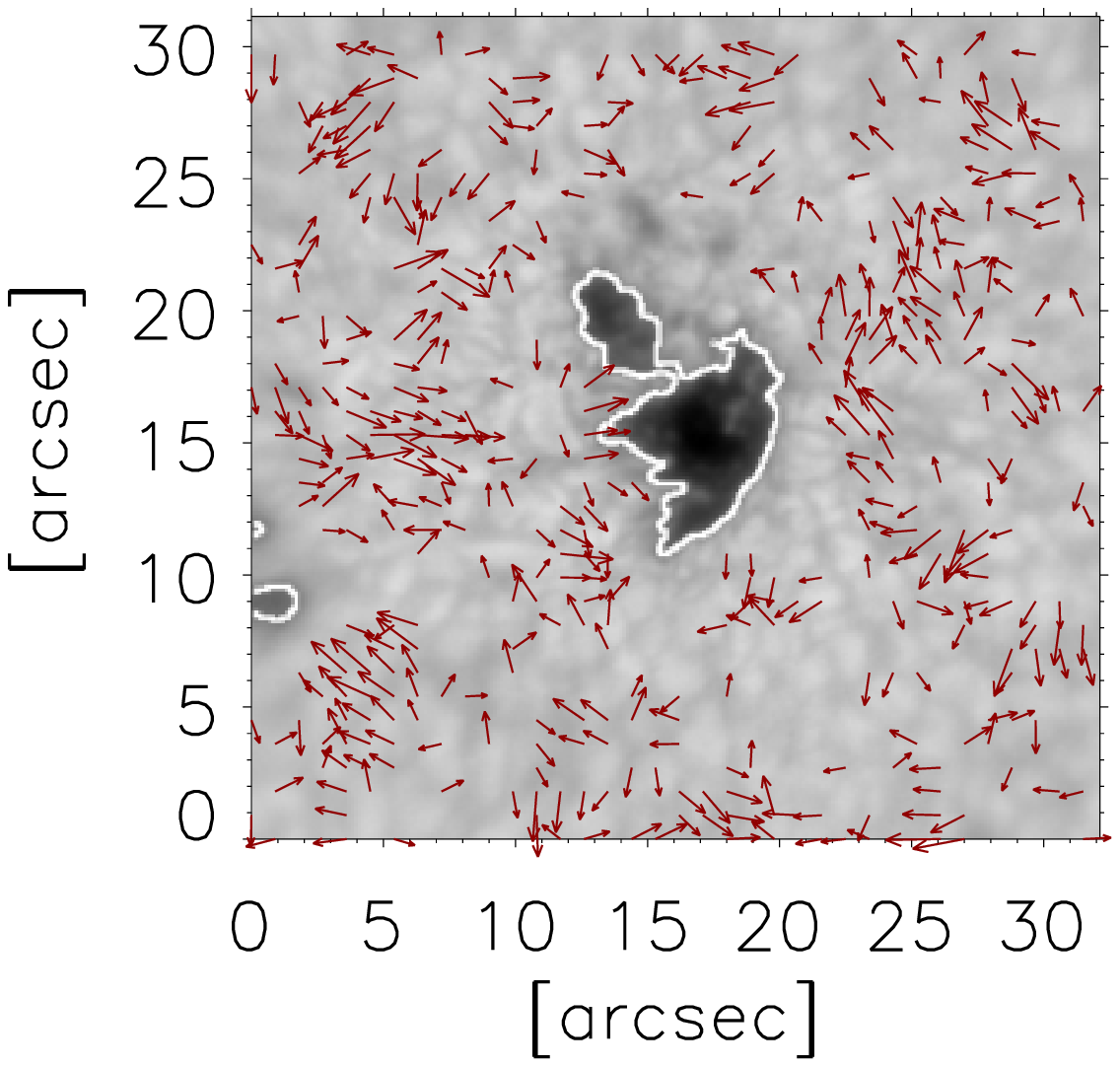}\\
\includegraphics[height=3.6cm,trim=1.5cm .5cm 4.35cm 1.1cm,clip=true,keepaspectratio=true]{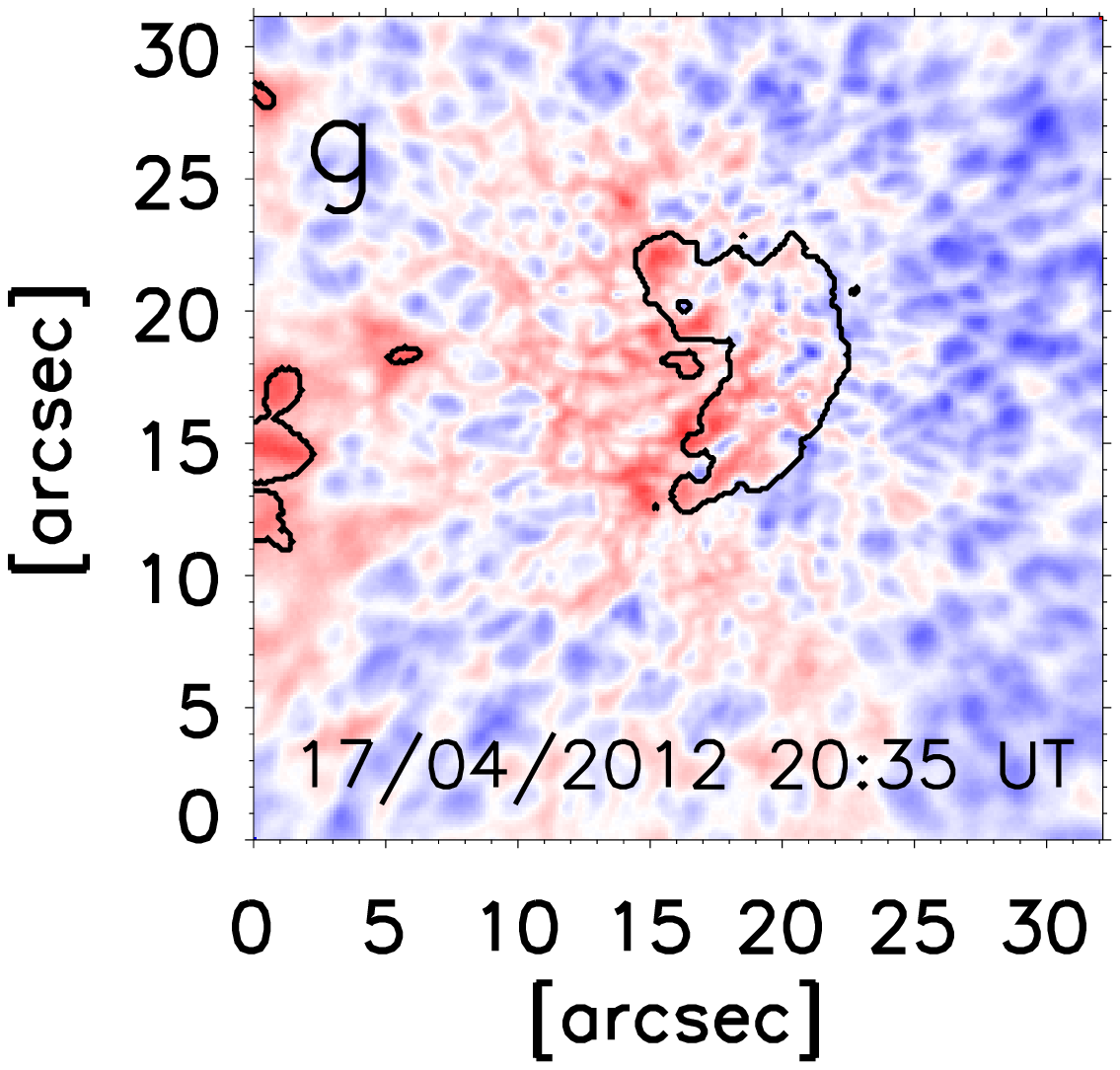}\includegraphics[height=3.6cm,trim=4.2cm .5cm 4.35cm 1.1cm,clip=true,keepaspectratio=true]{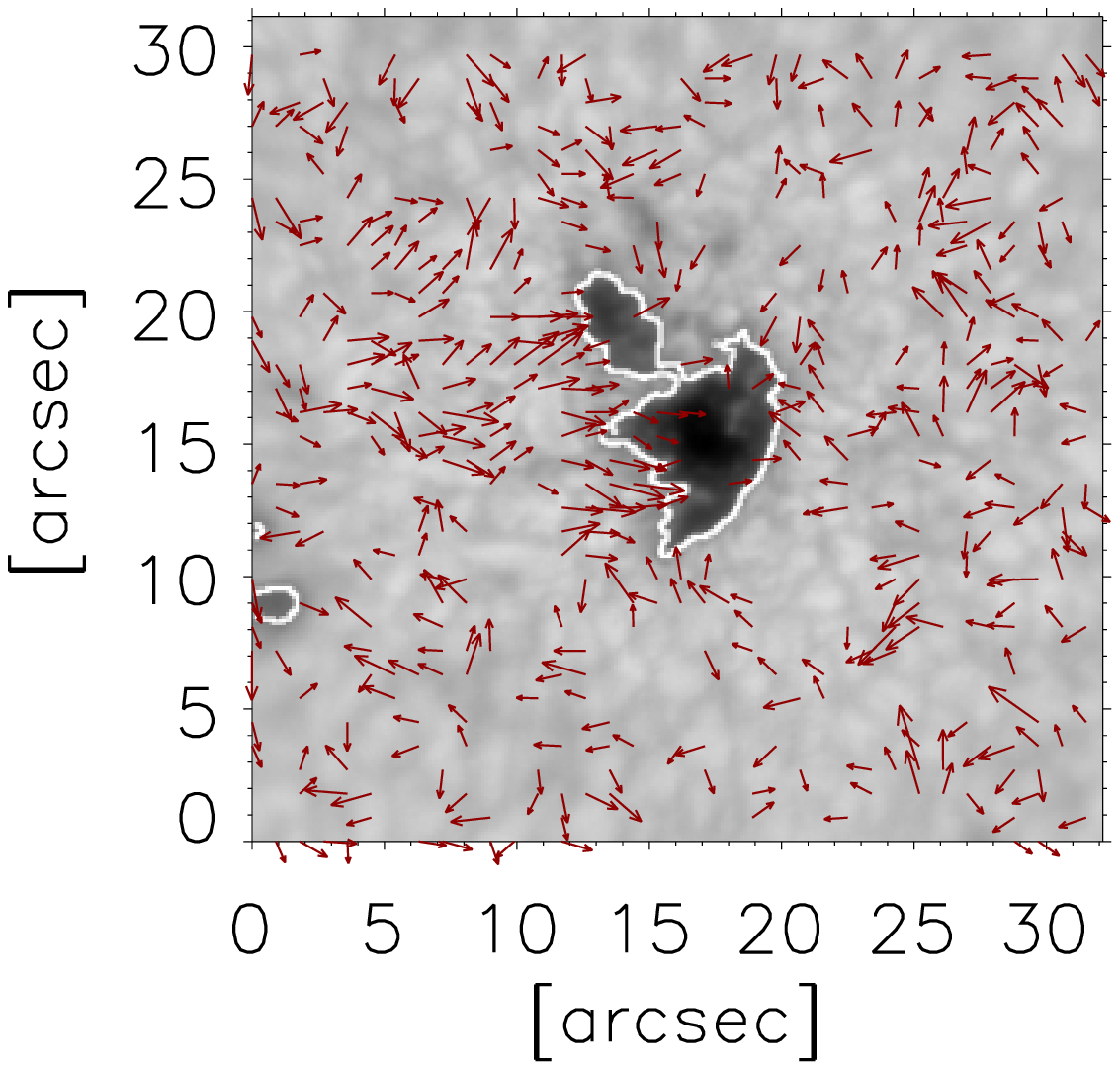}\\
}
\caption{\footnotesize{Example of  the v$_{LOS}$  (column 1) and v$_{H}$ (column 2) plasma velocity fields in the evolving region derived from the IBIS Fe I 617.3 nm  data at the seven stages (lines a-g) of the
pore formation shown in  Fig. \ref{f3} and specified in each panel of column 1.  
The blue (negative) and red (positive) v$_{LOS}$ indicate upflows and downflows, respectively.  
The intensity background in the v$_{H}$ maps shows the average image of the representative series. 
In panel 2a, the horizontal blue bar  indicates v$_{H}$ plasma velocity of 1 km/s. 
}}
\label{f5}
\end{figure}

Figure \ref{f5} shows examples of the v$_{LOS}$ 
and v$_{H}$ maps  derived from the IBIS  data at the same stages of the pore formation presented in Fig. \ref{f3}. 

The v$_{LOS}$ maps    (Fig. \ref{f5}, column 1) show    downflows  (corresponding to positive values in the velocity maps) characterized by velocities up to 2 km/s in the area of the evolving structure and its periphery.    %, in agreement with results presented in the literature, by e.g. \citet[][]{Sobotka_etal2012}. 
%These downflows are found to 
These downflows occur during the whole pore formation; as the feature evolves, they become stronger, which is likely as a consequence of the flux coalescence.  
They  are mostly located  at the northern edge of the feature during the initial and intermediate  evolutionary stages (Fig. \ref{f5}, lines a-d), and at its eastern and southern sides during the final stages (Fig. \ref{f5}, lines e-g). 
Upflows %(corresponding to negative values in the velocity maps) 
are suppressed in the evolving region during most of its evolution  (Fig. \ref{f5}, lines a-d), while they appear after the pore formation mostly at its western region (Fig. \ref{f5}, line g). 
%The photospheric   v$_{LOS}$ 
The maps also show convective upflows  around  the evolving region and localized downflows,  where magnetic flux accumulates (e.g. Fig. \ref{f5}, lines c, d, g, to the left side  of the map). These plasma motions are  also  characterized by  velocities   up to about  1-2 km/s.  It is worth noting that the positive (negative) values of plasma velocity  shown with red  (blue) colours  in the v$_{LOS}$ maps do not correspond to pure downflows (upflows) since they were derived from analyses of observations taken off disc centre.

The v$_{H}$ maps   (Fig. \ref{f5}, column 2)  show  plasma motions in agreement with those inferred from the analysis of the  SDO/HMI observations (Fig. \ref{f5b}). However, with respect to the latter maps,  those in Fig. \ref{f5}  more clearly represent the diverging motions of plasma seen  from the visual inspection of the available data in the close proximity of the evolving region.  
 At the initial stages of the pore formation, there are horizontal  motions  at  both sides of the elongated evolving structure (Fig. \ref{f5}, lines a-b), inwards  to its upper (northern)  edge and outwards beyond its lower (southern)  boundary; these  motions   push the evolving feature forward.  
At these stages, the v$_{H}$ maps  obtained from the  IBIS  data also clearly show the swirling motion of the plasma in the area of the evolving feature (see e.g. Fig. \ref{f5} lines a, d). 
After the pore formation, the maps show coherent motions  inwards directed around most of the eastward visible outline of the pore (Fig. \ref{f5}, line g).  After the formation of the funnel-shaped structure, the highest values in the v$_{H}$ maps are found to be located far from the visible outline of the evolving structure, at a distance of about  5-10 arcsec, as reported  for example by \citet[][]{Vargas_etal2010}.

\subsubsection{Temporal evolution} 

\begin{figure}%[ht!]
\centering{
\includegraphics[width=8.5cm,trim=1.1cm 0.3cm 0.7cm 0.3cm,clip=true,keepaspectratio=true]{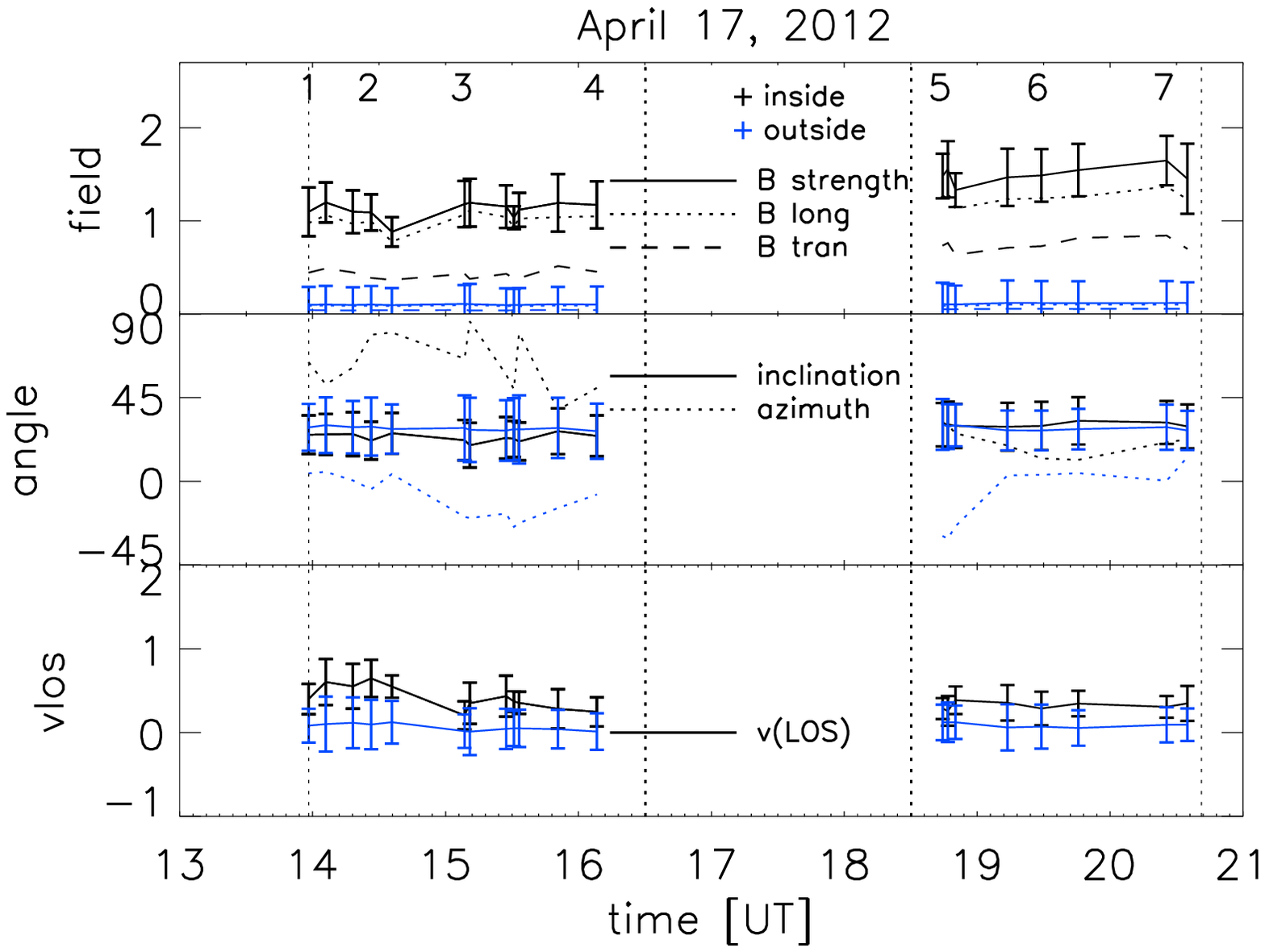}
}
\caption{\footnotesize{Variation of the magnetic field  strength (B strength), transverse (B tran), and longitudinal (B long) components in the LSF (top panel), field inclination, and azimuth in the LSF (middle panel), and  LOS plasma velocity v$_{LOS}$  (bottom panel) derived  from the IBIS photospheric   Fe I 617.3 nm line data during the pore formation, inside (black lines) and outside the evolving feature (blue lines). 
Each plot shows the mean and standard deviation of the values derived from the data inversion and other methods described in Sect. 2. Field, angle, and velocity values are given in kG, degree, and km/s units, respectively. For the sake of clarity, the standard deviation is shown only for  magnetic field strength, inclination, and v$_{LOS}$ estimates, inside and around the evolving structure. The vertical dotted lines show the time interval lacking IBIS observations. The 
numbers in the upper panel indicate the seven times corresponding to the evolutionary stages shown in Figs. \ref{f3}, \ref{f5}. }}
\label{f9a}
\end{figure}

Figure \ref{f9a}  shows the evolution of the physical quantities discussed above, as derived from  the SIR inversion and other methods applied to  the IBIS   data.  % and from the other methods applied to the same  data, at the twenty times selected for the SIR inversion of data taken at good, stable 
%taken at the twenty analysed times.  %shown in Figs. \ref{f2} to \ref{f5}. 
 Each panel represents the average and standard deviation of the various quantities estimated inside and around the evolving feature. %Tables \ref{table1} and  \ref{table2}  summarize the average and maximum values of the various  estimated quantities  at the seven stages of the pore formation shown in Figs. \ref{f3}, \ref{f5}. 

Inside the evolving feature,  $B$   varies  from about  1  kG to 2 kG, by  reaching average values larger than  1.5 kG  after the formation of the  funnel-shaped pore (Fig. \ref{f9a}, top panel). 
  The maximum  (not shown in Fig. \ref{f9a})  and average values of $B$ increase in time, as well as the same quantities for both the $B_t$  and $B_l$ components. %mean and standard deviation of measured values vary from xx$\pm$0.1 to xx$\pm$0.4 kG  over the studied time interval. 
Outside the forming pore, the average value of $B$  ranges between 0  kG and 0.1 kG,   but there are  regions in this area  with magnetic field strength $\ge$ 1 kG. The average value of $B$  over the whole area does not change significantly during the pore formation;  the same holds for both the mean value of $B$  over the stronger field elements and the maximum value of 
$B_t$ in the whole area, while the maximum 
values of $B_l$ and $B$ outside the evolving pore 
slightly decrease ($\le$ 10 \%) over time.  %The fields  are found to be  rather  vertical at the location of the evolving structure. % and inclined by 80-120$^{\circ}$ near the edge of the evolving region. 

 During the interval   of the IBIS observations, $\theta$ does not change significantly (Fig. \ref{f9a}, middle panel).  The  average   and the range of v$_{LOS}$ values measured  over   the magnetic region  slightly decrease after the pore formation, compared to those estimated in previous stages  (Fig. \ref{f9a}, bottom panel);  outside the evolving pore, the velocity of the plasma motions does  not change significantly over  the analysed period.

\section{Discussion and conclusions}

The results derived from our analysis agree  with the outcomes of former studies   of steady pores mentioned  above; 
%in Sects. 1 and 3,  %observed at high resolution with ground- and space based telescopes.  %. observed  at high spatial resolution and  at various spectral bands with ground- and space-based telescopes.
%In particular, 
specifically, with the  magnetic  field patterns and strengths presented, for example by \citet[][]{Sobotka_etal2012}, 
%from analysis of %average properties of magnetic field in 
%pores  observed  at high spatial resolution and  at various spectral bands with ground-based telescopes. Besides,  
and the %patterns and strengths of the 
  v$_{LOS}$  and v$_{H}$  reported   by  \citet[][]{Cho_etal2010},  \citet[][]{Sobotka_etal2012},  \citet[][]{Sobotka_etal2013}, and \citet[][]{Verma_etal2016}, for example.
  % from analysis of pores observed at high resolution with ground- and space-based telescopes. 
  However, unlike previous studies, our data  also allow us to investigate  the 
  properties of the photospheric plasma  in the evolving region  during the  pore formation, and thus provide further observational constraints to numerical models of the AR evolution.

Magnetic fields and fluid motions are  coupled according to the equations of the  magnetohydrodynamics in 3D space and time.  %3D MHD simulations allow to investigate this coupling.
%For example, 
\citet[][]{Cheung_etal2008} performed simulations  to investigate   the interaction between  convection and a  magnetic flux tube rising into the photosphere, and discussed the outcomes with respect to observations from the HINODE mission. 
They reported that the rising flux tube expands because of the strong stratification of the convective zone by forming  a magnetic sheet that  acts as a reservoir for small-scale flux emergence events occurring at the scale of granulation.  They also found that   the interaction of the convective downflows and the rising magnetic flux tube undulates it to form serpentine field lines that emerge into the photosphere.

Later on, from an in-depth progression of the above simulations to  the formation of an AR,  \citet[][]{Cheung_etal2010}  
 showed that the above serpentine fields gradually coalesce to form larger magnetic concentrations that eventually form a pair of opposite polarity spots.  They also pointed out  that
%, although the mean flow pattern in the vicinity of the developing spots is directed radially outward, 
correlations between the magnetic field and velocity field fluctuations allow the spots to accumulate flux by  Lorentz-force-driven, counter streaming motion of opposite polarity regions.

From recent 3D MHD investigations of the 
%recent simulations  by  \citet{Rempel_cheung2014}. These simulations aimed at investigating 
effects of inflows in the evolution of ARs, 
% produce observational signatures  consistent  with  the outcomes 
%results from 
%of the above  numerical works. % described above. % and observations analysed in our study and presented in other papers in the literature. 
%In particular, 
\citet{Rempel_cheung2014} reported that,  in their simulated photosphere, 
the flux appears organized on a granular scale with mostly mixed polarity with magnetic flux of the order 10$^{21}$ Mx and average value  lower than 0.1 kG.   
After emergence, the simulated dipoles undergo horizontal diverging flows,  reaching an amplitude of up to  2 km/s. In the numerical domain, these  motions produce a progressive separation of the polarity of the dipoles that migrate in the opposite direction, by moving the opposite polarity flux away from the emergence regions.

 The above simulations 
have contributed to a long series of numerical studies  aimed at the identification of the mechanisms responsible for the formation and evolution of solar magnetic structures  %follow previous  similar studies   
(see e.g.  \citet{Cameron_etal2007,Cameron_etal2011},  \citet{Cheung_etal2008,Cheung_etal2010}, \citet{Fang_etal2012a, Fang_etal2012b, Fang_etal2014}, \citet{Kitiashvili_etal2010}, \citet{Martinezsykora2012,Martinezsykora_etal2015}, \citet{Stein_etal2011,Stein_nordlund2012}, \citet{Toriumi_etal2012,Toriumi_etal2013},  to mention those that have been presented  during the last decade). 
Although these studies can strongly differ  in terms of 
the initial conditions  and scales of the simulated processes,  they have all  %all the above numerical simulations
reproduced  some flux evolution  signatures in agreement with  observations. 
Hence the question  arises   on which processes unveiled  by the numerical studies can be considered robust  with both the observations and model assumptions.

In this regard, the  data analysed in our study show several  observational facts that are  consistent with the outcomes of the  MHD simulations presented by \citet[][]{Rempel_cheung2014}.  In particular, we found as follows: %of a magnetic dipole emerging into the photosphere and interacting with plasma flows. 
%For example: 
\begin{itemize}
\item  At the initial stages of the AR evolution, the analysed observations display
mixed polarity flux patches  organized on a granular scale;  the flux patches in the trailing region of the forming AR are more clearly seen than those in the leading part (e.g.  Fig. \ref{f1b} and attached movie).
\item 
After about 24 hours, the patches in the leading part become stronger than those in the trailing region and form a filamentary, sheet-like, coherent structure (e.g.  movie available online and Figs. \ref{f1b}, \ref{f1}, \ref{f2});  these are the observed initial stages of pore formation.
\item At that time, the %data show  %: 
%, at the aryl satges of the pore formation they show:
%they 
%show 
small-scale mixed polarity  patches hold flux of the order 10$^{21}$ Mx in an evolving  region with  an average field below 0.1 kG (e.g. Figs. \ref{f1b}, \ref{f1},  \ref{f3}).    %as reported by \citet[][]{Rempel_cheung2014},
%Besides, 
%Moreover, 
%
%he pattern of the emerging flux regions, the cancellation of surface flux and associated high-speed downflows, the convective collapse of photospheric flux tubes, the appearance of anomalous darkenings, the formation of bright points, and the possible existence of transient kilogauss horizontal fields
%
\item %at the initial stages of the pore formation, our data  show
The flux then  shows an increase  of comparable magnitude in both polarity patches of the evolving region (e.g. Fig. \ref{f1}). %, as reported by \citet[][]{Rempel_cheung2014}. 
%, as by the simulated emerging loop.
%\end{itemize}
\item At a later time, the data clearly display 
small-scale opposite polarity features that counterstream, coalesce, and reinforce flux accumulation at distinct sites (e.g.  movie available online  and Fig. \ref{f5b}).    
\item  At that time,  elongated granules  appear in close proximity of the evolving feature, mostly located in the region the opposite polarities of the evolving AR facing each other (e.g. Fig. \ref{f2});
\item The pore increases in size, while the sites of flux accumulation   move away from each other with clear horizontal diverging   motions  and a rather small increase of the average field  in the forming pore (e.g. Figs. \ref{f5b}, \ref{f3}, \ref{f5}). %, as reported by 
\item  
Horizontal diverging  motions     %is in agreement with results from our observations as reported in Fig.  \ref{f5}.  In the simulations, such plasma flows  
  seem to produce further aggregation of field of the same polarity  (e.g.  movies available online and Figs.  \ref{f5b}, \ref{f5});   the plasma velocity is up to 0.4 km/s in the forming pore and up to 1 km/s outside it.  
  \item Strong downflows, with  plasma velocity  > 1.5 km/s,   appear  near the periphery of the forming pore and where magnetic flux accumulates (e.g. Fig.  \ref{f5}).  
\item
%Furthermore, the simulations show  that 
Most intense field concentrations   %with values reaching 3-4 kG   
 occur near the edges of the magnetic regions in evolution (e.g. Figs. \ref{f3},  \ref{f5}) as due to  the confinement of the field by the ambient plasma motions. 
\item The analysed data show that the pore formation in the leading region of the AR occurs rapidly (< 1 hour); the evolution of the flux patch in the leading part is faster (< 12 hour) than the evolution (20-30 hour) of the more diffuse and smaller scale flux patches in the trailing region (e.g. supplemental movies and Fig. \ref{f1b}).
\item At the final stages of the AR evolution,  about 48 hour after the pore formation,  
 the evolution of the region  leads to the formation of a large-scale AR  with a magnetic flux of the order of 10$^{22}$ Mx.   
\end{itemize}

%While  
   From analysis of  two relatively isolated ARs observed from the SDO mission, 
   \citet[][]{Centeno_2012} already reported some observational signatures of AR formation consistent with the  3D MHD numerical simulation of a rising-tube process,  
 % reported observational signatures consistent wit the results of the above simulations from her study of two relatively isolated ARs observed during 24 hour of their life from the SDO mission. %The analysed data were taken at the Fe I 617.3 nm spectral band, with a cadence of 12 minute and a spatial resolution of 1 arcsec. 
%For example, 
 %\citet[][]{Centeno_2012}  showed 
  such as evidence  of a connection between  horizontal field patches and strong upflows, elongated granulation around the evolving ARs, and a mass discharge process through magnetic reconnection, % that can allow magnetic arcades rising into the solar atmosphere, 
as envisaged in the simulations of   \citet[][]{Cheung_etal2010} and  \citet[][]{Rempel_cheung2014}, for example. 
In contrast, from a recent study of 
%spectro-polarimetric observations of a young dipolar subregion developing within an AR, by 
%focussing on 
%the distribution of the vertical and horizontal components of the magnetic and velocity fields derived from 
HINODE spectro-polarimetric observations of a young dipolar subregion developing within an AR, 
\citet[][]{Getling_etal2016} presented 
 observational results that are considered to conflict with the signatures expected by the emergence of a flux-tube loop. In particular, they  reported a fountain-like 3D magnetic  structure of the studied  features and lack of large-scale (horizontal and vertical) flows over the evolving area,  which are  seen as two pieces of evidence of magnetic region generation by local convective dynamo mechanisms as %; this is  
 % This alternative scenario of AR formation is 
 envisaged  in the 3D MHD simulations, for example by \citet[][]{Stein_nordlund2012}.

{\bf 
Compared to the HINODE and SDO observations analysed, for example by \citet[][]{Getling_etal2016}  and  \citet[][]{Centeno_2012},  
the IBIS data considered in our study have spatial and temporal resolution that are higher at least by factors 3.5 and 11, respectively. By analysing the available data, we found that the   simulations of the rising-tube process
successfully reproduce both the  average properties of the physical quantities estimated in the studied region and  the mechanisms driving  the observed pore formation.
 In particular, 
the studied  pore  seems to result  from the emergence into the photosphere of a strong field formed in the solar interior, with some amplification and structuring effects of the initially emerged %moderate and weak  
field by surface plasma motions, as  evinced from the simulations for example by  \citet[][]{Rempel_cheung2014}. 
The above simulations  
also describe the evolution of the studied region  at different spatial and temporal scales fairly well.    
The signatures observed in the studied region, in contrast, differ from those presented by \citet[][]{Getling_etal2016}, which  support the  scenario of pore and larger scale magnetic region generation by local convective dynamo mechanisms. 

}

\begin{acknowledgements}
The authors wish to thank Serena Criscuoli, Han Uitenbroek, and the whole DST staff for its support during the observing campaign and data reduction. This study
received funding from the European Unions Seventh Programme for Research, Technological
Development and Demonstration, under the Grant Agreements of the eHEROES
(n 284461, www.eheroes.eu), SOLARNET (n 312495, www.solarnet-east.eu), and SOLID
(n 313188, projects.pmodwrc.ch/solid/) projects.  This work was also supported by the Istituto Nazionale di Astrofisica (PRIN-INAF-2014) and Italian MIUR (PRIN-2012). We thank the anonymous referee for useful comments and suggestions.

The National Solar Observatory is operated by the Association of Universities for Research in Astronomy under a cooperative agreement with the National Science Foundation.
\end{acknowledgements}

\bibliographystyle{aa}
\bibliography{ibis2012_reply}

%\online
%\begin{appendix}
\appendix
\section{Movies}
%\end{appendix}

%\begin{appendix}
\section{Examples data inversion results}

\begin{figure*}%[ht!]
%  \centering{\hspace {1cm} 1                       \hspace{2.5cm}                                     2            \hspace{2.5cm}                                                 3     \hspace{2.5cm}                        4  \hspace{2.5cm}                        5 \hspace{2.5cm}                        6     \hspace {1cm}    }\\
%
\centering{
\hspace{6.5cm}
\includegraphics[height=0.68cm,trim=3.5cm 9.75cm 5.8cm 1.cm,clip=true,keepaspectratio=true]{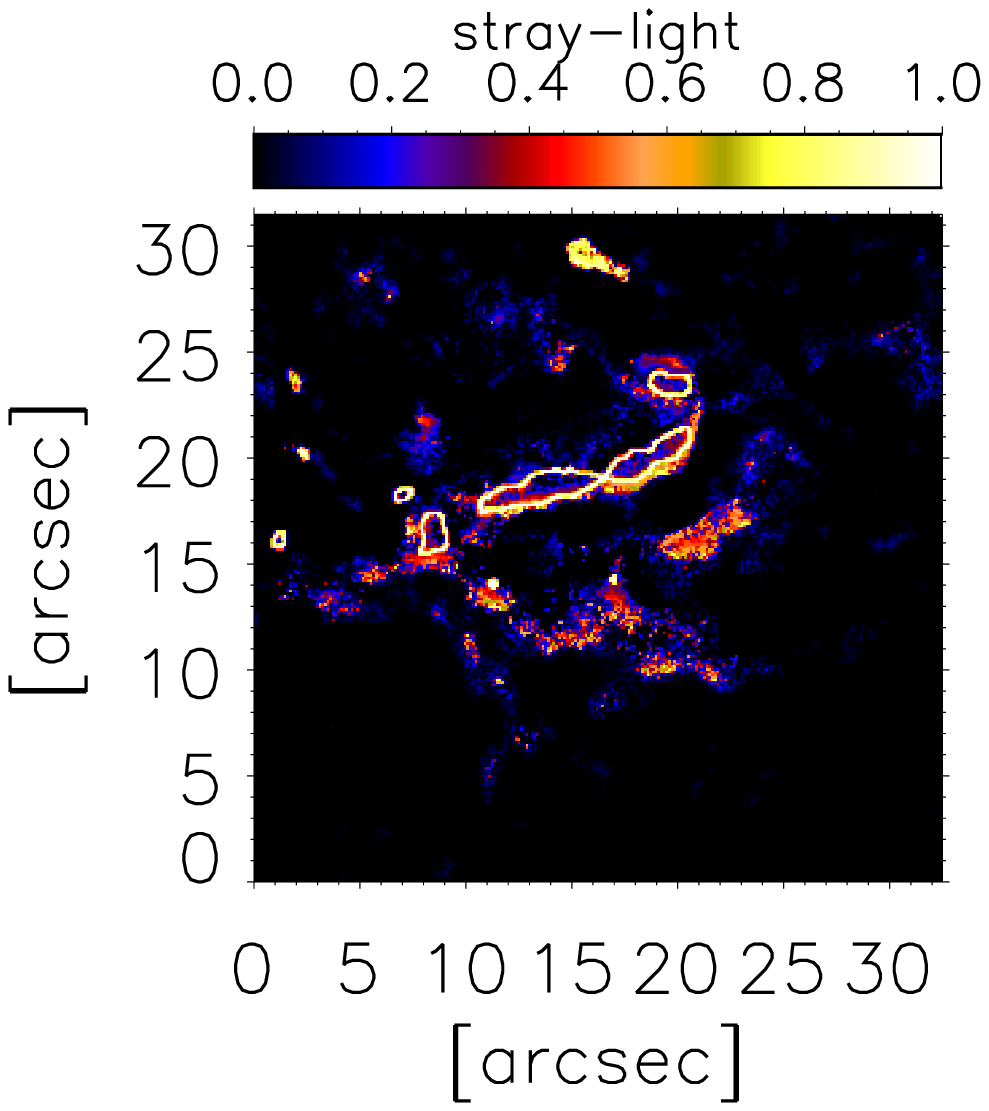}
\hspace{15.9cm}
}\\
\centering{
\includegraphics[height=3cm,trim=1.5cm 2.3cm 4.35cm 1.cm,clip=true,keepaspectratio=true]{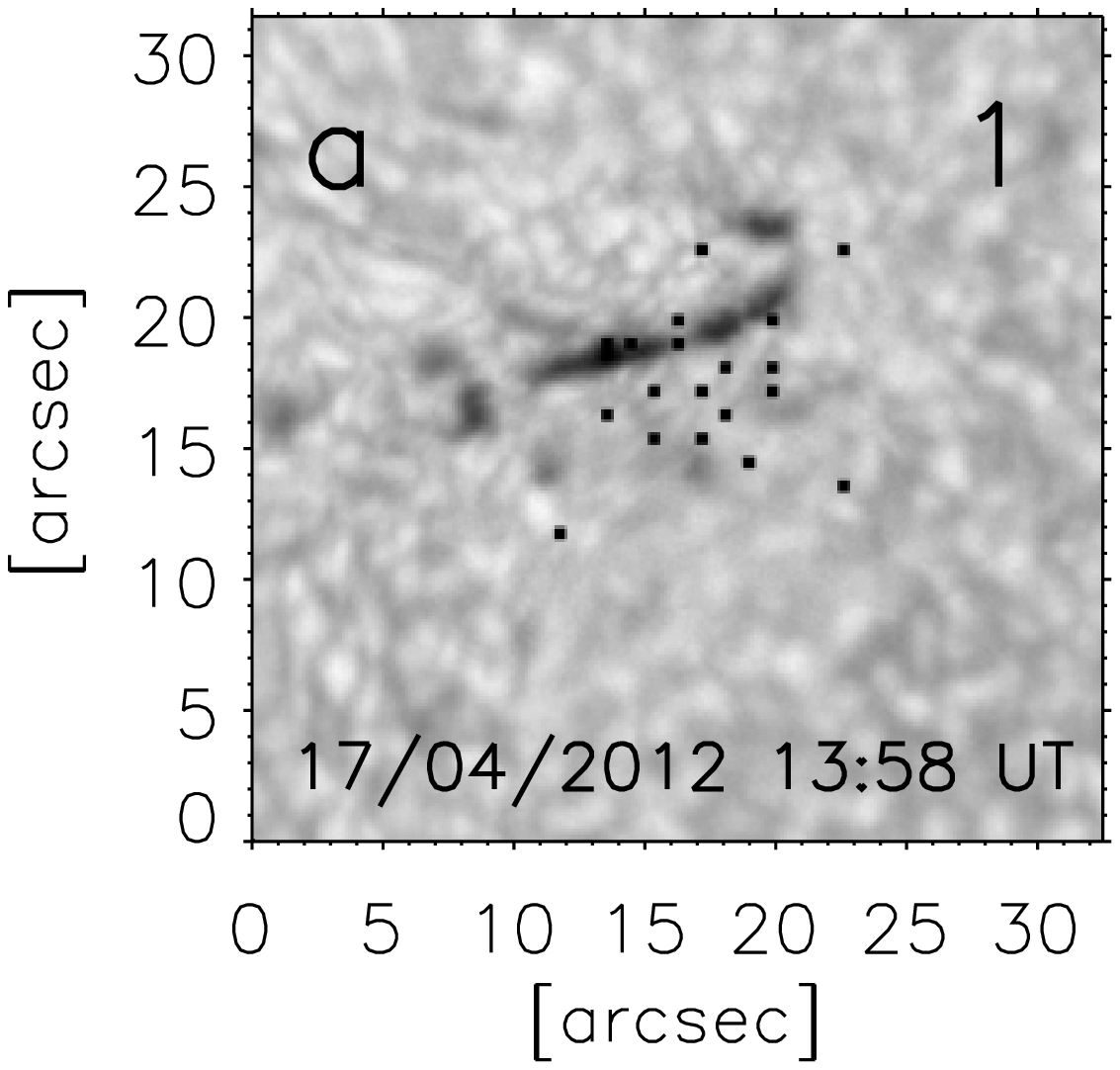}\includegraphics[height=3cm,trim=4.2cm 2.3cm 4.35cm 1.cm,clip=true,keepaspectratio=true]{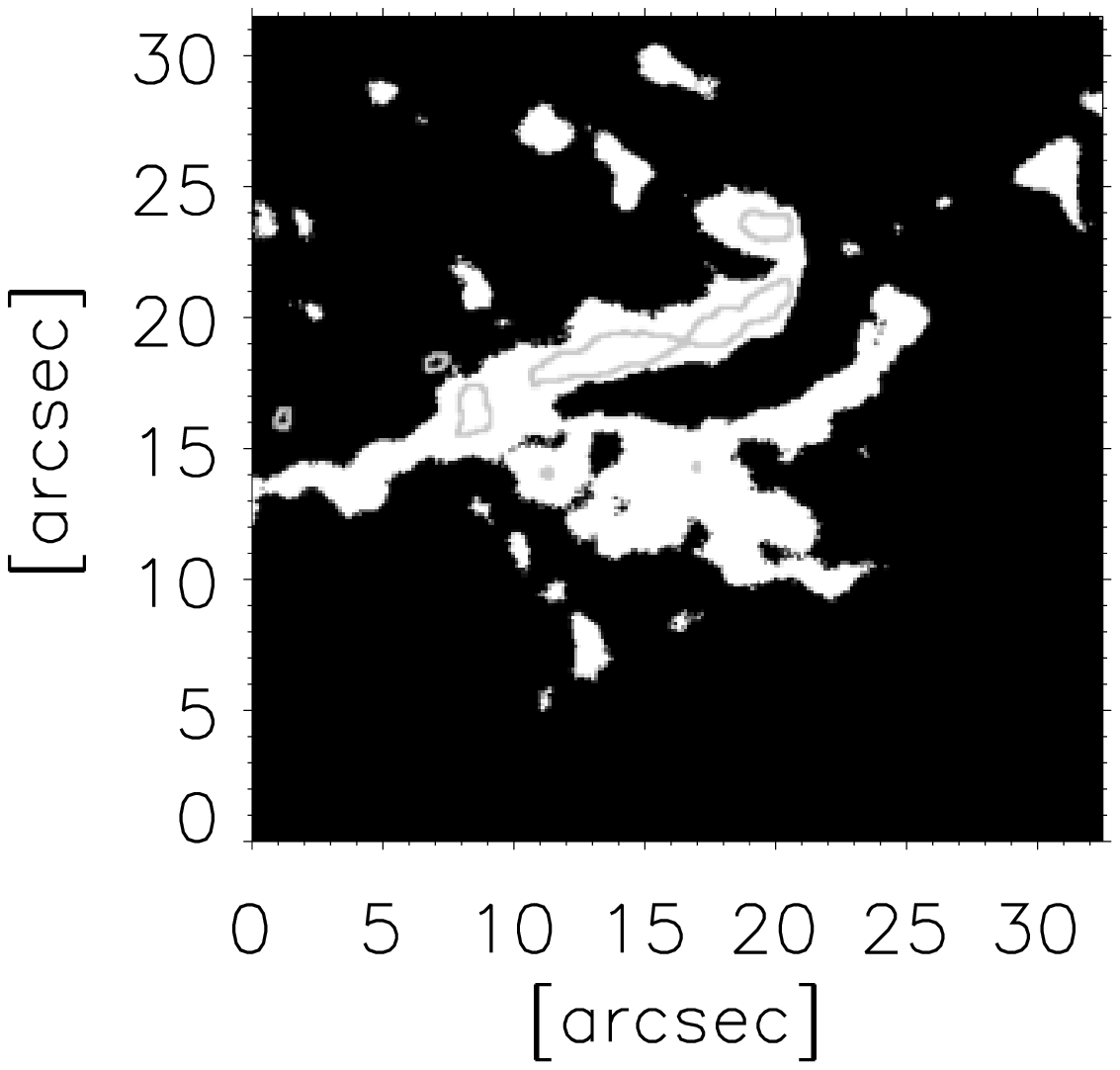}\includegraphics[height=3cm,trim=4.2cm 2.3cm 4.35cm 1.cm,clip=true,keepaspectratio=true]{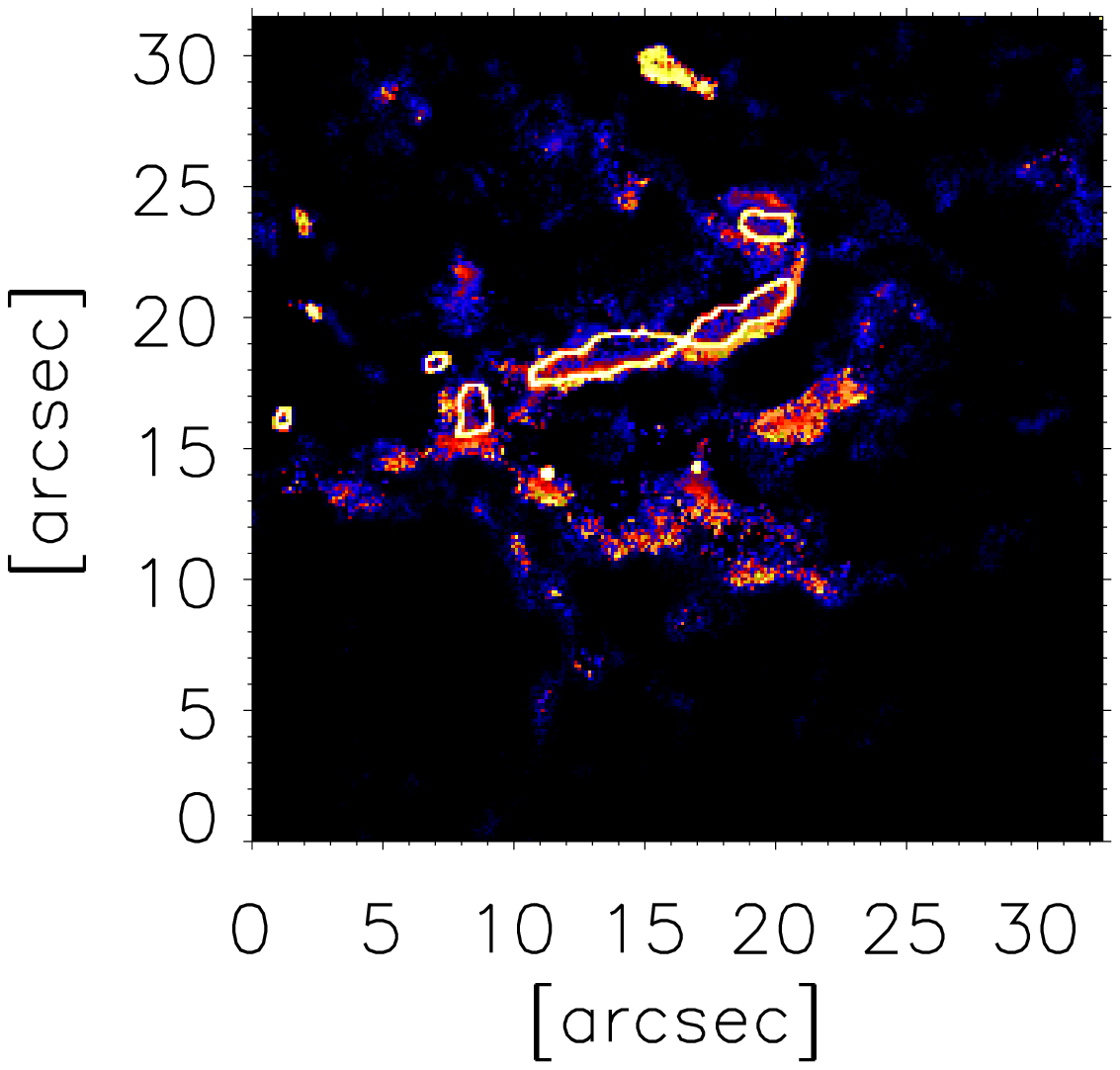}\\
\includegraphics[height=3cm,trim=1.5cm 2.3cm 4.35cm 1.cm,clip=true,keepaspectratio=true]{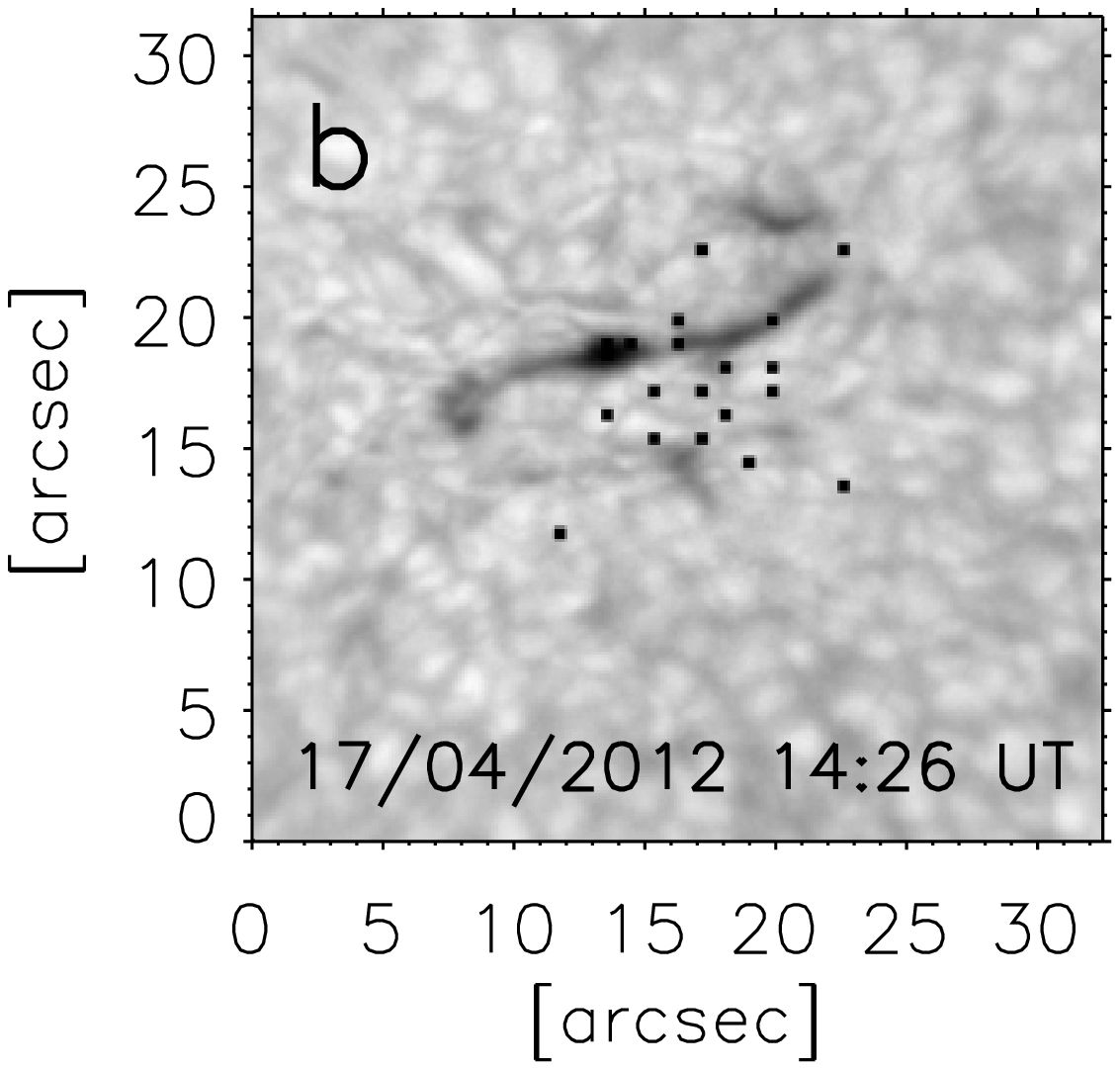}\includegraphics[height=3cm,trim=4.2cm 2.3cm 4.35cm 1.cm,clip=true,keepaspectratio=true]{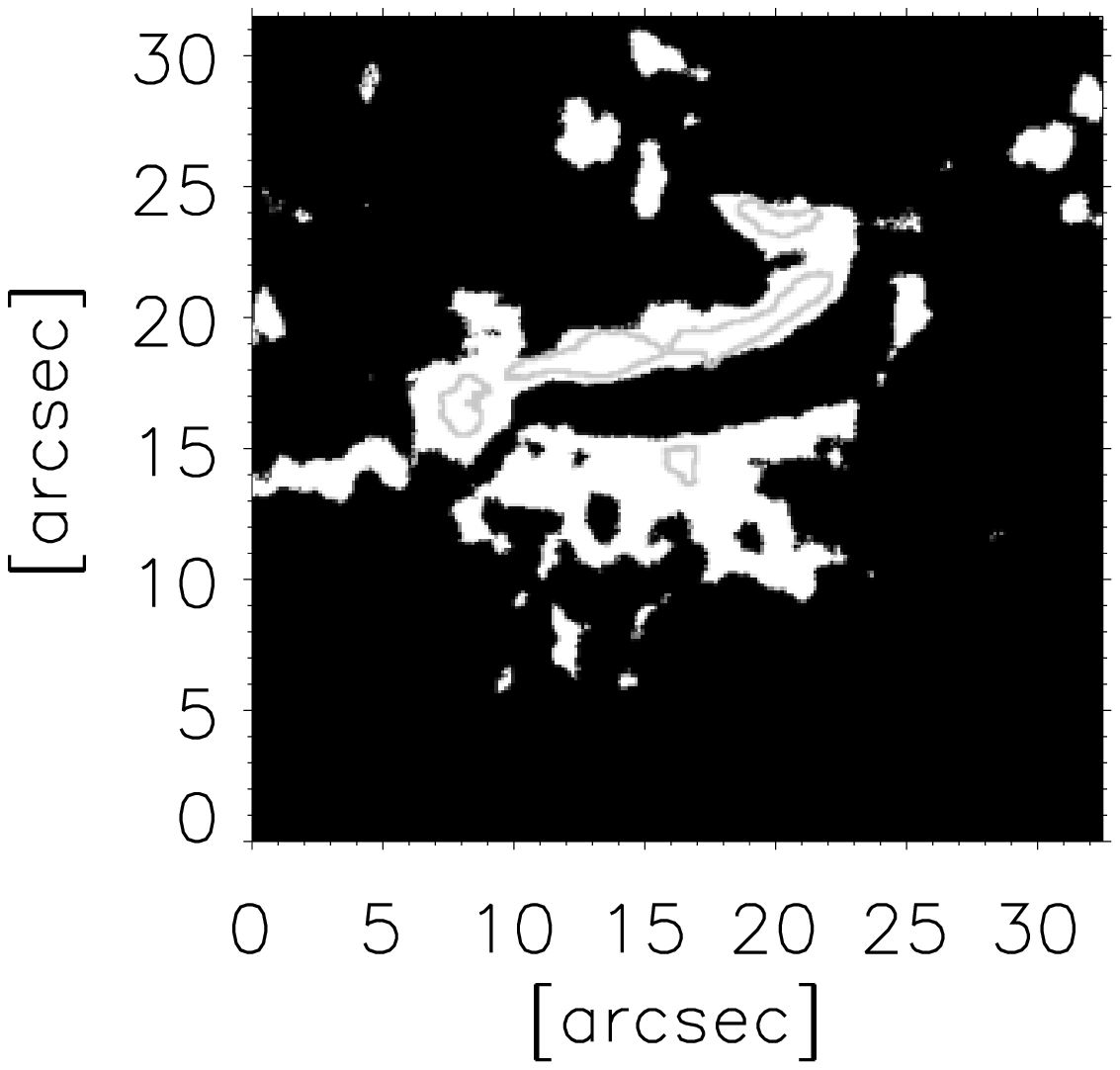}\includegraphics[height=3cm,trim=4.2cm 2.3cm 4.35cm 1.cm,clip=true,keepaspectratio=true]{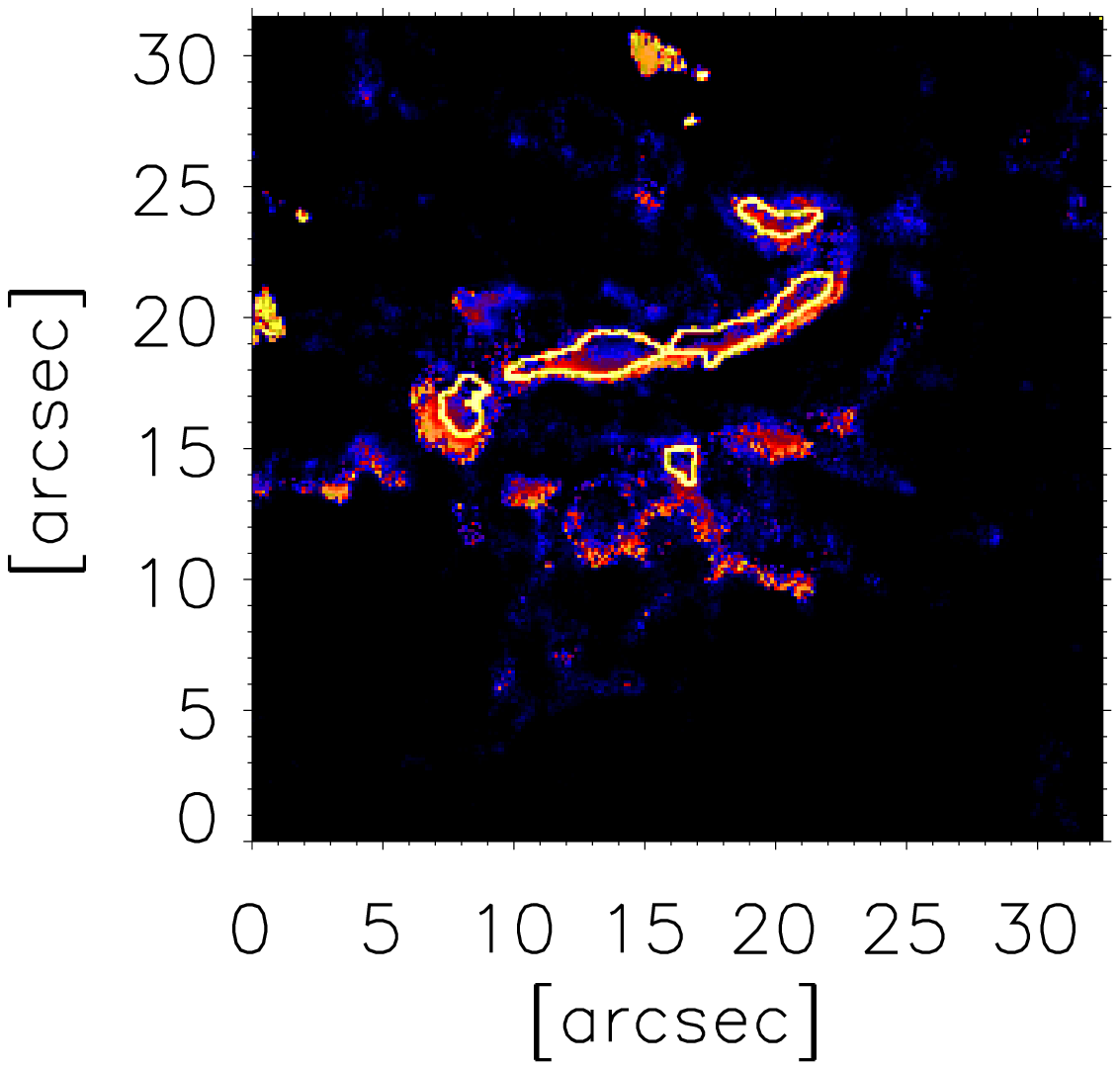}\\
\includegraphics[height=3cm,trim=1.5cm 2.3cm 4.35cm 1.cm,clip=true,keepaspectratio=true]{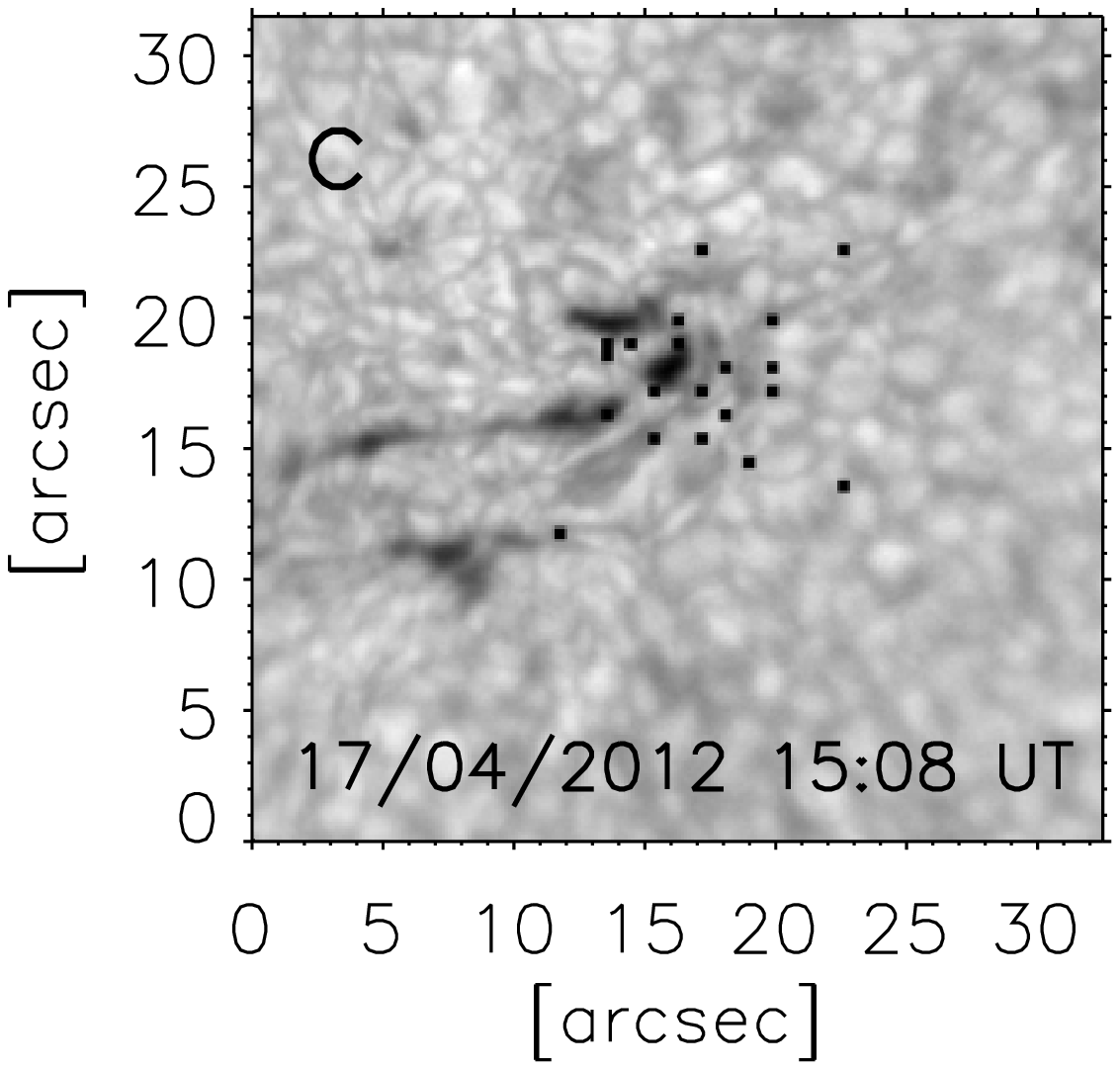}\includegraphics[height=3cm,trim=4.2cm 2.3cm 4.35cm 1.cm,clip=true,keepaspectratio=true]{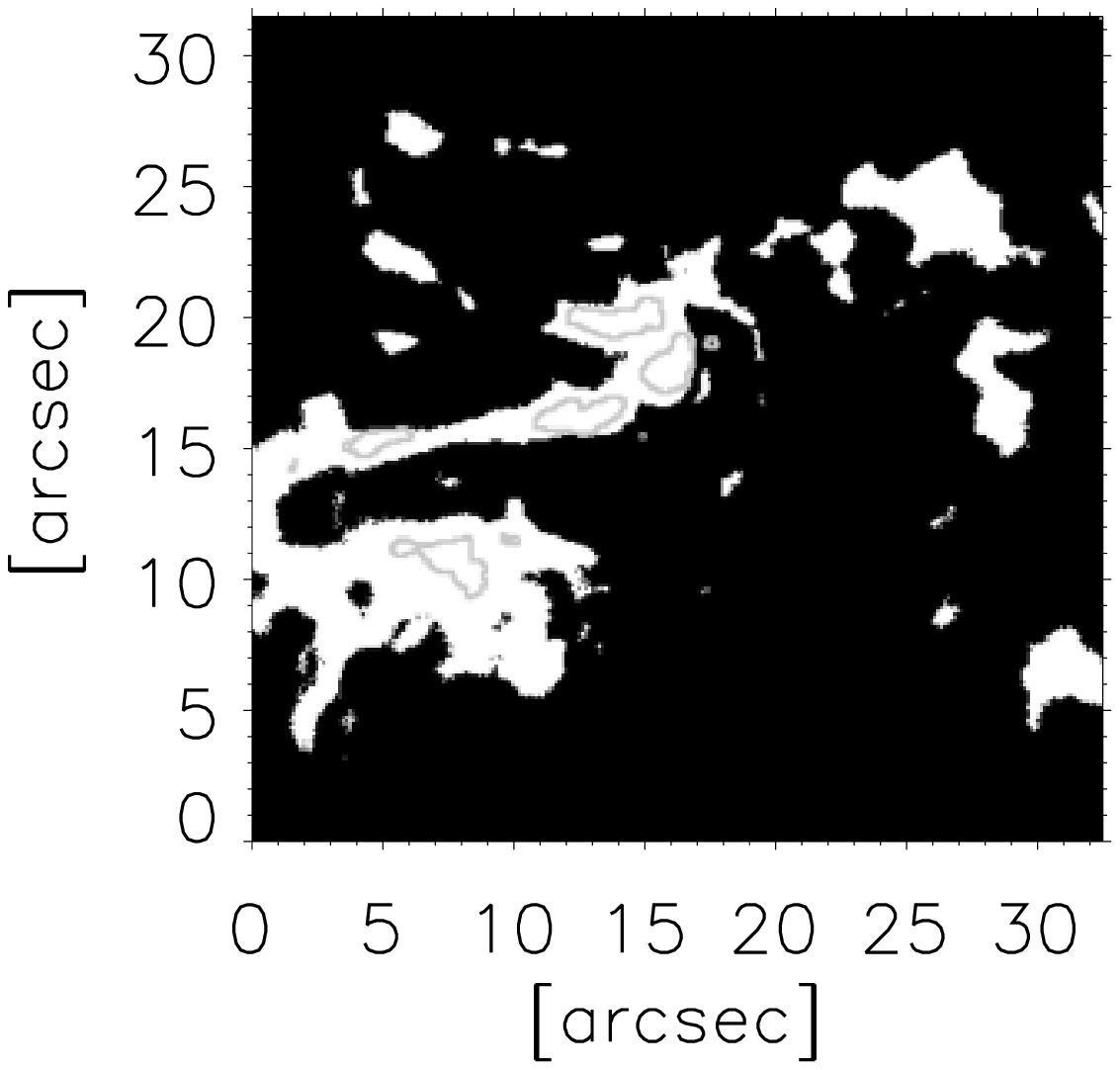}\includegraphics[height=3cm,trim=4.2cm 2.3cm 4.35cm 1.cm,clip=true,keepaspectratio=true]{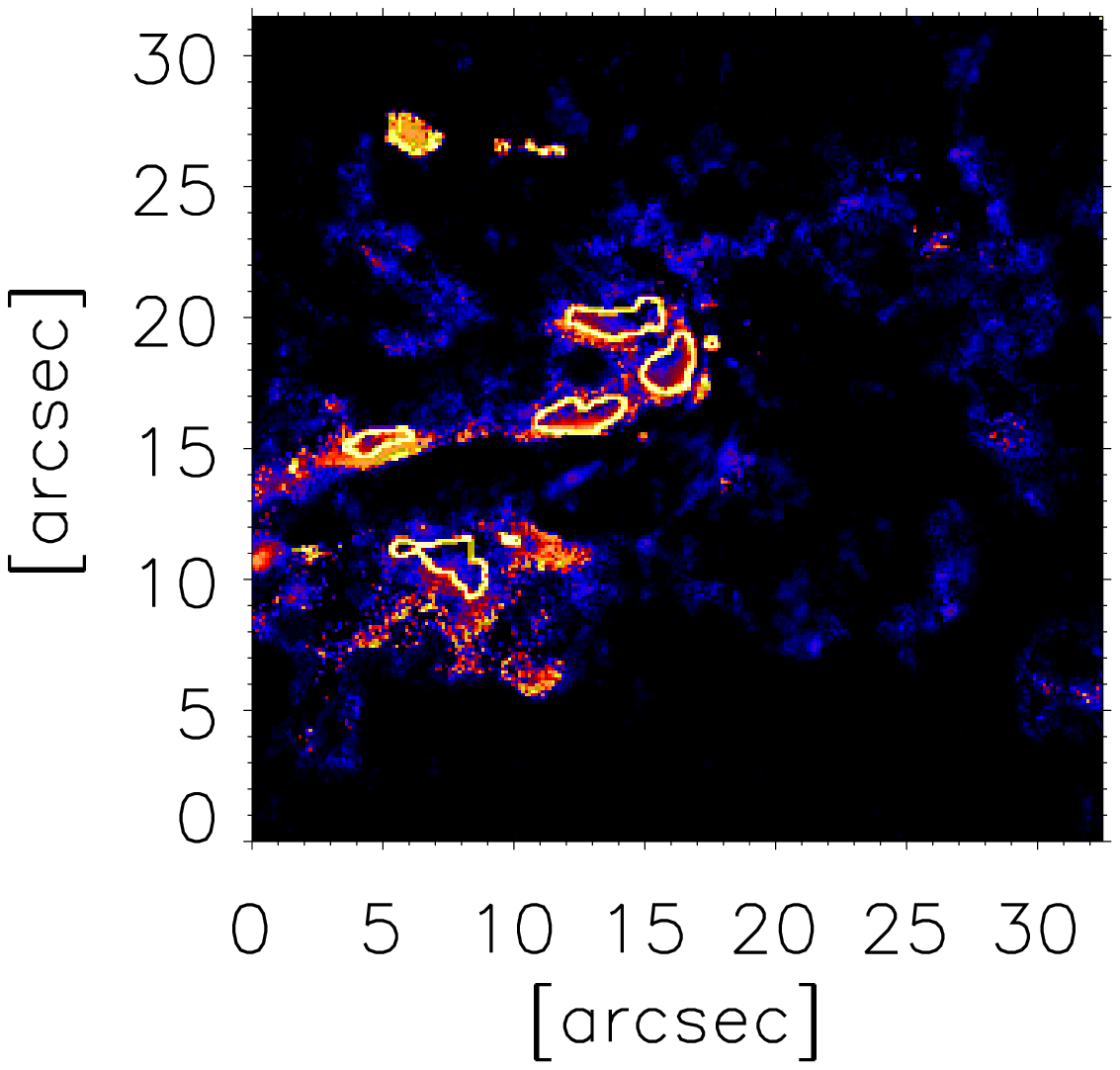}\\
\includegraphics[height=3cm,trim=1.5cm 2.3cm 4.35cm 1.cm,clip=true,keepaspectratio=true]{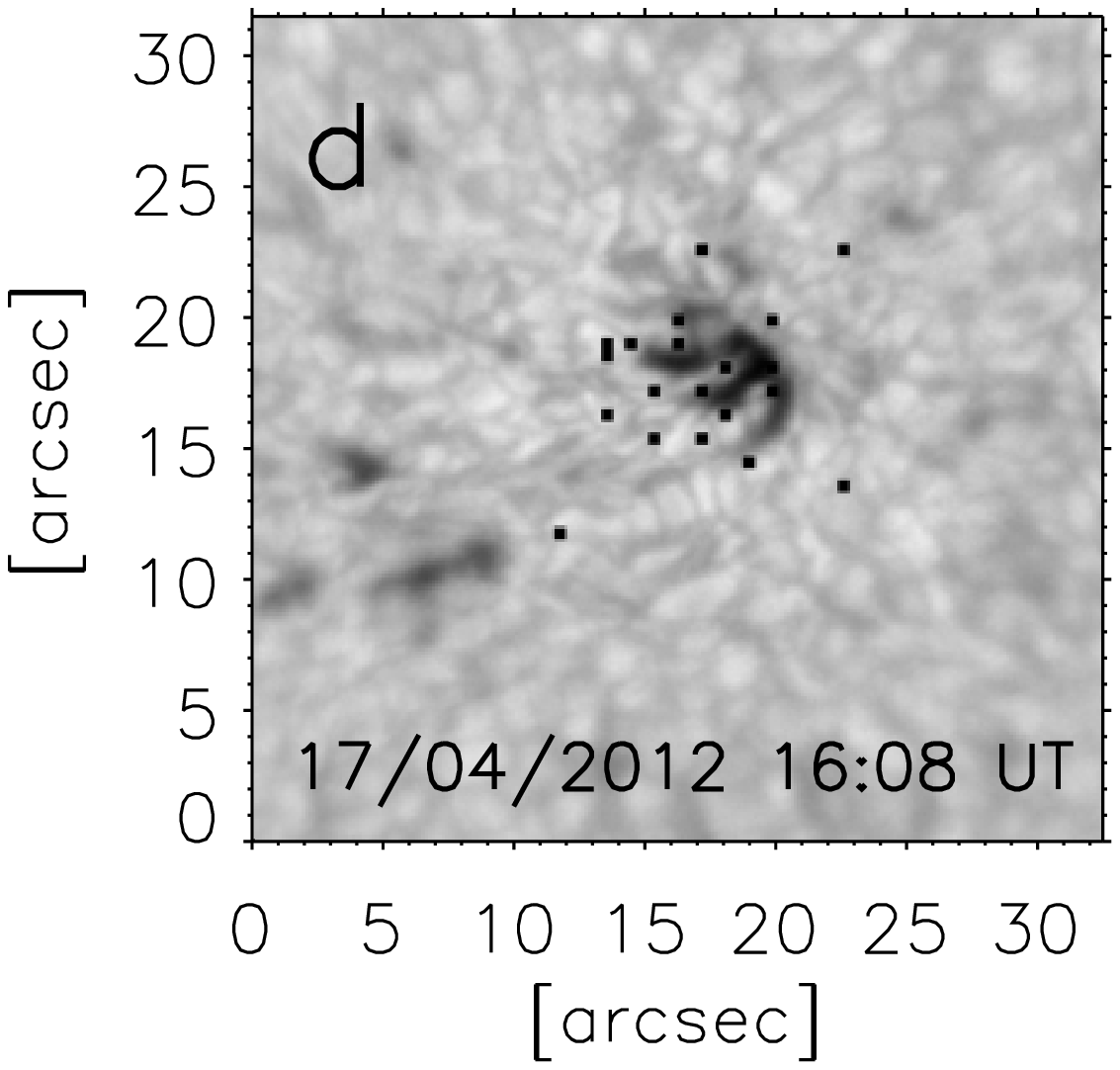}\includegraphics[height=3cm,trim=4.2cm 2.3cm 4.35cm 1.cm,clip=true,keepaspectratio=true]{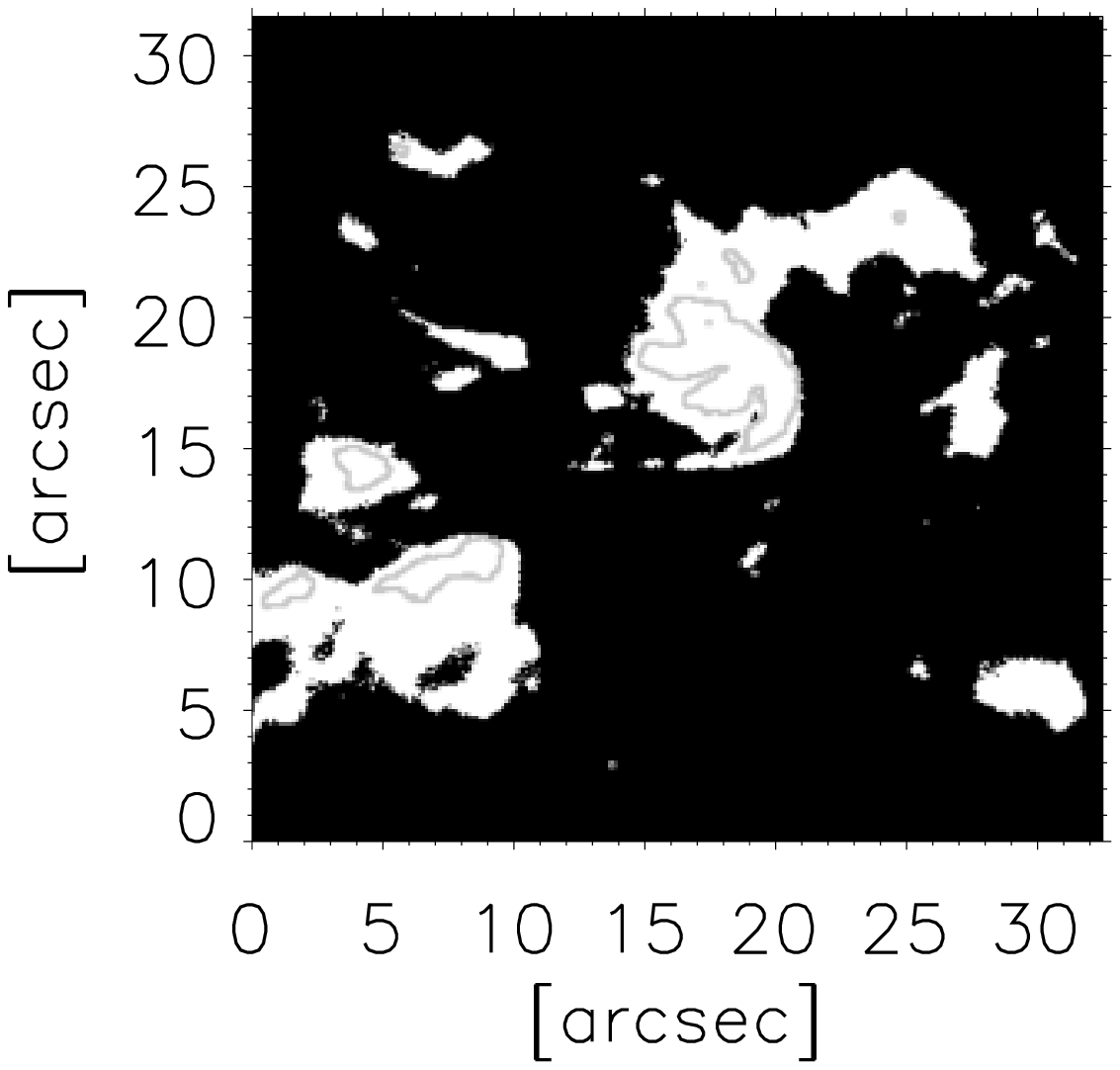}\includegraphics[height=3cm,trim=4.2cm 2.3cm 4.35cm 1.cm,clip=true,keepaspectratio=true]{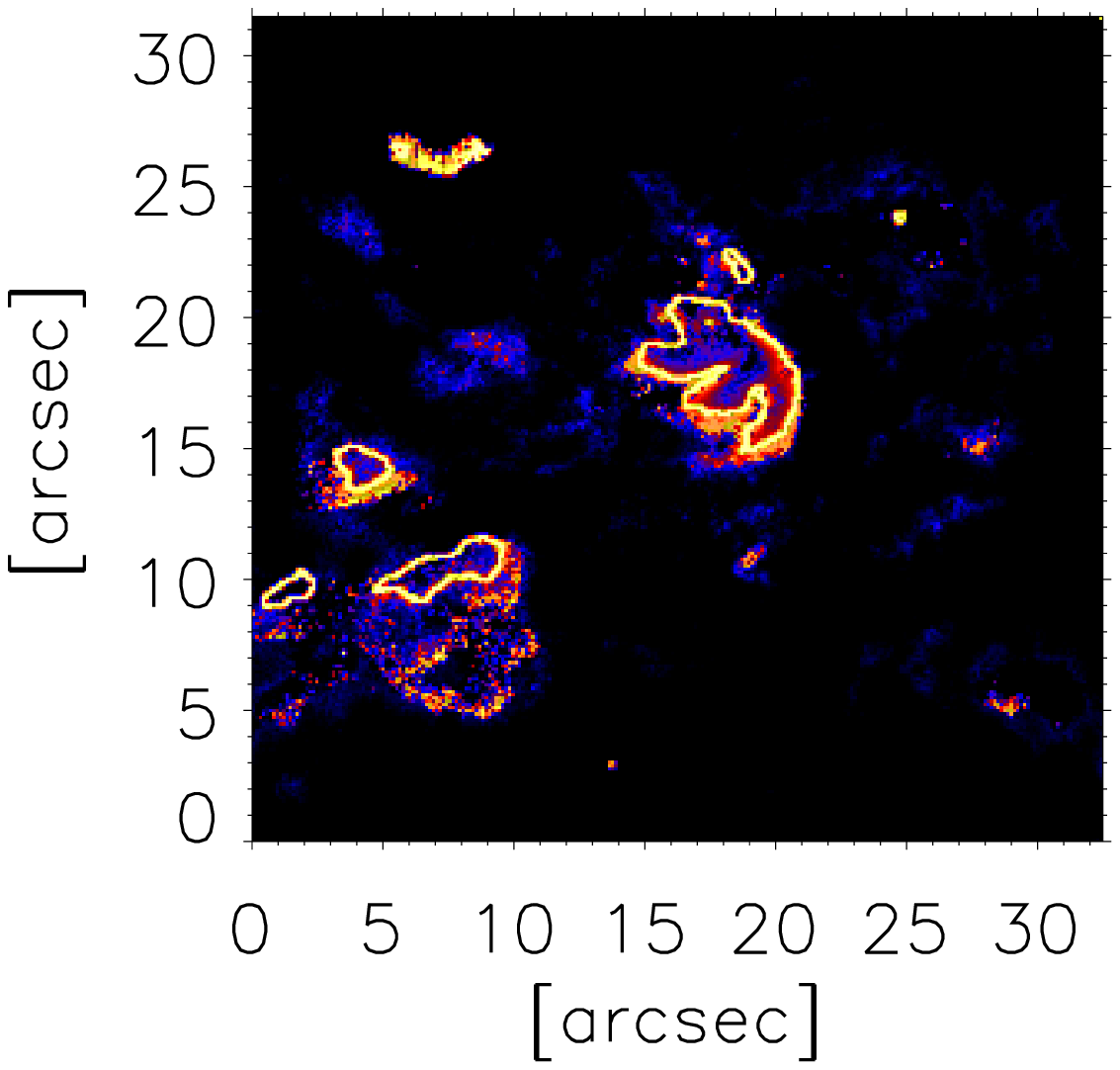}\\
\includegraphics[height=3cm,trim=1.5cm 2.3cm 4.35cm 1.cm,clip=true,keepaspectratio=true]{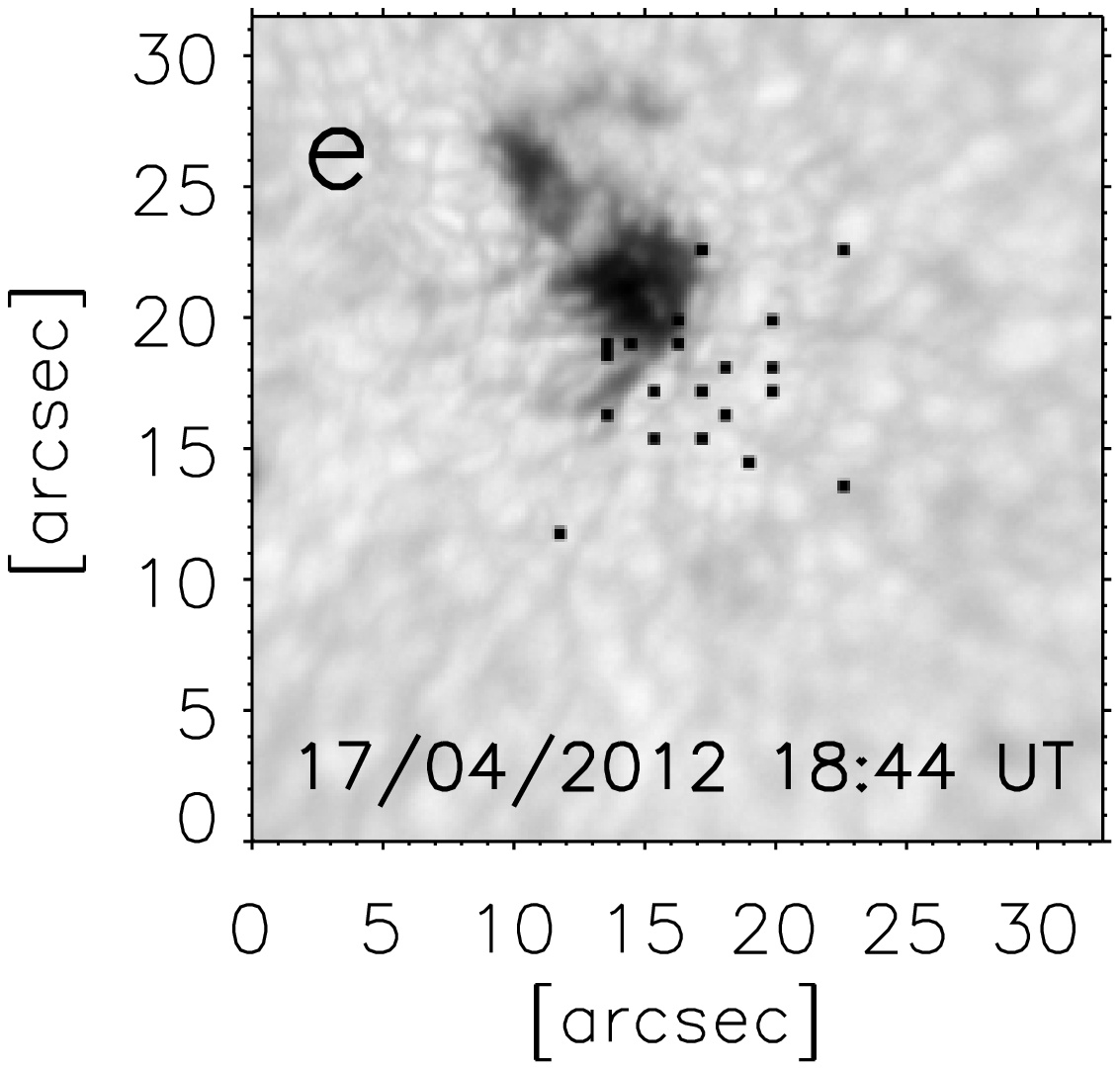}\includegraphics[height=3cm,trim=4.2cm 2.3cm 4.35cm 1.cm,clip=true,keepaspectratio=true]{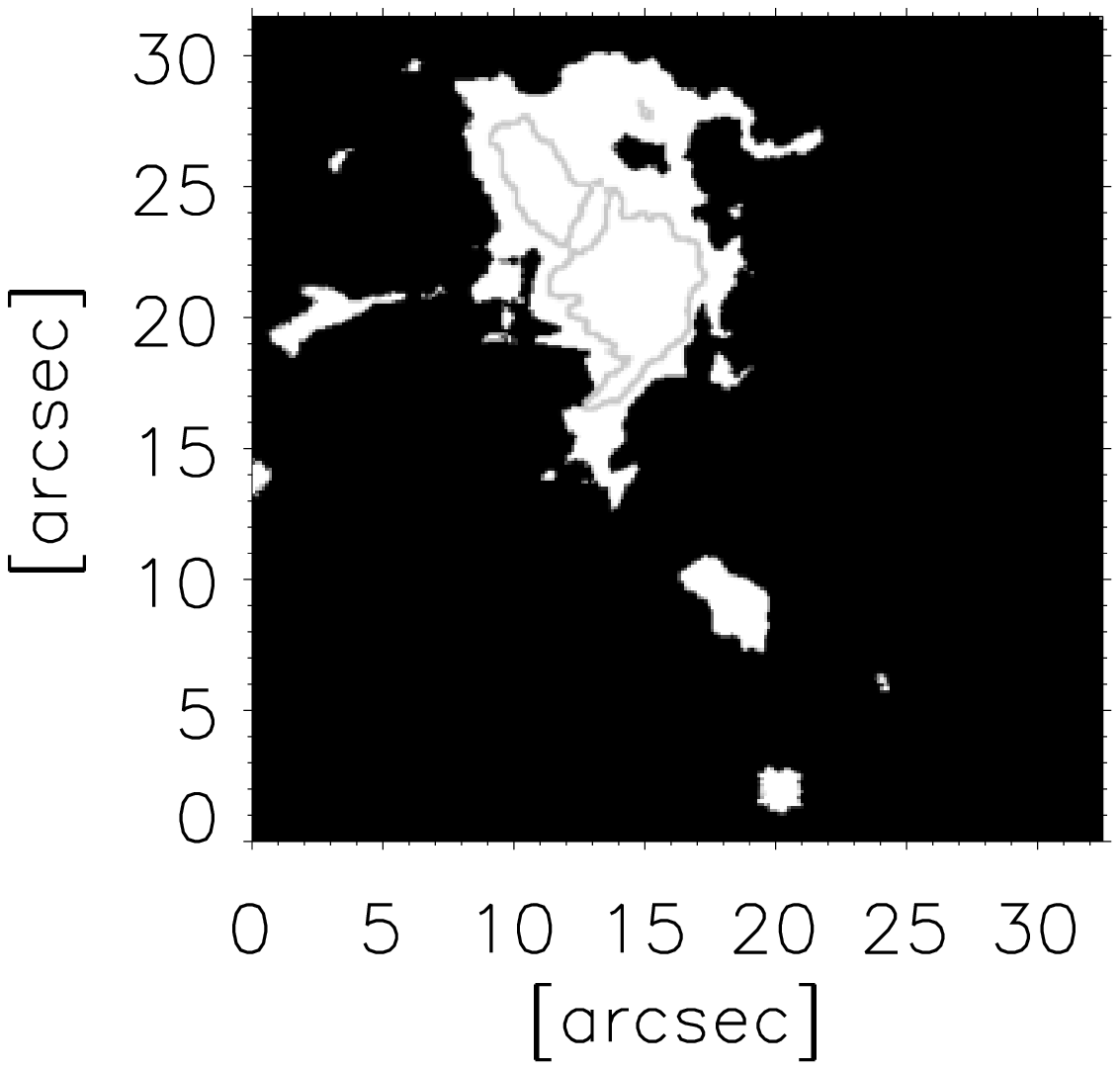}\includegraphics[height=3cm,trim=4.2cm 2.3cm 4.35cm 1.cm,clip=true,keepaspectratio=true]{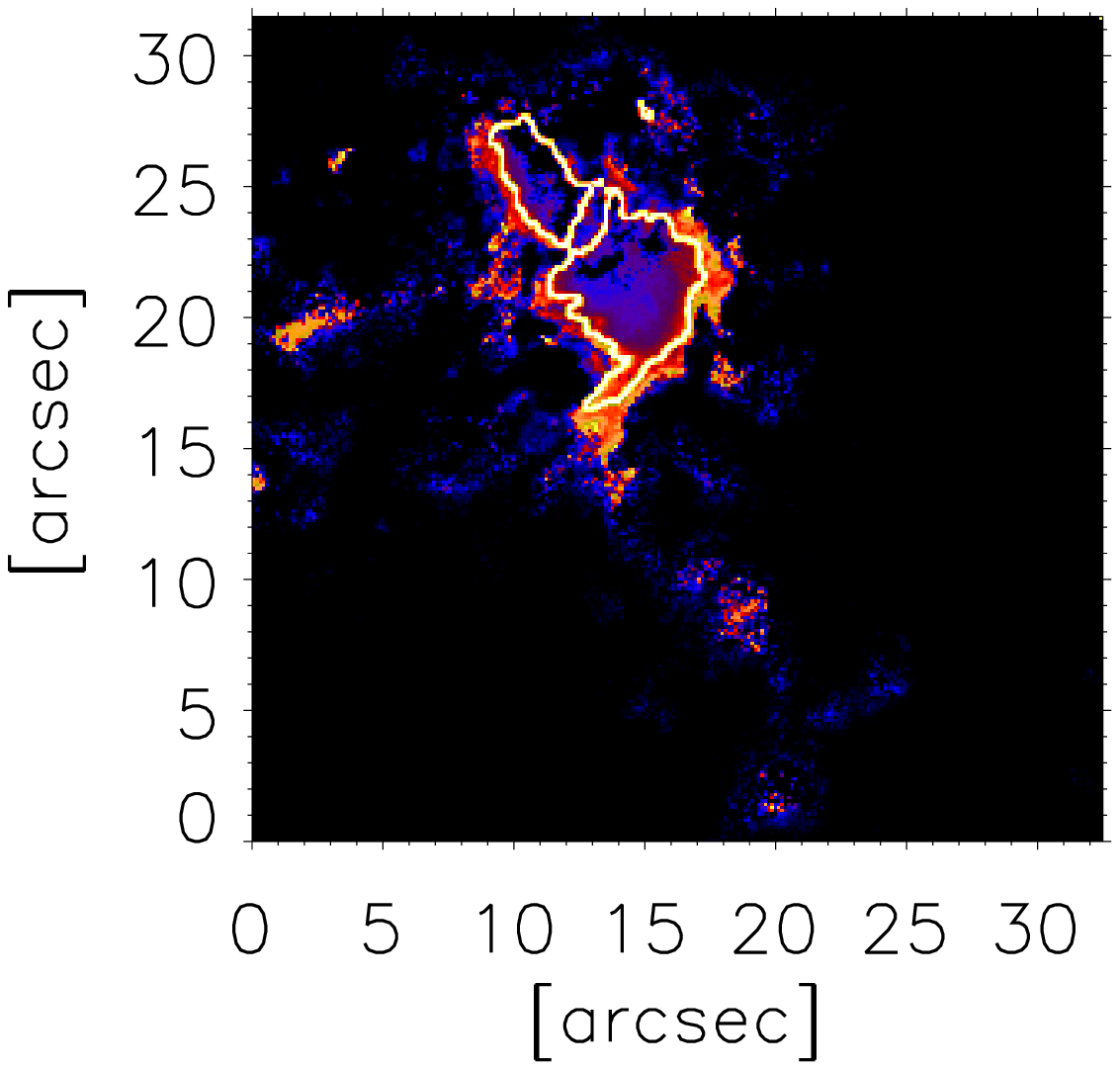}\\
\includegraphics[height=3cm,trim=1.5cm 2.3cm 4.35cm 1.cm,clip=true,keepaspectratio=true]{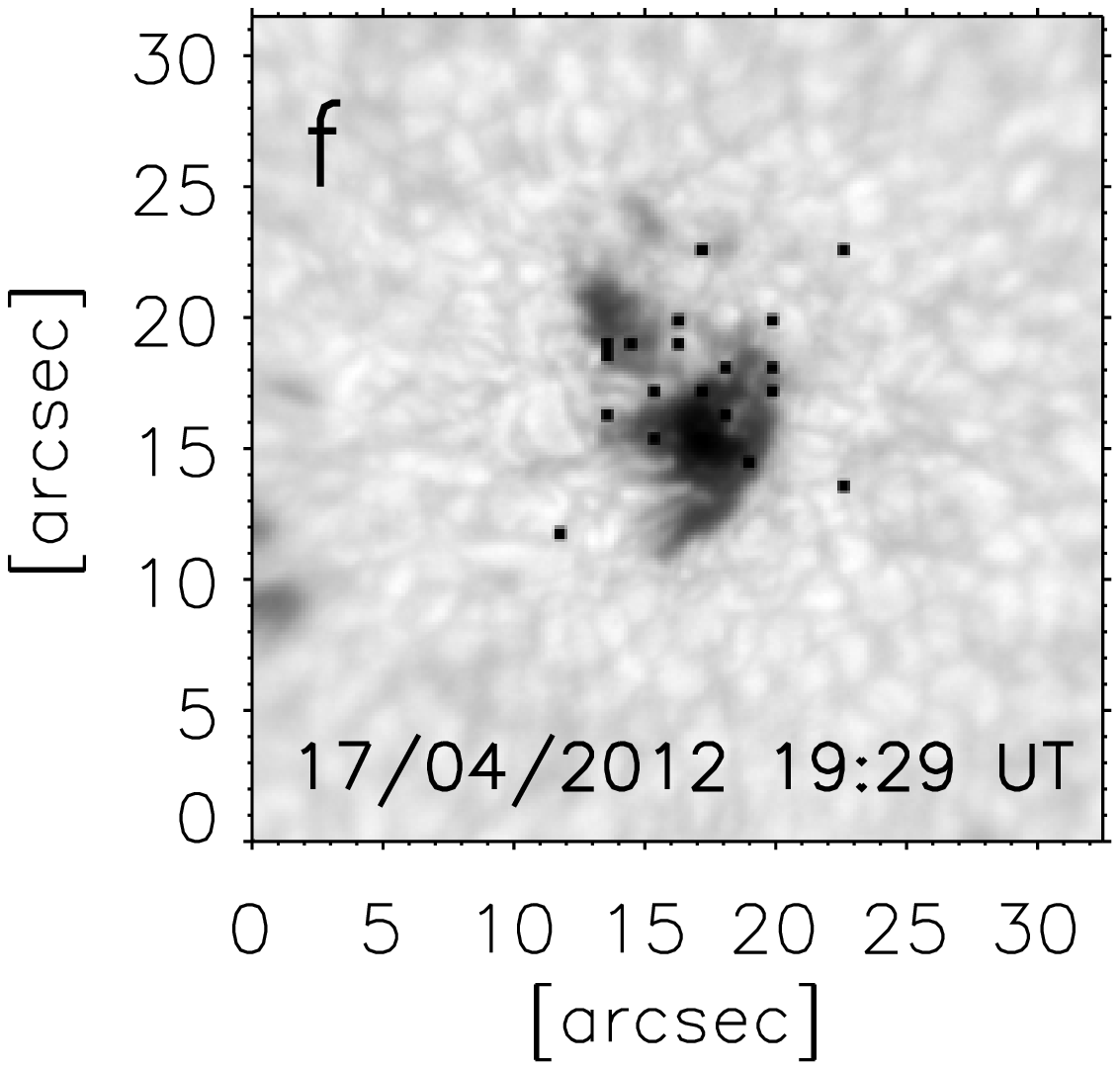}\includegraphics[height=3cm,trim=4.2cm 2.3cm 4.35cm 1.cm,clip=true,keepaspectratio=true]{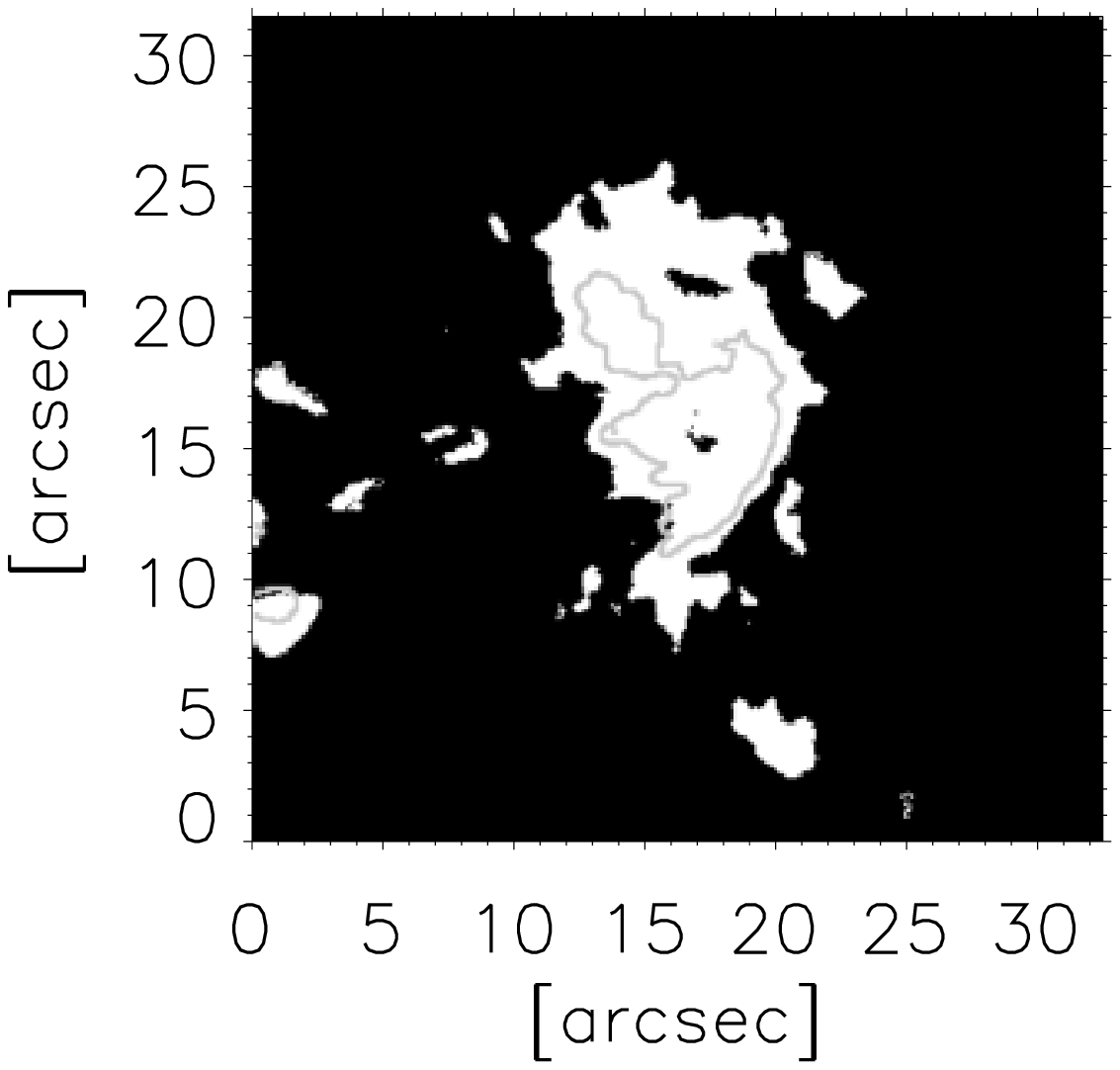}\includegraphics[height=3cm,trim=4.2cm 2.3cm 4.35cm 1.cm,clip=true,keepaspectratio=true]{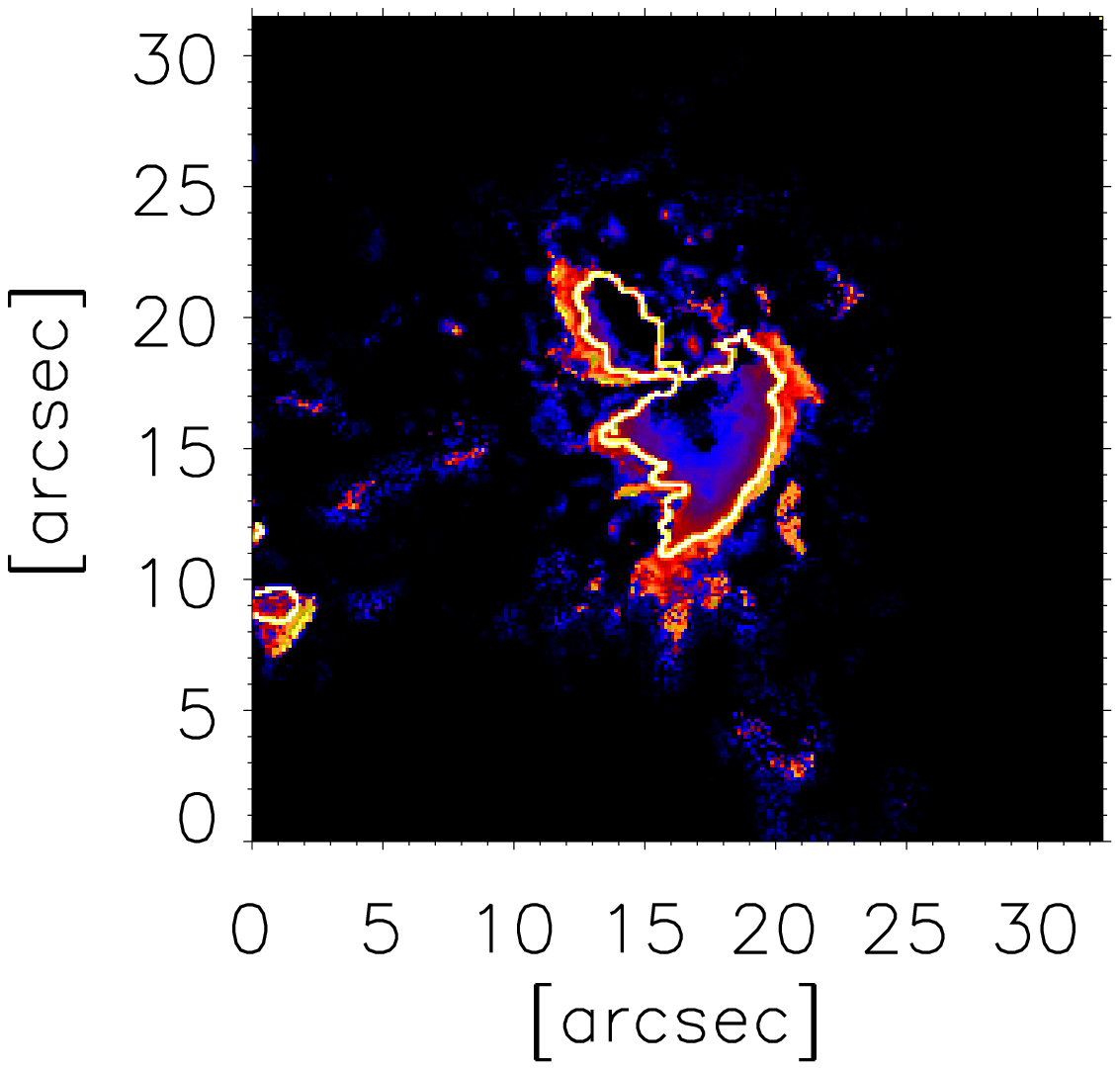}\\
\includegraphics[height=3.6cm,trim=1.5cm 0.3cm 4.35cm 1.cm,clip=true,keepaspectratio=true]{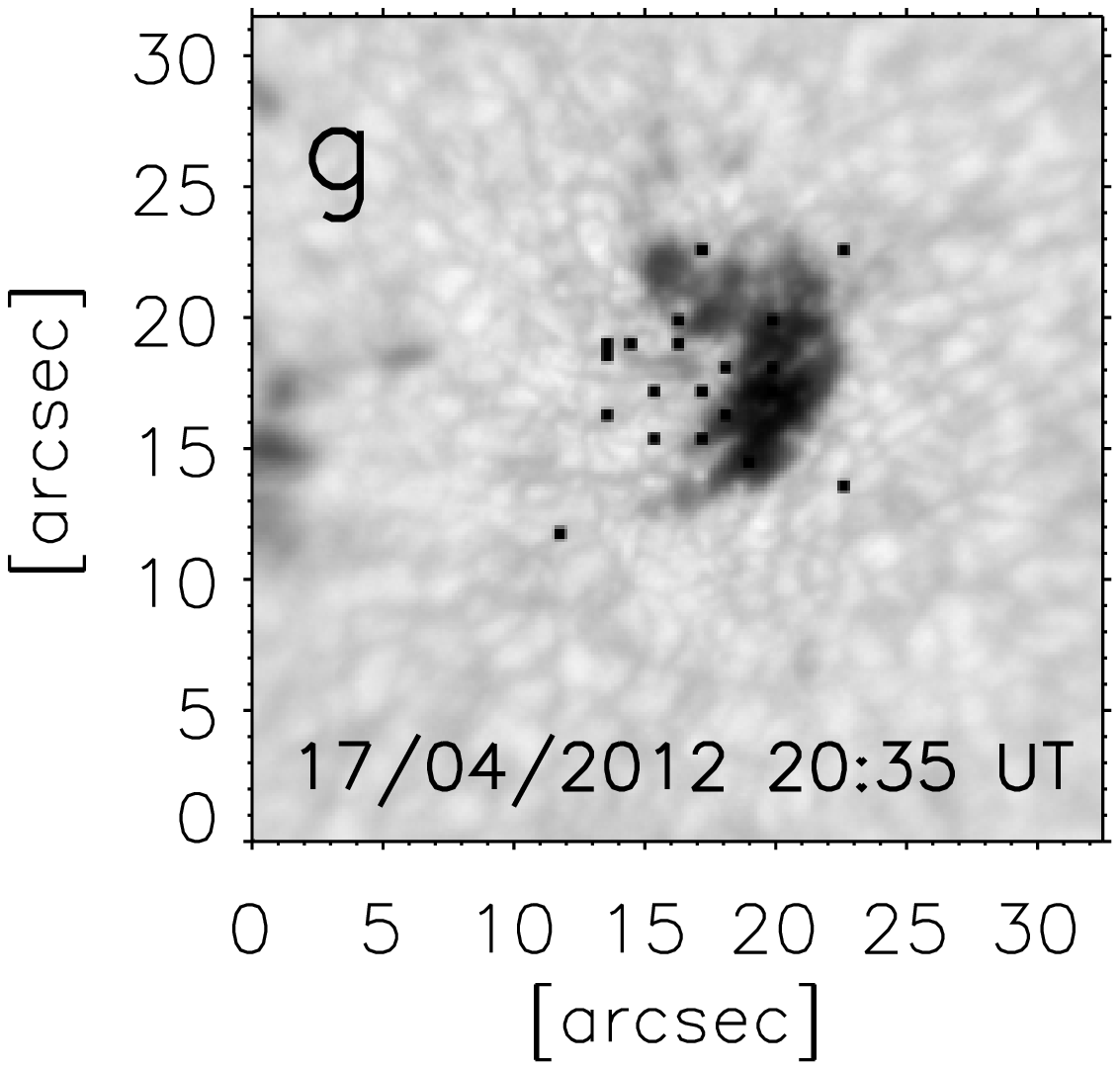}\includegraphics[height=3.6cm,trim=4.2cm 0.3cm 4.35cm 1.cm,clip=true,keepaspectratio=true]{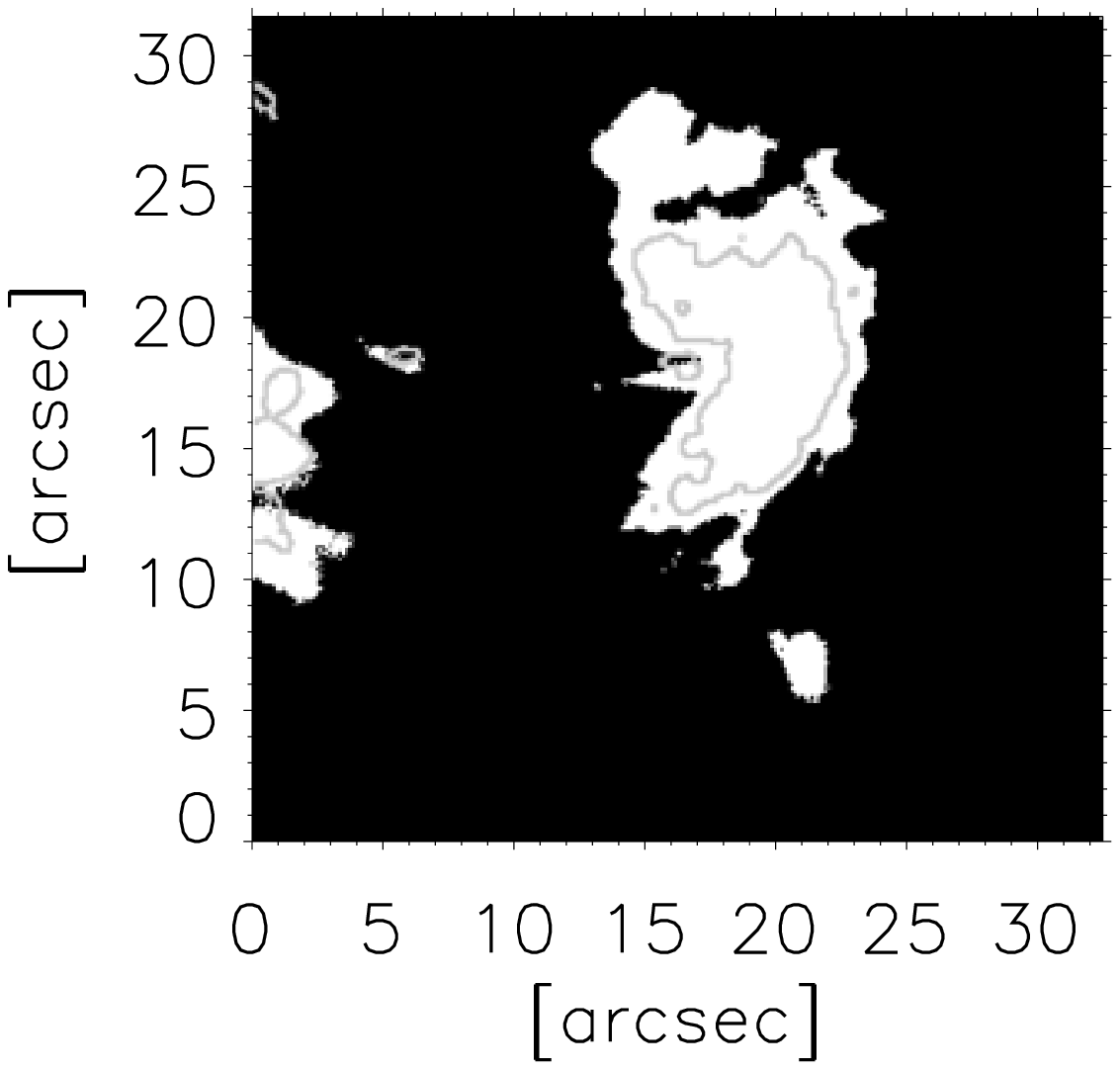}\includegraphics[height=3.6cm,trim=4.2cm 0.3cm 4.35cm 1.cm,clip=true,keepaspectratio=true]{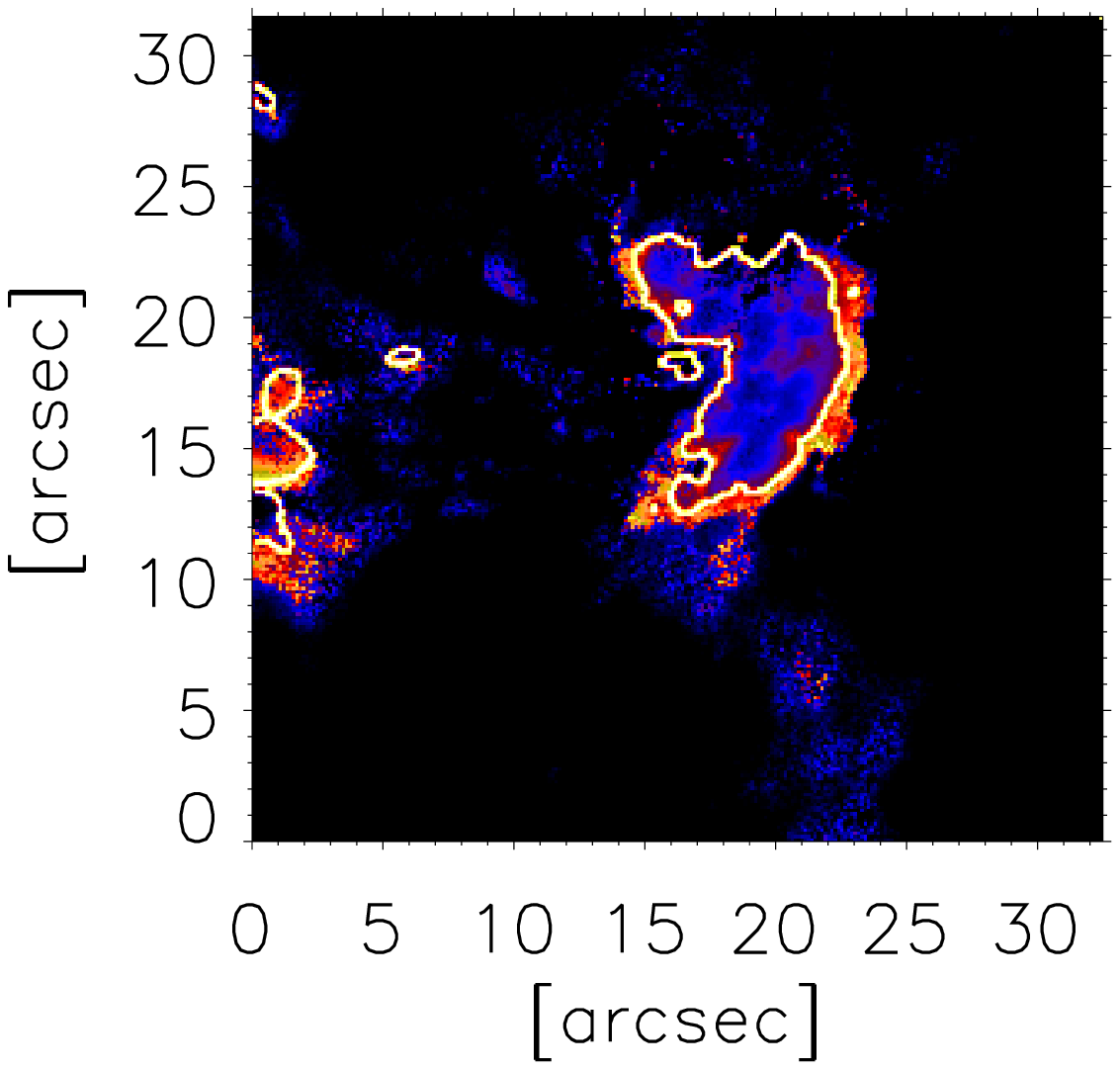}\\
}
\caption{\footnotesize{ Example of   continuum intensity images (left column),  mask of the magnetic regions identified in the data (middle column, white (black) shows the magnetic (quiet)  component) and stray-light fraction (right column)  derived  from the SIR inversion of the IBIS Fe I 617.3 nm line data, at seven stages of the pore formation at (a) 13:58 UT, (b) 14:26 UT, (c) 15:08 UT, (d) 16:08 UT, (e) 18:44 UT, (f) 19:29 UT, and (g) 20:35 UT. North is at the top; west is to the right. The small black squares in the continuum intensity images show the  20 positions  considered for the comparison between observed and inverted profiles  in Figs. \ref{a2}-\ref{a8}. The contour line in the panels of the  middle and right columns shows the location of the evolving structure singled out in the continuum data, as specified in Fig. \ref{f2}. 
}}
\label{a1}
\end{figure*}

\begin{figure*}%[ht!]
%  \centering{\hspace {1cm} 1                       \hspace{2.5cm}                                     2            \hspace{2.5cm}                                                 3     \hspace{2.5cm}                        4  \hspace{2.5cm}                        5 \hspace{2.5cm}                        6     \hspace {1cm}    }\\
%
\centering{
\hspace{.7cm}
\hspace{15.9cm}
}\\
\centering{
\includegraphics[height=17.5cm,trim=.3cm 0.3cm 0.5cm 0.cm,clip=true,keepaspectratio=true]{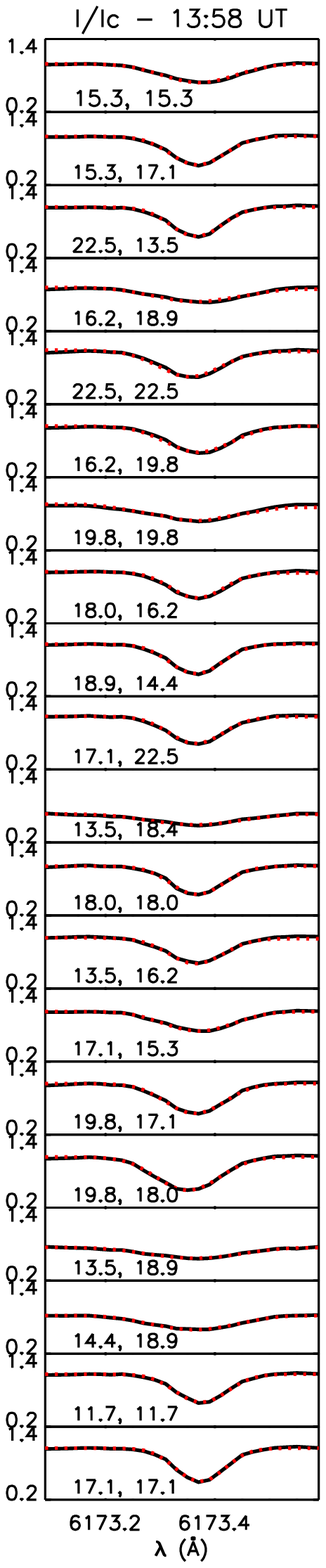}
\includegraphics[height=17.5cm,trim=.3cm 0.3cm 0.5cm 0.cm,clip=true,keepaspectratio=true]{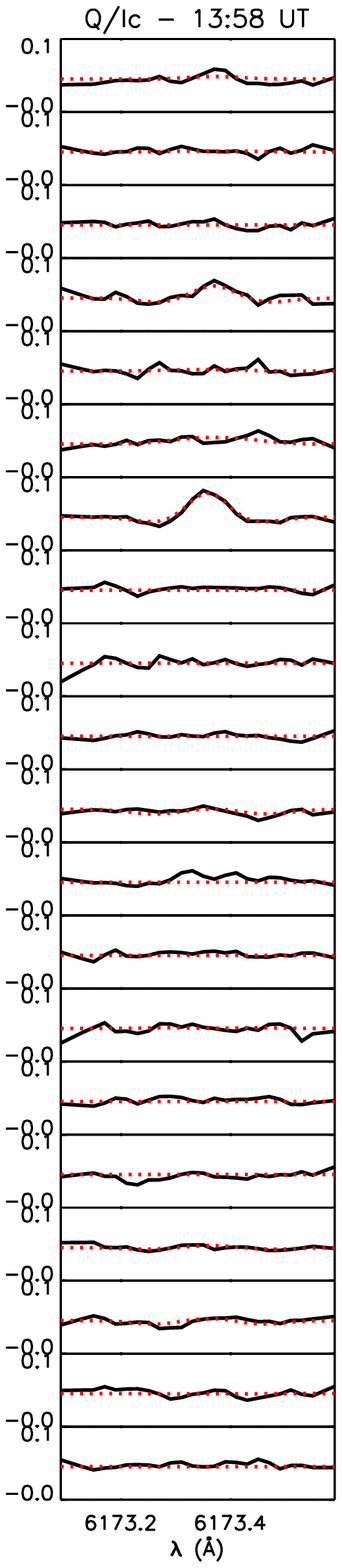}
\includegraphics[height=17.5cm,trim=.3cm 0.3cm 0.5cm 0.cm,clip=true,keepaspectratio=true]{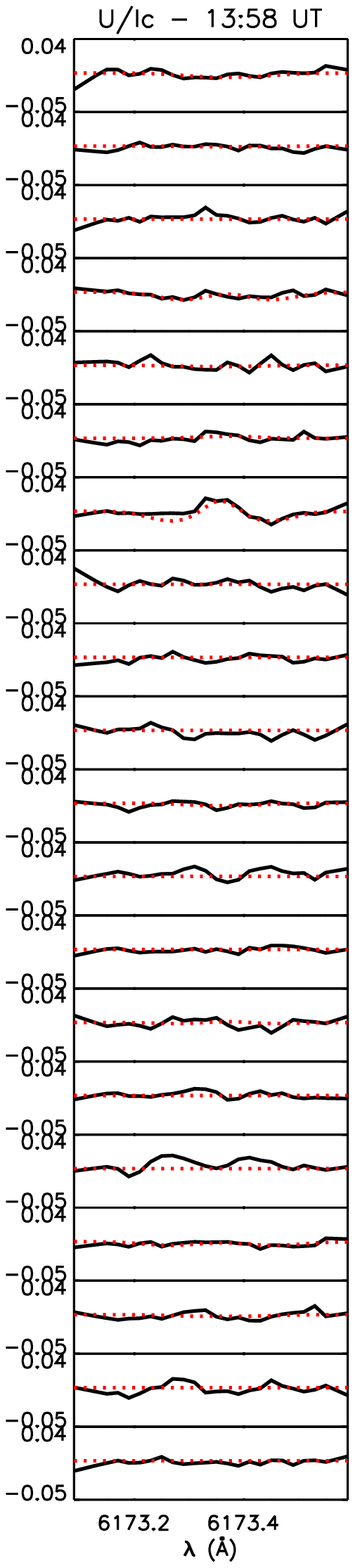}
\includegraphics[height=17.5cm,trim=.3cm 0.3cm 0.5cm 0.cm,clip=true,keepaspectratio=true]{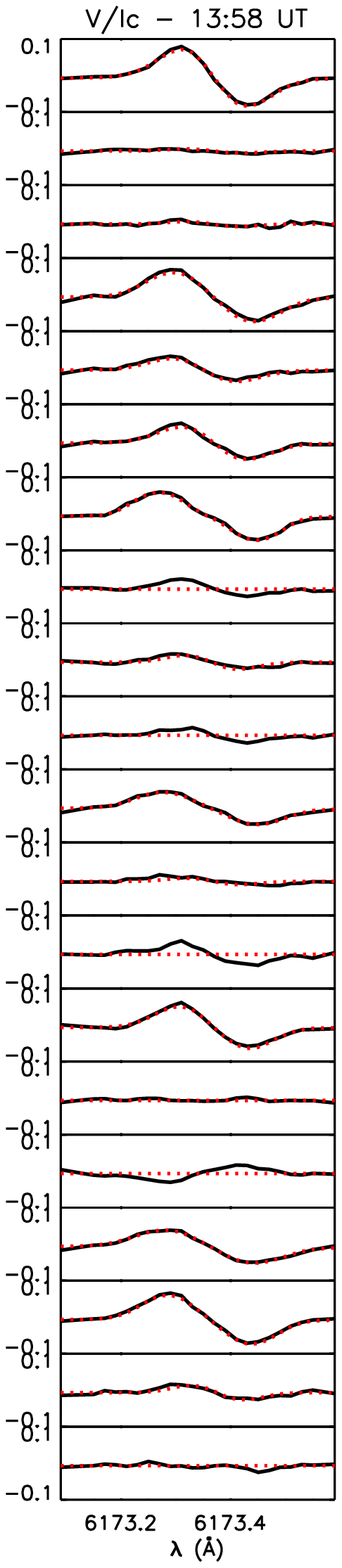}\\
}
\caption{\footnotesize{ From left to right: Examples of observed (black) and inverted (red) Stokes-I, -Q, -U, and -V  profiles at  the 20 positions on the sub-FOV  labelled (a) in Fig. \ref{a1}, taken during the pore formation at 13:58 UT. Spectra are normalized to the continuum intensity.
Numbers on the bottom left in each sub-panel of the Stokes-I figure indicate the position of the analysed pixel  expressed in arcsec with respect to the sub-FOV shown in Fig. \ref{a1}. Values of the stray-light fraction, LOS magnetic field strength (kG), field inclination and azimuth (degree), and LOS velocity (km/s)  derived  from our data analysis (at optical depth log$\tau _{500}=1$) at the same positions of the shown profiles are listed in Table \ref{tablea2};  the values are cut at 2 decimal digits. 
%Panels with Stokes-Q, -U, -V profiles show results at the same positions on the subFOV considered for the Stokes-I figure, by following the  order in that figure.
}}
\label{a2}
\end{figure*}

\begin{table*}
 \caption{ Summary of the plasma properties  estimated at the 20 positions considered in Figs. \ref{a2}. %: $\chi$ (eV) excitation potential, g$_{eff}$ effective Land\'e factor, estimated average line-formation height for the the line core, reference. %(1) Norton et al. 2006, (2) Faroubert, Ricort, Aime 2013, (3) Cauzzi et al. 2008.
} % title of Table
\label{tablea2} % is used to refer this table in the text
\centering % used for centering table
\begin{tabular}{r r r r r r r} % centered columns (4 columns)
\hline\hline % inserts double horizontal lines
position & position & stray-light  & field   &  field & field   & v$_{LOS}$  \\
x & y &  fraction & strength   &  inclination &  azimuth  &   \\
(arcsec) & (arcsec) &  &(kG) & (degree) & (degree)  & (km s$^{-1}$) \\
\hline
 15.3 & 15.3 &     0.00 &      0.67 &       25.33 &      -22.11 &      0.33\\
 15.3 & 17.1 &     0.00 &     0.02 &       20.33 &      -63.90 &     0.08\\
 22.5 & 13.5 &     0.00 &     0.03 &       19.88 &      -86.65 &      0.11\\
 16.2 & 18.9 &     0.18 &       1.10 &       11.15 &       161.89 &      0.46\\
 22.5 & 22.5 &     0.17 &      0.12 &       18.45 &      -87.51 &     -0.11\\
 16.2 & 19.8 &     0.00 &      0.41 &       23.97 &       20.41 &      0.17\\
 19.8 & 19.8 &     0.38 &       1.15 &       18.99 &       150.45 &      0.22\\
 18.0 & 16.2 &     0.16 &      0.11 &       24.10 &      -86.22 &      0.33\\
 18.9 & 14.4 &     0.00 &     0.07 &       19.68 &      -90.49 &     0.01\\
 17.1 & 22.5 &    0.05 &     0.06 &       16.13 &       106.53 &      0.16\\
 13.5 & 18.4 &    0.01 &       1.19 &       19.66 &       100.73 &      0.24\\
 18.0 & 18.0 &     0.00 &     0.07 &       35.82 &      -71.08 &      0.19\\
 13.5 & 16.2 &     0.11 &      0.18 &       18.92 &       55.18 &      0.13\\
 17.1 & 15.3 &     0.00 &      0.52 &       28.09 &      -69.31 &      0.34\\
 19.8 & 17.1 &     0.00 &     0.02 &       47.26 &      -100.35 &     0.01\\
 19.8 & 18.0 &     0.00 &      0.00 &       76.06 &      -105.08 &     -0.37\\
 13.5 & 18.9 &     0.00 &      0.84 &       36.02 &      -54.81 &      0.38\\
 14.4 & 18.9 &     0.00 &      0.82 &       34.25 &       17.01 &      0.16\\
 11.7 & 11.7 &     0.00 &     0.06 &       20.02 &       83.90 &      0.28\\
 17.1 & 17.1 &     0.00 &     0.02 &       25.53 &      -80.95 &      0.18\\
\hline
\end{tabular}
\end{table*}

\begin{figure*}%[ht!]
%  \centering{\hspace {1cm} 1                       \hspace{2.5cm}                                     2            \hspace{2.5cm}                                                 3     \hspace{2.5cm}                        4  \hspace{2.5cm}                        5 \hspace{2.5cm}                        6     \hspace {1cm}    }\\
%
\centering{
\hspace{.7cm}
\hspace{15.9cm}
}\\
\centering{
\includegraphics[height=17.5cm,trim=.3cm 0.3cm 0.5cm 0.cm,clip=true,keepaspectratio=true]{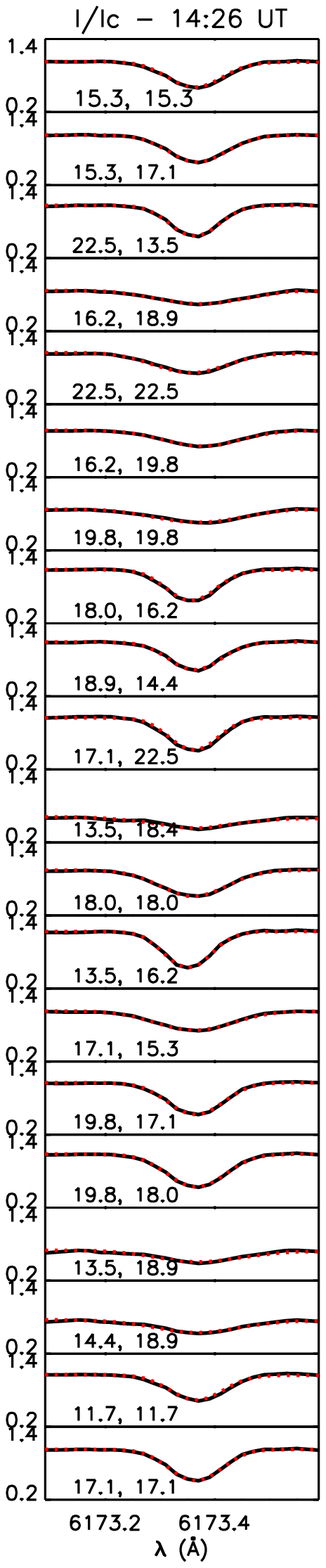}
\includegraphics[height=17.5cm,trim=.3cm 0.3cm 0.5cm 0.cm,clip=true,keepaspectratio=true]{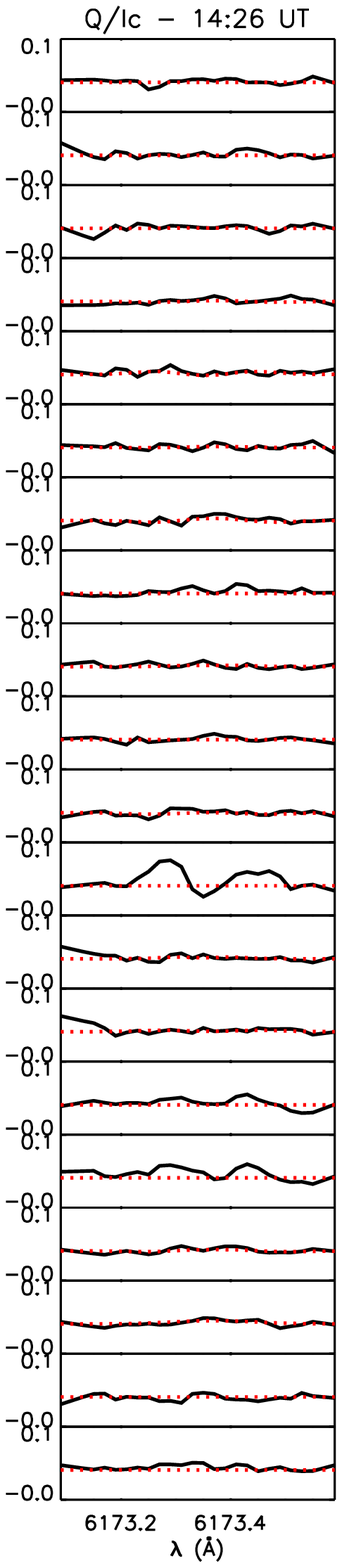}
\includegraphics[height=17.5cm,trim=.3cm 0.3cm 0.5cm 0.cmclip=true,keepaspectratio=true]{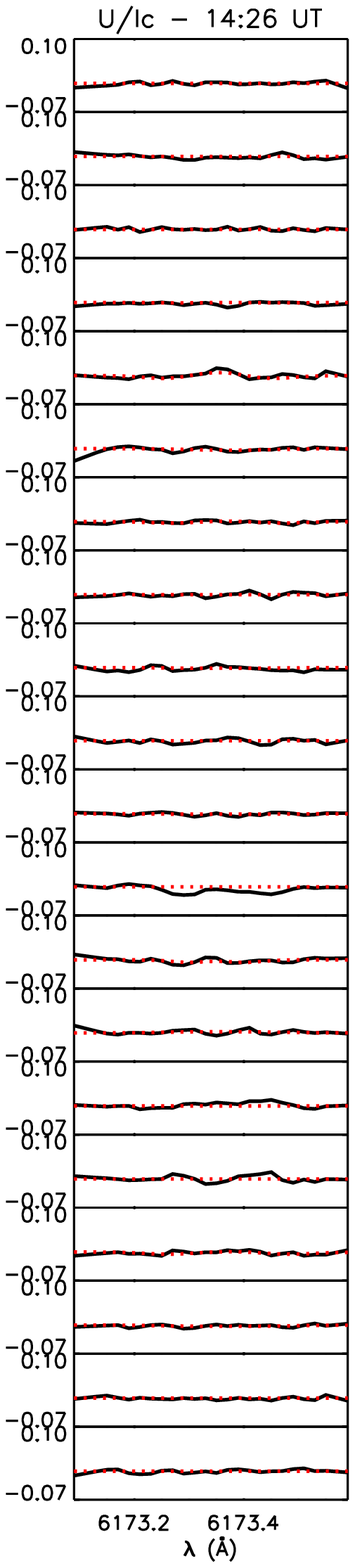}
\includegraphics[height=17.5cm,trim=.3cm 0.3cm 0.5cm 0.cm,clip=true,keepaspectratio=true]{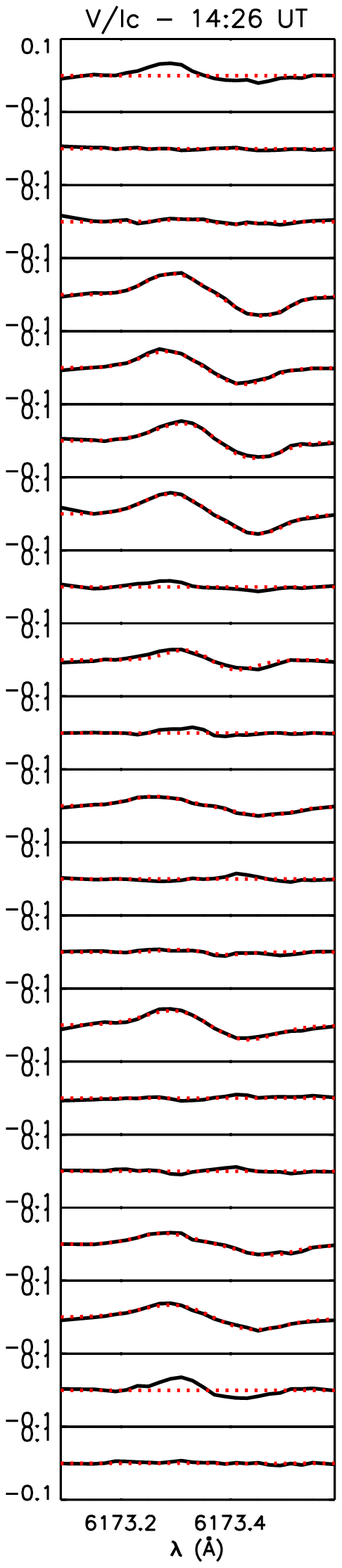}\\
}
\caption{\footnotesize{ As in Fig. \ref{a2} but for the data taken at 14:26 UT (stage (b)  of the pore formation shown in Fig. \ref{a1}). Values of the stray-light fraction, LOS magnetic field strength (kG), field inclination and azimuth (degree), and LOS velocity (km/s)  derived  from our data analysis (at optical depth log$\tau _{500}=1$) at the same positions of the shown profiles are listed in Table \ref{tablea3};  the values are cut at 2 decimal digits. }}
\label{a3}
\end{figure*}

 \begin{table*}
 \caption{Summary of the plasma properties  estimated at the 20 positions considered in Figs. \ref{a3}.  %: $\chi$ (eV) excitation potential, g$_{eff}$ effective Land\'e factor, estimated average line-formation height for the the line core, reference. %(1) Norton et al. 2006, (2) Faroubert, Ricort, Aime 2013, (3) Cauzzi et al. 2008.
} % title of Table
\label{tablea3} % is used to refer this table in the text
\centering % used for centering table
\begin{tabular}{r r r r r r r} % centered columns (4 columns)
\hline\hline % inserts double horizontal lines
position & position & stray-light  & field   &  field & field   & v$_{LOS}$  \\
x & y &  fraction & strength   &  inclination &  azimuth  &   \\
(arcsec) & (arcsec) &  &(kG) & (degree) & (degree)  & (km s$^{-1}$) \\
\hline
15.3 & 15.3 &    0.08 &      0.21 &       20.83 &      -71.28 &     0.39 \\
15.3 & 17.1 &    0.01 &     0.02 &       26.63 &       93.58 &      0.47 \\
22.5 & 13.5 &     0.00 &     0.03 &       20. 00&      -83.32 &     0.31 \\
16.2 & 18.9 &     0.00 &       1.05 &       10.36 &       142.37 &     0.77\\
22.5 & 22.5 &     0.11 &      0.59 &       25.39 &      -67.77 &     0.04\\
 16.2 & 19.8 &     0.00 &      0.63 &       20.08 &      -61.98 &      0.80\\
 19.8 & 19.8 &     0.00 &       1.11 &       12.01 &       158.17 &      0.77\\
 18.0 & 16.2 &    0.04 &     0.04 &       20.26 &      -81.93 &    -0.01\\
 18.9 & 14.4 &     0.00 &      0.23 &       20.84 &       119.02 &      0.37\\
 17.1 & 22.5 &     0.00 &     0.05 &       19.71 &      -30.36 &    -0.09\\
 13.5 & 18.4 &     0.15 &       1.16 &       16.09 &       53.05 &      0.72\\
 18.0 & 18.0 &     0.00 &     0.01 &       79.49 &       103.84&     -0.12\\
 13.5 & 16.2 &     0.00 &     0.06 &       20.17 &       101.17 &     -0.32\\
 17.1 & 15.3 &     0.00 &      0.45 &       26.86 &      -68.42 &      0.44\\
 19.8 & 17.1 &     0.00 &    0.01 &       104.87 &      -93.28 &      0.19\\
 19.8 & 18.0 &     0.00 &    0.03 &       119.86 &      -93.42 &     0.08\\
 13.5 & 18.9 &    0.09 &      0.74 &       32.33 &      -65.79 &      0.61\\
 14.4 & 18.9 &    0.07 &      0.67 &       37.87 &      -16.49 &      0.62\\
 11.7 & 11.7 &    0.07 &      0.19 &       18.68 &       66.81 &      0.44\\
 17.1 & 17.1 &     0.00 &     0.04 &       29.03 &      -84.17 &      0.17\\
\hline
\end{tabular}
\end{table*}
%\end{appendix}

\begin{figure*}%[ht!]
%  \centering{\hspace {1cm} 1                       \hspace{2.5cm}                                     2            \hspace{2.5cm}                                                 3     \hspace{2.5cm}                        4  \hspace{2.5cm}                        5 \hspace{2.5cm}                        6     \hspace {1cm}    }\\
%
\centering{
\hspace{.7cm}
\hspace{15.9cm}
}\\
\centering{
\includegraphics[height=17.5cm,trim=.3cm 0.3cm 0.5cm 0.cm,clip=true,keepaspectratio=true]{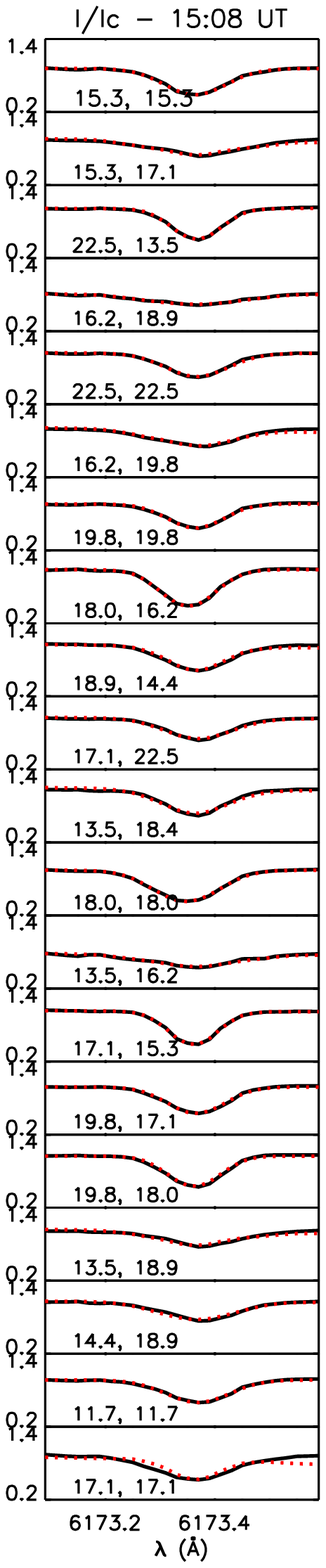}
\includegraphics[height=17.5cm,trim=.3cm 0.3cm 0.5cm 0.cm,clip=true,keepaspectratio=true]{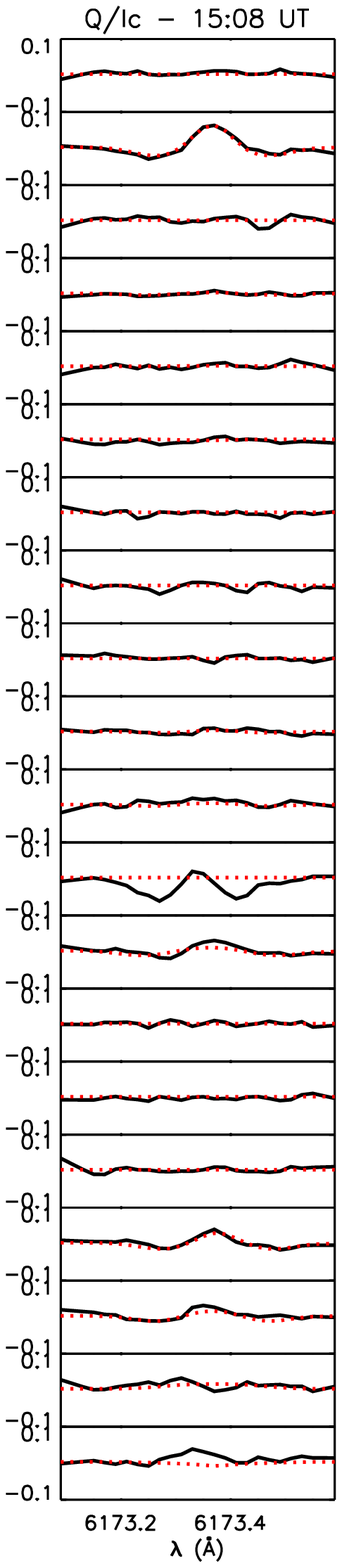}
\includegraphics[height=17.5cm,trim=.3cm 0.3cm 0.5cm 0.cmclip=true,keepaspectratio=true]{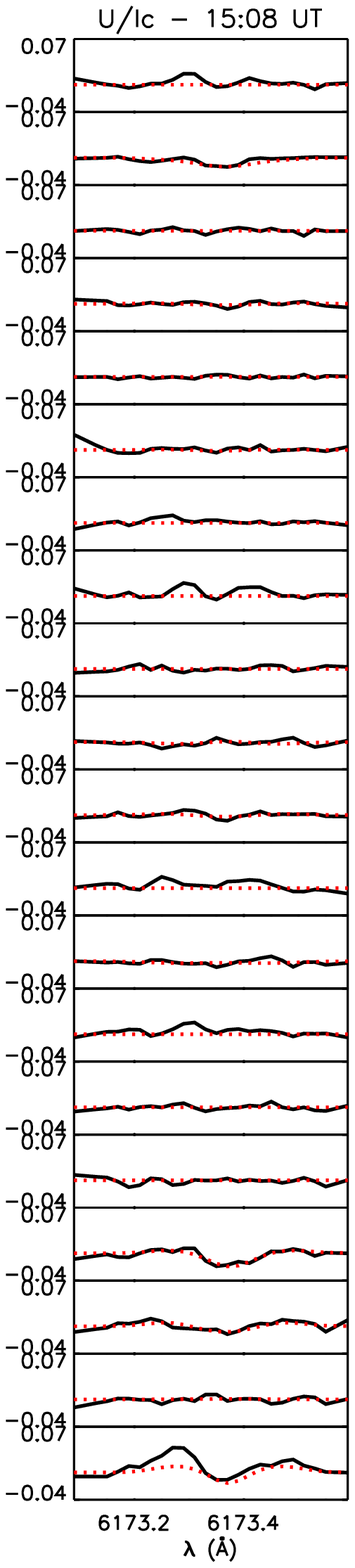}
\includegraphics[height=17.5cm,trim=.3cm 0.3cm 0.5cm 0.cm,clip=true,keepaspectratio=true]{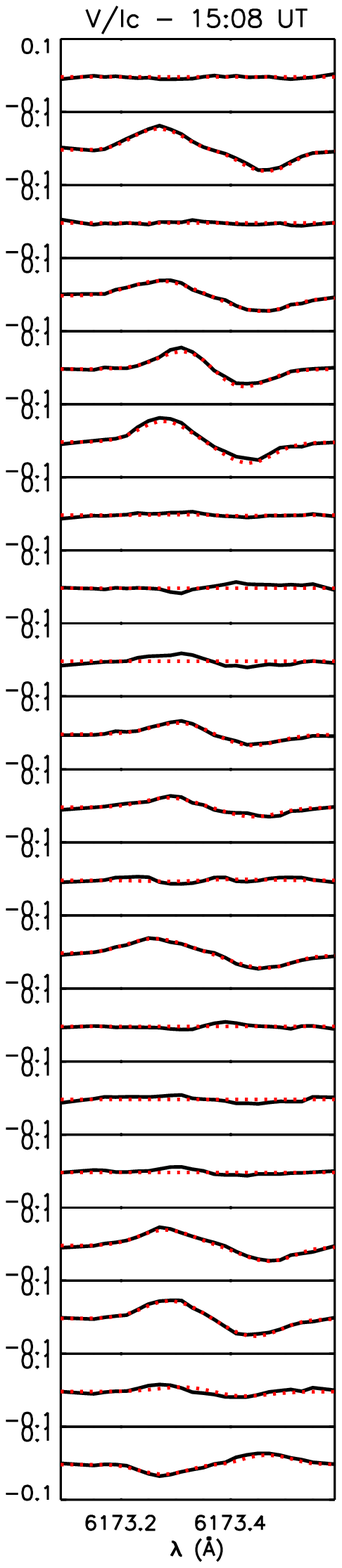}\\
}
\caption{\footnotesize{ As in Fig. \ref{a2} but for the data taken at 15:08 UT (stage (c)  of the pore formation shown in Fig. \ref{a1}). Values of the stray-light fraction, LOS magnetic field strength (kG), field inclination and azimuth (degree), and LOS velocity (km/s)  derived  from our data analysis (at optical depth log$\tau _{500}=1$) at the same positions of the shown profiles are listed in Table \ref{tablea4};  the values are cut at 2 decimal digits. }}
\label{a4}
\end{figure*}

 \begin{table*}
 \caption{ Summary of the plasma properties  estimated at the 20 positions considered in Figs. \ref{a4}. %: $\chi$ (eV) excitation potential, g$_{eff}$ effective Land\'e factor, estimated average line-formation height for the the line core, reference. %(1) Norton et al. 2006, (2) Faroubert, Ricort, Aime 2013, (3) Cauzzi et al. 2008.
} % title of Table
\label{tablea4} % is used to refer this table in the text
\centering % used for centering table
\begin{tabular}{r r r r r r r} % centered columns (4 columns)
\hline\hline % inserts double horizontal lines
position & position & stray-light  & field   &  field & field   & v$_{LOS}$  \\
x & y &  fraction & strength   &  inclination &  azimuth  &   \\
(arcsec) & (arcsec) &  &(kG) & (degree) & (degree)  & (km s$^{-1}$) \\
\hline
 15.3 & 15.3 &    0.010 &      0.00 &       106.61 &      -109.23 &     0.06\\
 15.3 & 17.1 &     0.37 &       1.30 &       17.66 &       49.68 &      0.37\\
 22.5 & 13.5 &     0.00 &     0.01 &       19.72 &      -91.58 &     0.06\\
 16.2 & 18.9 &     0.00 &       1.24 &       26.84 &      -69.30 &     0.25\\
 22.5 & 22.5 &    0.04 &      0.36 &       17.15 &       58.38 &      0.19\\
 16.2 & 19.8 &     0.47 &      0.84 &       17.13 &      -110.95 &     0.12\\
 19.8 & 19.8 &     0.00 &     0.03 &       23.16 &      -93.35 &     0.09\\
 18.0 & 16.2 &     0.00 &    0.02 &       115.11 &      -124.39 &     -0.28\\
 18.9 & 14.4 &     0.35 &     0.08 &       21.14 &      -109.74 &      0.39\\
 17.1 & 22.5 &     0.14 &      0.32 &       17.75 &      -95.71 &      0.27\\
 13.5 & 18.4 &     0.33 &      0.22 &       17.21 &       98.03 &    -0.01\\
 18.0 & 18.0 &     0.00 &      0.00 &       83.34 &      -158.86 &    -0.56\\
 13.5 & 16.2 &     0.12 &       1.35 &       8.02 &       164.21 &     0.22\\
 17.1 & 15.3 &     0.10 &    0.03 &       127.02 &      -108.06 &    0.07\\
 19.8 & 17.1 &     0.00 &     0.06 &       23.82 &      -90.13 &      0.17\\
 19.8 & 18.0 &     0.00 &     0.03 &       19.57 &      -99.72 &    -0.08\\
 13.5 & 18.9 &     0.38 &       1.16 &       12.59 &       125.59 &      0.66\\
 14.4 & 18.9 &     0.10 &      0.97 &       7.91 &     -0.98 &      0.34\\
 11.7 & 11.7 &     0.00 &      0.22 &       41.89 &      -72.44 &      0.13\\
 17.1 & 17.1 &     0.75 &      0.20 &       62.52 &      -122.73 &    0.09\\
\hline
\end{tabular}
\end{table*}

\begin{figure*}%[ht!]
%  \centering{\hspace {1cm} 1                       \hspace{2.5cm}                                     2            \hspace{2.5cm}                                                 3     \hspace{2.5cm}                        4  \hspace{2.5cm}                        5 \hspace{2.5cm}                        6     \hspace {1cm}    }\\
%
\centering{
\hspace{.7cm}
\hspace{15.9cm}
}\\
\centering{
\includegraphics[height=17.5cm,trim=.3cm 0.3cm 0.5cm 0.cm,clip=true,keepaspectratio=true]{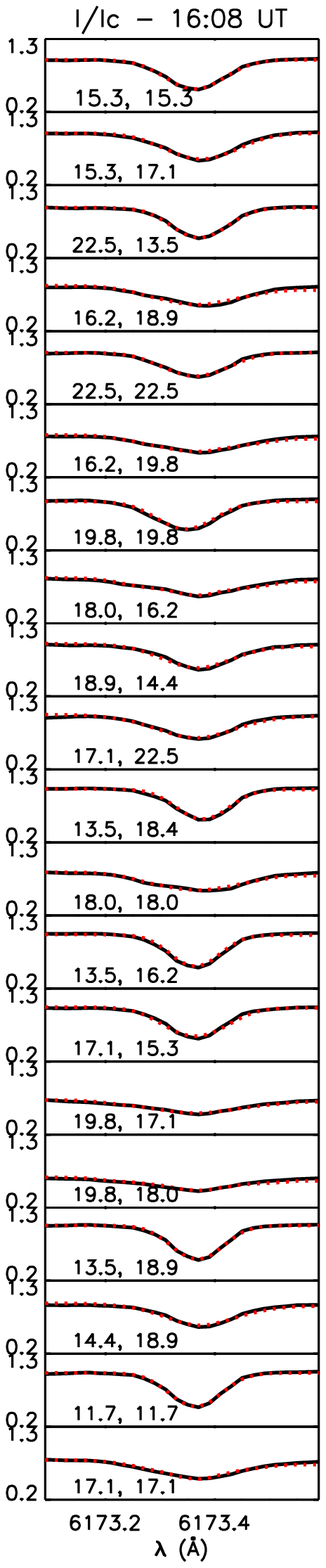}
\includegraphics[height=17.5cm,trim=.3cm 0.3cm 0.5cm 0.cm,clip=true,keepaspectratio=true]{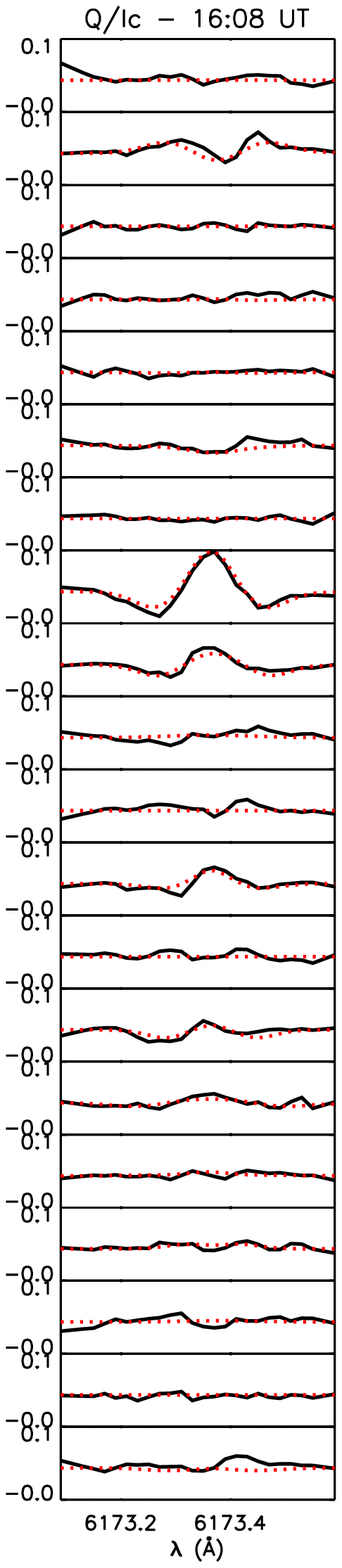}
\includegraphics[height=17.5cm,trim=.3cm 0.3cm 0.5cm 0.cmclip=true,keepaspectratio=true]{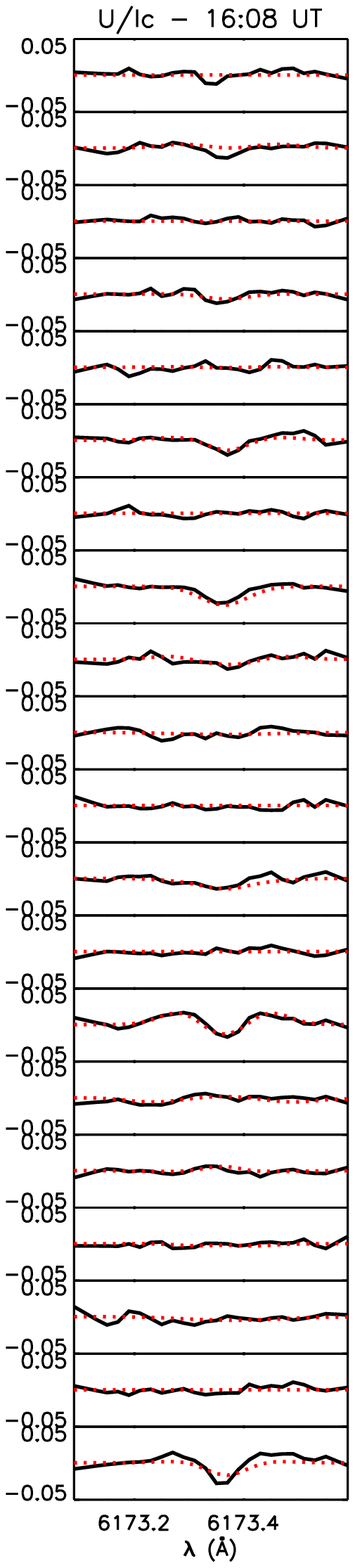}
\includegraphics[height=17.5cm,trim=.3cm 0.3cm 0.5cm 0.cm,clip=true,keepaspectratio=true]{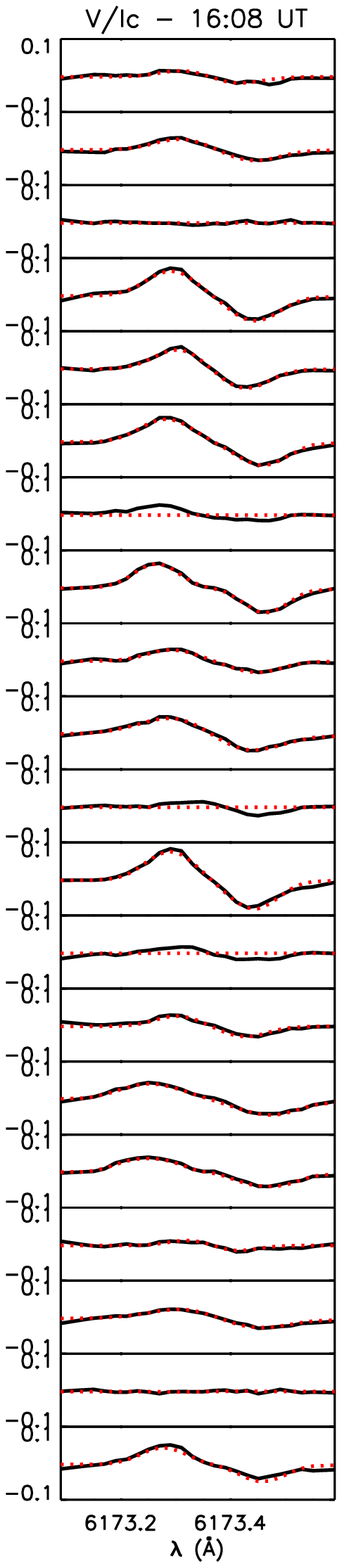}\\
}
\caption{\footnotesize{ As in Fig. \ref{a2} but for the data taken at  16:08 UT (stage (d)  of the pore formation shown in Fig. \ref{a1}). Values of the stray-light fraction, LOS magnetic field strength (kG), field inclination and azimuth (degree), and LOS velocity (km/s)  derived  from our data analysis (at optical depth log$\tau _{500}=1$) at the same positions of the shown profiles are listed in Table \ref{tablea5};  the values are cut at 2 decimal digits. }}
\label{a5}
\end{figure*}

 \begin{table*}
 \caption{Summary of the plasma properties  estimated at the 20 positions considered in Figs. \ref{a5}. %: $\chi$ (eV) excitation potential, g$_{eff}$ effective Land\'e factor, estimated average line-formation height for the the line core, reference. %(1) Norton et al. 2006, (2) Faroubert, Ricort, Aime 2013, (3) Cauzzi et al. 2008.
} % title of Table
\label{tablea5} % is used to refer this table in the text
\centering % used for centering table
\begin{tabular}{r r r r r r r } % centered columns (4 columns)
\hline\hline % inserts double horizontal lines
position & position & stray-light  & field   &  field & field   & v$_{LOS}$  \\
x & y &  fraction & strength   &  inclination &  azimuth  &   \\
(arcsec) & (arcsec) &  &(kG) & (degree) & (degree)  & (km s$^{-1}$) \\
\hline
 15.3 & 15.3 &     0.00 &     0.05 &       20.49 &       87.15 &     0.09 \\
 15.3 & 17.1 &     0.00 &      0.59 &       36.08 &       72.74 &      0.27\\
 22.5 & 13.5 &     0.00 &      0.00 &       51.10 &      -95.53 &      0.30\\
 16.2 & 18.9 &     0.47 &      0.77 &       35.52 &       64.74 &      0.58\\
 22.5 & 22.5 &     0.00 &      0.38 &       29.46 &      -59.62 &      0.22\\
 16.2 & 19.8 &     0.41 &      0.96 &       35.32 &       56.90 &      0.59\\
 19.8 & 19.8 &     0.00 &     0.02 &       17.96 &      -89.92 &     -0.39\\
 18.0 & 16.2 &     0.37 &       1.33 &       14.63 &      -20.10 &      0.39\\
 18.9 & 14.4 &     0.30 &      0.48 &       19.47 &       121.96 &      0.49\\
 17.1 & 22.5 &     0.14 &      0.56 &       23.63 &       69.23 &      0.48\\
 13.5 & 18.4 &    0.07 &     0.08 &       19.77 &       104.08 &      0.33\\
 18.0 & 18.0 &     0.37 &       1.17 &       4.90 &       163.44 &      0.39\\
 13.5 & 16.2 &    0.08 &     0.05 &       18.17 &       88.70 &    -0.04\\
 17.1 & 15.3 &    0.08 &      0.25 &       36.18 &       64.22 &      0.16\\
 19.8 & 17.1 &     0.11 &       1.50 &       12.76 &       145.21 &      0.36\\
 19.8 & 18.0 &     0.34 &       1.45 &       27.18 &      -70.36 &      0.18\\
 13.5 & 18.9 &     0.00 &     0.07 &       16.59 &       108.13 &     0.03\\
 14.4 & 18.9 &     0.23 &      0.34 &       20.94 &       93.83 &      0.62\\
 11.7 & 11.7 &     0.00 &      0.00 &       52.61 &      -14.82 &      0.00\\
 17.1 & 17.1 &     0.49 &      0.89 &       34.32 &       51.11 &      0.28\\
\hline
\end{tabular}
\end{table*}

\begin{figure*}%[ht!]
%  \centering{\hspace {1cm} 1                       \hspace{2.5cm}                                     2            \hspace{2.5cm}                                                 3     \hspace{2.5cm}                        4  \hspace{2.5cm}                        5 \hspace{2.5cm}                        6     \hspace {1cm}    }\\
%
\centering{
\hspace{.7cm}
\hspace{15.9cm}
}\\
\centering{
\includegraphics[height=17.5cm,trim=.3cm 0.3cm 0.5cm 0.cm,clip=true,keepaspectratio=true]{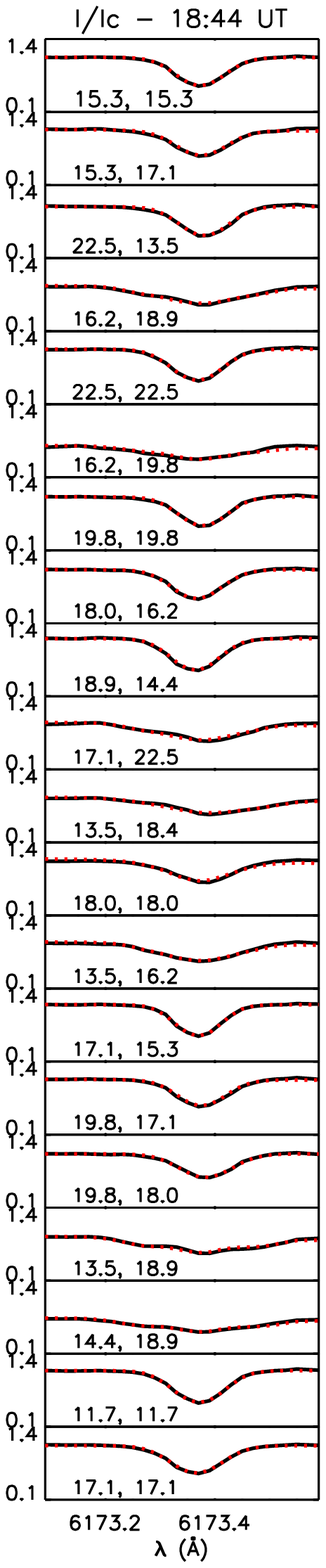}
\includegraphics[height=17.5cm,trim=.3cm 0.3cm 0.5cm 0.cm,clip=true,keepaspectratio=true]{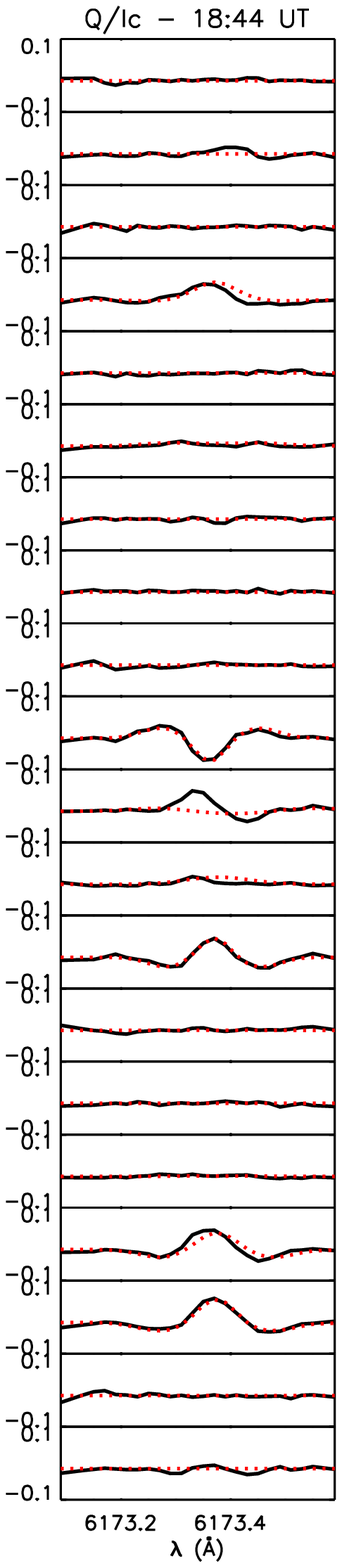}
\includegraphics[height=17.5cm,trim=.3cm 0.3cm 0.5cm 0.cmclip=true,keepaspectratio=true]{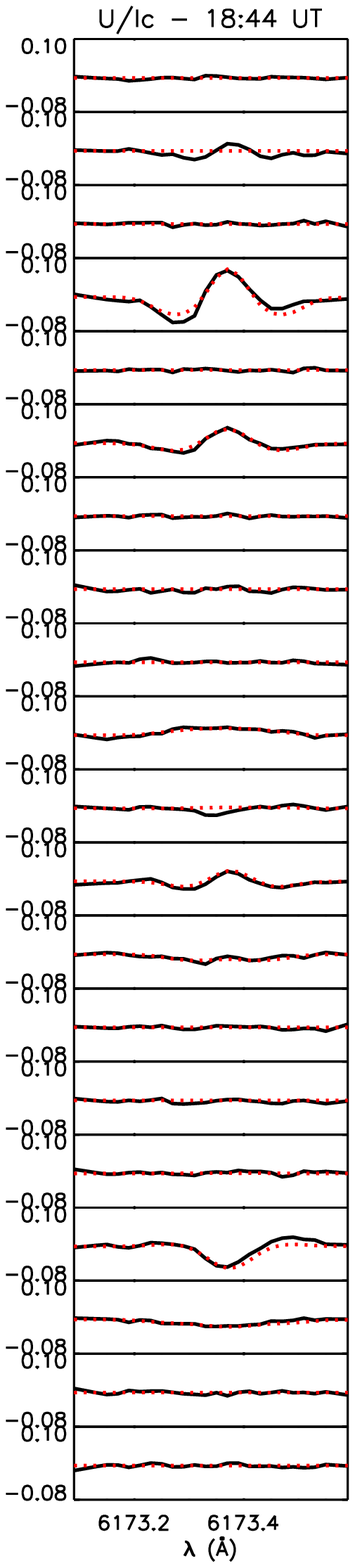}
\includegraphics[height=17.5cm,trim=.3cm 0.3cm 0.5cm 0.cm,clip=true,keepaspectratio=true]{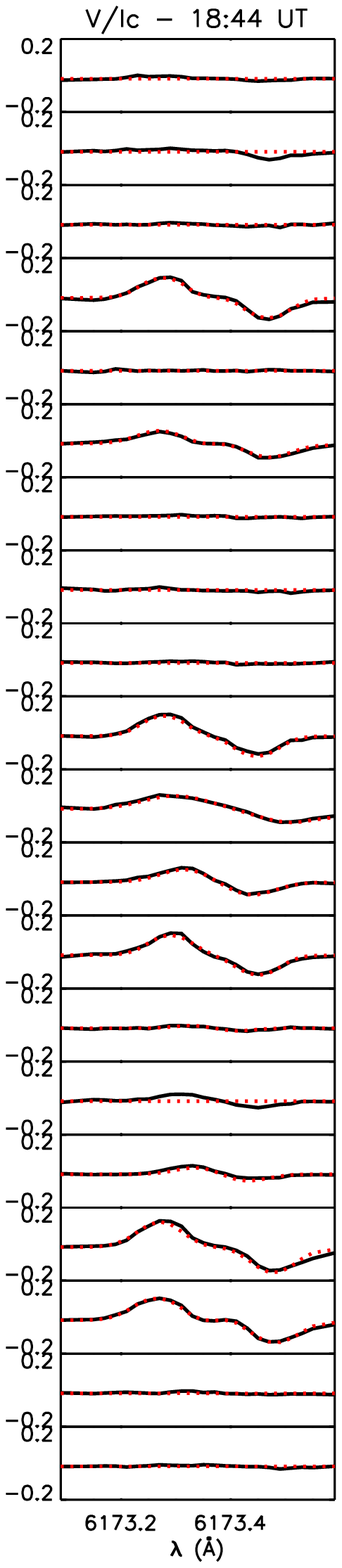}\\
}
\caption{\footnotesize{ As in Fig. \ref{a2} but for the data taken  at  18:44 UT (stage (e)  of the pore formation shown in Fig. \ref{a1}). Values of the stray-light fraction, LOS magnetic field strength (kG), field inclination and azimuth (degree), and LOS velocity (km/s)  derived  from our data analysis (at optical depth log$\tau _{500}=1$) at the same positions of the shown profiles are listed in Table \ref{tablea6};  the values are cut at 2 decimal digits. }}
\label{a6}
\end{figure*}

 \begin{table*}
 \caption{ Summary of the plasma properties  estimated at the 20 positions considered in Figs. \ref{a6}. %: $\chi$ (eV) excitation potential, g$_{eff}$ effective Land\'e factor, estimated average line-formation height for the the line core, reference. %(1) Norton et al. 2006, (2) Faroubert, Ricort, Aime 2013, (3) Cauzzi et al. 2008.
} % title of Table
\label{tablea6} % is used to refer this table in the text
\centering % used for centering table
\begin{tabular}{r r r r r r r} % centered columns (4 columns)
\hline\hline % inserts double horizontal lines
position & position & stray-light  & field   &  field & field   & v$_{LOS}$  \\
x & y &  fraction & strength   &  inclination &  azimuth  &   \\
(arcsec) & (arcsec) &  &(kG) & (degree) & (degree)  & (km s$^{-1}$) \\
\hline
 15.3 & 15.3 &    0.05 &     0.03 &       20.64 &      -86.64 &      0.34\\
 15.3 & 17.1 &     0.17 &      0.10 &       20.50 &       35.74 &      0.25\\
 22.5 & 13.5 &     0.00 &     0.01 &       27.04 &      -89.94 &      0.25\\
 16.2 & 18.9 &     0.34 &       1.27 &       25.51 &       49.70 &      0.54\\
 22.5 & 22.5 &     0.00 &     0.01 &       22.83 &      -91.06 &    -0.04\\
 16.2 & 19.8 &     0.29 &      0.87 &       56.23 &      -50.70 &      0.28\\
 19.8 & 19.8 &     0.00 &     0.02 &       19.52 &      -92.04 &      0.23\\
 18.0 & 16.2 &     0.00 &     0.01 &       21.06 &      -82.95 &      0.45\\
 18.9 & 14.4 &     0.00 &     0.02 &       21.55 &      -81.66 &     -0.20\\
 17.1 & 22.5 &     0.37 &      0.97 &       30.44 &      -95.63 &      0.38\\
 13.5 & 18.4 &     0.10 &       1.13 &       19.32 &       125.05 &       1.17\\
 18.0 & 18.0 &     0.50 &      0.57 &       22.98 &       126.65 &      0.50\\
 13.5 & 16.2 &     0.37 &      0.96 &       17.62 &       155.54 &      0.31\\
 17.1 & 15.3 &     0.00 &     0.03 &       22.18 &      -79.70 &      0.37\\
 19.8 & 17.1 &     0.10 &     0.08 &       25.69 &      -68.98 &     0.30\\
 19.8 & 18.0 &     0.00 &      0.12 &       21.31 &      -85.23 &      0.48\\
 13.5 & 18.9 &     0.36 &       1.50 &       14.60 &       112.04 &      0.63\\
 14.4 & 18.9 &     0.26 &       1.82 &       8.30 &       156.42 &      0.55\\
 11.7 & 11.7 &     0.00 &     0.03 &       26.85 &      -103.30 &      0.00\\
 17.1 & 17.1 &     0.00 &     0.01 &       18.60 &      -27.75 &      0.23\\
\hline
\end{tabular}
\end{table*}
%\end{appendix}

\begin{figure*}%[ht!]
%  \centering{\hspace {1cm} 1                       \hspace{2.5cm}                                     2            \hspace{2.5cm}                                                 3     \hspace{2.5cm}                        4  \hspace{2.5cm}                        5 \hspace{2.5cm}                        6     \hspace {1cm}    }\\
%
\centering{
\hspace{.7cm}
\hspace{15.9cm}
}\\
\centering{
\includegraphics[height=17.5cm,trim=.3cm 0.3cm 0.5cm 0.cm,clip=true,keepaspectratio=true]{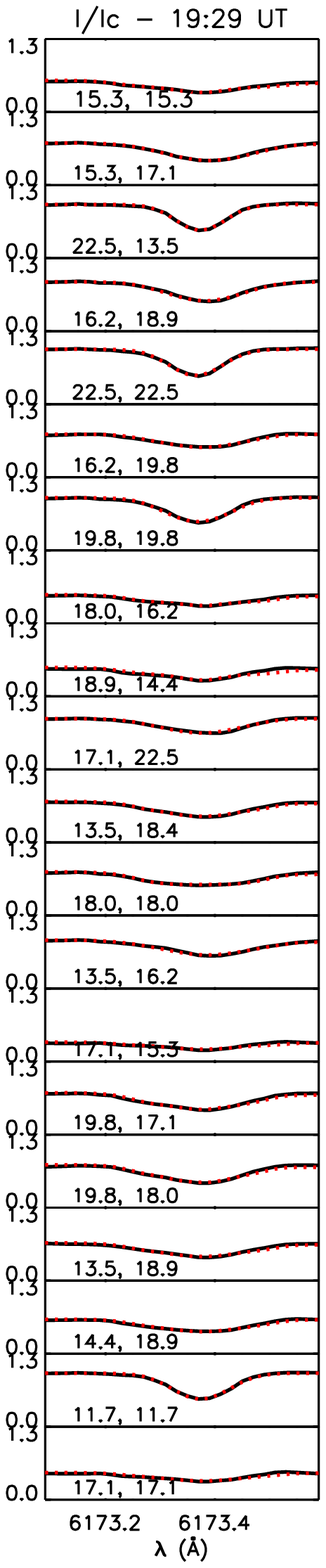}
\includegraphics[height=17.5cm,trim=.3cm 0.3cm 0.5cm 0.cm,clip=true,keepaspectratio=true]{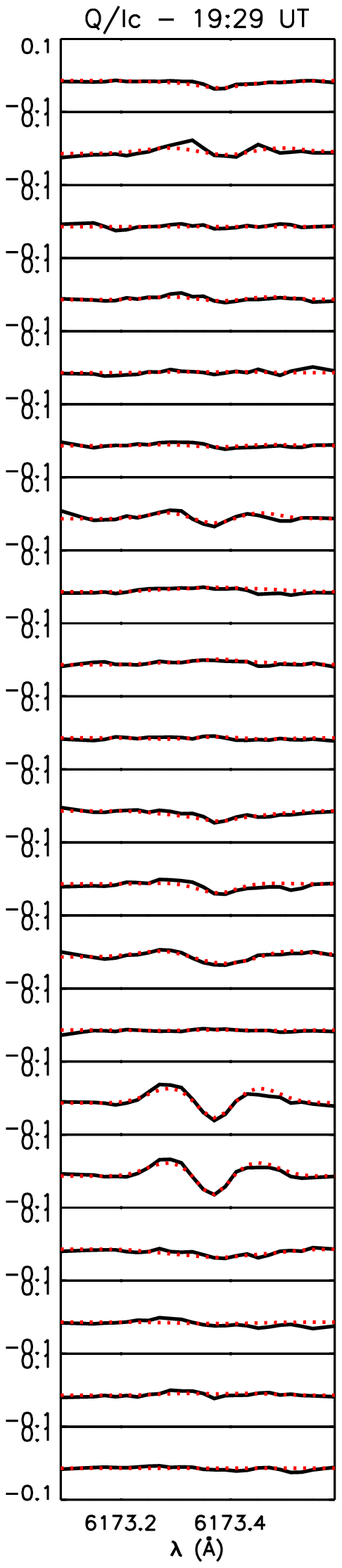}
\includegraphics[height=17.5cm,trim=.3cm 0.3cm 0.5cm 0.cmclip=true,keepaspectratio=true]{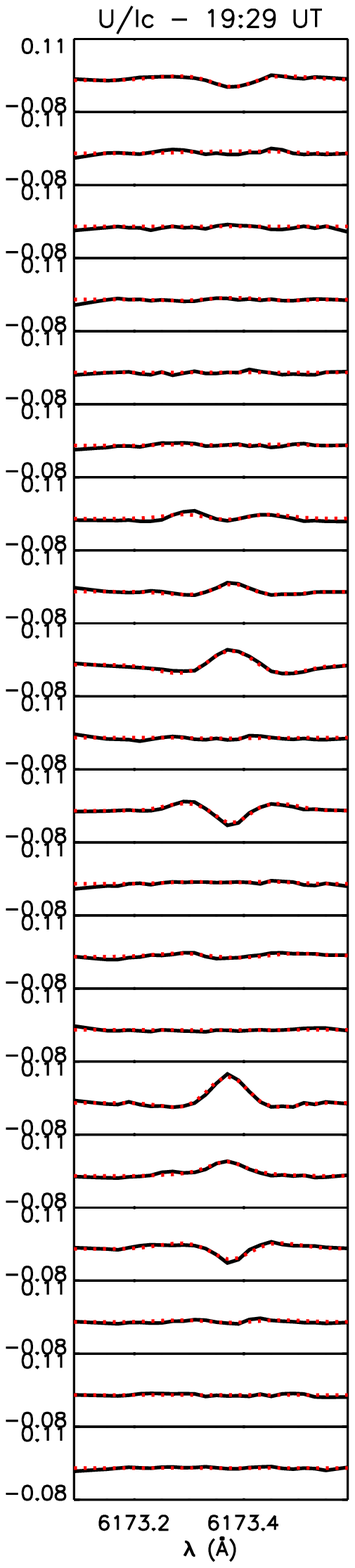}
\includegraphics[height=17.5cm,trim=.3cm 0.3cm 0.5cm 0.cm,clip=true,keepaspectratio=true]{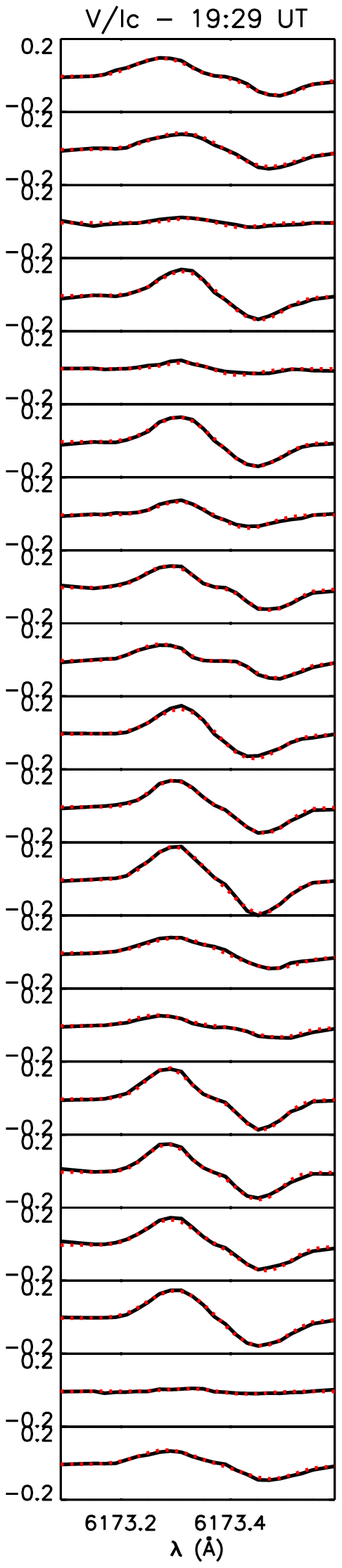}\\
}
\caption{\footnotesize{ As in Fig. \ref{a2} but for the data taken at 19:29 UT (stage (f)  of the pore formation shown in Fig. \ref{a1}). Values of the stray-light fraction, LOS magnetic field strength (kG), field inclination and azimuth (degree), and LOS velocity (km/s)  derived  from our data analysis (at optical depth log$\tau _{500}=1$) at the same positions of the shown profiles are listed in Table \ref{tablea7};  the values are cut at 2 decimal digits. }}
\label{a7}
\end{figure*}

 \begin{table*}
 \caption{ Summary of the plasma properties  estimated at the 20 positions considered in Figs. \ref{a7}. %: $\chi$ (eV) excitation potential, g$_{eff}$ effective Land\'e factor, estimated average line-formation height for the the line core, reference. %(1) Norton et al. 2006, (2) Faroubert, Ricort, Aime 2013, (3) Cauzzi et al. 2008.
} % title of Table
\label{tablea7} % is used to refer this table in the text
\centering % used for centering table
\begin{tabular}{r r r r r r r} % centered columns (4 columns)
\hline\hline % inserts double horizontal lines
position & position & stray-light  & field   &  field & field   & v$_{LOS}$  \\
x & y &  fraction & strength   &  inclination &  azimuth  &   \\
(arcsec) & (arcsec) &  &(kG) & (degree) & (degree)  & (km s$^{-1}$) \\
\hline
 15.3 & 15.3 &     0.23 &       1.45 &       30.54 &       39.10 &      0.52\\
 15.3 & 17.1 &    0.020 &      0.99 &       20.11 &       83.83 &      0.54\\
 22.5 & 13.5 &     0.00 &     0.07 &       22.64 &      -52.47 &      0.10\\
 16.2 & 18.9 &     0.00 &      0.70 &       33.60 &      -75.42 &      0.34\\
 22.5 & 22.5 &     0.00 &      0.11 &       23.24 &      -76.96 &     0.03\\
 16.2 & 19.8 &     0.00 &      0.94 &       28.88 &      -49.22 &      0.32\\
 19.8 & 19.8 &     0.00 &      0.37 &       37.93 &      -108.71 &     0.09\\
 18.0 & 16.2 &     0.17 &       1.39 &       37.88 &      -57.73 &      0.11\\
 18.9 & 14.4 &     0.24 &       1.00 &       56.16 &      -47.72 &    -0.03\\
 17.1 & 22.5 &     0.00 &      0.89 &       24.49 &      -14.89 &      0.43\\
 13.5 & 18.4 &     0.23 &      0.94 &       41.94 &       42.08 &      0.30\\
 18.0 & 18.0 &     0.26 &       1.33 &       19.86 &      -106.29 &      0.30\\
 13.5 & 16.2 &    0.07 &      0.95 &       31.87 &       75.94 &       1.06\\
 17.1 & 15.3 &    0.04 &      0.99 &       21.11 &      -67.65 &     0.09\\
 19.8 & 17.1 &     0.26 &      0.92 &       47.37 &      -77.50 &     0.08\\
 19.8 & 18.0 &     0.30 &      0.88 &       42.51 &      -88.23 &      0.10\\
 13.5 & 18.9 &     0.19 &       1.12 &       36.92 &       38.49 &      0.24\\
 14.4 & 18.9 &     0.00 &       1.06 &       36.38 &       37.72 &     0.08\\
 11.7 & 11.7 &     0.00 &     0.09 &       23.73 &       92.24 &      0.40\\
 17.1 & 17.1 &     0.00 &       1.07 &       17.76 &       86.69 &      0.27\\
\hline
\end{tabular}
\end{table*}

\begin{figure*}%[ht!]
%  \centering{\hspace {1cm} 1                       \hspace{2.5cm}                                     2            \hspace{2.5cm}                                                 3     \hspace{2.5cm}                        4  \hspace{2.5cm}                        5 \hspace{2.5cm}                        6     \hspace {1cm}    }\\
%
\centering{
\hspace{.7cm}
\hspace{15.9cm}
}\\
\centering{
\includegraphics[height=17.5cm,trim=.3cm 0.3cm 0.5cm 0.cm,clip=true,keepaspectratio=true]{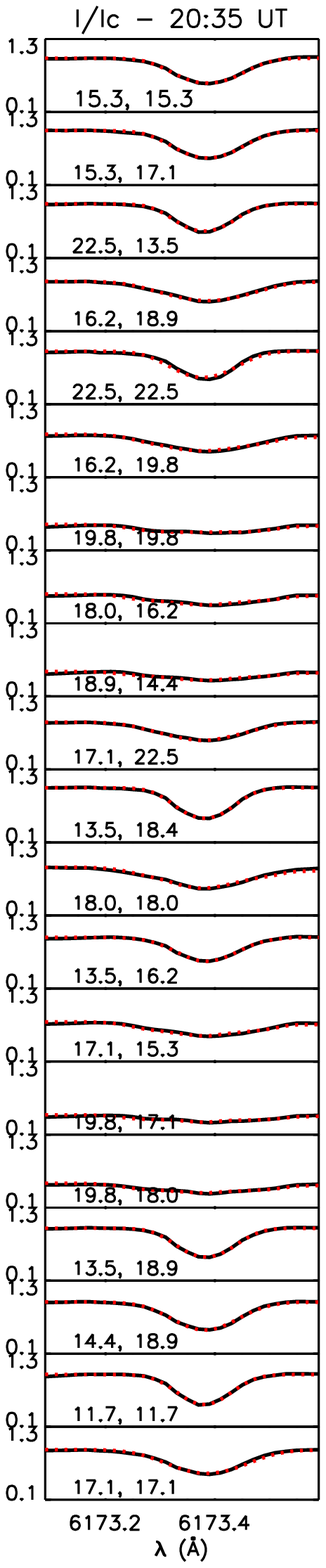}
\includegraphics[height=17.5cm,trim=.3cm 0.3cm 0.5cm 0.cm,clip=true,keepaspectratio=true]{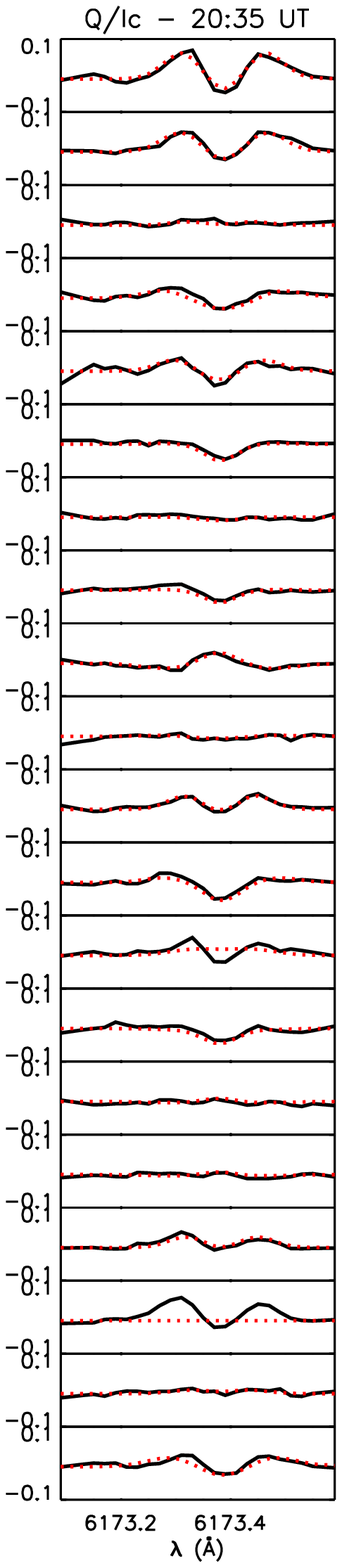}
\includegraphics[height=17.5cm,trim=.3cm 0.3cm 0.5cm 0.cmclip=true,keepaspectratio=true]{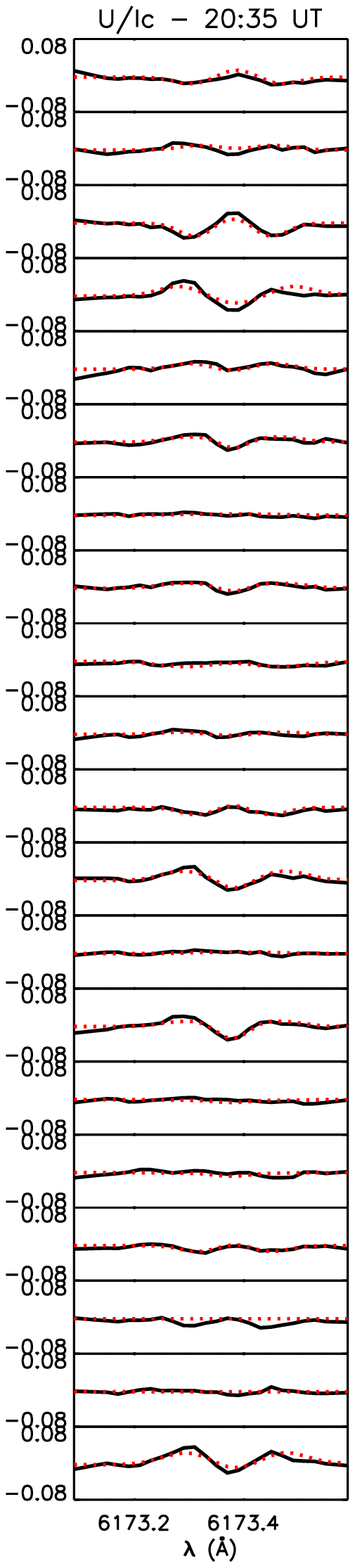}
\includegraphics[height=17.5cm,trim=.3cm 0.3cm 0.5cm 0.cm,clip=true,keepaspectratio=true]{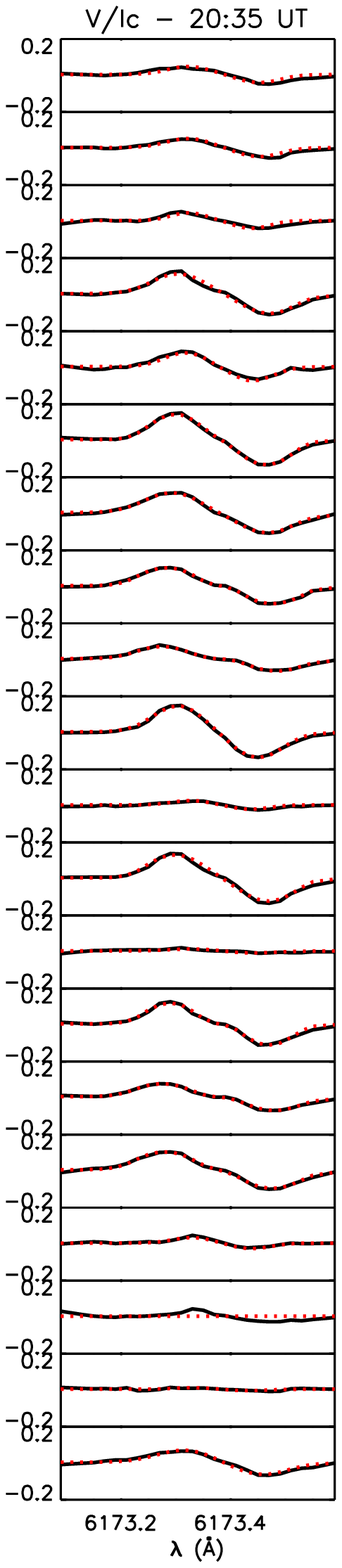}\\
}
\caption{\footnotesize{ As in Fig. \ref{a2} but for the data taken at  20:35 UT  (stage (g)  of the pore formation shown in Fig. \ref{a1}). Values of the stray-light fraction, LOS magnetic field strength (kG), field inclination and azimuth (degree), and LOS velocity (km/s)  derived  from our data analysis (at optical depth log$\tau _{500}=1$) at the same positions of the shown profiles are listed in Table \ref{tablea8};  the values are cut at 2 decimal digits. }}
\label{a8}
\end{figure*}

 \begin{table*}
 \caption{Summary of the plasma properties  estimated at the 20 positions considered in Figs. \ref{a8}. %: $\chi$ (eV) excitation potential, g$_{eff}$ effective Land\'e factor, estimated average line-formation height for the the line core, reference. %(1) Norton et al. 2006, (2) Faroubert, Ricort, Aime 2013, (3) Cauzzi et al. 2008.
} % title of Table
\label{tablea8} % is used to refer this table in the text
\centering % used for centering table
\begin{tabular}{r r r r r r r  } % centered columns (4 columns)
\hline\hline % inserts double horizontal lines
position & position & stray-light  & field   &  field & field   & v$_{LOS}$  \\
x & y &  fraction & strength   &  inclination &  azimuth  &   \\
(arcsec) & (arcsec) &  &(kG) & (degree) & (degree)  & (km s$^{-1}$) \\
\hline
 15.3 & 15.3 &     0.00 &      0.63 &       30.54 &       107.48 &     0.43\\
 15.3 & 17.1 &     0.00 &      0.59 &       33.45 &       88.06 &     0.35\\
 22.5 & 13.5 &     0.00 &      0.36 &       23.77 &       120.17 &      0.28\\
 16.2 & 18.9 &     0.00 &      0.75 &       42.66 &       65.46 &      0.30\\
 22.5 & 22.5 &     0.00 &      0.52 &       24.81 &      -101.15 &    -0.04\\
 16.2 & 19.8 &     0.34 &      0.93 &       43.09 &       55.84 &      0.47\\
 19.8 & 19.8 &     0.12 &       1.62 &       21.87 &       8.13 &      0.45\\
 18.0 & 16.2 &     0.16 &       1.43 &       33.70 &       43.96 &      0.56\\
 18.9 & 14.4 &    0.09 &       1.94 &       11.15 &       165.00 &      0.50\\
 17.1 & 22.5 &     0.00 &      0.78 &       36.15 &       45.12 &     0.06\\
 13.5 & 18.4 &     0.00 &      0.30 &       34.08 &       104.97 &      0.31\\
 18.0 & 18.0 &     0.39 &      0.89 &       40.03 &       61.39 &      0.23\\
 13.5 & 16.2 &     0.00 &     0.01 &       60.88 &       96.97 &      0.52\\
 17.1 & 15.3 &     0.30 &       1.02 &       44.47 &       47.94 &      0.66\\
 19.8 & 17.1 &     0.13 &       1.79 &       29.69 &      -71.89 &      0.54\\
 19.8 & 18.0 &     0.20 &       1.73 &       20.62 &      -85.48 &      0.20\\
 13.5 & 18.9 &     0.00 &      0.32 &       26.27 &       107.49 &      0.46\\
 14.4 & 18.9 &     0.00 &      0.40 &       45.63 &       104.67 &      0.27\\
 11.7 & 11.7 &     0.00 &     0.04 &       30.07 &       91.89 &      0.38\\
 17.1 & 17.1 &     0.00 &      0.74 &       31.65 &       68.66 &      0.23\\
\hline
\end{tabular}
\end{table*}

\end{document}